%% file: SMaSH.tex
\documentclass[a4paper,11pt]{article}

\usepackage{jheppub} 
\usepackage{braket}
\usepackage{amssymb}
\usepackage{amsmath}
\usepackage{amsthm}
\usepackage{mathrsfs}
\usepackage{slashed}
\usepackage{longtable}

\usepackage[makeroom]{cancel}
\usepackage{hyperref}
\usepackage{pifont}
\usepackage{float}

\usepackage{longtable}
\usepackage{booktabs}
\usepackage[utf8]{inputenc}

\hypersetup{colorlinks=true}
\hypersetup{linkcolor=violet}
\hypersetup{citecolor=red}
\hypersetup{urlcolor=violet}

\numberwithin{equation}{section}

\usepackage{datetime}

\usepackage{fancyhdr}
\include{SMaSHjheptexdefn}

\usepackage{framed}

\usepackage{tikz-cd}
\usepackage{tikz-feynman,contour}
\usepackage{tikz, pgf}
\usetikzlibrary{shapes.misc}

\usetikzlibrary{shapes}
\usetikzlibrary{fit}

\usepackage{tikz}
\usetikzlibrary{arrows,shapes}
\usetikzlibrary{trees}
\usetikzlibrary{matrix,arrows} 				
\usetikzlibrary{positioning}				
\usetikzlibrary{calc,through}				
\usetikzlibrary{decorations.pathreplacing}  
\usepackage{pgffor}							

\usetikzlibrary{decorations.pathmorphing}	
\usetikzlibrary{decorations.markings}

\tikzset{
	sugra/.style={decorate, decoration={snake}, draw=black},
	scalarphi/.style={dashed,draw=black, postaction={decorate},
	},
	scalarchi/.style={draw=brown}, 
	hwbou/.style={draw=blue, postaction={decorate}, ultra thick
	},
	vector/.style={draw=blue,decorate, decoration={snake}, draw},
	provector/.style={decorate, decoration={snake,amplitude=2.5pt}, draw},
	antivector/.style={decorate, decoration={snake,amplitude=-2.5pt}, draw},
	fermion/.style={draw=cyan, postaction={decorate},
		decoration={markings,mark=at position .55 with {\arrow[draw=black]{>}}}},
	fermionbar/.style={draw=cyan, postaction={decorate},
		decoration={markings,mark=at position .55 with {\arrow[draw=black]{<}}}},
	chspin/.style={draw=red, postaction={decorate},
		decoration={markings,mark=at position .55 with {\arrow[draw=black]{>}}}},
	chspinbar/.style={draw=red, postaction={decorate},
		decoration={markings,mark=at position .55 with {\arrow[draw=black]{<}}}},  
	fermionnoarrow/.style={draw=black},
	gluon/.style={decorate, draw=purple,
		decoration={coil, amplitude=4pt, segment length=5pt}},
	scalar/.style={dashed,draw=black, postaction={decorate},
		decoration={markings,mark=at position .55 with {\arrow[draw=black]{>}}}},
	scalarbar/.style={dashed,draw=black, postaction={decorate},
		decoration={markings,mark=at position .55 with {\arrow[draw=black]{<}}}},
	electron/.style={draw=black, postaction={decorate},
		decoration={markings,mark=at position .55 with {\arrow[draw=black]{>}}}},
	scalarnoarrow/.style={dashed, draw=black},
	electron/.style={draw=black, postaction={decorate},
		decoration={markings, mark=at position .55 with {\arrow[draw=black]{>}}}},
	bigvector/.style={decorate, decoration={snake, amplitude=4pt}, draw},
	photon/.style={draw=violet, decorate, decoration={snake}, draw},
	higgs/.style={dashed, draw=black, postaction={decorate},
	},	
	goldstone/.style={draw=brown, postaction={decorate},
	},    
	ghost/.style={dashed, draw=blue, postaction={decorate},
		decoration={markings, mark=at position .55 with {\arrow[draw=black]{>}}}
	},  
	antighost/.style={dashed, draw=blue, postaction={decorate},
		decoration={markings, mark=at position .55 with {\arrow[draw=black]{<}}}
	}, 
	scalartwo/.style={dashed,draw=brown, postaction={decorate},
		decoration={markings,mark=at position .55 with {\arrow[draw=black]{>}}}},
	scalarbartwo/.style={dashed,draw=brown, postaction={decorate},
		decoration={markings,mark=at position .55 with {\arrow[draw=black]{<}}}}, 
	fermiontwo/.style={draw=purple, postaction={decorate},
		decoration={markings,mark=at position .55 with {\arrow[draw=black]{>}}}},
	fermionbartwo/.style={draw=purple, postaction={decorate},
		decoration={markings,mark=at position .55 with {\arrow[draw=black]{<}}}},    
	realscalar/.style={draw=black}, 
	fakerealscalar/.style={draw=white}, 
	realscalarone/.style={ draw=black},
	realscalartwo/.style={draw=brown},    	    pseudoscalar/.style={draw=brown},
	mgluon/.style={decorate, draw=blue,
		decoration={coil,amplitude=4pt, segment length=5pt}},
	weylfermion/.style={draw=orange, postaction={decorate},
		decoration={markings,mark=at position .55 with {\arrow[draw=black]{>}}}},
	weylfermionbar/.style={draw=orange, postaction={decorate},
		decoration={markings,mark=at position .55 with {\arrow[draw=black]{<}}}}, 
	majorana/.style={draw=cyan, postaction={decorate},
		decoration={markings,mark=at position .55 with {\arrow[draw=black]{><}}}},
	majoranabar/.style={draw=cyan, postaction={decorate},
		decoration={markings,mark=at position .55 with {\arrow[draw=black]{><}}}},    
	wboson/.style={draw=blue,decorate, decoration={snake,amplitude=4pt}, draw},  
	zboson/.style={draw=violet, decorate, decoration={snake}, draw},   
	lepton/.style={draw=black, postaction={decorate},
		decoration={markings, mark=at position .55 with {\arrow[draw=black]{>}}}},
	leptonbar/.style={draw=black, postaction={decorate},
		decoration={markings, mark=at position .55 with {\arrow[draw=black]{<}}}}, 
	clepton/.style={draw=cyan, postaction={decorate},
		decoration={markings, mark=at position .55 with {\arrow[draw= black]{>}}}},
	cleptonbar/.style={draw=cyan, postaction={decorate},
		decoration={markings, mark=at position .55 with {\arrow[draw=black]{<}}}},   
	nlepton/.style={draw=orange, postaction={decorate},
		decoration={markings, mark=at position .55 with {\arrow[draw=black]{>}}}},
	nleptonbar/.style={draw=orange, postaction={decorate},
		decoration={markings, mark=at position .55 with {\arrow[draw=black]{<}}}},              
	graviton/.style={draw=blue, decorate, decoration={snake, amplitude=4pt}, draw},  
	spinj/.style={draw=red, decorate, decoration={snake, amplitude=4pt}, draw},  
	bgraviton/.style={draw=blue, decorate, decoration={snake, amplitude=4pt}, draw},  
	gravitino/.style={draw=red, postaction={decorate}, 
		decoration={snake,  markings, mark=at position .55 with {\arrow[draw=black]{><}}}},
	gravitinobar/.style={draw=red, postaction={decorate},
		decoration={snake, markings, mark=at position .55 with {\arrow[draw=black]{><}}} },  
	phir/.style={draw=blue, postaction={decorate},},
	phil/.style={dashed,draw=blue,},
	phiav/.style={draw=cyan, postaction={decorate},},
	phidif/.style={dashed,draw=cyan,},  
	chir/.style={draw=red, postaction={decorate},},
	chil/.style={dashed,draw=red,},  
}

\def\iimg{{\bf i}}

\usepackage{xcolor}
\usepackage{lipsum} 

\usepackage[draft]{todonotes}   

\def\redbox#1{\begin{tcolorbox}[
	colback=white!5,
	colframe=gray!60!gray,
	boxrule=1pt,
	arc=3mm,
	left=6pt,right=6pt,top=3pt,bottom=3pt
	]	
#1
\end{tcolorbox}}

\usepackage[most]{tcolorbox}
\usepackage{xparse}

\makeatletter
\def\@pullargument#1\tt#2#3\@nil{#2}

\newcommand{\pullargument}[1]{\@pullargument#1\tt{#1}\@nil}
\makeatother

\newenvironment{documentationboxWL}[2]
{%
	\begin{tcolorbox}[
    enhanced,
    breakable,
    colback=blue!3,
    colframe=orange!35!black,
    boxrule=1pt,
    arc=2mm,
    left=8pt,
    right=8pt,
    top=8pt,
    bottom=8pt
  ]
  \noindent
  \textcolor{blue!65!black}{\bfseries #1}%
  \if\relax\detokenize{#2}\relax
  \else
    \quad \textcolor{blue!65!black}{\vspace{15pt}\\ \normalfont #2}%
  \fi
  \par\smallskip
  \normalcolor
}
{%
  \end{tcolorbox}
}

\NewDocumentEnvironment{documentationbox}{m O{}}
{%
	\label{dcmnt:\pullargument{#1}}
	\begin{documentationboxWL}{#1}{#2}
}
{%
  	\end{documentationboxWL}
}

\NewDocumentEnvironment{examplebox}{m}
{%
  \begin{tcolorbox}[
    enhanced,
    breakable,
    colback=orange!5,
    colframe=blue!12,
    boxrule=0pt,
    arc=2mm,
    left=8pt,
    right=8pt,
    top=8pt,
    bottom=8pt
  ]
  \noindent\textcolor{orange!45!black}{\bfseries Examples for #1}
  \par\smallskip
  \normalcolor
}
{%
  \end{tcolorbox}
}

\NewDocumentEnvironment{conceptbox}{m}
{%
 \par\noindent
 {\bfseries #1}
 \par\smallskip
}
{%
 \par
}

\NewDocumentEnvironment{morematerialbox}{O{}}
{%
 \par\noindent
 \if\relax\detokenize{#1}\relax
 \else
   {\bfseries #1}
   \par\smallskip
 \fi
}
{%
 \par
}

\definecolor{warningbrown}{RGB}{145,95,45}
\definecolor{warningfill}{RGB}{255,247,235}

\NewDocumentEnvironment{warningbox}{m}
{%
  \begin{tcolorbox}[
    enhanced,
    breakable,
    colback=warningfill,
    colframe=warningbrown,
    boxrule=1pt,
    arc=2mm,
    left=8pt,
    right=8pt,
    top=8pt,
    bottom=8pt
  ]
  \noindent\textcolor{warningbrown}{\bfseries Warning: #1}
  \par\smallskip
  \normalcolor
}
{%
  \end{tcolorbox}
} 

\newenvironment{commandtable}{
	\begin{longtable}{@{} l p{10cm} c @{}}
	\toprule
	\textbf{Command} & \textbf{Description} & \textbf{Page} \\
	\midrule
	\endfirsthead

	\toprule
	\textbf{Command} & \textbf{Description} & \textbf{Page} \\
	\midrule
	\endhead

	\bottomrule
	\endfoot

	\bottomrule
	\endlastfoot
}{\end{longtable}}

\newenvironment{basicobjectstable}{
	\begin{longtable}{@{} l  l p{10cm} c @{}}
	\toprule
	\textbf{Command} & \textbf{Shortcuts} & \textbf{Description} & \textbf{Page} \\
	\midrule
	\endfirsthead

	\toprule
	\textbf{Command} & \textbf{Shortcuts} & \textbf{Description} & \textbf{Page} \\
	\midrule
	\endhead

	\bottomrule
	\endfoot

	\bottomrule
	\endlastfoot
}{\end{longtable}}

\def\tt#1{\texttt{#1}}
\def\ttr#1{\texttt{\color{red}{#1}}}

\def\epsilond{{{\epsilon \textrm{d}}}}
\def\epsilonlg{{{\epsilon \textrm{lg}}}}
\def\deltad{{{\delta \textrm{d}}}}
\def\deltalg{{{\delta \textrm{lg}}}}

\title{\boldmath \tt{SMaSH} : Simplify Massive Spinor Helicity}

\vspace{10mm}

\author{Aakash Kumar,}
\author{ Arnab Rudra \&}
\author{ Rahul Shaw}
\author[]{\\ }
\vspace*{4.0ex} 
\vspace{10mm}
\affiliation[]{Indian Institute of Science Education and Research Bhopal,\\
	Bhopal Bypass Road, Bhauri, Madhya Pradesh 462066, India.\\ }
\vspace*{4.0ex} 
\vspace*{2.0ex}
\emailAdd{aakash.kumar.physics@gmail.com}
\emailAdd{rudra@iiserb.ac.in} 
\emailAdd{rahulshaw.phy@gmail.com}

\abstract{We present \tt{SMaSH}, a \tt{Mathematica} package to do spinor helicity computations in four spacetime dimensions(\href{https://github.com/aakash-kmr/SMaSH}{github link}). It can handle massive spinor helicity computations with explicit little group indices which is a novel feature. It can also handle massless as well as off-shell spinor helicity variables. It is designed to compute perturbative computations; it comes with predefined three point amplitudes and propagators for any masses and spins \cite{Arkani-Hamed:2017jhn}. It can implement the high energy limit over an expression, check the discrete $\tt{C,P,T}$ transformations, compute contact terms and impose gauge invariance for any scattering process. We have shown the usage of such functions for computing gauge invariant Weinberg minimal amplitudes\cite{Kumar:2025znu,Kumar:2025juz}. 
	
The package can also generate both real and complex numerical kinematics for any $n$-point scattering for arbitrary masses and energy scales by implementing the \tt{RAMBO} algorithm. It is also rich with basic spinor helicity manipulations like Schouten simplification, Clifford algebra manipulation, conversion between spinor helicity and Lorentz vectors, derivative w.r.t. spinors and their scalars, helicity scaling etc.}
\begin{document}
	\maketitle
	
	\newpage

    \input{Introduction}
	\input{DeclareLegs}

	\input{BasicObjects}

	\input{PolynomialRules}

	\input{ScatteringAmplitudes}

	\input{UsefulTools}

	\input{Conclusion}

	\newpage 
	\appendix
	
	\input{App_Load}

	\input{App_CPT}

	\input{App_RAMBO}

	\input{App_MoreCommands}
	\input{App_SectionWiseSummary}

\bibliographystyle{unsrt}
\bibliography{SMaSHbiblio}

\end{document}

%% file: SMaSHjheptexdefn.tex
\usepackage{braket}
\usepackage{amssymb}
\usepackage{amsmath}
\usepackage{amsthm}
\usepackage{mathrsfs}
\usepackage{skak}
\usepackage{framed}
\usepackage{hyperref}
\usepackage{slashed}
\usepackage{array}

\hypersetup{colorlinks=true}
\hypersetup{linkcolor=violet}
\hypersetup{citecolor=violet}
\hypersetup{urlcolor=violet}

\numberwithin{equation}{section}

\usepackage
{datetime}
\usepackage{fancyhdr}
\usepackage{tikz, pgf}
\usetikzlibrary{shapes.misc}
\usetikzlibrary{shapes}
\usetikzlibrary{fit}
\usetikzlibrary{arrows}

\usetikzlibrary{decorations.pathmorphing}	
\usetikzlibrary{decorations.markings}
 \usetikzlibrary{patterns}

\tikzset{
    sugra/.style={decorate, decoration={snake}, draw=black},
    scalarphi/.style={dashed,draw=black, postaction={decorate},
        },
    scalarchi/.style={draw=brown}, 
    hwbou/.style={draw=blue, postaction={decorate}, ultra thick
        },
    vector/.style={draw=blue,decorate, decoration={snake}, draw},
	provector/.style={decorate, decoration={snake,amplitude=2.5pt}, draw},
	antivector/.style={decorate, decoration={snake,amplitude=-2.5pt}, draw},
   	 fermion/.style={draw=cyan, postaction={decorate},
        decoration={markings,mark=at position .55 with {\arrow[draw=black]{>}}}},
    fermionbar/.style={draw=cyan, postaction={decorate},
        decoration={markings,mark=at position .55 with {\arrow[draw=black]{<}}}},
    chspin/.style={draw=red, postaction={decorate},
        decoration={markings,mark=at position .55 with {\arrow[draw=black]{>}}}},
    chspinbar/.style={draw=red, postaction={decorate},
        decoration={markings,mark=at position .55 with {\arrow[draw=black]{<}}}},  
    fermionnoarrow/.style={draw=black},
    gluon/.style={decorate, draw=purple,
        decoration={coil, amplitude=4pt, segment length=5pt}},
    scalar/.style={dashed,draw=black, postaction={decorate},
        decoration={markings,mark=at position .55 with {\arrow[draw=black]{>}}}},
    scalarbar/.style={dashed,draw=black, postaction={decorate},
        decoration={markings,mark=at position .55 with {\arrow[draw=black]{<}}}},
    electron/.style={draw=black, postaction={decorate},
        decoration={markings,mark=at position .55 with {\arrow[draw=black]{>}}}},
    scalarnoarrow/.style={dashed, draw=black},
    electron/.style={draw=black, postaction={decorate},
        decoration={markings, mark=at position .55 with {\arrow[draw=black]{>}}}},
	bigvector/.style={decorate, decoration={snake, amplitude=4pt}, draw},
    photon/.style={draw=violet, decorate, decoration={snake}, draw},
    higgs/.style={dashed, draw=black, postaction={decorate},
        },	
        goldstone/.style={draw=brown, postaction={decorate},
        },    
          ghost/.style={dashed, draw=blue, postaction={decorate},
        decoration={markings, mark=at position .55 with {\arrow[draw=black]{>}}}
        },  
          antighost/.style={dashed, draw=blue, postaction={decorate},
        decoration={markings, mark=at position .55 with {\arrow[draw=black]{<}}}
        }, 
            scalartwo/.style={dashed,draw=brown, postaction={decorate},
        decoration={markings,mark=at position .55 with {\arrow[draw=black]{>}}}},
    scalarbartwo/.style={dashed,draw=brown, postaction={decorate},
        decoration={markings,mark=at position .55 with {\arrow[draw=black]{<}}}}, 
    fermiontwo/.style={draw=purple, postaction={decorate},
        decoration={markings,mark=at position .55 with {\arrow[draw=black]{>}}}},
    fermionbartwo/.style={draw=purple, postaction={decorate},
        decoration={markings,mark=at position .55 with {\arrow[draw=black]{<}}}},    
        realscalar/.style={draw=black}, 
        fakerealscalar/.style={draw=white}, 
        realscalarone/.style={ draw=black},
    	realscalartwo/.style={draw=brown},    	    pseudoscalar/.style={draw=brown},
        mgluon/.style={decorate, draw=blue,
        	decoration={coil,amplitude=4pt, segment length=5pt}},
         weylfermion/.style={draw=orange, postaction={decorate},
        decoration={markings,mark=at position .55 with {\arrow[draw=black]{>}}}},
         weylfermionbar/.style={draw=orange, postaction={decorate},
        decoration={markings,mark=at position .55 with {\arrow[draw=black]{<}}}}, 
    majorana/.style={draw=cyan, postaction={decorate},
        decoration={markings,mark=at position .55 with {\arrow[draw=black]{><}}}},
    majoranabar/.style={draw=cyan, postaction={decorate},
        decoration={markings,mark=at position .55 with {\arrow[draw=black]{><}}}},    
   	wboson/.style={draw=blue,decorate, decoration={snake,amplitude=4pt}, draw},  
    zboson/.style={draw=violet, decorate, decoration={snake}, draw},   
    lepton/.style={draw=black, postaction={decorate},
        decoration={markings, mark=at position .55 with {\arrow[draw=black]{>}}}},
    leptonbar/.style={draw=black, postaction={decorate},
        decoration={markings, mark=at position .55 with {\arrow[draw=black]{<}}}}, 
    clepton/.style={draw=cyan, postaction={decorate},
        decoration={markings, mark=at position .55 with {\arrow[draw= black]{>}}}},
    cleptonbar/.style={draw=cyan, postaction={decorate},
        decoration={markings, mark=at position .55 with {\arrow[draw=black]{<}}}},   
   nlepton/.style={draw=orange, postaction={decorate},
        decoration={markings, mark=at position .55 with {\arrow[draw=black]{>}}}},
    nleptonbar/.style={draw=orange, postaction={decorate},
        decoration={markings, mark=at position .55 with {\arrow[draw=black]{<}}}},              
        graviton/.style={draw=blue, decorate, decoration={snake, amplitude=4pt}, draw},  
        spinj/.style={draw=red, decorate, decoration={snake, amplitude=4pt}, draw},  
        bgraviton/.style={draw=blue, decorate, decoration={snake, amplitude=4pt}, draw},  
        gravitino/.style={draw=red, postaction={decorate}, 
        decoration={snake,  markings, mark=at position .55 with {\arrow[draw=black]{><}}}},
    	gravitinobar/.style={draw=red, postaction={decorate},
        decoration={snake, markings, mark=at position .55 with {\arrow[draw=black]{><}}} },  
    phir/.style={draw=blue, postaction={decorate},},
   phil/.style={dashed,draw=blue,},
     phiav/.style={draw=cyan, postaction={decorate},},
   phidif/.style={dashed,draw=cyan,},  
    chir/.style={draw=red, postaction={decorate},},
   chil/.style={dashed,draw=red,},  
}

\newcommand{\treelevelthreepointskel}[4]{
\begin{scope}[shift={(0,0)}, rotate=#4]
	\draw[#1] (0,0)--(1.5,0);    
	\draw[#2][rotate=120] (0,0)--(1.5,0);  
	\draw[#3][rotate=-120](0,0)--(1.5,0);  

\end{scope}	 
}

\newcommand{\labeltreelevelthreepoint}[4]{
\begin{scope}[shift={(0,0)}, rotate=#4]

\node at (2,0) {#1}; 
\begin{scope}[shift={(0,0)},rotate=120]	
\node at (2,0) {#2};
\end{scope}	
\begin{scope}[shift={(0,0)},rotate=240]	
	\node at (2,0) {#3};
\end{scope}	
	
\end{scope}	 
} 

\newcommand{\treefourpointexchange}[6]{
\begin{scope}[rotate=#6]

\begin{scope}[shift={(-.75,0)} ]

\treelevelthreepointskel{fakerealscalar}{#1}{#2}{0}	
\begin{scope}[shift={(1.5,0)}]	
\treelevelthreepointskel{#5}{#3}{#4}{180}	
\end{scope}	

\end{scope}	 
\end{scope}	 
}

\newcommand{\treefourpointcrossexchange}[6]{
\begin{scope}[rotate=#6]

\begin{scope}[shift={(-.75,0)} ]

	\draw[#5] (0,0)--(1.5,0);    
	\draw[#2][rotate=-120](0,0)--(1.5,0); 

	\draw[#4][rotate=30](0,0)--(2.6,0); 
	\filldraw[white] (.75,.45) circle (.2);  

\begin{scope}[shift={(1.5,0)}, rotate=0]
	\draw[#3][rotate=-60] (0,0)--(1.5,0);
	
	\draw[#1][rotate=150](0,0)--(2.6,0);        
\end{scope}	 

\end{scope}	 
\end{scope}	 
}

\newcommand{\labeltreefourpointexchange}[6]{
\begin{scope}[rotate=#6]

\begin{scope}[shift={(-.75,0)} ]

\node at (.75,-.5) {#5}; 

\labeltreelevelthreepoint{}{#1}{#2}{0} 
\begin{scope}[shift={(1.5,0)}]	
\labeltreelevelthreepoint{}{#3}{#4}{180}
\end{scope} 
	
\end{scope}	 
\end{scope}	 
}

\usepackage{comment}
\definecolor{cobalt}{rgb}{0.0, 0.28, 0.67}

\specialcomment{rudranote}{%
  \begingroup \color{cobalt}}{%
  \endgroup}

\tikzstyle{block} = [draw, rectangle, 
    minimum height=3em, minimum width=6em]

 \definecolor{ao(english)}{rgb}{0.0, 0.5, 0.0}
\usepackage{pifont}

\usepackage{comment}

 \specialcomment{rudradetails}{%
  \begingroup \color{blue}}{%
  \endgroup}

%% file: Introduction.tex
\section{Introduction}
The spinor helicity (SH) formalism has become extremely useful in the computations of scattering amplitudes. The foundation for the SH formalism was laid down by Weyl-van-der-Waerden in the book  ``The Theory of Groups and Quantum Mechanics" \cite{WeylvanderWaerden} in the year 1950 (see \cite{Dittmaier_1998} for a review). Much later, in the year 1981, the technique was introduced by Berends, Kleiss, Causmaecker, Gastmans and Tsun Wu to compute the single and multiple Bremsstrahlung process in QED \cite{Berends:1981rb,Berends:1981uq}. It was further used to compute the mutliple Bremsstrahlung in non-abelian gauge theories \cite{Xu:1986xb} by Xu, Zhang and Chang in the year 1986. This work also introduced much of the modern formalism. The major advancements in the formalism came in the same year through the works of Parke and Taylor \cite{Parke:1986gb}. They showed its true strength by writing down single term formula for the color ordered MHV amplitudes of n-gluons.
\begin{align}
	\mathtt{A_n(1^{-}, 2^{-}, 3^{+}, \cdots, n^{+})= \frac{\langle 1\,2\rangle ^4}{ \langle 1\,2\rangle\langle 2\,3\rangle \cdots   \langle n\,1\rangle }}
\end{align}
where $+,-$ denotes the helicity of the gluons. The same amplitude when computed using the usual Feynman diagrammatic technique grows very fast with the number of gluons. For $n=4$, there are 4 diagrams, for $n=5$ there are 25, and for $n=6$ there are $220$ diagrams. The diagrams grow to around one million for $n=8$. The power of massless SH formalism to simplify huge expressions down to single term has made it the standard language for the modern day amplitudes community. Beyond the extreme simplifications, the formalism also keeps the physics manifest and makes many properties of the amplitudes much more transparent. There are many review articles and books for the massless SH formalism\cite{LDixonSH,cheung2017tasilecturesscatteringamplitudes,Srednicki:2007qs,Elvang:2015rqa,Schwartz:2013pla}. We follow the notations and conventions from the books by Srednicki \cite{Srednicki:2007qs} and Elvang-Huang \cite{Elvang:2015rqa}. We give a brief review of the conventions in the next section.

In the modern times, the formalism has also been extended to massive particles \cite{Arkani-Hamed:2017jhn,Aoude:2020onz} relying on the foundations laid by Hermann Weyl \cite{WeylvanderWaerden}. The massive SH formalism has become the backbone to the state of the art computations of classical black-holes scattering amplitudes using the higher spin point particle amplitudes \cite{Vines:2018gqi,  Guevara:2019fsj, Maybee:2019jus, Bautista:2021wfy, Bautista:2022wjf, Guevara:2017csg, Chung:2018kqs,Vines:2017hyw,Guevara:2018wpp, Arkani-Hamed:2019ymq, Aoude:2020onz, Ochirov:2018uyq, Chiodaroli:2021eug, Johansson:2019dnu}. It has been shown that the classical black holes scattering can be explained as the large mass and large spin limit of the point particle amplitudes. Much of the recent results in this area have become possible due to the extreme simplifications provided by the SH formalism.

Although the spinor-helicity (SH) formalism greatly streamlines amplitude computations, it also introduces certain practical complications. One such complication is that helicity spinors are contracted using antisymmetric bilinears, rather than the symmetric bilinears familiar from Lorentz-vector notation. As a result, by-hand manipulations can easily lead to sign errors. Another important technical feature of the formalism is the repeated use of Schouten identities, which play a role analogous to Jacobi identities among spinor variables. These identities are extremely powerful: they allow one to simplify large expressions and to recognize the equivalence of expressions that look different but are kinematically identical. However, applying them recursively by hand is cumbersome and error-prone.

The main purpose of the SH package is to eliminate the sign mistakes and simplify expressions as much as possible. This is done using the automation of Schouten identities supplemented by the on-shell conditions like momentum conservation and equations of motion. It should be emphasized that these difficulties are largely confined to the simplification stage of the calculation, and they have already been addressed to a significant extent by several \tt{Mathematica} packages available in the literature \cite{S@M,spinorhelicity4d,spinorsextras}. More recently, Schouten simplification using machine-learning algorithms has also been explored in the context of gluon scattering \cite{SchwartzSpinorHelicity}. In addition to the features introduced in the packages \cite{S@M,spinorhelicity4d,spinorsextras}, we provide various novel features in a completely new environment which is suitable for handling massive little group indices. The key features of the package are discussed below.

\subsection{Key features}
The main features of \tt{SMaSH} are as follows:

\begin{enumerate}
	\item Handling massive little group and spinor indices,
	
	\item Computing scattering amplitudes, checking gauge invariance, and
	constructing contact terms,
	
	\item Converting a SH expression to component form and generating real as well as complex numerical kinematic data,
	
	\item Analyzing high-energy limits, discrete symmetries, and conversions from Lorentz vector to SH variables,
	
	\item Differentiation w.r.t. SH variables, little group scaling, and replacing \& scaling momenta as well as spinors.
	
\end{enumerate}

\subsection{Comparison with other publicly available packages}

Here we outline the similarities and differences between \tt{SMaSH} and other available packages. A complete list of all the public functions and their shortcuts is stored in the variables \tt{\$SMaSHFunctions} and \tt{\$SMaSHShortcuts} and is also summarized in appendix \ref{app:sectionwisesummary}.
\begin{enumerate}
	\item  This is the only public package which can handle explicit massive little group indices without reducing them into massless spinors. Thus, it allows to do massive SH computations keeping the little group covariance manifest at every step. The usual decomposition to massless spinors can be done in a simple step using the command \tt{ToMassless}. The result can also be copy-pasted to other packages by simple replacements rules of the basic objects.
	
	\item The package can generate the numeric for any scattering process including arbitrary number of external particles. It uses the \tt{RAMBO} algorithm \cite{RAMBO1} which is reviewed in appendix \ref{app:RAMBO}. As compared to the other packages, the numerics can be generated for both massive and massless particles. It also comes with multiple options like numerics in COM frame, numerics for particular mass values etc. The package also allows to generate complex numerics for all massless particle scattering.
	
	\item The package allows to declare the variables as massive or massless to do specific on-shell computations where mass of the particle is an important parameter. A massive or massless variable can be any positive number or variable\footnote{Any negative variable is considered as the variable for negative momenta following the properties \eqref{negativespinors}, symbol as well as any composite variable.}. Declaring the variables as numbers has not been possible in the other packages like \tt{SpinorHelicity4D} \cite{spinorhelicity4d}. Of-course, one doesn't need to declare the variables as massive or massless to do off-shell computations where the mass of the variables is not needed.
	
	\item In many existing packages, operations involving index contraction are heavily automated. This package is designed in a way that it can switch between explicitly showing the contracted indices and putting it in scalar form through simple commands like \tt{NoScalar}, \tt{PutScalar} and \tt{ContractBilinears} at any step of the computation. This makes the computations more transparent and allows a more complete platform for a beginner to play around with the basic objects. At the same time, the package is optimized to be computationally efficient and easy to use, enabling an experienced user to also use it seamlessly.
	
	\item The package can handle the canonicalization of the little group as well as spinor indices. Two uncontracted expressions which are same but carry different contracted indices can be set as same using the \tt{CanonicalizeIndices} command. There are also more specific commands which we will discuss later which can either canonicalize left handed or the right handed indices.
	
	\item There are commands which can convert any spinor or their scalars into its component form i.e. in terms of the spherical/Cartesian coordinates. These are highly relevant for scattering cross-section computations.
	
	\item The package makes a distinction between real and complex momenta. It has well defined complex conjugation rules. It comes with commands like \tt{ImposeReality} and \tt{RemoveReality} to impose or remove reality conditions over the spinor variables. This is a novel feature which can be extremely useful in amplitudes computations.
	
	\item  It has commands to impose the action of charge conjugation, parity and time reversal on both massive and massless spinors. These actions are also written in a little group covariant way. The complete derivation of these rules in the helicity basis is also provided in the appendix \ref{app:CPT}.
	
	\item The package also contains inbuilt three point amplitudes for any spins/helicities and masses\cite{Arkani-Hamed:2017jhn} . It has inbuilt functions for any higher spin propagator. The package can be used to compute higher spin amplitudes of any n-point scattering involving both massive and massless particles. It has simpler commands like \tt{PureGaugeTransformations}, \tt{FindContactTerm} and \tt{ManifestGI} to impose gauge invariance and find the relevant contact terms\cite{Kumar:2025znu,Kumar:2025juz}. 
	
	\item The package has other basic operations like differentiation wrt spinors, helicity scaling operators, replacing spinors with other spinors, basis transformations, conversions from Lorentz to SH etc. It comes equipped with basic Schouten identities over angles, boxes and bilinears. However, it can also do more high end Schouten simplifications using the \tt{SimplifySchouten} command which has been adapted from the package \tt{SpinorHelicity4D} \cite{spinorhelicity4d}.
\end{enumerate}

\subsection{A minimal review of massless and massive Spinor Helicity}\label{sec:BriefReviewofSH}

We begin with a brief review of the massless and massive spinor-helicity
formalism. The purpose of this section is not to give a pedagogical
introduction to the formalism; readers seeking such an introduction may consult
the standard references \cite{Elvang:2015rqa, Arkani-Hamed:2017jhn}. Instead,
our goal is to state the notation and sign conventions adopted in the package,
so that the subsequent documentation can be read unambiguously.

The spinor helicity formalism is very sensitive to signs conventions. However the literature is divided into various conventions. In the QFT literature, the mostly used spacetime convention is mostly negative, whereas in the GR literature it has become a norm to use the mostly positive metric convention. We follow the convention where the spacetime metric is mostly positive and the Clifford algebra has a negative sign 
\begin{align}
	\mathtt{\eta^{\mu\nu}=(-1,+1,+1,+1)\quad,\quad \{\Gamma^{\mu},\Gamma^{\nu}\}=-2\eta^{\mu\nu}}\label{metricandcliffordalgebra}
\end{align}
This is same as the convention followed in the textbook \cite{Elvang:2015rqa} by Elvang and Huang. The Clifford matrices in the Weyl representation are given by
\begin{align}
	\mathtt{\Gamma}^\mu=\begin{pmatrix}
		\tt{0}&\mathtt{\sigma}^{\mu}\\\bar{\mathtt{\sigma}}^{\mu}&\tt{0}
	\end{pmatrix}\quad\text{where}\quad \mathtt{\sigma}^\mu_{\alpha\dot{\beta}}=(\mathtt{\sigma}^\tt{0},\vec{\sigma}),\quad (\bar{\mathtt{\sigma}}^{\mu})^{\dot{\beta}\alpha}=(\mathtt{\sigma}^\tt{0},-\vec{\mathtt{\sigma}})
\end{align}
The undotted index $\alpha$ transforms under the left handed $\mathtt{SU(2)_L}$, whereas the dotted index transforms under the right handed $\mathtt{SU(2)_R}$.
In this convention, the sigma matrices satisfy the following properties
{\begin{align}
		\sigma^{\mu}_{\alpha\dot{\beta}}\bar{\sigma}^{\nu\dot{\beta}\alpha}&=-\tt{2}\eta^{\mu\nu},\label{sigmamatrixidentity1} \\
		\left(\bar{\sigma }_{\mu }\right){}^{\dot{\beta }\beta } \left(\bar{\sigma }^{\mu }\right)^{\dot{\alpha }\alpha }&=-\tt{2}\epsilon^{\alpha\beta}\epsilon^{\dot{\alpha}\dot{\beta}}\label{sigmamatrixidentity2}
\end{align}}
We chose the following convention for the invariant Levi-Civita tensor of $\mathtt{SU(2)_L}$ and $\mathtt{SU(2)_R}$
\begin{align}
	\epsilon^{\alpha\beta}=-\epsilon_{\alpha\beta}=\epsilon^{\dot{\alpha}\dot{\beta}}=-\epsilon_{\dot{\alpha}\dot{\beta}}&=\begin{pmatrix}
		\tt{0}&\tt{1}\\-\tt{1}&\tt{0}
	\end{pmatrix}\label{bilinearprop1}\\
	\epsilon^{\alpha\beta}\epsilon_{\gamma\delta}=\delta^{\alpha}{}_{\gamma}\delta^{\beta}{}_{\delta}-\delta^{\alpha}{}_{\delta}\delta^{\beta}{}_{\gamma}\quad&,\quad \epsilon^{\dot{\alpha}\dot{\beta}}\epsilon_{\dot{\gamma}\dot{\delta}}=\delta^{\dot{\alpha}}{}_{\dot{\gamma}}\delta^{\dot{\beta}}{}_{\dot{\delta}}-\delta^{\dot{\alpha}}{}_{\dot{\delta}}\delta^{\dot{\beta}}{}_{\dot{\gamma}}\label{bilinearprop2}
\end{align}
The sign in \eqref{sigmamatrixidentity1} follows from the negative sign in the Clifford algebra \eqref{metricandcliffordalgebra}, whereas the sign in \eqref{sigmamatrixidentity2} is solely due to the mostly positive metric convention. The contracted Lorentz index makes it easy to see that the sign flips with the change of metric convention. Changing the conventions flips these signs. Using the sigma matrices, one can construct a $\mathtt{2\times 2}$ matrix representation of the momentum $\mathtt{p^\mu}$ of a particle which transforms as a vector under the $\mathtt{\mathtt{SU(2)_L}\otimes \mathtt{SU(2)_R}}$ representation\footnote{Since $\mathtt{p^0=-p_0}$, this matches with the convention in Elvang and Huang.}.
\begin{align}
	\mathtt{p}_{\alpha\dot{\alpha}}:=\mathtt{p}_{\mu}\sigma^{\mu}_{\alpha\dot{\alpha}}=\begin{pmatrix}
		\mathtt{p_0+p_3}&\mathtt{p_1-\iimg p_2} \\\mathtt{ p_1+\iimg p_2 }& \mathtt{p_0-p_3}
	\end{pmatrix},\label{Vectortomatrixrepresentation}
\end{align}
The determinant of the momentum matrix $\tt{p}^{\dot{\alpha}\alpha}$ gives the magnitude of the Lorentz vector as $\mathtt{Det(p)=-p^2}$. One can use \eqref{sigmamatrixidentity2} in \eqref{Vectortomatrixrepresentation} and derive the inverse relation.
\begin{align}
	\mathtt{p}_\mu&=-\frac{1}{2}\mathtt{p}_{\alpha\dot{\alpha}}\bar{\sigma}_{\mu}^{\dot{\alpha}\alpha} \label{matrixtovectorrepresentation}
\end{align}
The negative sign is again due to the mostly positive convention. In case of massless particles, $\mathtt{p^2=0}$, the momentum matrix has determinant zero which means it is a rank $\tt{1}$ matrix which can be written as a tensor product of a row and a column vector. Whereas for massive momenta i.e. $\mathtt{Det(p)=m^2}$, the momentum matrix is a written as a sum of two rank $\tt{1}$ matrices.
\begin{align}
	\text{Massless}~:~ \tt{p}_{\alpha\dot{\alpha}}\mathtt{=-|p]}_{\alpha}\langle \mathtt{p|}_{\dot{\alpha}}\quad;\quad \text{Massive}~:~ \mathtt{p}_{\alpha\dot{\alpha}}\mathtt{=-|p^{J}]}_{\alpha}\langle \mathtt{p_{J}|}_{\dot{\alpha}}
\end{align}
The massive spinors transform under the fundamental $\mathtt{2\times 2}$ representation of the $\mathtt{SU(2)}$ little group. The index $\mathtt{J}$ is the $\mathtt{SU(2)}$ fundamental tensor index. The little group index is raised and lowered using the antisymmetric invariant bilinear $\epsilon^{\tt{JK}},\epsilon_{\tt{JK}}$ which satisfies the same properties \eqref{bilinearprop1} and \eqref{bilinearprop2} as the left handed or the right handed invariant bilinear. For negative momenta $\mathtt{-p}_{\alpha\dot{\alpha}}$, the spinors are defined such that
\begin{align}
	\mathtt{|-p\rangle := - |p\rangle \quad,\quad |-p]:= |p]}\label{negativespinors}
\end{align}

One can decompose any expression in the Lorentz vector formalism in the spinor helicity notation. The on-shell massive and massless polarizations can be written in the following way
\begin{align}
	\text{Massless}\quad&:\quad\mathtt{\varepsilon_{i}(+)=-\frac{| i] \langle r_i|}{\sqrt{2} \langle i\,r_i\rangle }\quad,\quad \varepsilon_{i}(-)=\frac{| r_i] \langle i|}{\sqrt{2} [ i\,r_i] }},\\
	\text{Massive}\quad&:\quad \zeta_\tt{i}^{\ttr{I}\ttr{J}}=\frac{\left| \tt{i}^{(\ttr{I}}\right] \left\langle \tt{i}^{\ttr{J})}\right| }{\mathtt{ \sqrt{2} m_i}}
\end{align}
where in case of massless polarizations, the $\mathtt{r_i}$ denotes the reference spinors which are null and satisfy $\mathtt{\langle i \, r_i \rangle \neq 0}$. The variation of the reference spinors can be used to check for the gauge invariance of an amplitude. One can check that the polarizations are manifestly transverse to their momenta. One can also write down higher spin massive polarizations as the copies of spin $\tt{1}$ polarizations with the little group indices symmetrized. We color the massive polarization's little group indices in \textcolor{red}{red} to make a difference with the summed over little group indices.

Under complex conjugation, the left and the right handed representations flip. For hermitian momentum, the angle and box spinors are related to each other by complex-conjugation
\begin{align}
	(\tt{p}_{\alpha\dot{\alpha}})^\dagger=\tt{p}_{\alpha\dot{\alpha}}\implies \begin{cases}
		\left(\mathtt{|p]}_{\alpha}\right)^*=\langle \mathtt{p|}_{\dot{\alpha}}\quad,\quad \left(\mathtt{|p^J]}_{\alpha}\right)^*=\langle \mathtt{p_J|}_{\dot{\alpha}}\quad,\quad \left(\langle \mathtt{p^J|}_{\dot{\alpha}}\right)^*\mathtt{=-|p_J]}_{\alpha}
	\end{cases}
\end{align}
The extra negative sign in the conjugation of angles is due to lowering the $\mathtt{\mathtt{SU(2)_{LG}}}$ index on LHS and raising it on the RHS.
For a non-hermitian momentum, the angle and box spinors need not be related by conjugation and are in general independent of each other. In such a case, we denote the new conjugated spinors (which now has opposite handedness) with a tilde above it.
\begin{align}
	(\tt{p}_{\alpha\dot{\alpha}})^\dagger\neq \tt{p}_{\alpha\dot{\alpha}}~:~\begin{cases}
		\left(\mathtt{|p]}_{\alpha}\right)^*\equiv\langle \mathtt{\tilde{p}}|_{\dot{\alpha}}\quad&,\quad \left(\langle \tt{p}|_{\dot{\alpha}}\right)^*\equiv|\mathtt{\tilde{p}}]_{\alpha}\\
		\left(|\mathtt{p^J}]_{\alpha}\right)^*\equiv\langle \mathtt{\tilde{p}}_\tt{J}|_{\dot{\alpha}}\quad&,\quad \left(\langle \mathtt{p^J}|_{\dot{\alpha}}\right)^*\equiv-|\mathtt{\tilde{p}_J}]_{\alpha}
	\end{cases}
\end{align} 
To impose the reality condition over the momentum, we simply remove the tilde i.e. we set $\mathtt{\tilde{p}=p}$. The angle and box spinors satisfies the following Dirac equations of motion
\begin{align}
	\mathtt{	p^2=0}\qquad&:\qquad \begin{cases}
		\tt{p}^{\dot{\beta}\alpha}\langle \tt{p}|_{\dot{\beta}}=\tt{0}\quad,\quad  \tt{p}^{\dot{\beta}\alpha}| \tt{p} ]_{\alpha}=\tt{0}
	\end{cases},\\
	\mathtt{p^2=-m_p^2}\qquad&:\qquad \begin{cases}
		\tt{p}^{\dot{\beta}\alpha}\langle \mathtt{p^{J}} |_{\dot{\beta}}=-\mathtt{m_p}\, [\mathtt{p^{J}}|^{\alpha} \quad&,\quad  \tt{p}^{\dot{\beta}\alpha}| \mathtt{p^{J}} ]_{\alpha}=\mathtt{m_{p}}\, |\mathtt{p^{J}}\rangle^{\dot{\beta}},\\
		\tt{p}_{\alpha\dot{\beta}}| \mathtt{p^{J}} \rangle^{\dot{\beta}}=\mathtt{m_p}\, |\mathtt{p^{J}}]_{\alpha} \quad&,\quad  \tt{p}_{\alpha\dot{\beta}}[ \mathtt{p^{J}} |^{\alpha}=-\mathtt{m_{p}}\, \langle \mathtt{p^{J}}|_{\dot{\beta}}
	\end{cases}
\end{align}

\subsection{A quick guide for a reader}
Before jumping to the detailed description, we guide the user briefly through the contents of the upcoming sections. 
\begin{enumerate}
	\item Section \ref{sec:DeclareLegs} introduces the commands to declare and undeclare massive and massless legs. 
	
	Many useful features in the package rely on whether a number/variable is declared as massive or massless leg label. So this section serves as the starting point to understand the functioning of the package.

	\item Section \ref{sec:BasicObjects} introduces the basic spinor-helicity objects used in \texttt{SMaSH}:
	
	\( |\tt{i}\rangle_{\alpha} \), \( |\tt{i}]^{\dot\alpha} \), \( \langle \tt{i}\,\tt{j}\rangle \), \( [\tt{i}\,\tt{j}] \), \( \langle \tt{i}|{\tt{j}}|\tt{k}] \),  \( [\tt{i}|{\tt{j}}|\tt{k}\rangle \), \( \epsilon_{\alpha\beta} \), \( \epsilon_{\dot\alpha\dot\beta} \), \( \epsilon_{IJ} \).
	
	This section may be viewed as learning the {\it spinor-helicity alphabets of} \texttt{SMaSH}: it fixes the elementary objects from which all later expressions are built. A summary of the all the objects introduced in this section can be found in \ref{sec:Summaryofbasicobjects}.   
	
	\item Section \ref{sec:PolynomialRules} deals with various operations to apply on the polynomials of SH variables. 
	
	These simplification tools are designed to put SH expressions into a canonical and computable form. They handle the signs arising from antisymmetric spinor contractions, contract repeated spinor, little-group, and Lorentz indices, translate between indexed notation and scalar-bracket notation, impose on-shell relations, and simplify expressions using Schouten identities. A summary the rules introduced in this section can be found in sec. \ref{sec:SummaryofrulesforSimplifyingSpinorHelicityPolynomials}.

	\item Section \ref{sec:computescamp} discusses how to use \texttt{SMaSH} to compute scattering amplitudes. 
	
	This section introduces the commands for the inbuilt three point amplitudes, higher spin propagators, finding contact terms and imposing gauge invariance. A list of commands could be found in \ref{sec:SummaryOfSACommands}
	
	\item Section \ref{sec:usefultools} discusses several useful tools for analyzing scattering amplitudes. These include utilities for studying little group scaling, discrete symmetries, computing high-energy limits, and conversion from Lorentz vector to SH formalism. Summary of all the commands for this can be found in \ref{sec:SummaryOfUsefulTools}
	
	This section serves as the classification of all the commands as per the usage. It also serves as the quick summary of the what every command does. It contains all the commands which are in the main section as well as in the appendices.
	
	\item In appendix \ref{sec:load}, we describe how to load the package.
	
	\item Appendix \ref{app:helicitybasis} gives the basics of discrete $\tt{C,P,T}$ transformations.
	
	This section reviews the spin and helicity basis. It derives the relation between the two basis and the \tt{C,P,T} transformations in bot the basis. Together with the derivation, this section serves to clear the confusion between the two basis.
	
	\item Appendix \ref{app:RAMBO} reviews the basics of \tt{RAMBO} algorithm\cite{RAMBO1,RAMBO2} which is important for generating numerical kinematics.
	\item Appendix \ref{app:MoreCommands} describes all the sub-commands whose main commands have been discussed in the main sections \ref{sec:DeclareLegs}, \ref{sec:PolynomialRules}, \ref{sec:computescamp} and \ref{sec:usefultools}. 
	
	\item Appendix \ref{app:sectionwisesummary} has the summary of all the commands introduced in various sections.
\end{enumerate}

The details of the commands are organized in \textcolor{blue}{blue} boxes (for example \pageref{dcmnt:DeclareLegs}), with their examples given below them.

Along with this document, attached are two example \tt{Mathematica} notebooks, one named \tt{Example.nb} and \tt{ComptonAmplitudeExample.nb}. The first one contains the same examples that are presented in this document separated into sections. The second one computes the spin-s compton amplitude using this package.

%% file: DeclareLegs.tex
\section{Declaring massive and massless legs}\label{sec:DeclareLegs}
The basic objects of the package are spinor-helicity variables. Spinor-helicity
variables associated with a null vector will be called massless SH variables,
while those associated with a time-like vector will be called massive SH
variables. For example, if the null vector is \(\mathtt{k_3}\), the corresponding spinors
may be denoted either by
\begin{align}
	|\mathtt{3}\rangle^{\dot{\alpha}},
	\qquad
	\mathtt{[3|}^{\beta},
\end{align}
or equivalently by
\begin{align}
	|\mathtt{k_3}\rangle^{\dot{\alpha}},
	\qquad
	\mathtt{[k_3|}^{\beta}.
\end{align}
In scattering applications, these null or time-like vectors are usually the
momenta of external particles/legs. In \texttt{SMaSH}, the leg labels are used as the basic identifiers. The
momenta and other kinematic variables are then built by attaching appropriate
letters to the leg label. For instance, for a massless leg \(\mathtt{i}\), the
corresponding momentum is denoted by \(\mathtt{ki}\). The legs can be declared and undeclared using the command \tt{DeclareLegs} and \tt{UndeclareLegs[]}.

\begin{conceptbox}{}
	
\end{conceptbox}

\begin{documentationbox}{\tt{DeclareLegs}}{\tt{DeclareLegs[Mslegs\_][Mllegs\_]}}
	
	Massive and massless legs can be declared simultaneously using this single
	command. The first argument is the list of massive legs, while the second argument is
	the list of massless legs. A leg label can either be a
	\href{https://reference.wolfram.com/language/ref/Symbol.html?q=Symbol}{\tt{Symbol}}
	or an element of
	\href{https://reference.wolfram.com/language/ref/PositiveIntegers.html}{\tt{PositiveIntegers}}.
	Negative labels are reserved for negative momenta; see
	\eqref{negativespinors}.
	
	This commands declare the default kinematic variables associated with each
	leg. For a massive leg \tt{i}, the command assigns $\tt{ki}$ as the momentum, $\zeta \tt{i}$ as the polarization. For a massless leg \tt{i}, the command assigns $\tt{ki}$ as the momentum, $\varepsilon\tt{i}$ as the polarization, \tt{Bi} as the field strength, \tt{ri} as the reference spinor and $+1$ as its helicity.
\end{documentationbox}

\begin{examplebox}{\tt{DeclareLegs}}
	
	\redbox{\includegraphics[scale=0.5]{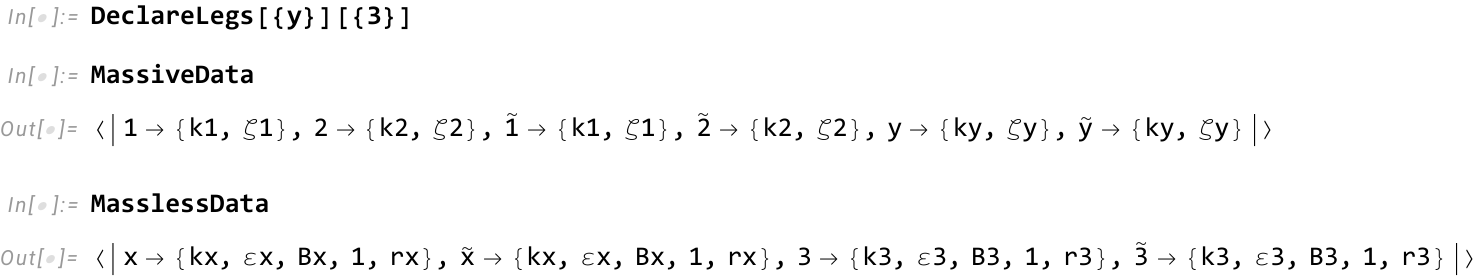}}
	
\end{examplebox}

\begin{morematerialbox}{}
	
\textit{Complex-conjugate legs}: For any leg \tt{i}, the declaration commands also define an association with
	the same \tt{Values} for its complex-conjugate leg, denoted by a tilde
	\(\mathtt{\tilde{i}}\). The notation for complex conjugation was introduced in
	section \ref{sec:BriefReviewofSH}. For a Hermitian momentum, one has
	\begin{align}
		\mathtt{\tilde{i}=i}.
	\end{align}
	The conjugation rules and the implementation of the corresponding reality
	conditions will be discussed in section \ref{sec:ComplexConjugation}.
	
\end{morematerialbox}

\begin{documentationbox}{\tt{UndeclareLegs}}{\tt{UndeclareLegs[Mslegs\_][Mllegs\_]}}
	
	Undeclares a previously declared leg. The arguments are the list of massive and massless leg labels (positive numbers or symbols) respectively.
	
\end{documentationbox}

\begin{examplebox}{\tt{UndeclareLegs}}
	
	\redbox{\includegraphics[scale=0.5]{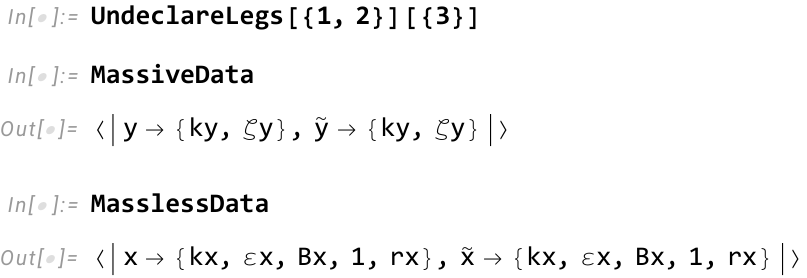}}
	
\end{examplebox}

\begin{morematerialbox}{} 
	We have also provided more related commands which also allows \textit{editing the kinematic data}. The details of these can be found in sec \ref{subsec:moreondeclaring massive and massless legs}.
	
	These commands are introduced to make amplitude computations leg-aware. Once
	the kinematic data of each external leg is declared, \texttt{SMaSH} can
	automatically recognize the corresponding momentum, polarization, helicity,
	field strength, and reference spinor. This avoids repeatedly specifying the
	same data and makes Lorentz-to-SH conversion, on-shell simplification,
	gauge-invariance checks, and frame-dependent evaluations unambiguous.

\end{morematerialbox}

\begin{conceptbox}{}
	One can check whether a variable is massive or massless using the commands
	\footnote{
		These commands check whether the \tt{BaseHead} of the expression has been
		declared as massive or massless. Here \tt{BaseHead} is a custom command that
		extracts the \tt{leg} label from a compound expression of the form
		\(\tt{leg[\,][\,]$\cdots$[\,]}\).} \tt{MsQ},	\tt{MlQ}.
\end{conceptbox}

\begin{documentationbox}{\tt{MsQ}}{\tt{MsQ[expr\_]}}
	
	returns \tt{True} for an expression if and only if its
	underlying leg label has been declared massive; otherwise, it returns
	\tt{False}.
	
\end{documentationbox}

\begin{documentationbox}{\tt{MlQ}}{\tt{MlQ[expr\_]}}
	
	returns \tt{True} for an expression if and only if its
	underlying leg label has been declared massless; otherwise, it returns
	\tt{False}.
	
\end{documentationbox}

\begin{morematerialbox}{}
	
	\textit{Off-shell variables}: Compound variables that are neither declared massive nor massless are treated
	as off-shell variables.
	
\end{morematerialbox}

\begin{examplebox}{\tt{MsQ} and \tt{MlQ}}
	
	\redbox{
		\begin{minipage}{0.5\linewidth}
			\includegraphics[scale=0.5]{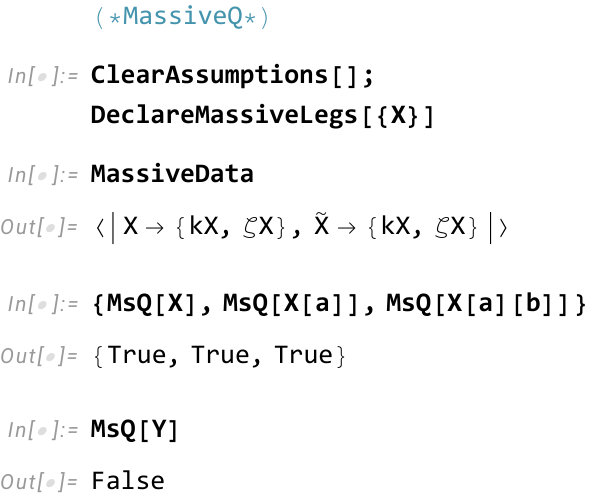}
		\end{minipage}
		\begin{minipage}{0.5\linewidth}
			\includegraphics[scale=0.5]{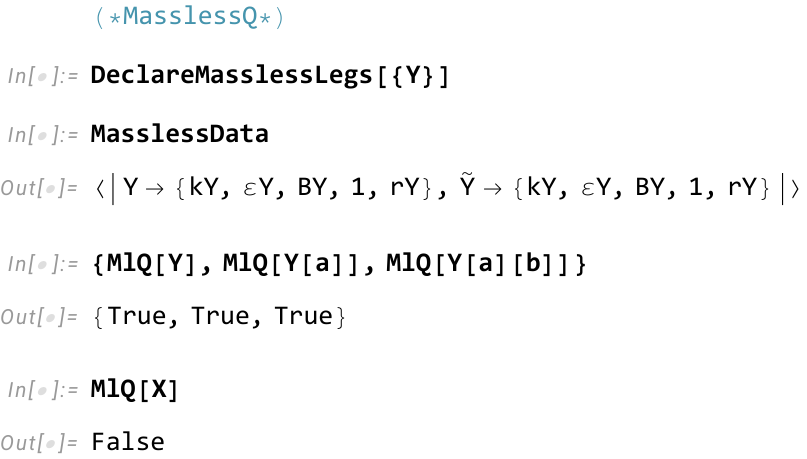}
		\end{minipage}
	}
	
\end{examplebox}

\begin{conceptbox}{Little group indices}
	Since we will be dealing with both massless legs and massive legs with little group indices, we adopt the following notation for the leg label:
	\begin{eqnarray}
		\tt{leg\_}
		&=&
		\tt{particlelabel\_}
		\qquad\qquad \qquad\qquad \qquad\qquad\qquad\qquad \text{for massless particles},
		\nonumber	
		\\
		&=&
		\tt{particlelabel\_[massivelittlegroupindex\_]}
		\qquad\quad \text{formassive particles}.
		\nonumber	
	\end{eqnarray}
	where \tt{particlelabel} is declared using \tt{DeclareLegs}. The full label \(\tt{leg}\) denotes the external leg.  In the massive case,
	the same object also carries the corresponding massive little-group index. A more elaborate discussion on little group indices is discussed in appendix \ref{app:LGcriteria}.
\end{conceptbox}

%% file: BasicObjects.tex
\section{Basic objects}
\label{sec:BasicObjects}
This section will introduce the readers how to write basic spinor helicity variables in \tt{SMaSH}.

\subsection{Basic spinor-helicity variables}
\begin{conceptbox}{}
	
	The basic objects in the spinor-helicity formalism are two-component
	spinors transforming under the subgroups of the Lorentz group
	\(\mathtt{SU(2)_L}\) and \(\mathtt{SU(2)_R}\).  We denote them schematically by
	\begin{eqnarray}
		|i\rangle^{\dot{\alpha}},\qquad
		\langle i|_{\dot{\alpha}},\qquad
		|i]_{\alpha},\qquad
		[i|^{\alpha} .
	\end{eqnarray}
	For massive particles, these spinors also carry an index transforming in the
	fundamental representation of the massive little group \(SU(2)\).  For
	massless particles, they instead carry a definite weight under the massless
	little group \(U(1)\). 
\end{conceptbox}

In the package, these spinors are represented by the elementary objects
\textbf{\tt{SHA}} and \textbf{\tt{SHB}}.  Here \textbf{\tt{SHA}} stands for
{\it {\bf S}pinor {\bf H}elicity {\bf A}ngle}, while \textbf{\tt{SHB}} denotes
the corresponding box spinor.

\begin{documentationbox}{\tt{SHA}}{\tt{SHA[leg\_, $\alpha$\_]} 
	}
	
	The object \(\tt{SHA}\) represents an angle spinors: $\mathtt{\langle leg |}_{\dot{\alpha}}$ and $\mathtt{| leg\rangle}^{\dot{\alpha}}$.  The right angle spinor
	and the left angle spinor are distinguished by the sign convention assigned
	to the spinor index: the right angle spinor carries an overall positive sign,
	whereas the left angle spinor carries an overall negative sign.
	
	\paragraph{Keyboard shortcut}
	\(\boxed{\tt{Esc}}\tt{<}\boxed{\tt{Esc}}\), \qquad
	\(\boxed{\tt{Esc}}\tt{>}\boxed{\tt{Esc}}\).
	
\end{documentationbox}

\begin{documentationbox}{\tt{SHB}}{\tt{SHB[leg\_,$\alpha$\_]} 
	}
	
	The object \(\tt{SHB}\) represents a box spinors: $\mathtt{| leg ]}_{{\alpha}}$ and $\mathtt{[ leg|}^{{\alpha}}$.  The right box spinor and
	the left box spinor are distinguished by the sign convention assigned to the
	spinor index: the right box spinor carries an overall negative sign, whereas
	the left box spinor carries an overall positive sign.
	
	\paragraph{Keyboard shortcut}
	\(\boxed{\tt{Esc}}\tt{[}\boxed{\tt{Esc}}\), \qquad
	\(\boxed{\tt{Esc}}\tt{]}\boxed{\tt{Esc}}\).
	
\end{documentationbox}

\begin{examplebox}{\tt{SHA} and \tt{SHB}}

	\redbox{\begin{minipage}{0.5\linewidth}
			\includegraphics[scale=0.5
			]{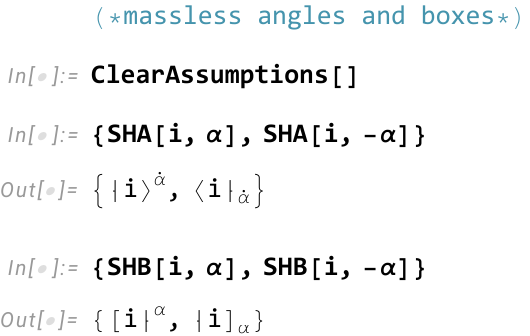}
		\end{minipage}
		\begin{minipage}{0.5\linewidth}
			\includegraphics[scale=0.5
			]{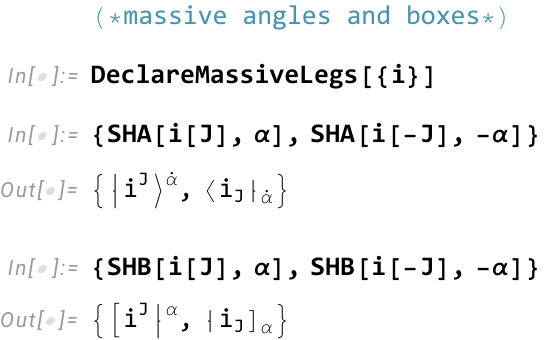}
		\end{minipage}
	}
	
\end{examplebox}

\begin{morematerialbox}
	\textbf{Note}: As discussed before, if a leg is declared to be massive, namely if
	$
	\tt{MsQ[leg] == True},
	$
	then one must supply an additional massive little-group index by writing
	$
	\tt{leg[LGIndex]} .
	$
	If the sign of \(\tt{LGIndex}\) is positive, the little-group index is placed
	as a superscript.  As with Lorentz spinor indices, the little-group index can
	be lowered by putting a minus sign in front of the index.
	
	The syntax also works when the legs have not been declared.  In that case,
	the input variable \(\tt{leg}\) can be any valid expression and need not be
	declared as long as one is working only with the basic spinor-helicity	
	objects.  As we will see in the next section, declaring the legs becomes
	important when performing on-shell manipulations, such as using equations of
	motion.
	
	If the leg variable is of the form \(\tt{leg[Index]}\), then \(\tt{Index}\)
	is interpreted as a massive little-group index only when
	$
	\tt{MsQ[leg] == True}.
	$
	This convention has the advantage that legs can also be equipped with other
	types of labels, such as charge, flavour, or species indices, without forcing
	them to be interpreted as little-group indices.
	
	\redbox{
		\includegraphics[scale=0.5]{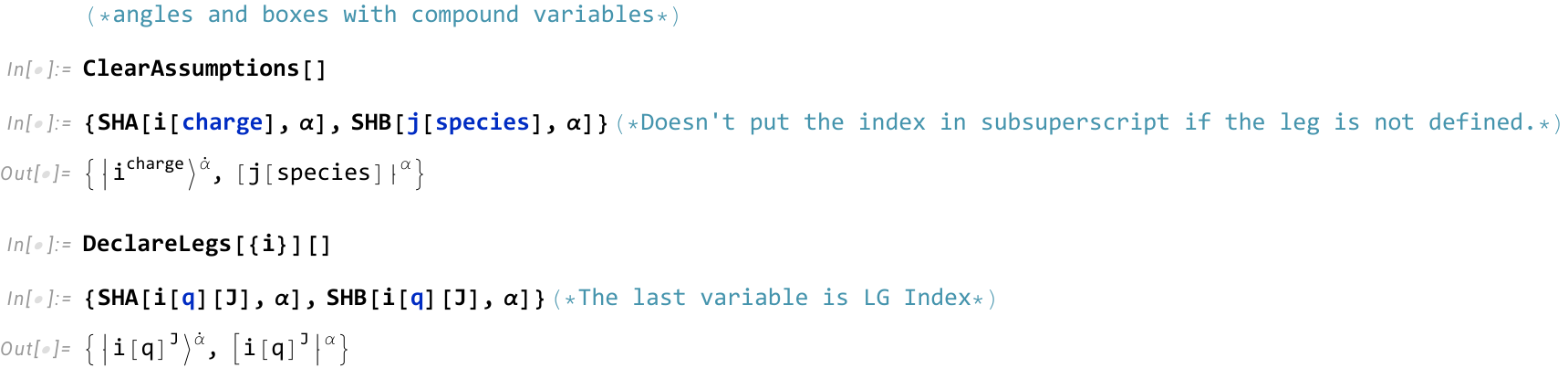}
	}
	
\end{morematerialbox}

\subsection{Scalars}
\begin{conceptbox}{}
	
	The scalar contractions of spinor-helicity variables are denoted by
	\(\tt{SHAA[]}\) and \(\tt{SHBB[]}\).  Here \(\tt{SHAA}\) stands for
	{\it {\bf S}pinor {\bf H}elicity {\bf A}ngle {\bf A}ngle}, while
	\(\tt{SHBB}\) stands for {\it {\bf S}pinor {\bf H}elicity {\bf B}ox {\bf B}ox}.
	They represent Lorentz-scalar contractions in the two spinor sectors,
	respectively:
	\begin{eqnarray}
		\tt{SHAA}
		&\sim&
		\mathtt{SU(2)_L}\text{-spinor contraction},
		\nonumber\\
		\tt{SHBB}
		&\sim&
		\mathtt{SU(2)_R}\text{-spinor contraction}.
	\end{eqnarray}

\end{conceptbox}

\begin{documentationbox}{\tt{SHAA} and \tt{SHBB}}{\tt{SHAA[leg1\_, leg2\_]} and \tt{SHBB[leg1\_, leg2\_]}}

These objects represent the $\mathtt{\langle leg1, leg2\rangle }$ and $\mathtt{[leg1, leg2]}$ scalars respectively.

	\paragraph{Keyboard shortcuts}
	\(\boxed{\tt{Esc}}\tt{<>}\boxed{\tt{Esc}}\), \qquad
	\(\boxed{\tt{Esc}}\tt{[]}\boxed{\tt{Esc}}\).
	
\end{documentationbox}

\begin{morematerialbox}{}
	
The legs \(\tt{leg1}\) and \(\tt{leg2}\) need not be declared \ref{dcmnt:DeclareLegs}. When they are declared, the rules for their output form of the leg labels are the same as for the basic spinors \tt{SHA[leg,$\alpha$]} and \tt{SHB[leg,$\alpha$]}.

\end{morematerialbox}

\begin{examplebox}{\tt{SHAA} and \tt{SHBB}}
	
	\redbox{\includegraphics[scale=0.5
		]{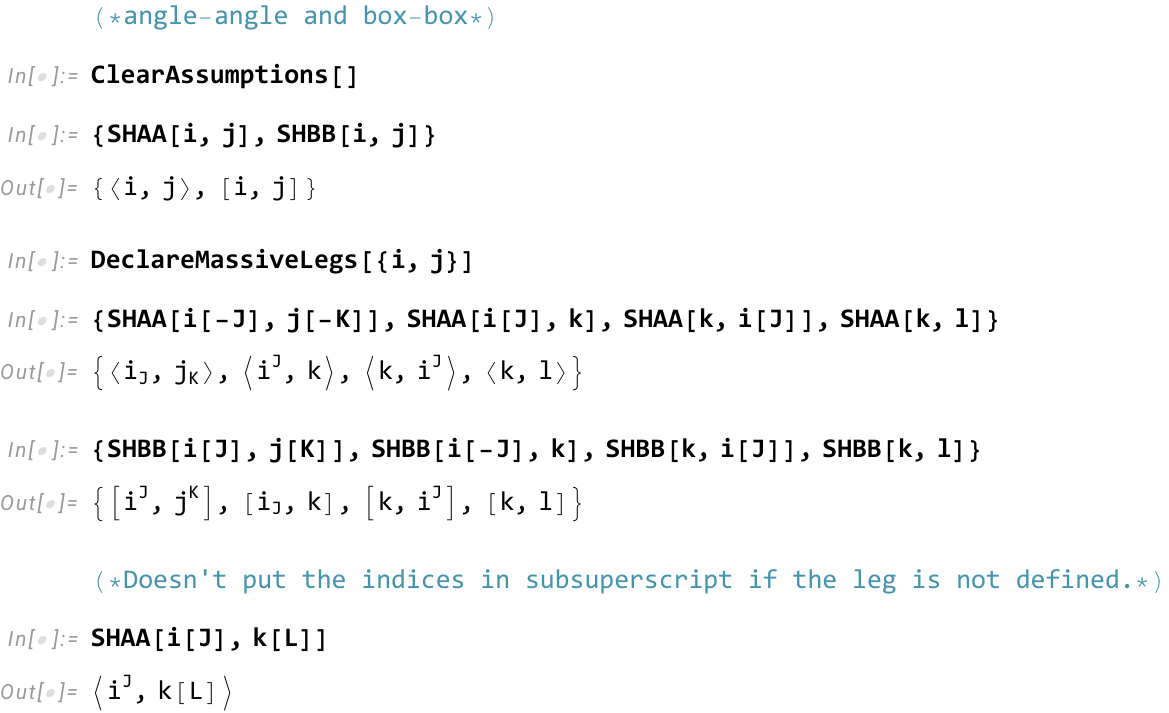}}
	
\end{examplebox}

\begin{conceptbox}{Mixed spinor-helicity scalars}
	
	In addition to the purely angle and purely box contractions, one can also form
	mixed scalar contractions involving both angle and box spinors.  These objects
	are denoted by \(\tt{SHAB}\) and \(\tt{SHBA}\).  Here \(\tt{SHAB}\) stands
	for {\it {\bf S}pinor {\bf H}elicity {\bf A}ngle {\bf B}ox}, while
	\(\tt{SHBA}\) stands for {\it {\bf S}pinor {\bf H}elicity {\bf B}ox {\bf A}ngle}.
	
	These contractions are Lorentz scalars: the \(\mathtt{SU(2)_L}\) and
	\(\mathtt{SU(2)_R}\) spinor indices are contracted.  The mixed scalar has
	three leg inputs because the middle leg supplies the momentum insertion that
	connects the angle and box spinor sectors.  Schematically, these objects
	correspond to contractions of the form
	\begin{eqnarray}
		\langle i|j|k],
		\qquad
		[i|j|k\rangle .
	\end{eqnarray}
	If the middle leg is massive, its little-group indices are also contracted
	internally.  Therefore the resulting scalar does not carry a free
	little-group index associated with the middle leg.
	
\end{conceptbox}

\begin{documentationbox}{\tt{SHAB} and \tt{SHBA}}{\tt{SHAB[leg1\_, leg2\_, leg3\_]} or \tt{SHBA[leg1\_, leg2\_, leg3\_]}}
	
These objects represents the little group scalars $\langle \mathtt{leg1|leg2|leg3} \tt{]}$ and $\mathtt{[\mathtt{leg1|leg2|leg3}\rangle}$.

	\paragraph{Keyboard shortcuts}
	\(\boxed{\tt{Esc}}\tt{<]}\boxed{\tt{Esc}}\), \qquad
	\(\boxed{\tt{Esc}}\tt{[>}\boxed{\tt{Esc}}\).
	
\end{documentationbox}

\begin{morematerialbox}{}
	
Here \(\tt{leg2}\) can be either massive or massless.  If \(\tt{leg2}\) is
	massive, its massive little-group indices are contracted internally.  Thus
	the resulting object does not transform under little-group transformations
	acting on \(\tt{leg2}\).	

\end{morematerialbox}

\begin{examplebox}{\tt{SHAB} and \tt{SHBA}}
	\redbox{\includegraphics[scale=0.5
		]{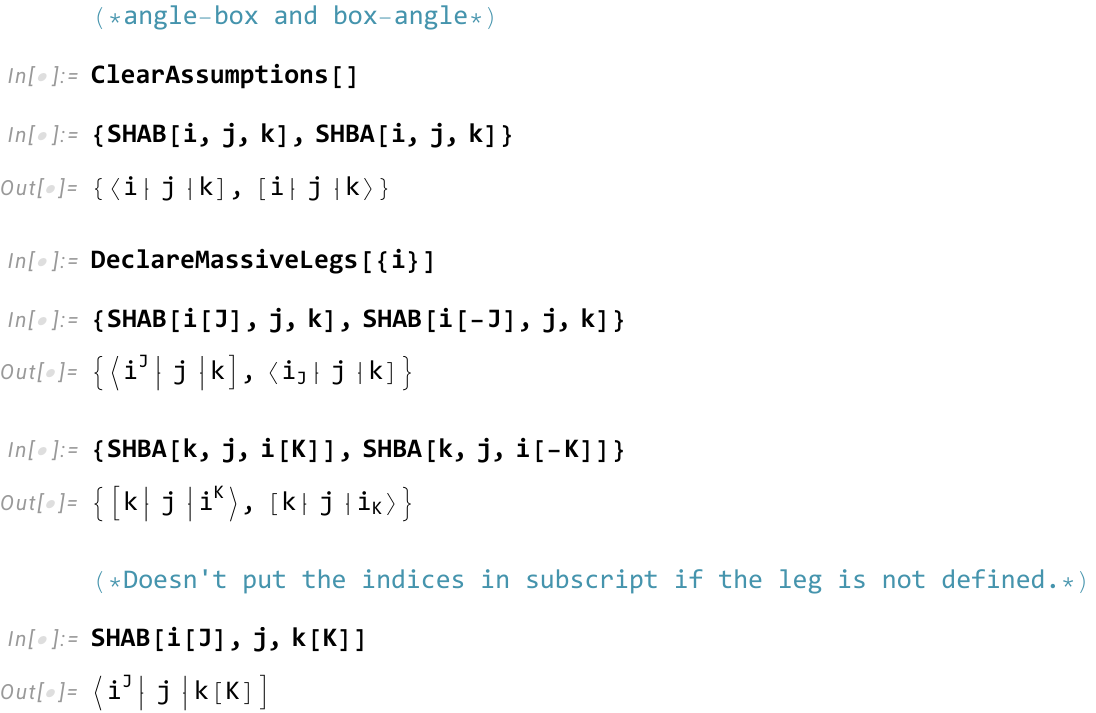}}
	
\end{examplebox}

\subsection{Bilinears}

\begin{conceptbox}{}
	
	There are three antisymmetric invariant bilinears that appear repeatedly in
	spinor-helicity computations.  These are the invariant tensors of
	\(\mathtt{SU(2)_L}\), \(\mathtt{SU(2)_R}\), and the massive little group
	\(\mathtt{SU(2)_{LG}}\):
	\begin{eqnarray}
		\epsilon^{\alpha\beta},
		\qquad
		\epsilon^{\dot\alpha\dot\beta},
		\qquad
		\epsilon^{IJ}.
	\end{eqnarray}
	In addition to these antisymmetric invariant tensors, we also use the
	corresponding identity tensors
	\begin{eqnarray}
		\delta^\alpha{}_\beta,
		\qquad
		\delta^{\dot\alpha}{}_{\dot\beta},
		\qquad
		\delta^I{}_J .
	\end{eqnarray}
	Although these six bilinears are distinct objects, their index conventions are
	handled uniformly in the package.  Raising and lowering of indices is
	implemented by changing the sign of the corresponding index.  Thus an index
	with positive sign is displayed as an upper index, while an index with
	negative sign is displayed as a lower index.
	
\end{conceptbox}

\begin{documentationboxWL}{\tt{$\epsilon$}, \tt{$\epsilond$}, \tt{$\epsilon$lg}}{\tt{$\epsilon$[$\alpha$\_, $\beta$\_]}, \tt{$\epsilond$[$\alpha$\_, $\beta$\_]}, \tt{$\epsilon$lg[I\_, J\_]}}
	\label{dcmnt:epsilon}	

	Both indices must have the same sign. 
	\paragraph{Keyboard shortcuts}
	\(\boxed{\tt{Esc}}\tt{ep}\boxed{\tt{Esc}}\), \qquad
	\(\boxed{\tt{Esc}}\tt{epd}\boxed{\tt{Esc}}\), \qquad
	\(\boxed{\tt{Esc}}\tt{eplg}\boxed{\tt{Esc}}\).
	
\end{documentationboxWL}

\begin{documentationboxWL}{\tt{$\delta$}, \tt{$\deltad$}, \tt{$\delta$lg}}{\tt{$\delta$[$\alpha$\_, -$\beta$\_]}, \tt{$\deltad$[$\alpha$\_, -$\beta$\_]}, \tt{$\delta$lg[I\_, -J\_]}}
	\label{dcmnt:delta}
	
	The two indices must have opposite signs.
	
	\paragraph{Keyboard shortcuts}
	\(\boxed{\tt{Esc}}\tt{dl}\boxed{\tt{Esc}}\), \qquad
	\(\boxed{\tt{Esc}}\tt{dld}\boxed{\tt{Esc}}\), \qquad
	\(\boxed{\tt{Esc}}\tt{dlg}\boxed{\tt{Esc}}\).
\end{documentationboxWL}

\begin{examplebox}{$\mathtt{\epsilon},\epsilond,\epsilonlg$ and $\mathtt{\delta,\deltad,\deltalg}$}

	\redbox{\begin{minipage}{0.3\linewidth}
			\includegraphics[scale=0.4
			]{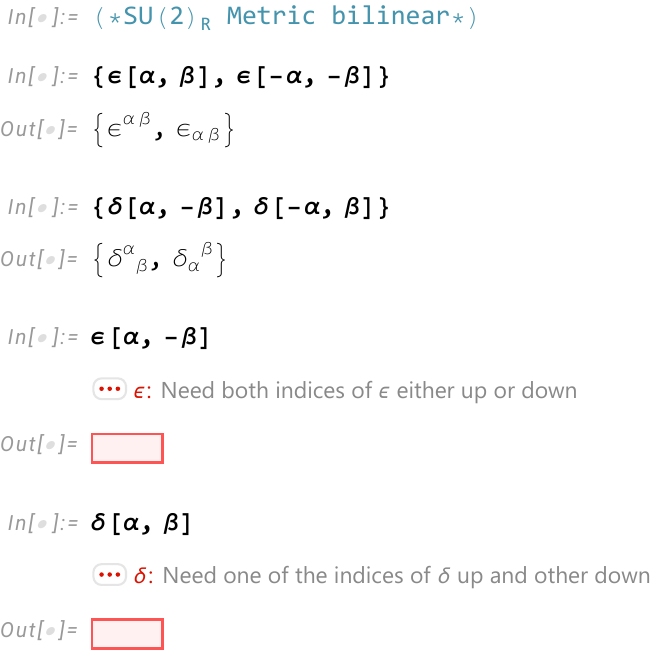}
		\end{minipage}\begin{minipage}{0.3\linewidth}
			\includegraphics[scale=0.4
			]{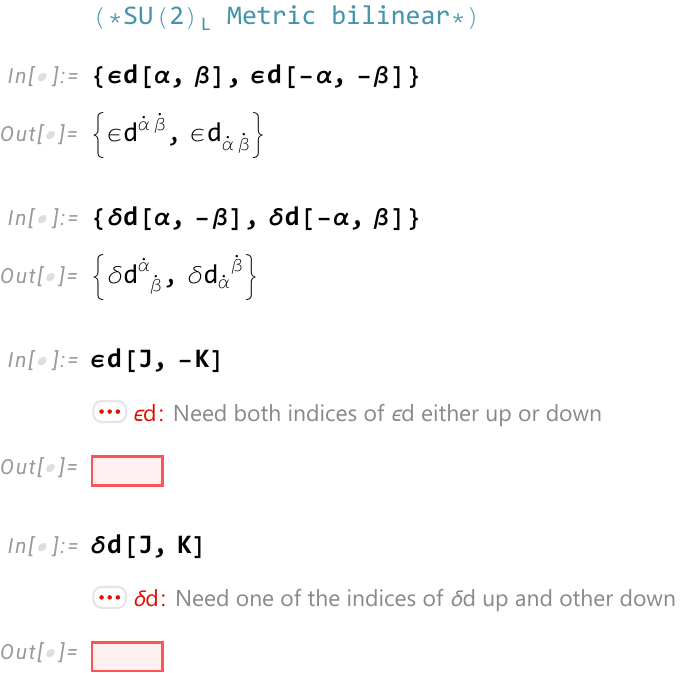}
		\end{minipage}\begin{minipage}{0.3\linewidth}
			\includegraphics[scale=0.4
			]{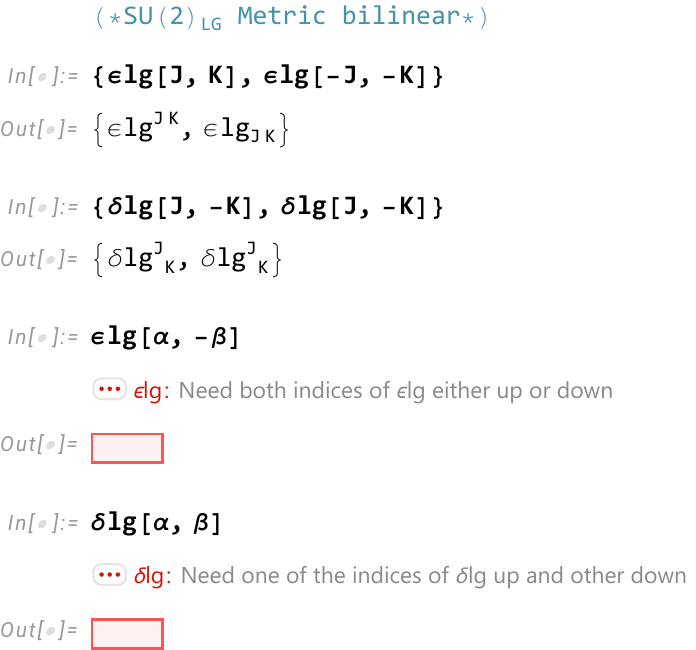}
	\end{minipage}}
	
\end{examplebox}

\begin{warningbox}{}
	The package warns the user (as shown in the example above) when the index structure is incompatible with the
	type of bilinear.  In particular, invariant epsilon tensors require indices
	with the same sign, while identity tensors require indices with opposite
	signs.
\end{warningbox}

\begin{morematerialbox}
	
	\textcolor{black}{\bfseries Note:}
	
	The lowering and raising of indices is controlled entirely by the signs of the
	indices.  For example, changing
	\[
	\tt{$\epsilon$[$\alpha$\_, $\beta$\_]}
	\qquad \textrm{to} \qquad
	\tt{$\epsilon$[-$\alpha$\_, -$\beta$\_]}
	\]
	changes the output from \(\epsilon^{\alpha\beta}\) to
	\(\epsilon_{\alpha\beta}\).  Similarly,
	\[
	\tt{$\delta$[$\alpha$\_, -$\beta$\_]}
	\qquad \textrm{and} \qquad
	\tt{$\delta$[-$\alpha$\_, $\beta$\_]}
	\]
	represent the two possible index placements for the identity tensor.
	
\end{morematerialbox}

\begin{documentationbox}{\tt{ep}, \tt{epd}, and \tt{eplg}}{\tt{ep[$\alpha$\_, $\beta$\_]}, \tt{epd[$\alpha$\_, $\beta$\_]}, \tt{eplg[I\_, J\_]}}
	
	The invariant bilinear and the identity bilinear can also be called using the
	single ep-like commands. These commands decide whether to return the antisymmetric invariant bilinear or the identity bilinear by inspecting the signs of the two indices.
	
	\begin{eqnarray}
		\tt{ep[$\alpha$\_, $\beta$\_]}
		\longrightarrow
		\epsilon^{\alpha\beta},
		\qquad
		\tt{epd[$\alpha$\_, $\beta$\_]}
		\longrightarrow
		\epsilon^{\dot\alpha\dot\beta},
		\qquad
		\tt{eplg[I\_, J\_]}
		\longrightarrow
		\epsilon^{IJ}.
	\end{eqnarray}
	If the two indices have opposite signs, the output is the corresponding
	identity tensor:
	\begin{eqnarray}
		\tt{ep[$\alpha$\_, -$\beta$\_]}
		\longrightarrow
		\delta^\alpha{}_\beta,
		\qquad
		\tt{epd[$\alpha$\_, -$\beta$\_]}
		\longrightarrow
		\delta^{\dot\alpha}{}_{\dot\beta},
		\qquad
		\tt{eplg[I\_, -J\_]}
		\longrightarrow
		\delta^I{}_J .
	\end{eqnarray}
	
\end{documentationbox}

\begin{examplebox}{$\mathtt{ep,epd,eplg}$}

	\redbox{\includegraphics[scale=0.55]{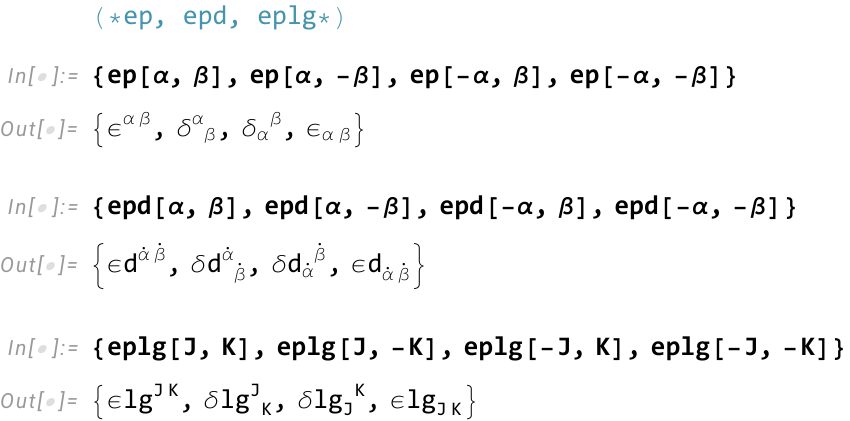}}
\end{examplebox}

\subsection{Lorentz metric}
\begin{documentationboxWL}{\tt{$\eta$}}{\tt{$\eta$[$\mu$\_, $\nu$\_]}}
	\label{dcmnt:eta}

	As in the previous cases, the position of an index is controlled by its sign.
	An index with positive sign is placed as a superscript, while an index with a
	negative sign is placed as a subscript.  Thus the metric with lowered indices
	can be entered by putting negative signs in front of both indices:
	\begin{eqnarray}
		\tt{$\eta$[-$\mu$\_, -$\nu$\_]} .
	\end{eqnarray}
	
	\paragraph{Keyboard shortcut}{$\eta$}
	
\end{documentationboxWL}

\begin{examplebox}{\tt{$\eta$}}

	\redbox{\includegraphics[scale=0.5
		]{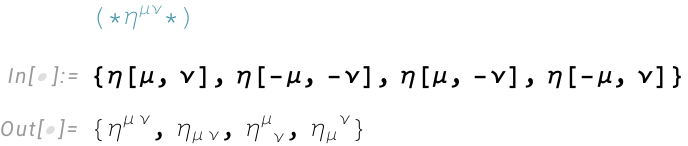}}
	
\end{examplebox}

\noindent
We will discuss the rules involving the contractions using the Lorentz metric in the next section.

\subsection{Clifford algebra}

\begin{conceptbox}{}
	
	Pauli matrices provide the bridge between Lorentz-vector notation and
	spinor-helicity notation.  A Lorentz vector index can be converted into a pair
	of spinor indices by contracting with the sigma matrices.  Schematically, this
	conversion is of the form
	\begin{eqnarray}
		V_\mu
		\quad \longleftrightarrow \quad
		V_{\alpha \dot\alpha}
		\sim
		V_\mu \sigma^\mu_{\alpha\dot\alpha}.
	\end{eqnarray}
	Thus the sigma matrices carry one Lorentz-vector index and one spinor index
	from each of the two Lorentz-spinor sectors.
	
\end{conceptbox}

In the package, the sigma matrices are represented by \(\tt{$\sigma$mat[]}\).
They are used whenever expressions written in Lorentz-vector notation are
converted into spinor-helicity variables.

\begin{documentationboxWL}{\tt{$\sigma$mat}}{\tt{$\sigma$mat[LorentzVecIndex\_][SpinorIndexR\_, SpinorIndexL\_]}}
	\label{dcmnt:sigma}
	
	An index with positive sign is placed as a superscript, while an index
	with negative sign is placed as a subscript.  When both spinor indices are
	raised, the output form is written using the barred Pauli matrix,
	\(\bar\sigma\).

	\paragraph{Keyboard shortcut}
	\(\boxed{\tt{Esc}}\sigma\boxed{\tt{Esc}}\).
	
\end{documentationboxWL}

\begin{examplebox}{\tt{$\sigma$mat}}

	\redbox{\includegraphics[scale=0.5
		]{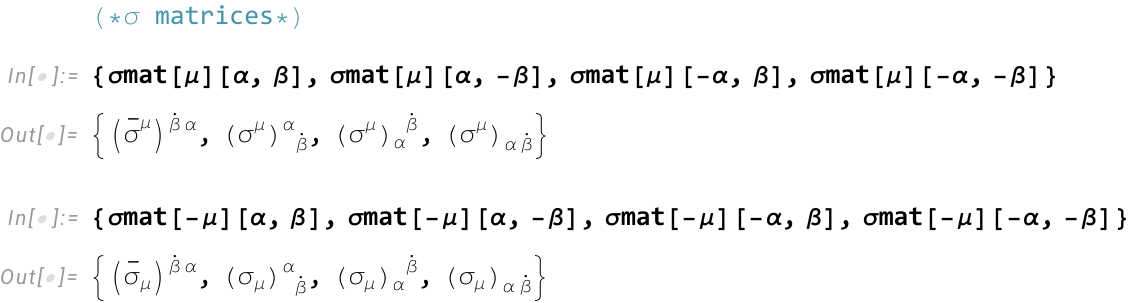}}
	
\end{examplebox}

A summary of the all the objects introduced in this section can be found in \ref{sec:Summaryofbasicobjects}. 

\subsection{Mandelstam Variables}
\begin{conceptbox}{}
	
	For an \(\mathtt{N}\)-point scattering process, with all external momenta
	taken to be incoming, the Mandelstam variables are defined by
	\begin{align}
		\mathtt{\textit{S}_{i_1\cdots i_n}}
		&=
		\mathtt{(k_{i_1}+\cdots+k_{i_n})^2},
		&
		\mathtt{\textit{S}_{\dots i\dots j\dots}}
		&=
		\mathtt{\textit{S}_{\dots j\dots i\dots}} .
	\end{align}
	The number of possible variables in $\mathtt{\textit{S}_{i_1\cdots i_n}}$ grows with the number of external legs.  It is therefore useful to introduce the smaller two-index Mandelstam variables
	\begin{align}
		\mathtt{\textit{s}_{ij}}
		&=
		\mathtt{k_i\cdot k_j}
		=
		\mathtt{\frac{1}{2}\langle i|j|i]},
		&
		\mathtt{\textit{s}_{ij}}
		&=
		\mathtt{\textit{s}_{ji}},
		&
		\mathtt{\textit{s}_{ii}}
		&=
		\mathtt{-m_i^2}.
	\end{align}
	The larger Mandelstam variables are linearly related to the smaller ones.
	For example,
	\begin{align}
		\mathtt{\textit{S}_{123}}
		&=
		\mathtt{
			2\textit{s}_{12}
			+2\textit{s}_{23}
			+2\textit{s}_{13}
			-(m_1^2+m_2^2+m_3^2)
		}.
	\end{align}
	
\end{conceptbox}

\begin{documentationbox}{\tt{SM}}{\tt{SM[\(\cdots\)]}} 
	
	It gives Mandelstam variables of the form
	\begin{align}
		\mathtt{\textit{S}_{i_1\cdots i_n}}
		=
		\mathtt{(k_{i_1}+\cdots+k_{i_n})^2}.
	\end{align}
	The arguments of \(\tt{SM}\) can be any sequence of external-leg labels.
	
\end{documentationbox}

\begin{documentationbox}{\tt{sm}}{\tt{sm[i\_, j\_]}}

	It gives the the two-index Mandelstam variable
	\begin{align}
		\mathtt{\textit{s}_{ij}}
		=
		\mathtt{k_i\cdot k_j}.
	\end{align}
	
	\paragraph{Keyboard shortcut}
	\boxed{\tt{Esc}}\ \tt{sm}\ \boxed{\tt{Esc}} .
	
\end{documentationbox}

\begin{examplebox}{\tt{SM} and \tt{sm}}
	
	\redbox{
		\begin{minipage}{0.5\linewidth}
			\includegraphics[scale=0.5]{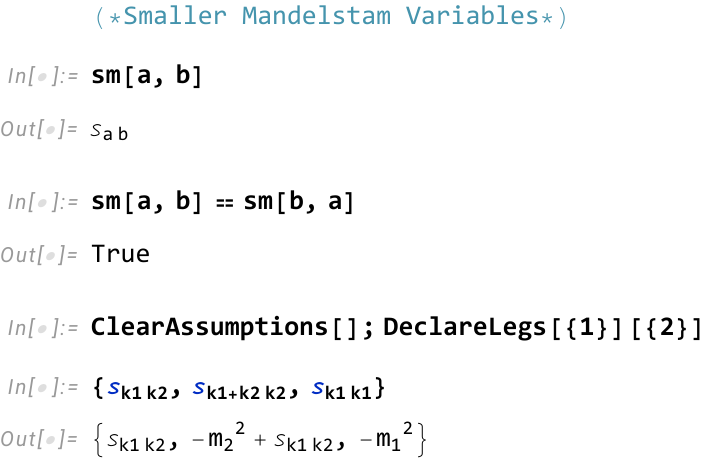}
		\end{minipage}
		\begin{minipage}{0.5\linewidth}
			\includegraphics[scale=0.5]{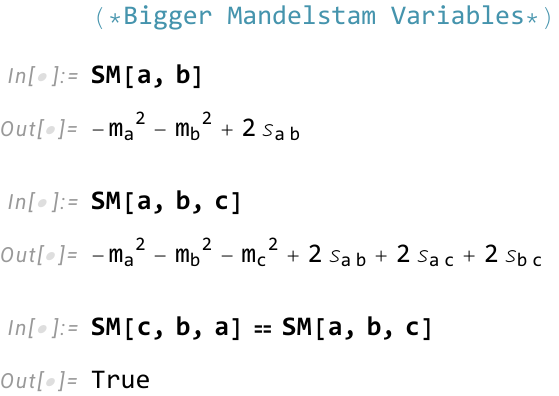}
		\end{minipage}
	}
	
\end{examplebox}

%% file: PolynomialRules.tex
\section{Rules for Simplifying SH Polynomials}
\label{sec:PolynomialRules}

Physical quantities in quantum field theory are typically expressed as
rational functions of spinor-helicity variables.  The numerator and denominator
are polynomials built from spinor brackets, mixed contractions, invariant
bilinears, and momentum insertions.  In this section, we introduce the basic
rules used by the package to simplify such spinor-helicity polynomials.

\subsection{Canonical Ordering}
\begin{conceptbox}{}
	
	Several basic objects in the package are antisymmetric under the exchange of
	their arguments.  This includes the invariant bilinears
	\(\epsilon\), \(\epsilond\), and \(\epsilonlg\), as well as the scalar
	contractions \(\tt{SHAA}\), \(\tt{SHBB}\), \(\tt{SHAB}\), and \(\tt{SHBA}\).
	Hence these objects vanish whenever the relevant arguments coincide. To avoid writing the same expression in several sign-equivalent forms, the
	package uses the following order
	\begin{enumerate}
		\item two massless legs, or two massive legs, are ordered lexicographically;
		\item if one leg is massive and the other is massless, the massive leg is
		placed first.
		\item 
		There is also a canonical convention for contracted massive little-group
		indices between an angle spinor and a box spinor.  The angle-spinor
		little-group index is placed downstairs, while the box-spinor little-group
		index is placed upstairs:
		\begin{eqnarray}
			|Y_{K}]\langle X^{K}|
			=
			-|Y^{K}]\langle X_{K}| .
		\end{eqnarray}
	\end{enumerate}
\end{conceptbox}

\begin{documentationbox}{\tt{PutCanonicalOrder}}{\tt{PutCanonicalOrder[expr\_]}}
	
	The command \(\tt{PutCanonicalOrder}\) rewrites an expression in the canonical
	ordering convention.  
\end{documentationbox}

\begin{examplebox}{\tt{PutCanonicalOrder}}
	
	\redbox{\begin{minipage}{0.57\linewidth}
		\includegraphics[scale=0.425]{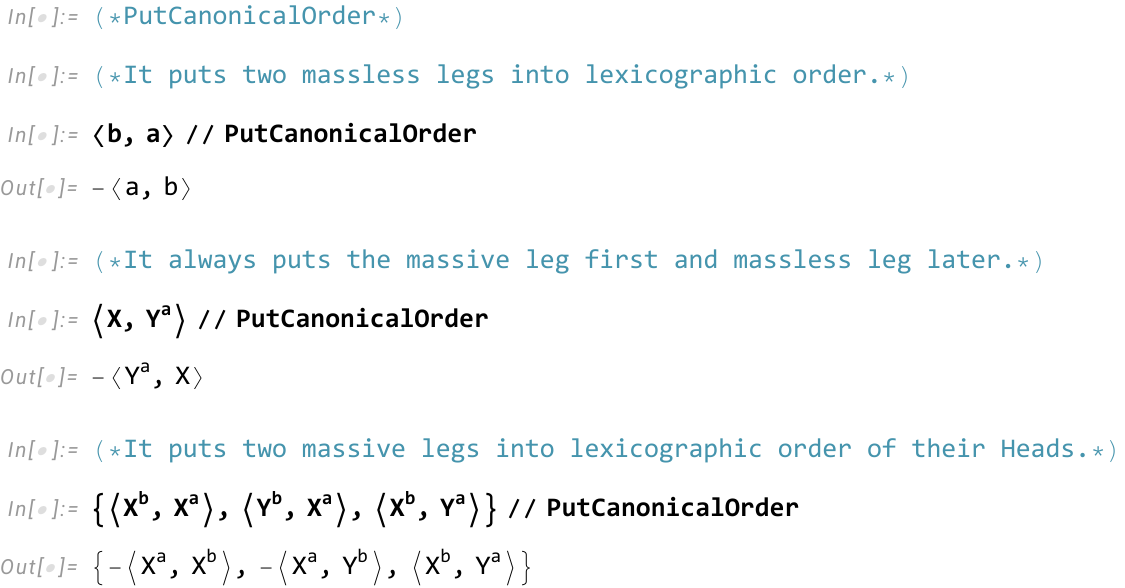}
	\end{minipage}
	\begin{minipage}{0.45\linewidth}
		\includegraphics[scale=0.425]{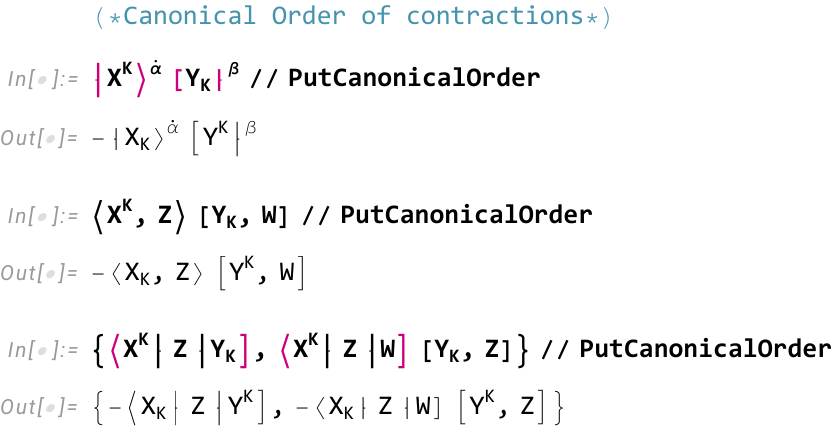}
\end{minipage}}

\end{examplebox}

\begin{conceptbox}{}
Instead of applying the \tt{PutCanonicalOrder} command after every evaluation, the canonical order can be made automatic by running \tt{SetDefaultOrdering[]}.
\end{conceptbox}

\begin{documentationbox}{\tt{SetDefaultOrdering}}{\tt{SetDefaultOrdering[]}}
	
	This command sets the canonical ordering rules as
	the default simplification rules.  After this command is run, the package
	automatically applies the canonical ordering convention to the relevant
	antisymmetric objects and contracted massive little-group structures.

\end{documentationbox}

\begin{examplebox}{\tt{SetDefaultOrdering}}
	
	\redbox{\includegraphics[scale=0.475]{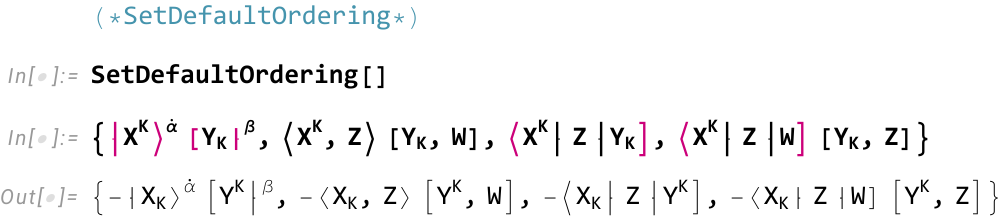}}

\end{examplebox}

\subsection{Contract Bilinears}

\begin{conceptbox}{}
	
	Spinor-helicity expressions often contain repeated spinor, little-group, or
	Lorentz indices.  Such repeated indices should be contracted using the
	corresponding invariant bilinears.  The command \(\tt{ContractBilinears}\)
	performs these contractions automatically, whenever repeated indices are
	present in the expression.
	
\end{conceptbox}

\begin{documentationbox}{\tt{ContractBilinears}}{\tt{ContractBilinears[expr\_]}}
	
	The command \(\tt{ContractBilinears}\) contracts all repeated indices associated
	with the spinor, little-group, and Lorentz metrics.  
	
\end{documentationbox}

\begin{examplebox}{\tt{ContractBilinears}}
	
	The following example shows the action of the full \(\tt{ContractBilinears}\)
	command:
	
	\redbox{
		\includegraphics[scale=0.5]{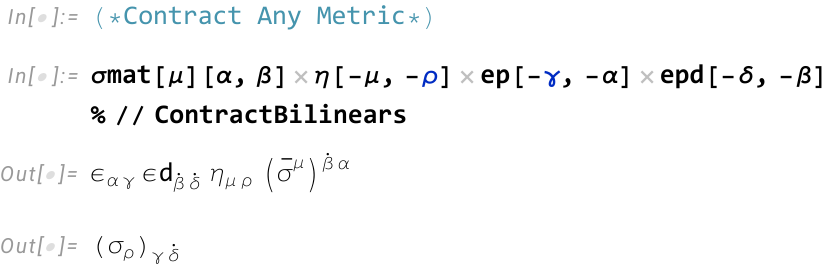}
	}

\end{examplebox}
\begin{morematerialbox}
	
	\(\tt{ContractBilinears}\) is a combined contraction command.  It applies the
	four sector-wise rules
	\[
	\tt{ContractLBilinear},\qquad
	\tt{ContractRBilinear},\qquad
	\tt{ContractLGBilinear},\qquad
	\tt{ContractLorMetric},
	\]
	which contract repeated \(\mathtt{SU(2)_L}\), \(\mathtt{SU(2)_R}\),
	massive little-group, and Lorentz indices, respectively. These individual commands are often useful in computations. These component commands are described in more detail in \ref{subsec:Moreoncontractmetric}.

\end{morematerialbox}

\begin{conceptbox}{Clifford algebra identities}
	
	When Lorentz indices are contracted, their conversion often requires Clifford
	algebra identities such as
	\begin{align}
		\label{eq:cliffrule1}
		(\bar{\sigma}^{\mu})^{\dot{\beta}\alpha}
		(\bar{\sigma}_{\mu})^{\dot{\delta}\gamma}
		&=
		-2\epsilon^{\alpha\gamma}
		\epsilon^{\dot{\beta}\dot{\delta}},
		\\
		\label{eq:cliffrule2}
		(\bar{\sigma}^{\mu})^{\dot{\beta}\alpha}
		(\bar{\sigma}^{\nu})_{\alpha\dot{\beta}}
		&=
		-2\eta^{\mu\nu}.
	\end{align}
	All such rules are included in the command $\tt{\(\sigma\)matRule}$.
	
\end{conceptbox}
\begin{documentationboxWL}{\tt{\(\sigma\)matRule}}{\tt{\(\sigma\)matRule[expr\_]}}
	\label{dcmnt:sigmamatrule}

	The command \(\tt{\(\sigma\)matRule}\) applies the Clifford algebra identities given in \eqref{eq:cliffrule1} and \eqref{eq:cliffrule2}.
\end{documentationboxWL}

\begin{examplebox}{Clifford algebra rules}
	
	\redbox{
		\includegraphics[scale=0.5]{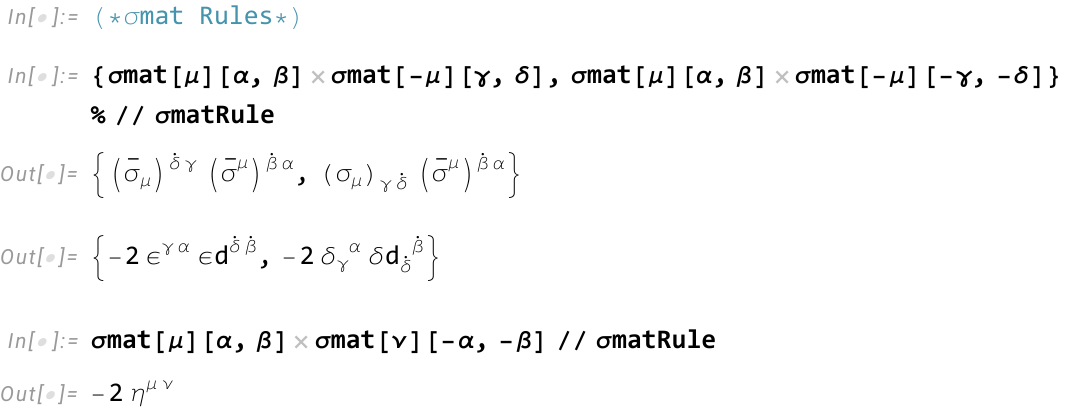}
	}
	
\end{examplebox}

\subsection{Make and Break Scalars}
\subsubsection{$SL(2,\mathbb{C})$ Scalars}

\begin{conceptbox}{}
	
	Products of two spinors with contracted \(SL(2,\mathbb C)\) indices can be
	rewritten as scalar spinor-helicity brackets.  The two basic contractions are
	\begin{equation}
		\langle p|_{\dot\alpha}|q\rangle^{\dot\alpha}
		=
		\langle p,q\rangle,
		\qquad
		[p|^{\alpha}|q]_{\alpha}
		=
		[p,q].
	\end{equation}
	In the package these scalar brackets are represented by \(\tt{SHAA}\) and
	\(\tt{SHBB}\).
	
	There are two complementary operations.  The command
	\(\tt{PutSL2CScalar}\) replaces contracted spinor products by scalar
	brackets.  The command \(\tt{NoSL2CScalar}\) performs the inverse operation:
	it expands scalar brackets back into products of spinors with contracted
	spinor indices.  The latter command also chooses dummy spinor indices so that
	indices are not accidentally repeated unless they are contracted within the
	same object.
	
\end{conceptbox}

\begin{documentationbox}{\tt{PutSL2CScalar}}{\tt{PutSL2CScalar[expr\_]}}
	
	The command \(\tt{PutSL2CScalar}\) replaces contracted
	\(SL(2,\mathbb C)\) spinor products by the corresponding scalar
	spinor-helicity brackets. 
\end{documentationbox}

\begin{examplebox}{\tt{PutSL2CScalar}}
	\redbox{
		\includegraphics[scale=0.5]{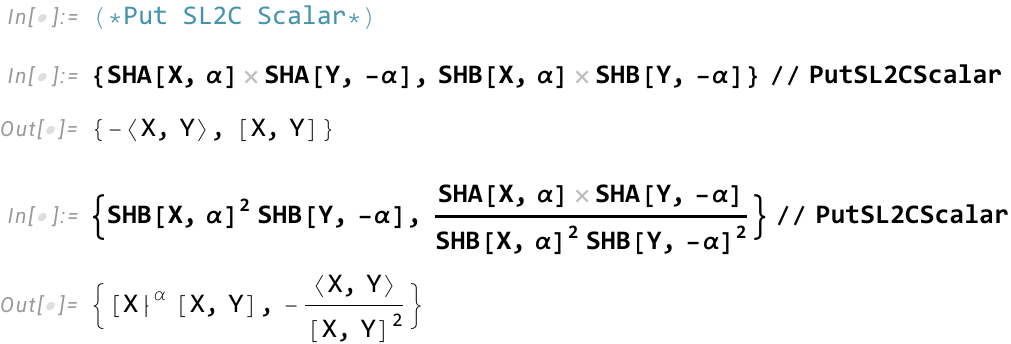}
	}
\end{examplebox}

\begin{morematerialbox}
	In many computations, it is useful to form the
	\(\mathtt{SU(2)_L}\) and \(\mathtt{SU(2)_R}\) scalar contractions separately.
	For this purpose, the package provides the sector-wise commands
	\(\tt{PutSU2LScalar}\) and \(\tt{PutSU2RScalar}\).  These commands are
	described in detail in \ref{subsec:Moreonputsltwocscalar}.
	
\end{morematerialbox}

\begin{documentationbox}{\tt{NoSL2CScalar}}{\tt{NoSL2CScalar[expr\_]}}
	
	The command \(\tt{NoSL2CScalar}\) expands scalar spinor-helicity brackets into
	products of spinors with contracted \(SL(2,\mathbb C)\) indices. 
\end{documentationbox}

\begin{examplebox}{\tt{NoSL2CScalar}}
	\redbox{
		\includegraphics[scale=0.5]{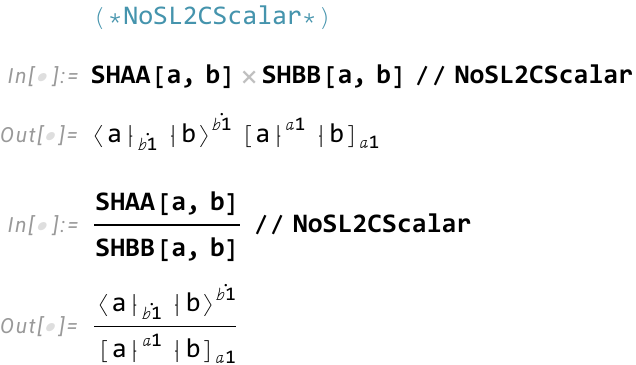}
	}
\end{examplebox}

\begin{morematerialbox}
	
	In many computations, it is also useful to split the
	\(\mathtt{SU(2)_L}\) and \(\mathtt{SU(2)_R}\) scalar contractions separately.
	For this purpose, the package provides the sector-wise inverse commands
	\(\tt{NoSU2LScalar}\) and \(\tt{NoSU2RScalar}\).  These commands are
	described in detail in \ref{subsec:Moreonnosltwocscalar}.
	
\end{morematerialbox}

\subsubsection{Little-group scalars}

\begin{conceptbox}{}
	
	The mixed scalar \(\tt{SHAB}\) is obtained by contracting an angle bracket and
	a box bracket through a common leg.  Schematically, the contraction has the
	form
	\begin{equation}
		\langle a,b\rangle [c,b]
		\quad \longrightarrow \quad
		\langle a|b|c].
	\end{equation}
	For massive legs, the same operation contracts the massive little-group
	indices:
	\begin{equation}
		\langle a,b_J\rangle [c,b^J]
		\quad \longrightarrow \quad
		\langle a|b|c].
	\end{equation}
	
	The package provides two complementary operations.  The command
	\(\tt{PutLGScalar}\) converts products such as
	\(\tt{SHAA[]SHBB[]}\) into the little-group scalar \(\tt{SHAB[]}\).  The
	command \(\tt{NoLGScalar}\) performs the inverse operation and expands
	\(\tt{SHAB[]}\) back into a product of \(\tt{SHAA[]}\) and \(\tt{SHBB[]}\).
	
\end{conceptbox}

\begin{documentationbox}{\tt{PutLGScalar}}{\tt{PutLGScalar[expr\_]}}
	
	It converts products of \(\tt{SHAA[]}\) and \(\tt{SHBB[]}\) into
	\(\tt{SHAB[]}\) for both massive and massless middle legs.
	
\end{documentationbox}

\begin{examplebox}{\tt{PutLGScalar}}
	\redbox{
		\includegraphics[scale=0.5]{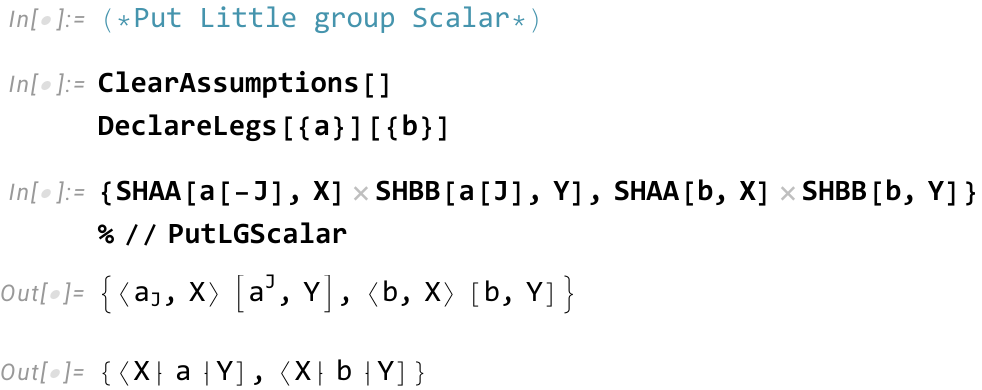}
	}
	
\end{examplebox}

\begin{morematerialbox}
	The package also allows one to form little-group scalars separately for
	massless and massive legs.  The command \(\tt{PutMasslessLGScalar}\) acts on
	non-massive middle legs, while \(\tt{PutMassiveLGScalar}\) acts on declared
	massive legs.  See sec.~\ref{Subsec:moreonputlgscalar} for details.
	
\end{morematerialbox}

\begin{documentationbox}{\tt{NoLGScalar}}{\tt{NoLGScalar[expr\_]}}
	
	It expands \(\tt{SHAB}\) into products of \(\tt{SHAA}\) and
	\(\tt{SHBB}\) for both massive and massless middle legs.
	
\end{documentationbox}

\begin{examplebox}{\tt{NoLGScalar}}
	\redbox{
		\includegraphics[scale=0.5]{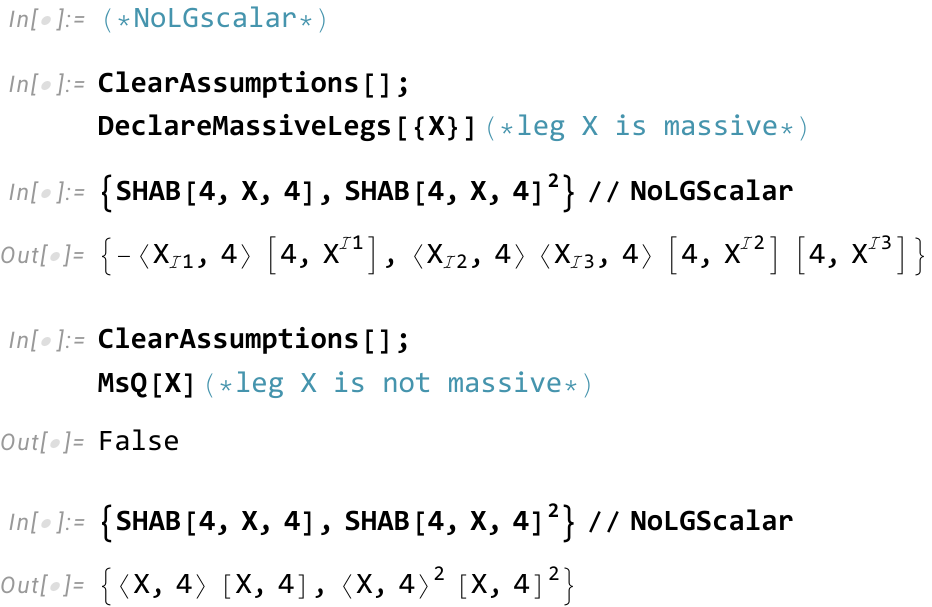}
	}
	
\end{examplebox}

\begin{morematerialbox}
	
	The package also allows one to split little-group scalars separately for
	massless and massive legs.  The command \(\tt{NoMasslessLGScalar}\) acts on
	non-massive middle legs, while \(\tt{NoMassiveLGScalar}\) acts on declared
	massive legs.  See sec.~\ref{Subsec:moreonnolgscalar} for details.
	
\end{morematerialbox}

\subsection{On-shell Rules}

\begin{conceptbox}{}
	
	All the commands discussed above works without declaring the mass of the particles. This allows the user enough flexibility to handle the spinor variables from a completely off-shell approach. In order to do the on-shell computations, we introduce the on-shell rules on top of the basic scalars. For example:	
	\begin{eqnarray}
		&&\mathtt{		[X_{I1}|^{\alpha}[X^{I1}|^{\beta}	=-m_X \epsilon^{\alpha\beta}
			\qquad,\qquad
			|X_{I1}\rangle}^{\dot \alpha}\mathtt{|X^{I1}\rangle}^{\dot \beta}\mathtt{=-m_X\epsilon}^{\dot{\alpha}\dot{\beta}}
		\\
		&&\mathtt{\langle X^{I1}X^{I2}\rangle=	 -m_X\epsilonlg^{I1\,I2}
			\qquad,\qquad
			[ X^{I1}X^{I2}]= m_X\epsilonlg^{I1\,I2}}
		\\
		&&\mathtt{		\langle X^{I1}|X|Y] =-m_X[X^{I1},Y]\quad,\quad 	\langle Y|X|X^{I1}] =-m_X\langle X^{I1},Y\rangle}
		\\
		&&\mathtt{[X^I,a][X_I,b]=-m_X[a,b]}	
	\end{eqnarray}
	The above identity is true for all $a$, $b$ (respective whether it is massless or massive).It does the same thing for angle variables.

	The function \tt{PutOnShell} puts the on-shell relations for leg $X$ with mass $m_X$. The command is designed to be applied after the basic spinor-helicity
	expression has been constructed.  Thus one may first manipulate expressions
	off shell, and then impose the on-shell conditions only when needed.
\end{conceptbox}

\begin{documentationbox}{\tt{PutOnShell}}{\tt{PutOnShell[expr\_]}}
	
	It acts on spinor-helicity polynomials by replacing contracted massive
	spinor-helicity structures with their on-shell values.
	
\end{documentationbox}

\begin{examplebox}{\tt{PutOnShell}}

	\redbox{
		\includegraphics[scale=0.5]{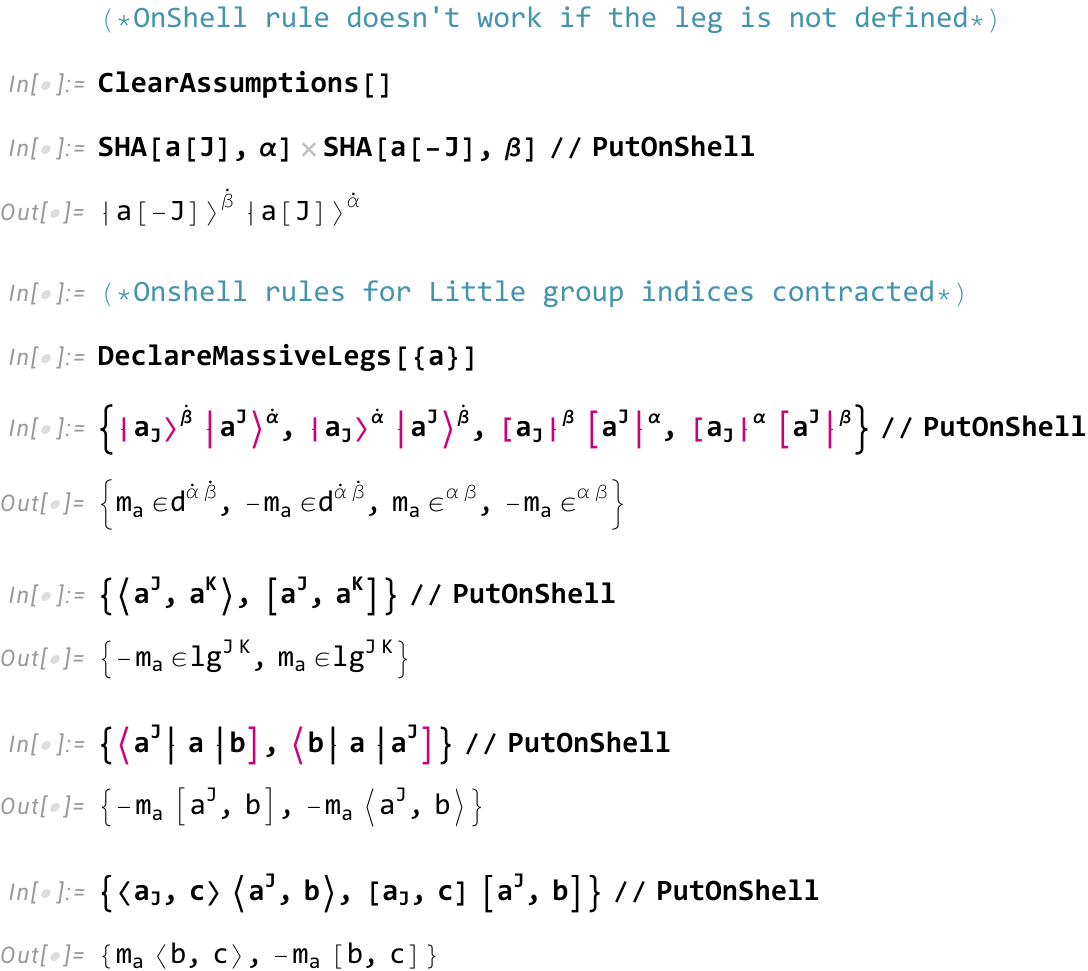}
	}
	
\end{examplebox}

\subsection{Convert to Mandelstam variables}

\begin{documentationbox}{\tt{ToMandelstam}}{\tt{ToMandelstam[expr\_]}}
	
	converts spinor-helicity and Lorentz products into Mandelstam variables wherever possible. 
	
\end{documentationbox}

\begin{examplebox}{\tt{ToMandelstam}}
	
	\redbox{
					\includegraphics[scale=0.5]{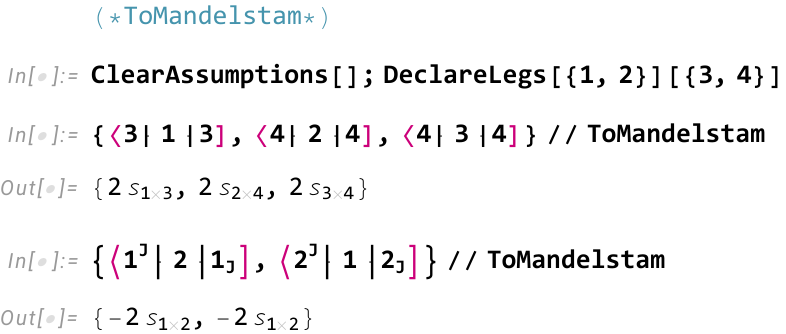}
	}
	
\end{examplebox}
\begin{morematerialbox}{}
	We have also defined more focused conversion commands. For example,
	\(\tt{MandelstamtoSH}\) and \(\tt{MandelstamtoLor}\) convert Mandelstam
	variables to spinor-helicity and Lorentz notation, respectively.  The inverse
	conversions are performed by \(\tt{SHToMandesltam}\) and
	\(\tt{LorToMandesltam}\).  See sec.~\ref{subsec:moreonmandelstamvar} for a
	detailed discussion.
	
\end{morematerialbox}

\subsection{Schouten identities}

\begin{conceptbox}{}
	
	Schouten identities express the fact that a two-dimensional spinor space has
	only two independent spinor directions.  Equivalently, any completely
	antisymmetric object with three spinor indices vanishes.  These identities are
	therefore useful for rewriting products of spinors, epsilon tensors, and
	spinor brackets.
	
	The basic identities used in the package are
	\begin{align}
		\epsilon^{\alpha\beta}\epsilon^{\gamma\delta}
		&=
		\epsilon^{\alpha\gamma}\epsilon^{\beta\delta}
		+
		\epsilon^{\alpha\delta}\epsilon^{\gamma\beta},
		&
		\epsilon^{\dot\alpha\dot\beta}
		\epsilon^{\dot\gamma\dot\delta}
		&=
		\epsilon^{\dot\alpha\dot\gamma}
		\epsilon^{\dot\beta\dot\delta}
		+
		\epsilon^{\dot\alpha\dot\delta}
		\epsilon^{\dot\gamma\dot\beta},
		\\
		[a|^{\alpha}[b|^{\beta}
		&=
		[a|^{\beta}[b|^{\alpha}
		-
		[a,b]\epsilon^{\alpha\beta},
		&
		|a\rangle^{\dot\alpha}|b\rangle^{\dot\beta}
		&=
		|a\rangle^{\dot\beta}|b\rangle^{\dot\alpha}
		+
		\langle a,b\rangle\epsilon^{\dot\alpha\dot\beta},
		\\
		[a,b][c,d]
		&=
		[a,c][b,d]
		-
		[a,d][b,c],
		&
		\langle a,b\rangle\langle c,d\rangle
		&=
		\langle a,c\rangle\langle b,d\rangle
		-
		\langle a,d\rangle\langle b,c\rangle .
	\end{align}
	The package applies these identities efficiently on the expressions.
	
\end{conceptbox}

\subsubsection{\tt{SplitSchouten}}

\begin{documentationbox}{\tt{SplitSchouten}}{\tt{SplitSchouten[\{a\_, b\_\},\{c\_, d\_\}]}}
	
	The command \(\tt{SplitSchouten}\) applies the Schouten identity to four
	angle-sector or four box-sector objects labelled by \(a,b,c,d\).  
	
	Here \(a,b,c,d\) may denote either particle labels or spinor indices,
	depending on the expression being simplified.
\end{documentationbox}

\begin{examplebox}{\tt{SplitSchouten}}
	
	\redbox{
		\includegraphics[scale=0.5]{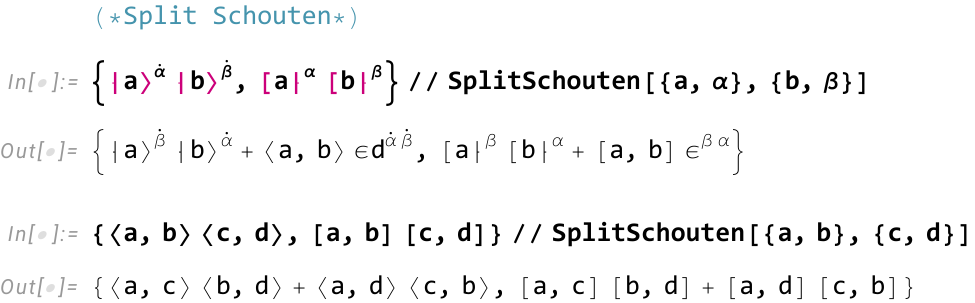}
	}

\end{examplebox}

\begin{morematerialbox}
	
	The package also allows Schouten identities to be applied separately in the
	angle and box sectors.  The commands \(\tt{SplitSchoutenA}\) and
	\(\tt{SplitSchoutenB}\) act on angle-sector and box-sector structures,
	respectively.  See sec.~\ref{subsec:MoreonSplitSchouten} for details.

\end{morematerialbox}

\subsubsection{\tt{SimplifySchouten}}

\begin{conceptbox}{}

	Schouten identities are not unique rewriting rules.  The same expression can
	often be transformed in several different ways, and a single application of a
	Schouten identity may not be enough to reveal the simplest form of the
	polynomial.  For this reason, it is useful to have an automated command which
	tries Schouten transformations recursively and searches for a simpler
	representative of the same expression.
	
	The command \(\tt{SimplifySchouten}\) (adapted from \cite{spinorhelicity4d}) performs this task.  It applies
	Schouten identities recursively in order to simplify spinor-helicity
	polynomials.  It uses Mathematica's built-in simplification framework together
	with Schouten transformations as additional
	\(\tt{TransformationFunctions}\).
	
\end{conceptbox}

\begin{documentationbox}{\tt{SimplifySchouten}}{\tt{SimplifySchouten[expr\_]}}
	
	The command \(\tt{SimplifySchouten}\) applies Schouten identities recursively
	in order to simplify spinor-helicity polynomials.

\end{documentationbox}

\begin{examplebox}{\tt{SimplifySchouten}}
	\redbox{
		\includegraphics[scale=0.335]{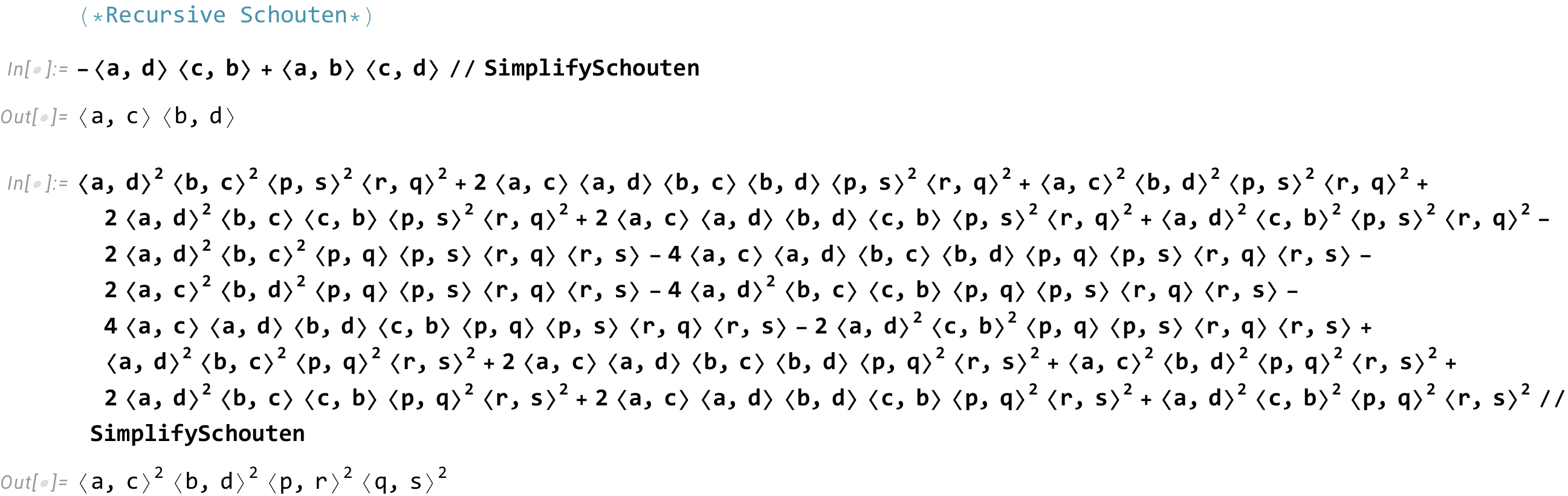}
	}
\end{examplebox}

\subsection{Canonicalize indices}\label{CanonicalIndices}

\begin{conceptbox}{}
	
	A difficult part of spinor-helicity computations is the simplification of long
	expressions using Schouten identities.  The command \(\tt{SimplifySchouten}\)
	solves part of this problem, but it is often not enough when the simplification
	also involves momentum-conservation or on-shell rules.
	
	A useful intermediate step is to put the same little-group indices on
	contracted massive legs. For example, consider the following sum of two terms with $SL(2,\mathbb{C})$ indices contracted
	\begin{align}
\langle \tt{a}| _{\dot{\alpha }} \langle \tt{b}| _{\dot{\beta }} | \tt{c}\rangle ^{\dot{\alpha }} | \tt{d}\rangle ^{\dot{\beta }}+\langle \tt{a}| _{\dot{\text{$\alpha
			$\tt{1}}}} \langle \tt{b}| _{\dot{\text{$\beta $\tt{1}}}} | \tt{c}\rangle ^{\dot{\text{$\alpha $\tt{1}}}} | \tt{d}\rangle ^{\dot{\text{$\beta $\tt{1}}}}
	\end{align}
	Both the terms are same but the SH variables carry different spinor indices. To put uniform indices across the same terms, one can use the \tt{CanonicalizeIndices} command.
\end{conceptbox}	
	\begin{documentationbox}{\tt{CanonicalizeIndices}}{\tt{CanonicalizeIndices[expr\_]}}
		
		Puts uniform spinor and little group indices across the same terms in the expression \tt{expr}.
		
	\end{documentationbox}
	
	\begin{examplebox}{\tt{CanonicalizeIndices}}
		
		\redbox{\begin{minipage}{0.53\linewidth}
\includegraphics[scale=0.39]{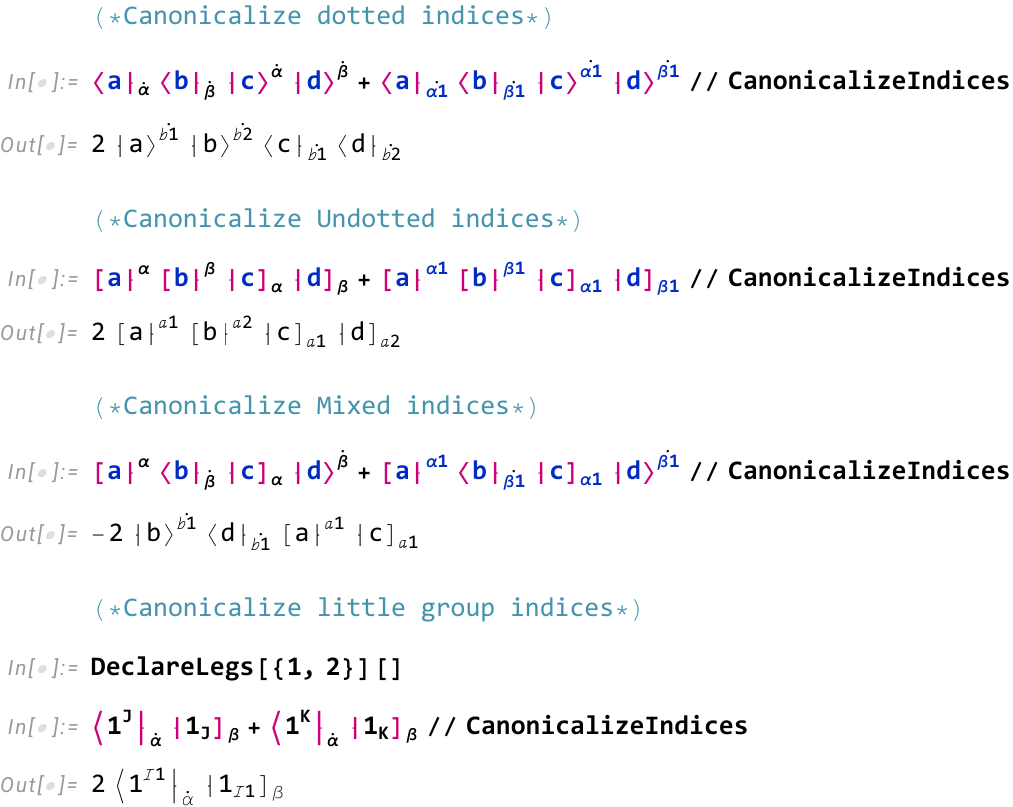}
		\end{minipage}\begin{minipage}{0.5\linewidth}
		\includegraphics[scale=0.39]{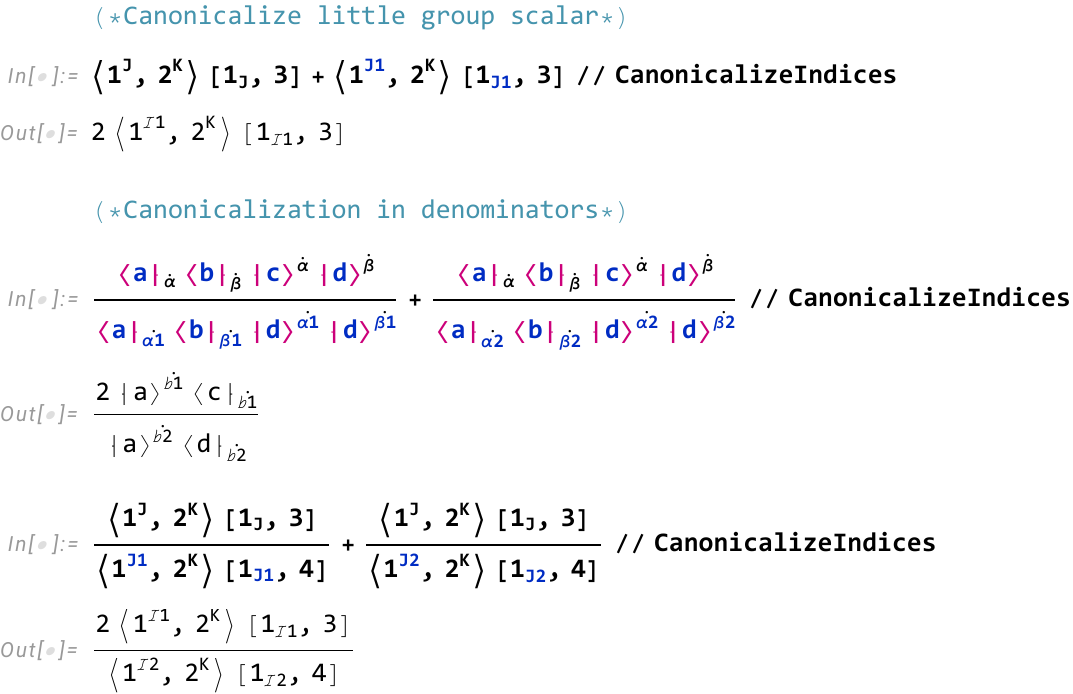}
	\end{minipage}}
		
	\end{examplebox}
	
	More related commands which are specific to canonicalization of left or right handed scalars are discussed in appendix \ref{app:moreoncanonicalizeindices}.

\subsection{Simplify Polynomial}

\begin{conceptbox}{}
	
	In practical computations, spinor-helicity expressions usually contain several
	different kinds of structures at the same time: bilinear contractions,
	\(SL(2,\mathbb C)\) scalar contractions, massive little-group contractions,
	and on-shell relations.  Applying each simplification rule by hand can be
	tedious.
	
	For this reason, the package combines the basic polynomial simplification
	rules into a single function \tt{SimplifyPolynomialFunction}, defined in \ref{subsec:moreonsimplifypolynomial}.  It combines \(\tt{ContractBilinears}\), \(\tt{PutSL2CScalar}\),  \(\tt{PutOnShell}\), \(\tt{PutLGScalar}\).  The command \(\tt{SimplifyPolynomial}\) applies the \tt{SimplifyPolynomialFunction} repeatedly until the expression stops changing.  
	
	This command performs a single simplification pass.  
	
\end{conceptbox}

\begin{documentationbox}{\tt{SimplifyPolynomial}}{\tt{SimplifyPolynomial[expr\_]}}
	
	uses all the simplification repeatedly until the expression stops changing.  
	
\end{documentationbox}

\begin{examplebox}{\tt{SimplifyPolynomial}}

	\redbox{
		\includegraphics[scale=0.5]{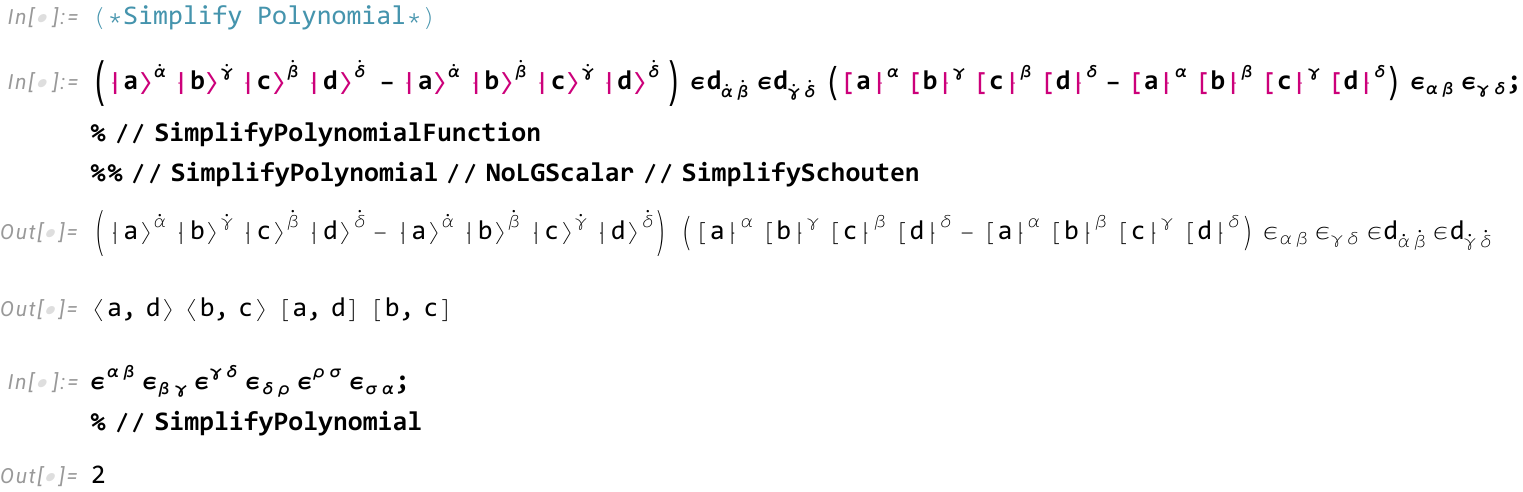}
	}
	
\end{examplebox}
A summary the rules introduced in this section can be found in sec. \ref{sec:SummaryofrulesforSimplifyingSpinorHelicityPolynomials}.

%% file: ScatteringAmplitudes.tex
\section{Computing scattering amplitudes}
\label{sec:computescamp}
In this section, we show how to compute scattering amplitude using \tt{SMaSH}. A typically scattering process takes the following form:
\begin{eqnarray}
	X_1+\cdots +X_{n_i}\longrightarrow Y_1+\cdots +Y_{n_f}
\end{eqnarray} 
Here $\{X_i\}$ are incoming particles and $\{Y_i\}$ are out-going particles. So, computing scattering involves the following steps:
\begin{enumerate}
	\item Specifying the kinematical data for the external legs.
	
	\item Writing down the Feynman rules 
	
	\item Compute the contributions from the Feynman diagrams
	
	\item Simplification and consistency checks like gauge invariance, little group scaling etc.
\end{enumerate}

In the following subsections, we will discuss the features of \tt{SMaSH} that are useful for each of the above steps. After that we will show a full example of computing a scattering amplitude using \tt{SMaSH}. One can choose to directly read the example section and come back to the following subsections when required. 

\subsection{Utilities for amplitude}
In this sub-section, we write down some useful utilities which will be used in computing scattering amplitudes. In particular, we introduce the commands for adding and removing assumptions, defining the momenta and polarizations for massive and massless legs in the SH notation.
\subsubsection{Assumptions}\label{sec:Assumptions}

\begin{morematerialbox}{}
	
	Declaring a leg automatically assigns the following mass assumptions. For every massive leg \tt{i}, the package assumes
	\begin{align}
		\mathtt{\{m_i\neq 0,\ m_i\in \href{https://reference.wolfram.com/language/ref/Reals.html?q=Reals}{Reals}\}} .
	\end{align}
	For every massless leg \tt{i}, the package assumes
	\begin{align}
		\mathtt{\{m_i==0\}} .
	\end{align}
	All such assumptions are automatically appended to
	\href{https://reference.wolfram.com/language/ref/Assumptions.html}{\tt{\$Assumptions}},
	which is used by \texttt{Mathematica} as the default assumption set in
	simplification and refinement operations.
	
\end{morematerialbox}

\begin{conceptbox}{}
	In addition to the already defined assumptions, it is always helpful to add the add new assumptions to \$\tt{Assumptions}. For example, one may add the assumption that two external particles have equal
	masses, impose an ordering among masses or Mandelstam variables, declare a
	symbol to be real or positive, or specify that a particular denominator is
	non-vanishing. Such assumptions are often useful when simplifying amplitudes,
	taking special kinematic limits, or comparing two expressions that are
	equivalent only after using additional physical conditions. The same
	assumptions can later be removed using \tt{RemoveAssumptions}.	
\end{conceptbox}

\begin{documentationbox}{\tt{AddAssumptions}, \tt{RemoveAssumptions}}{\tt{AddAssumptions[assumptions\_List]}, \tt{RemoveAssumptions[assumptions\_List]}}
	
	The user can add new assumptions to \tt{\$Assumptions} or remove existing ones
	using these. The argument in each case is a list of assumptions. 
\end{documentationbox}

\begin{examplebox}{\tt{AddAssumptions} \& \tt{RemoveAssumptions}}
	
	\redbox{
	\includegraphics[scale=0.5]{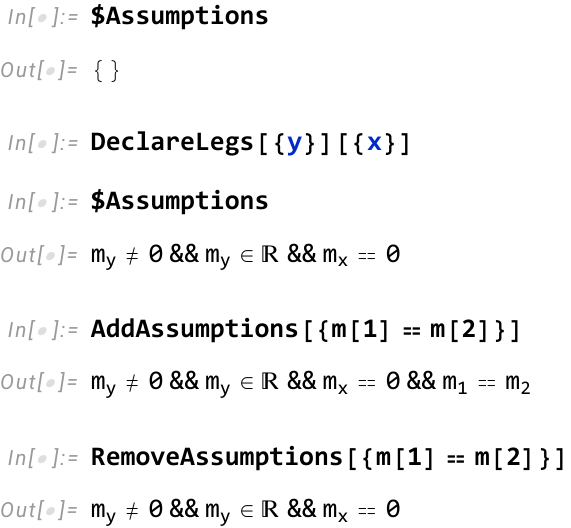}
	}
	
\end{examplebox}

\begin{conceptbox}{}
	It is often required to start with a clean palette without quitting the \tt{Kernel}. To do that, one can use the \tt{ClearAssumptions[]} command. 
\end{conceptbox}

\begin{documentationbox}{\tt{ClearAssumptions} }{\tt{ClearAssumptions[]}}
	
	This command removes the assumptions stored in \tt{\$Assumptions} and also
	clears the massive and massless leg data.
\end{documentationbox}

\begin{examplebox}{\tt{ClearAssumptions} }
	
	\redbox{
		\includegraphics[scale=0.5]{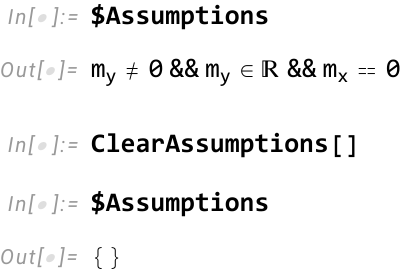}
	}
	
\end{examplebox}

\subsubsection{Massless and massive SH momentum}\label{Mom}

\begin{conceptbox}{}
	
	In the spinor-helicity formalism, a massless momentum is characterized by the
	leg label and a pair of left- and right-handed spinor indices.  For massive
	particles, the left- and right-handed spinors also carry little-group indices,
	which are summed over.
	
	We use a single syntax for both massless and massive momenta:
	\begin{align}
		\tt{Mom[leg\_, undotIndex\_, dotIndex\_]} .
	\end{align}
	This gives the massive momentum if
	\begin{align}
		\tt{MsQ[leg]==True},
	\end{align}
	and gives the massless momentum otherwise.
	
\end{conceptbox}

\begin{documentationbox}{\tt{Mom}}{\tt{Mom[leg\_, undotIndex\_, dotIndex\_]}}
	
	returns the spinor-helicity momentum associated with the specified leg and
	spinor indices. 
	
\end{documentationbox}

\begin{documentationbox}{\tt{DressMom}}{\tt{DressMom[Mom\_, undotIndex\_, dotIndex\_]}}

	distributes spinor indices over a momentum expression.  The first argument may
	also be a linear combination of momentum labels.
	
\end{documentationbox}

\begin{examplebox}{\tt{Mom} and \tt{DressMom}}
	
	\redbox{
		\begin{minipage}{0.5\linewidth}
			\includegraphics[scale=0.5]{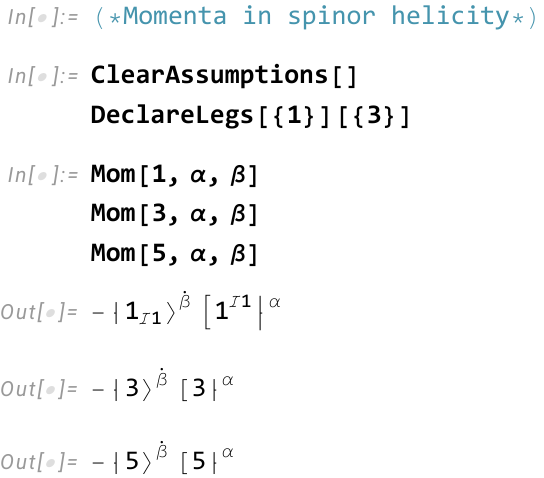}
		\end{minipage}
		\begin{minipage}{0.6\linewidth}
			\includegraphics[scale=0.5]{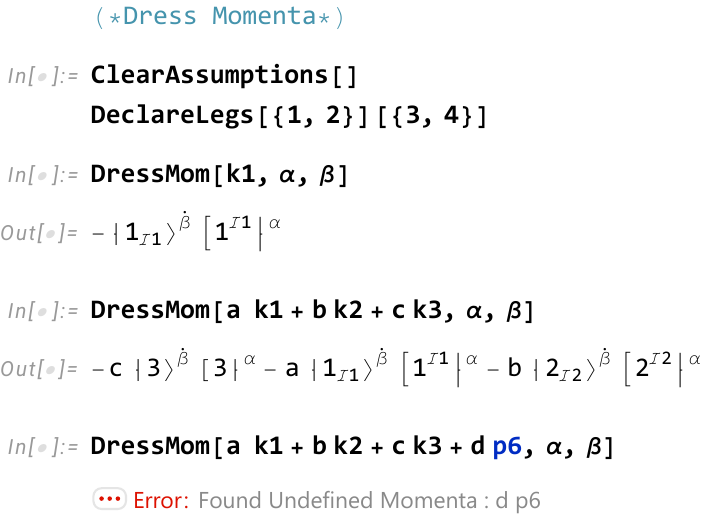}
		\end{minipage}
	}
	
\end{examplebox}

\begin{warningbox}{}

Since the coefficients of the linear combination may be arbitrary, the \tt{DressMom} command
throws an error if any term contains a momentum that has not been declared as
a leg.

\end{warningbox}

\subsubsection{Massive polarizations}\label{MsPol}

\begin{conceptbox}{}
	
	Massive polarizations can be written in spinor-helicity variables.  A massive
	spin-\(1\) polarization transforming as a Lorentz vector is characterized by
	its leg label, one undotted spinor index, one dotted spinor index, and two
	fundamental little-group indices.  The two little-group indices are
	symmetrized.
	
\end{conceptbox}

\begin{documentationbox}{\tt{MsPol}}{	\tt{MsPol[legIndex\_, undotIndex\_, dotIndex\_, LG1Index\_:\ttr{J1},LG2Index\_:\ttr{J2}]} .}

	returns a massive spin-\(1\) polarization.  The default little-group indices
	are	\ttr{J1},	\ttr{J2}, which are also symmetrized by default.
	
\end{documentationbox}

\begin{documentationbox}{\tt{MsPolNonSymm}}{\tt{MsPolNonSymm[legIndex\_, undotIndex\_, dotIndex\_, LG1Index\_, LG2Index\_]}}
	
	The automatic symmetrization of little-group indices may make computations
	unnecessarily long.  In such cases, one can use the non-symmetrized
	polarization.

	One may also use \(\tt{SetSameMsIndices}\) to set all symmetrized
	little-group indices to be the same.
	
\end{documentationbox}

\begin{examplebox}{\tt{MsPol} and \tt{MsPolNonSymm}}
	
	\redbox{
		\begin{minipage}{0.5\linewidth}
			\includegraphics[scale=0.5]{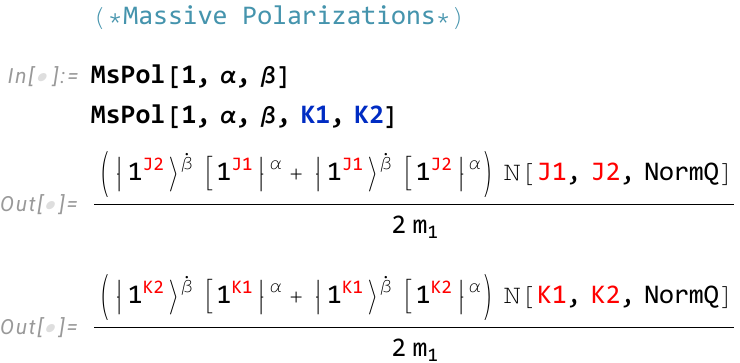}
		\end{minipage}
		\begin{minipage}{0.5\linewidth}
			\includegraphics[scale=0.5]{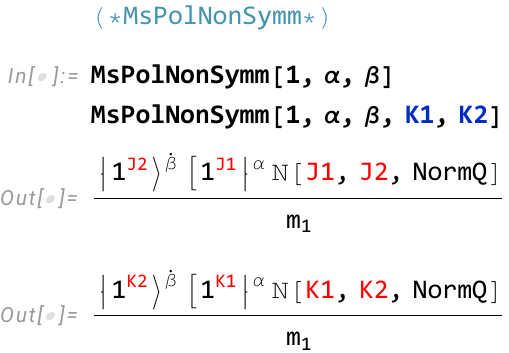}
		\end{minipage}
	}
	
\end{examplebox}

\begin{morematerialbox}{}

	\textit{Little-group index:} Little-group indices appearing on the polarization are colored in red in order to distinguish them from summed little-group indices. A little-group index \(\ttr{Ji}\) has the input form
	\begin{align}
		\tt{MsIndex[J,i]} \quad\quad 
	\end{align}
\textbf{Short Form}:\, \tt{MsI}, \textbf{keyboard Shortcut}: \,	\boxed{\tt{Esc}}\tt{msi}\boxed{\tt{Esc}}.
	
	\textit{Normalization factor}: The default little-group indices are \(\ttr{J1}\) and \(\ttr{J2}\).  The
	factor
	\begin{align}
		\tt{N[\ttr{J1},\ttr{J2},NormQ]}
	\end{align}
	is the normalization factor.  Its value depends on whether \(\tt{NormQ}\) is
	set to \(\tt{True}\) or \(\tt{False}\).  In general, one may set
	\(\tt{NormQ=False}\) after loading the package in order to drop the
	normalization factor.

\end{morematerialbox}

\begin{examplebox}{Normalization factor}
	
	\redbox{\includegraphics[scale=0.5]{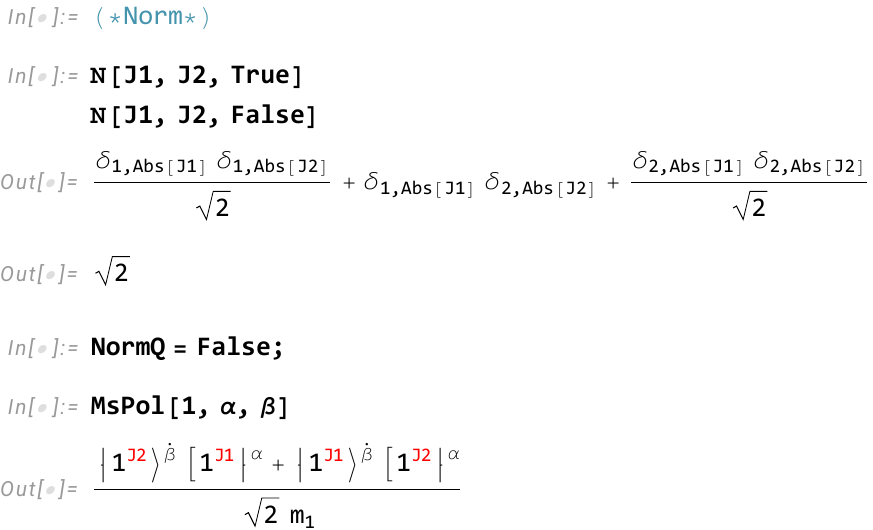}}
	
\end{examplebox}

\begin{documentationbox}{\tt{HsPol}}{\tt{HsPol[S\_, leg\_, PrefixUndot\_, PrefixDot\_, LGIndex\_:\ttr{J1}]}}
	
	Provides  higher integer-spin polarizations. The syntax is the same as \(\tt{MsPol}\), with an additional first argument
	specifying the spin \(\tt{S}\).
	
	Note: This command always puts the same little-group index on the polarization.  This
	is possible {but gives a special case} because the little-group indices are symmetrized.
	
\end{documentationbox}

\begin{examplebox}{\tt{HsPol}}
	
	\redbox{\includegraphics[scale=0.5]{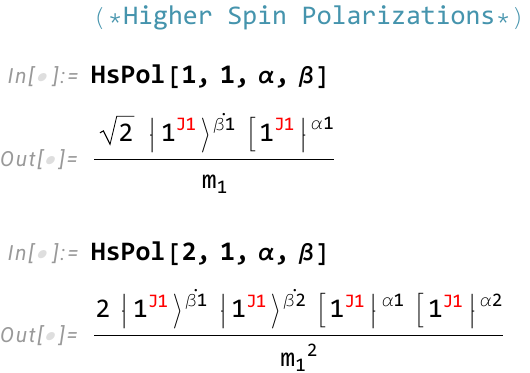}}
	
\end{examplebox}

\subsubsection{Massless polarizations}\label{MlPol}

\begin{conceptbox}{}
	
	Massless polarizations, as discussed in Sec.~\ref{sec:BriefReviewofSH}, are
	characterized by the leg label, helicity, reference spinor, and spinor indices.
	
\end{conceptbox}

\begin{documentationbox}{\tt{MlPol}}{\tt{MlPol[leg\_, undotIndex\_, dotIndex\_, helicity\_, Reference\_]}}
	
	returns the massless polarization for a declared massless leg with helicity and reference spinors specified. The default value of helicity
and reference spinor are taken from the data specified when the leg was
declared.
	
\end{documentationbox}

\begin{examplebox}{\tt{MlPol}}
	
	\redbox{\includegraphics[scale=0.5]{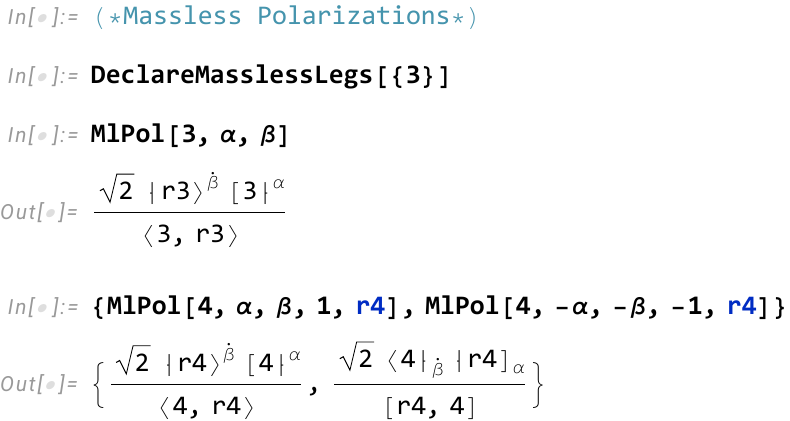}}
	
\end{examplebox}

\begin{examplebox}{Transversality and gauge transformation of \tt{MlPol}}

	The massless polarization satisfies the transversality relation.  For example,
	\begin{align}
		\tt{MlPol[3,1,r3,\(\alpha\),\(\beta\)] Mom[3,-\(\alpha\),-\(\beta\)]}
		&=0,
	\end{align}
	because by default
	\begin{align}
		\tt{SHAA[3,3]=0},
		\qquad
		\tt{SHBB[3,3]=0}.
	\end{align}
	One can also check the gauge transformation of the massless polarization by
	changing the reference spinor, for example from \(\tt{r3}\) to \(\tt{r33}\).
	The difference between the two polarizations is proportional to the
	corresponding momentum.

	\redbox{\includegraphics[scale=0.5]{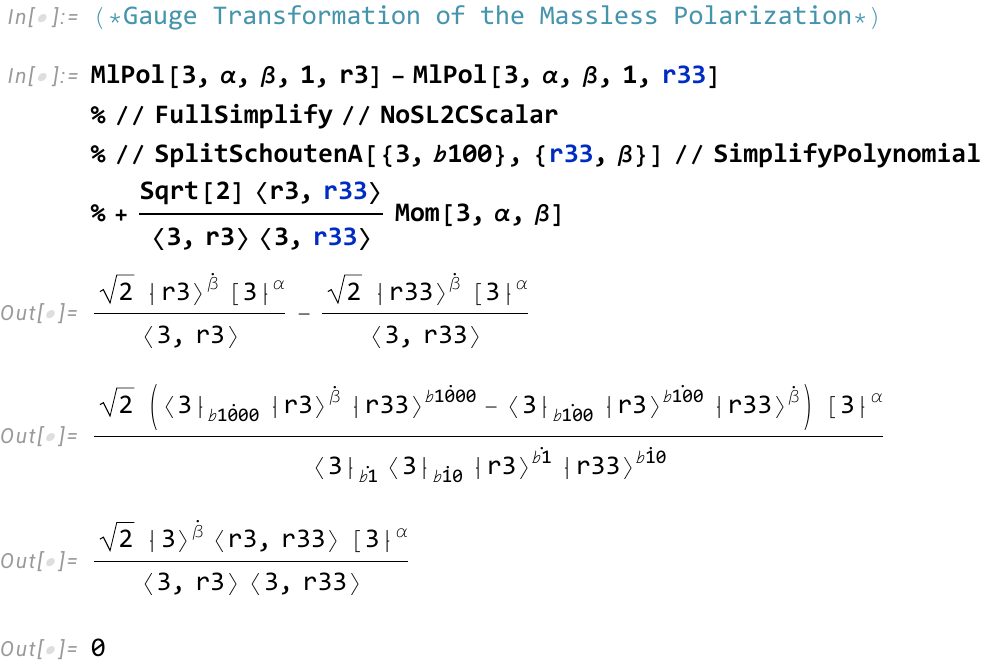}}
	
\end{examplebox}

\subsection{Feynman rules}  
\subsubsection{Three-point amplitudes}
Three-point amplitudes involving at least one massless leg are highly
constrained by three-point kinematics. In spinor-helicity variables, these
constraints are strong enough to fix the allowed structures up to coupling
constants and possible factors of Mandelstam invariants
\cite{Arkani-Hamed:2017jhn,Conde_2016}. 

The situation is different when all three particles are massive. In that case,
three-point kinematics alone does not restrict the amplitude to a finite set of
structures. For this reason, \texttt{SMaSH} provides built-in commands only for
the cases in which at least one of the three legs is massless.

\begin{conceptbox}{Three massless particles}
	For three massless particles with helicities
	\(\mathtt{h_1,h_2,h_3}\), the three-point amplitude is fixed by little-group
	scaling up to an overall factor. It is given by the square-bracket branch if
	\(\mathtt{h_1+h_2+h_3\geq 0}\), and by the angle-bracket branch if
	\(\mathtt{h_1+h_2+h_3<0}\):
	\begin{align}
		\mathtt{h_1+h_2+h_3\geq 0}
		\quad &:\quad
		\mathtt{
			[1\,2]^{h_1+h_2-h_3}
			[2\,3]^{h_2+h_3-h_1}
			[3\,1]^{h_1+h_3-h_2}
		},
		\\
		\mathtt{h_1+h_2+h_3<0}
		\quad &:\quad
		\mathtt{
			\langle 1\,2\rangle^{-h_1-h_2+h_3}
			\langle 2\,3\rangle^{-h_2-h_3+h_1}
			\langle 3\,1\rangle^{-h_1-h_3+h_2}
		}.
	\end{align}
	
\end{conceptbox}

\begin{documentationbox}{\tt{ThreeMasslessAmplitude}}{\tt{ThreeMasslessAmplitude[\{leg1\_,leg2\_,leg3\_\},\{h1\_,h2\_,h3\_\}]}}
	
	returns the three-point massless amplitude fixed by the helicities of the three external legs. 
\end{documentationbox}

\begin{examplebox}{\tt{ThreeMasslessAmplitude}}
	
	\redbox{
		\begin{minipage}{0.5\linewidth}
			\includegraphics[scale=0.5]{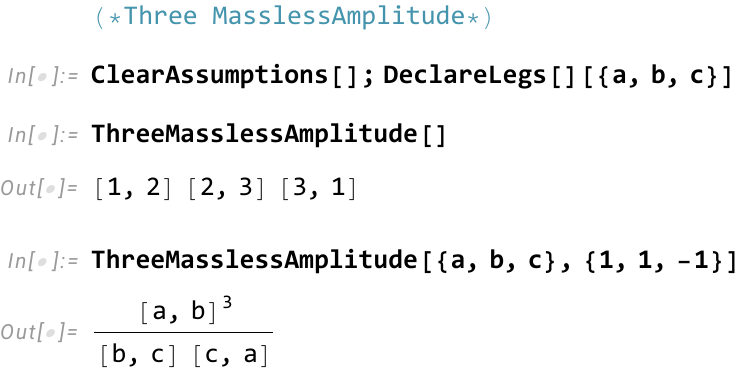}
		\end{minipage}
		\begin{minipage}{0.3\linewidth}
			\includegraphics[scale=0.5]{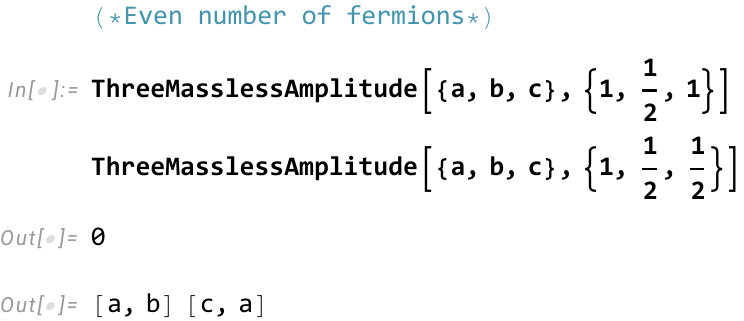}
		\end{minipage}
	}
	
\end{examplebox}
\begin{conceptbox}{Two massless particles and one massive particle}
	
	Consider two massless particles with helicities \(\mathtt{h_1,h_2}\) and one
	massive particle of spin \(\mathtt{S}\). In this case, the three-point
	kinematics fixes a unique structure. With the massive spin indices stripped, the amplitude can be written schematically as
	\begin{align}
		\mathtt{
			\frac{1}{m^{2S+h_1+h_2}}
			\left(
			\langle a|^{S+h_1-h_2}
			\langle b|^{S+h_2-h_1}
			\right)
		}_{\{\dot{\alpha}_1\cdots \dot{\alpha}_{2S}\};
			\{\dot{\beta}_1\cdots \dot{\beta}_{2S}\}}
		\mathtt{
			[a\,b]^{S+h_1+h_2}
		}.
	\end{align}
	
	For equal helicities, \(\mathtt{h_1=h_2}\), the amplitude vanishes for odd
	massive spin because of spin-statistics. For opposite helicities,
	\(\mathtt{h_1=-h_2}\), the amplitude vanishes if
	\begin{align}
		\mathtt{|h_1|>\frac{S}{2}},
	\end{align}
	because the powers of the massive spinors must be non-negative.
	
\end{conceptbox}

\begin{documentationbox}{\tt{TwoMasslessOneMassiveAmplitude}}{\tt{TwoMasslessOneMassiveAmplitude[\{masslessleg1\_,masslessleg2\_,leg3\_\},\{h1\_,h2\_,S\_\}]}}
	
	returns the three-point amplitude involving two massless particles and one massive particle. 
	
	
\end{documentationbox}

\begin{examplebox}{\tt{TwoMasslessOneMassiveAmplitude}} 
	
	\redbox{
		\begin{minipage}{0.5\linewidth}
			\includegraphics[scale=0.5]{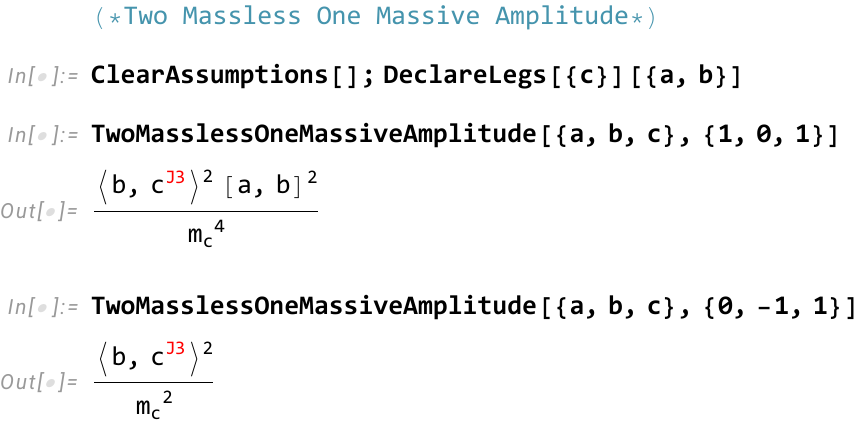}
		\end{minipage}
		\begin{minipage}{0.3\linewidth}
			\includegraphics[scale=0.5]{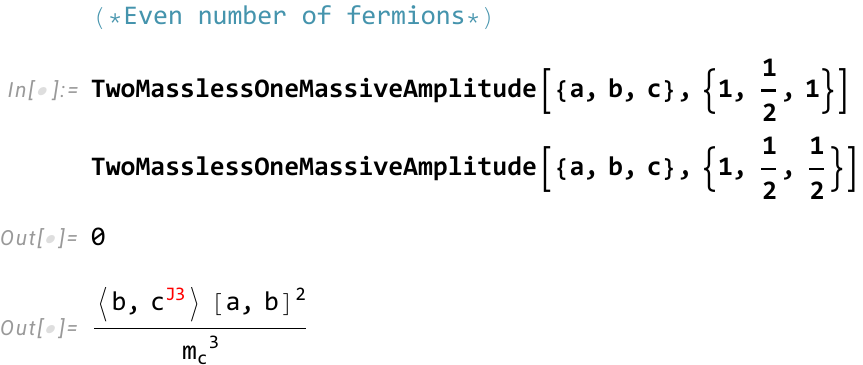}
		\end{minipage}
	}
	
	\redbox{
		\begin{minipage}{0.5\linewidth}
			\includegraphics[scale=0.5]{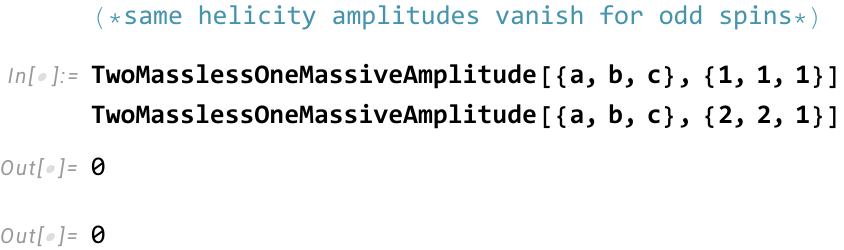}
		\end{minipage}
		\begin{minipage}{0.3\linewidth}
			\includegraphics[scale=0.5]{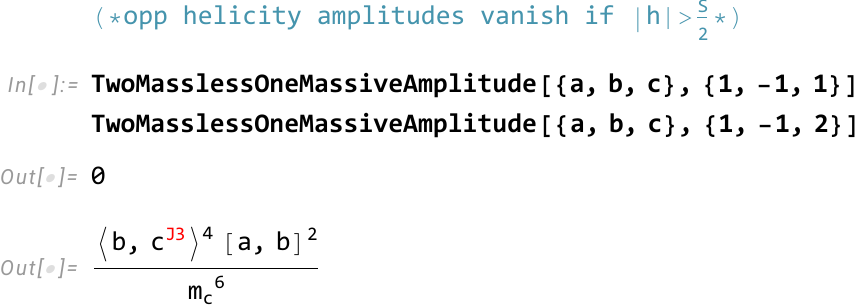}
		\end{minipage}
	}
	
\end{examplebox}

\begin{conceptbox}{Two massive particles and one massless particle}
	
	Unlike the previous two cases, the amplitude is not unique. Instead, it is spanned by a finite number of independent three-point structures. The precise form of the basis depends on whether the two massive particles have equal or unequal masses \cite{Arkani-Hamed:2017jhn}. We have reviewed them for the readers in \ref{subsec:moreonthreepointfunctions}.
	
\end{conceptbox}

\begin{documentationbox}{\tt{TwoMassiveOneMasslessAmplitude}}{\tt{TwoMassiveOneMasslessAmplitude[\{massiveleg1\_,massiveleg2\_,leg3\_\},\{S1\_,S2\_,h\_\}]}}
	
	The command \tt{TwoMassiveOneMasslessAmplitude} returns the most general
	three-point amplitude involving two massive particles and one massless
	particle. 
	
	The command distinguishes between the unequal-mass and equal-mass cases.
	
\end{documentationbox}

\begin{examplebox}{\tt{TwoMassiveOneMasslessAmplitude}}
	
	\redbox{\includegraphics[scale=0.5]{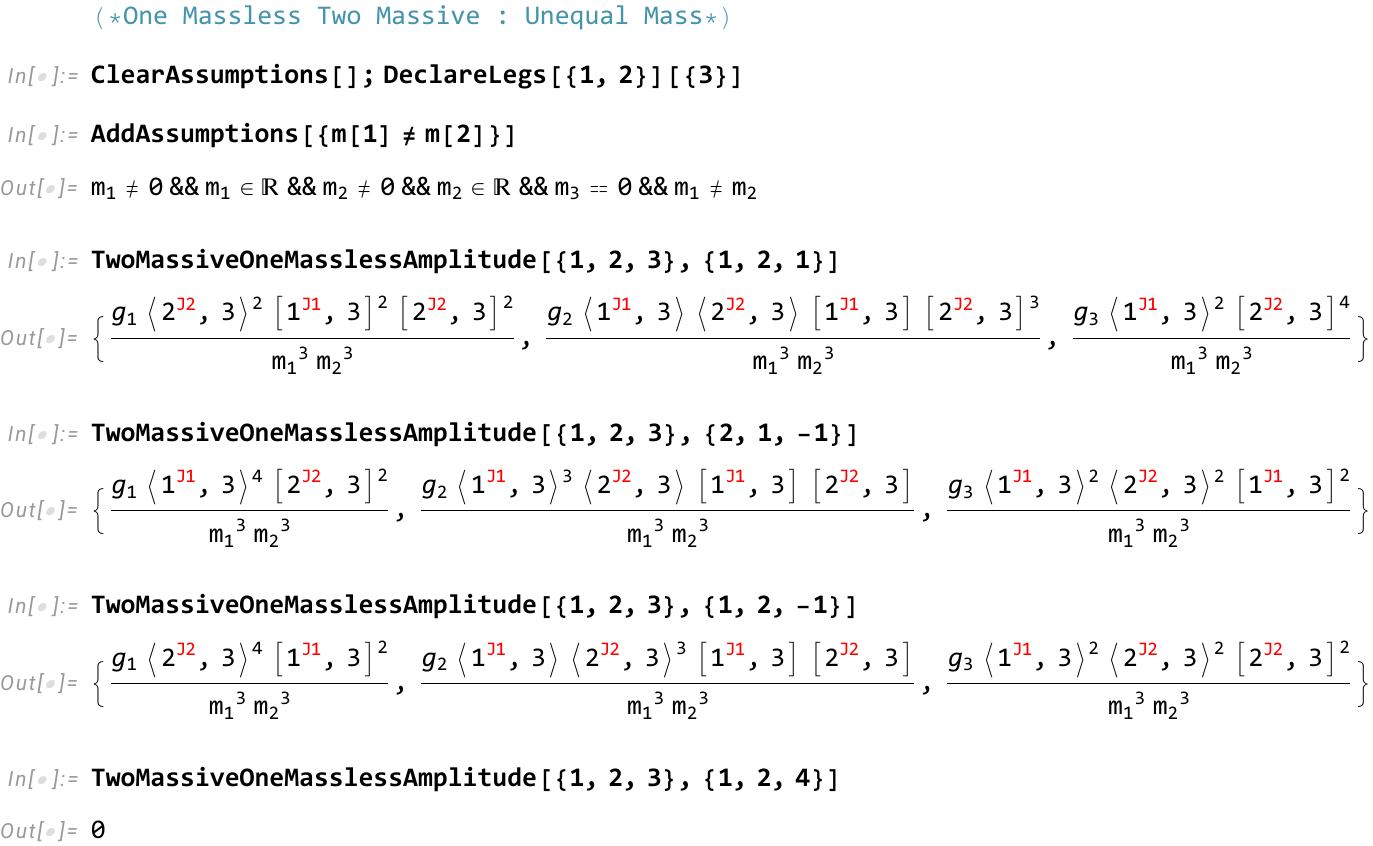}}

	Here the coupling constants can be accessed using the command \tt{Coupling} described in appendix \ref{subsec:moreonthreepointfunctions}
\end{examplebox}

\begin{examplebox}{\tt{TwoMassiveOneMasslessAmplitude}: equal masses}
	
	\redbox{\includegraphics[scale=0.49]{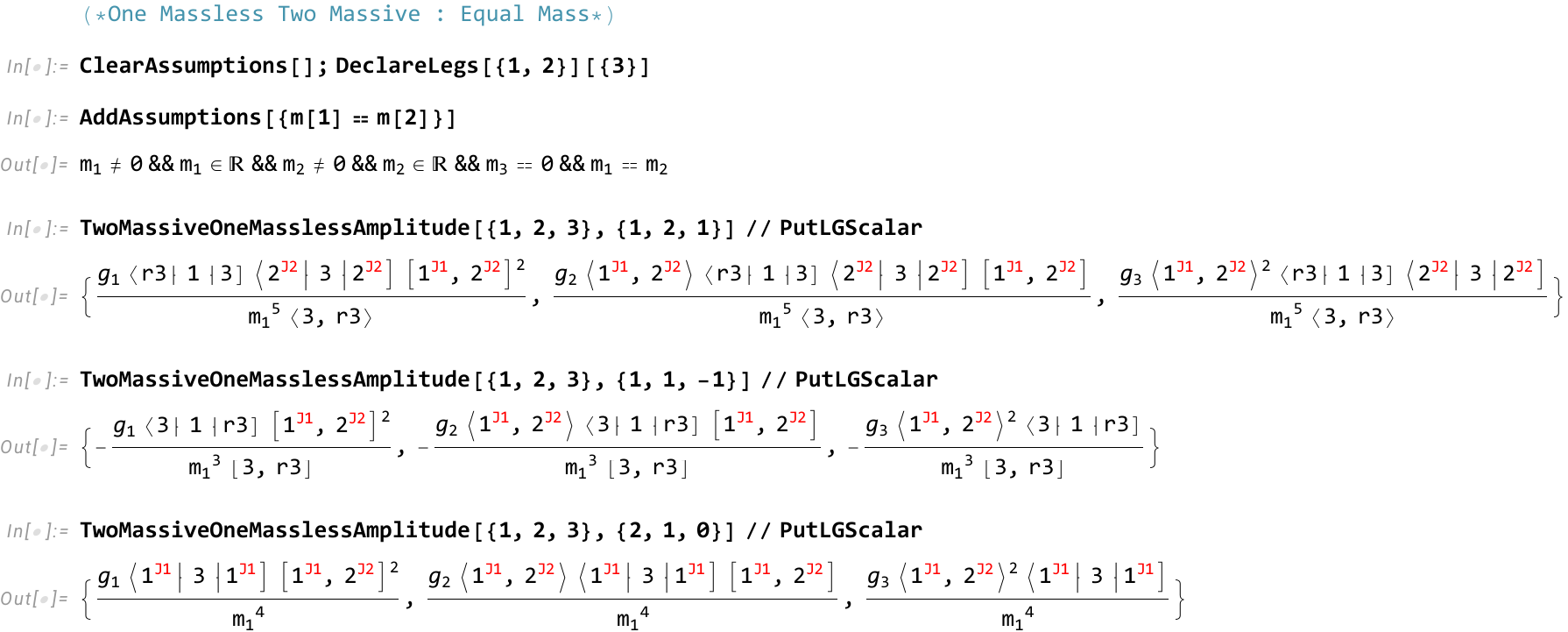}}
	
\end{examplebox}

\subsubsection{Higher-spin propagators}\label{subsubsec:higherspinpropagators}

In this section, we describe the commands for massless and massive
higher-spin propagators.  The higher-spin propagators are constructed from
the spin-\(1\) projector.

\begin{conceptbox}{Spin-\(1\) projector}
	
	The basic spin-\(1\) projector in the \(\mathtt{R_\xi}\) gauge is
	\begin{align}
		\mathtt{\Theta}^{\{\alpha\dot{\alpha}\};\{\beta\dot{\beta}\}}
		\mathtt{(k,\xi)}
		&=
		-2\,\epsilon^{\alpha\beta}\epsilon^{\dot{\alpha}\dot{\beta}}
		+
		\frac{(\xi-1)\,
			\mathtt{k}^{\dot{\alpha}\alpha}
			\mathtt{k}^{\dot{\beta}\beta}}
		{\mathtt{k^2+\xi m^2}},
		\qquad
		\mathtt{k^2=k^\mu k_\mu}.
	\end{align}
	This projector transforms as a rank-\(2\) tensor of
	\(\mathtt{SU(2)_L}\) and a rank-\(2\) tensor of
	\(\mathtt{SU(2)_R}\).
	
	One may also define mixed projectors by trading dotted indices for
	undotted indices, or vice versa, using momentum contractions of the form
	\(\frac{1}{\mathtt m}\mathtt{k}_{\alpha\dot{\alpha}}\).  For example, a
	rank-\(4\) projector of \(\mathtt{SU(2)_L}\) is
	\begin{align}
		\mathtt{\Theta}^{\{\alpha\gamma\};\{\beta\delta\}}
		\mathtt{(k,\xi)}
		&\equiv
		\frac{1}{\mathtt m}
		\mathtt{k}^{\gamma}{}_{\dot{\alpha}}\,
		\mathtt{\Theta}^{\{\alpha\dot{\alpha}\};\{\beta\dot{\beta}\}}
		\mathtt{(k,\xi)}
		\frac{1}{\mathtt m}
		\mathtt{k}^{\delta}{}_{\dot{\beta}} .
	\end{align}
	On shell, \(\mathtt{k^2=-m^2}\), this reduces to
	\begin{align}
		\mathtt{\Theta}^{\{\alpha\gamma\};\{\beta\delta\}}
		\mathtt{(k,\xi)}
		&=
		-2\,\epsilon^{\alpha\beta}\epsilon^{\gamma\delta}.
	\end{align}
	
\end{conceptbox}

\begin{documentationboxWL}{$\mathtt{\Theta}$}{\tt{$\mathtt{\Theta}$[k\_, m\_, $\xi$\_, leftlist\_, rightlist\_]}}
	\label{dcmn:Spin1Projector}	

	returns the spin-\(1\) projector.
	
	The default values of the left and right index lists are
	\begin{equation}
		\mathtt{\{\alpha,\alpha\}},
		\qquad
		\mathtt{\{\beta,\beta\}} .
		\nonumber
	\end{equation}

\end{documentationboxWL}

\begin{examplebox}{$\mathtt{\Theta}$}
	
	\redbox{\includegraphics[scale=0.5]{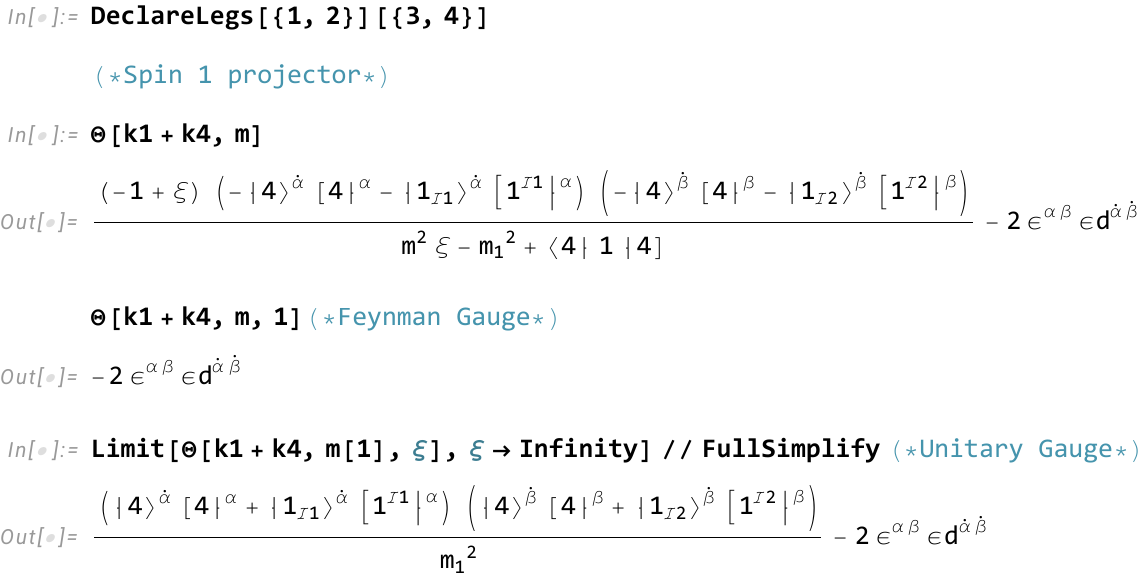}}
	
	\textit{Unitary and Feynman gauge Spin \tt{1} projector}: The spin \tt{1} projector in the unitary and Feynman gauge can also be directly obtained using the commands $$\mathtt{UG\Theta[k\_,m\_,leftlist\_,rightList\_]}\quad\text{and}\quad \mathtt{FG\Theta[k\_,m\_,leftlist\_,rightList\_]}$$ respectively.
\end{examplebox}

\begin{conceptbox}{Spin-\(\tt{S}\) propagator}
	The unitary gauge propagator for the higher spin particles is known from the following works\cite{Ingraham:1974un,Kumar:2025znu,Kumar:2025juz}. For any spin \tt{S}, the propagator is a rank-\(\mathtt{2S}\) tensor
	in both dotted and undotted spinor indices.  In terms of the spin-\(1\)
	projector $\mathtt{UG\Theta}$ at $\xi\rightarrow \infty$, it is given by
	\begin{align}
		\mathtt{\mathcal{P}}^{\{\alpha\dot{\alpha}\};\{\beta\dot{\beta}\}}
		\mathtt{(k)}
		&\equiv
		\frac{-\iimg}{\mathtt{k^2+m^2+\iimg\epsilon}}
		\sum_{a=0}^{\lfloor S/2\rfloor}
		\mathtt{A(S,a)}
		\bigg[
		\mathtt{UG\Theta}^{\alpha_{\tt 1}\dot{\alpha}_{\tt 1},
			\alpha_{\tt 2}\dot{\alpha}_{\tt 2}}
		\cdots
		\mathtt{UG\Theta}^{\alpha_{\tt{2a-1}}\dot{\alpha}_{\tt{2a-1}},
			\alpha_{\tt{2a}}\dot{\alpha}_{\tt{2a}}}
		\nonumber\\
		&\hspace{2.8cm}\times
		\mathtt{UG\Theta}^{\beta_{\tt 1}\dot{\beta}_{\tt 1},
			\beta_{\tt 2}\dot{\beta}_{\tt 2}}
		\cdots
		\mathtt{UG\Theta}^{\beta_{\tt{2a-1}}\dot{\beta}_{\tt{2a-1}},
			\beta_{\tt{2a}}\dot{\beta}_{\tt{2a}}}
		\nonumber\\
		&\hspace{2.8cm}\times
		\mathtt{UG\Theta}^{\alpha_{\tt{2a+1}}\dot{\alpha}_{\tt{2a+1}},
			\beta_{\tt{2a+1}}\dot{\beta}_{\tt{2a+1}}}
		\cdots
		\mathtt{UG\Theta}^{\alpha_{\tt{2S}}\dot{\alpha}_{\tt{2S}},
			\beta_{\tt{2S}}\dot{\beta}_{\tt{2S}}}
		\bigg]_{\tt{sym}(\alpha,\dot{\alpha},\beta,\dot{\beta})}.
	\end{align}\label{UGPropagator}
	The coefficients \(\mathtt{A(S,a)}\) are evaluated at
	\(\mathtt{d=4}\).  More generally, they are given by
	\begin{align}
		\mathtt{A(S,a,d)}
		&=
		\mathtt{
			\frac{
				\left(-\frac{1}{2}\right)^a S!
				(-2a+d+2S-5)\text{!!}
			}{
				a!(S-2a)!(d+2S-5)\text{!!}
			}
		}.
	\end{align}
	
	Although the higher-spin analogue of the \(\mathtt{R_\xi}\) gauge is not
	known in full generality, this expression reproduces several standard
	gauge choices.  In particular, it gives the unitary-gauge propagator in
	the limit \(\mathtt{\xi\rightarrow\infty}\), and the Feynman-gauge
	propagator for \(\mathtt{\xi=1}\).
	
\end{conceptbox}

\begin{documentationbox}{\tt{PropCoeff}}{\tt{PropCoeff[S\_, a\_, d\_]}}
	
	The coefficient \(\mathtt{A(S,a,d)}\) appearing in the spin-\(\mathtt S\)
	propagator in 3+1 dimensions. 
	
\end{documentationbox}

\begin{documentationbox}{\tt{UGPropagator}}{\tt{UGPropagator[S\_, k\_, m\_, LeftPrefix\_, RightPrefix\_,OptionsPattern[]]}} 
	
	returns the spin-\(\mathtt S\) propagator in unitary gauge where the input variables have the following:

	\paragraph{Arguments:}
	\begin{itemize}
		\item \(\tt{S}\) is the spin.  Its default value is \(\mathtt{1}\).
		\item \(\tt{k}\) is the momentum.  Its default value is \(\tt{k1}\).
		This argument may also be a linear combination of momenta.
		\item \(\tt{m}\) is the mass appearing in the pole.  Its default value
		is \(\mathtt{m_1}\).
		\item \(\tt{LeftPrefix}\) and \(\tt{RightPrefix}\) are the lists of
		prefixes for the left and right spinor indices.  Their default values
		are
		\begin{align}
			\mathtt{\{\alpha,\alpha\}},
			\qquad
			\mathtt{\{\beta,\beta\}} .
		\end{align}
	\end{itemize}
	
	\paragraph{Options}
	\begin{itemize}
\item \(\tt{LeftSymm}\) and \(\tt{RightSymm}\) specify whether the
corresponding spinor indices should be symmetrized. The indices are
symmetrized when these arguments are set to \(\tt{True}\). Since the propagator indices are usually contracted with already
symmetrized tensors, it is not necessary to symmetrize them beforehand. The default value for both these options is \tt{False}.
	\end{itemize}

	\paragraph{Related command:}
	The spin-\(\mathtt S\) projector has the same syntax, with the head \tt{UGPropagator} replaced by
	\(\tt{UGProjector}\).
	
\end{documentationbox}

\begin{examplebox}{\tt{UGPropagator}}
	
	\redbox{\includegraphics[scale=0.42]{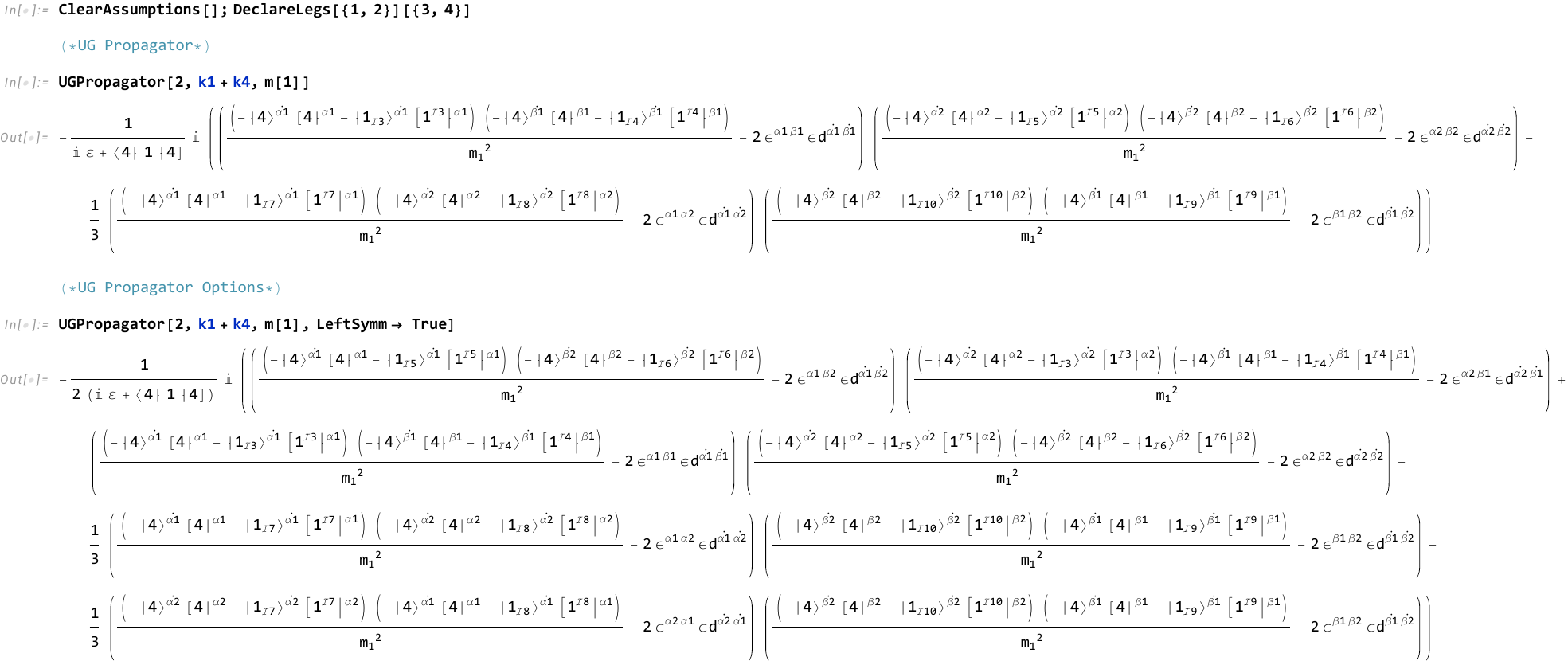}}
	
\end{examplebox}
\begin{morematerialbox}
	In \cite{Kumar:2025juz, Kumar:2025znu}, the authors ad-hocly defined a general $\xi$ propagator for any integer spin \tt{S} whose $\xi\rightarrow \infty$ gives the unitary-gauge propagator \tt{UGPropagator}. We have given commands for that propagator in appendix \ref{subsec:moreonpropagator} where we also discuss the Feynman gauge analogue for higher spins.
	
\end{morematerialbox}

\subsection{Computing Feynman Diagrams}
\subsubsection{Strip off polarizations}

\begin{conceptbox}{}
	
	The \(\tt{Differentiate}\) command can be used to strip off massive or
	massless polarization factors from an amplitude.  For higher-spin massive
	polarizations, we also provide dedicated commands that strip off the
	polarization tensors carrying explicit \(\tt{MsIndex}\) indices.
	
\end{conceptbox}

\begin{documentationbox}{\tt{StripPol}}{\tt{StripPol[leg\_, SpinIndexB\_, SpinIndexA\_]}}

	strips off the massive polarization associated with the specified leg.  The
	first argument is the leg label.  The labels \(\tt{A}\) and \(\tt{B}\) denote
	the prefixes for the dotted and undotted spinor indices, whose default values
	are \(\beta\) and \(\alpha\), respectively.
	
\end{documentationbox}

\begin{examplebox}{\tt{StripPol}}
	
	\redbox{
		\begin{minipage}{0.5\linewidth}
			\includegraphics[scale=0.5]{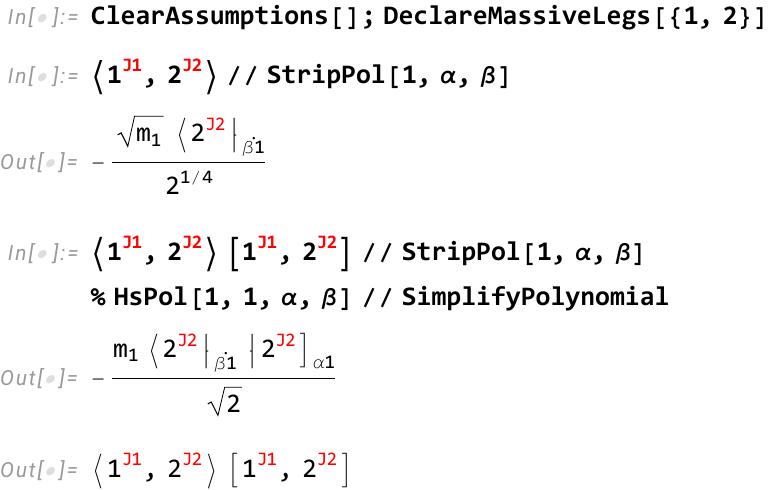}
		\end{minipage}
		\begin{minipage}{0.5\linewidth}
			\includegraphics[scale=0.5]{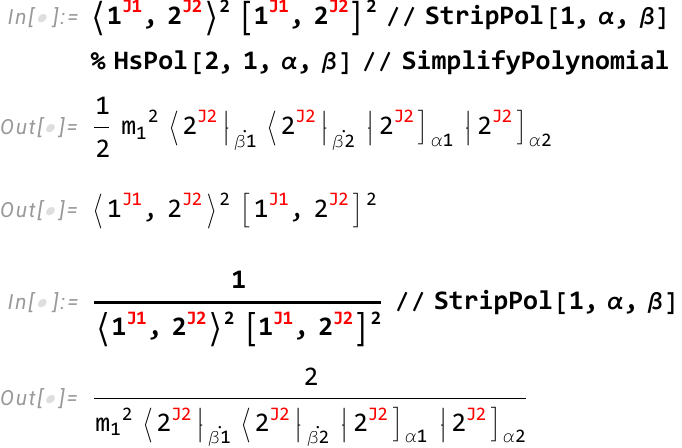}
		\end{minipage}
	}
	
\end{examplebox}

\subsubsection{Declaring momentum conservation}

\begin{conceptbox}{}
	
	To compute scattering amplitudes, one must also impose momentum conservation. Momentum conservation allows us to put constraints on the Mandelstam variables
	\(\mathtt{\textit{s}_{ij}}\), and hence to simplify expressions substantially.
\end{conceptbox}

\begin{documentationbox}{\tt{DeclareMomentumConservation}}{\tt{DeclareMomentumConservation[\(\cdots\)]}} 
	
	declares the momentum-conservation identity for the external momenta.
		
	The input is the sum of external momenta.  For example, for a three-point
	amplitude with all momenta incoming, one should use
	\begin{align}
		\tt{DeclareMomentumConservation[k1+k2+k3]} .
	\end{align}
	If a momentum is outgoing, it appears with a minus sign.  For example, if
	\(\mathtt{k_3}\) is outgoing, one should use
	\begin{align}
		\tt{DeclareMomentumConservation[k1+k2-k3]} .
	\end{align}
	
\end{documentationbox}

\begin{morematerialbox}
	
	When \(\tt{DeclareMomentumConservation}\) is evaluated, these rules are automatically generated 
	\begin{enumerate}
		\item \tt{momconsRule} : returns a replacement rule that can be used to replace the leading momentum in terms of the remaining momenta. 	\begin{align}
			\tt{k1+k2+k3=0}
			\qquad\Longrightarrow\qquad
			\tt{k1 -> -k2-k3}.
		\end{align}
		\item \tt{MandelstamRules}: returns the rules that express dependent Mandelstam variables in terms of the
		independent Mandelstam variables
	\end{enumerate}
	The output of \(\tt{DeclareMomentumConservation}\) is the list of independent
	Mandelstam variables.  For a three-point amplitude there are no free
	Mandelstam variables, and hence the output is
	\begin{align}
		\tt{\{\}} .
	\end{align}
	\tt{ClearMomentumConservation} can be used to clear the momentum-conservation rules. For sake of brevity, we have put those documentation in \ref{subsec:moreondeclaringmomconv}
	
\end{morematerialbox}
\begin{warningbox}{}

If an undefined momentum is used inside \(\tt{DeclareMomentumConservation}\), the evaluation throws an error.  This is because momentum conservation can only be declared for momenta that have already been introduced using \(\tt{DeclareLegs[]}\).

\end{warningbox}

\begin{morematerialbox}
	
	By default, \(\tt{DeclareMomentumConservation}\) chooses the independent
	Mandelstam variables automatically using Mathematica's
	\href{https://reference.wolfram.com/language/ref/Solve.html}{\tt{Solve}}.

	\tt{SMaSH} also allows to specify a preferred list of independent Mandelstam variables. See \ref{subsec:moreondeclaringmomconv} for documentation and examples. 
	
\end{morematerialbox}

\subsection{Consistency Checks and contact-term completion}
\subsubsection{Gauge invariance}

\begin{conceptbox}{Gauge transformations}
	
	One important aspect of amplitudes is their behaviour under gauge
	transformations.  For massless particles, gauge transformations are controlled
	by changes of reference spinors.  Recall that declaring massless legs assigns a
	unique reference spinor to each leg.  We can use this to impose gauge
	transformations on any expression.
	
	If one changes the reference spinor from \(\mathtt{r_i}\) to \(\mathtt{r_{ii}}\) in the expression then it amounts to performing a gauge transformation. 
\end{conceptbox}

\begin{documentationbox}{\tt{PureGaugeTransformation}}{\tt{PureGaugeTransformation[i\_][expr\_]}}

	performs a reference-spinor shift for the massless leg \(\tt{i}\).  
	
	can be used to test the gauge variation of spinor-helicity expressions involving massless external particles.
	
\end{documentationbox}

\begin{examplebox}{\tt{PureGaugeTransformation}}
	
	\redbox{
		\includegraphics[scale=0.45]{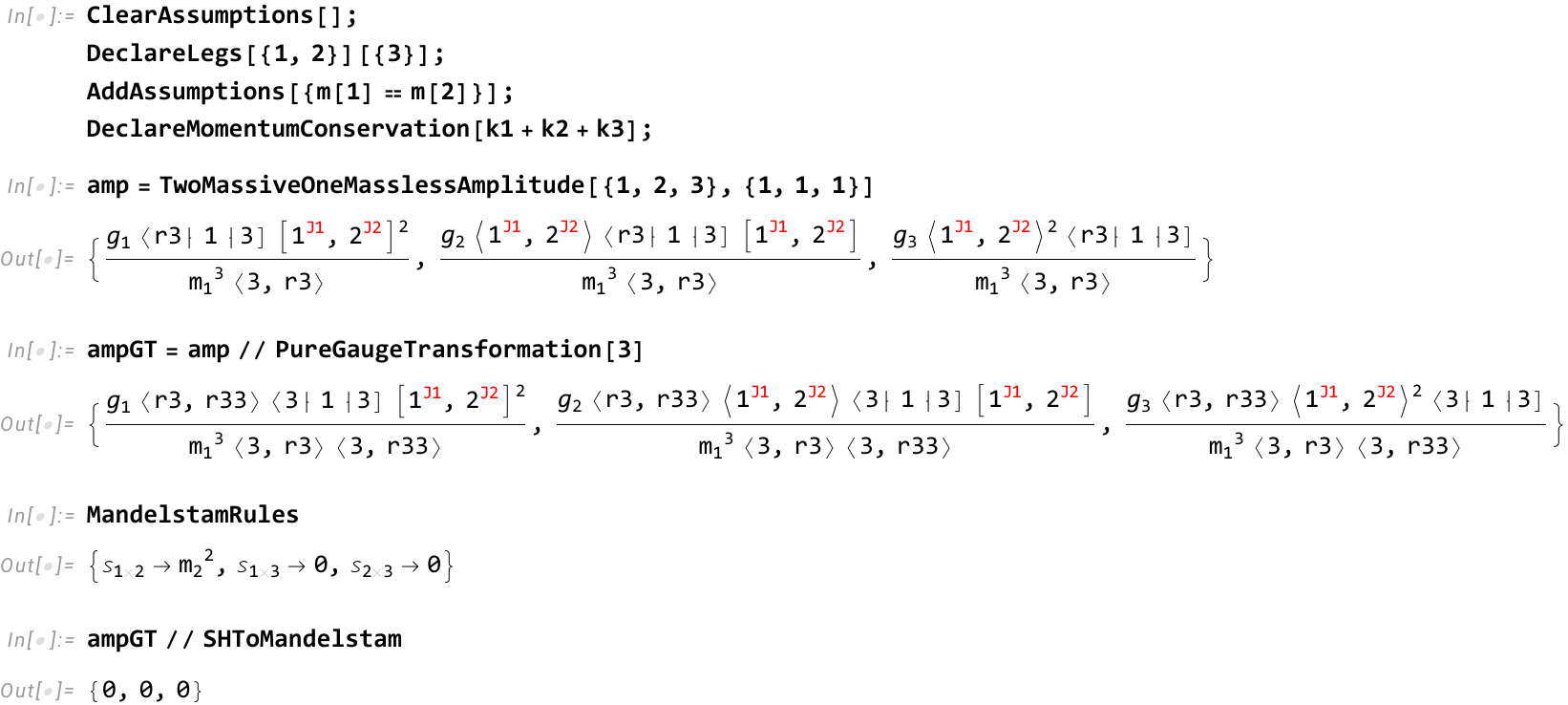}
	}
	
\end{examplebox}

\begin{documentationbox}{\tt{GIQ}}{\tt{GIQ[leg\_][expr\_]}}
	
	checks whether an expression is gauge invariant with respect to the massless
	leg \(\tt{leg}\) and returns True/False. 
	
\end{documentationbox}

\begin{examplebox}{\tt{GIQ}}
	
	\redbox{
		\includegraphics[scale=0.45]{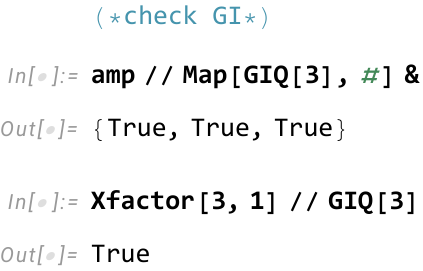}
	}
	
\end{examplebox}

\begin{conceptbox}{\tt{Manifestly gauge-invariant form}}

	An expression written in terms of the field strength is gauge invariant by
	construction, because the field strength is antisymmetric. Therefore, checking
	the gauge invariance of such expressions does not require using any on-shell
	relations. In the spinor-helicity language, gauge invariance should be reflected
	in the absence of dependence on the reference spinor. Thus, whenever an
	expression is gauge invariant, it should be possible to rewrite it in a form in
	which the reference spinor drops out completely. We call such a representation
	a \tt{manifestly gauge-invariant form}.

\end{conceptbox}

\begin{documentationbox}{\tt{ManifestGI}}{\tt{ManifestGI[basistransf\_][expr\_]}}
	
	If an expression is independent of reference spinors, then \tt{ManifestGI}  puts it to a manifestly gauge-invariant form. 
	
\end{documentationbox}

\begin{examplebox}{\tt{ManifestGI}}
	
	\redbox{
		\includegraphics[scale=0.435]{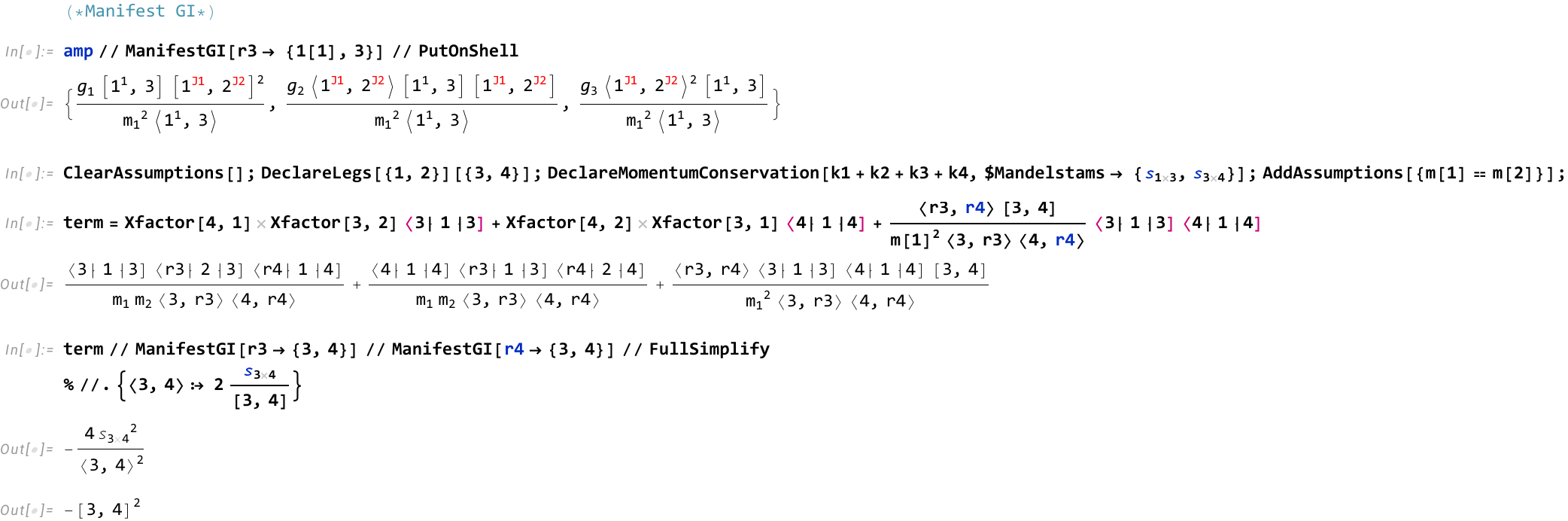}
	}
	
\end{examplebox}

\begin{conceptbox}{Dependence on momentum-conservation convention}
	
	The command \(\tt{ManifestGI}\) doesn't depend on the convention used for momentum conservation, as declared by \(\tt{DeclareMomentumConservation}\).
	For example, if the momentum-conservation relation is
	\begin{align}
		\mathtt{k_1+k_4-k_2-k_3=0}
	\end{align}
	instead of the all-incoming convention
	\begin{align}
		\mathtt{k_1+k_2+k_3+k_4=0},
	\end{align}
	the command can still simplify expressions involving reference spinors to give the correct answer.
	
\end{conceptbox}

\begin{examplebox}{Changing the momentum-conservation convention}
	
	\redbox{
		\includegraphics[scale=0.45]{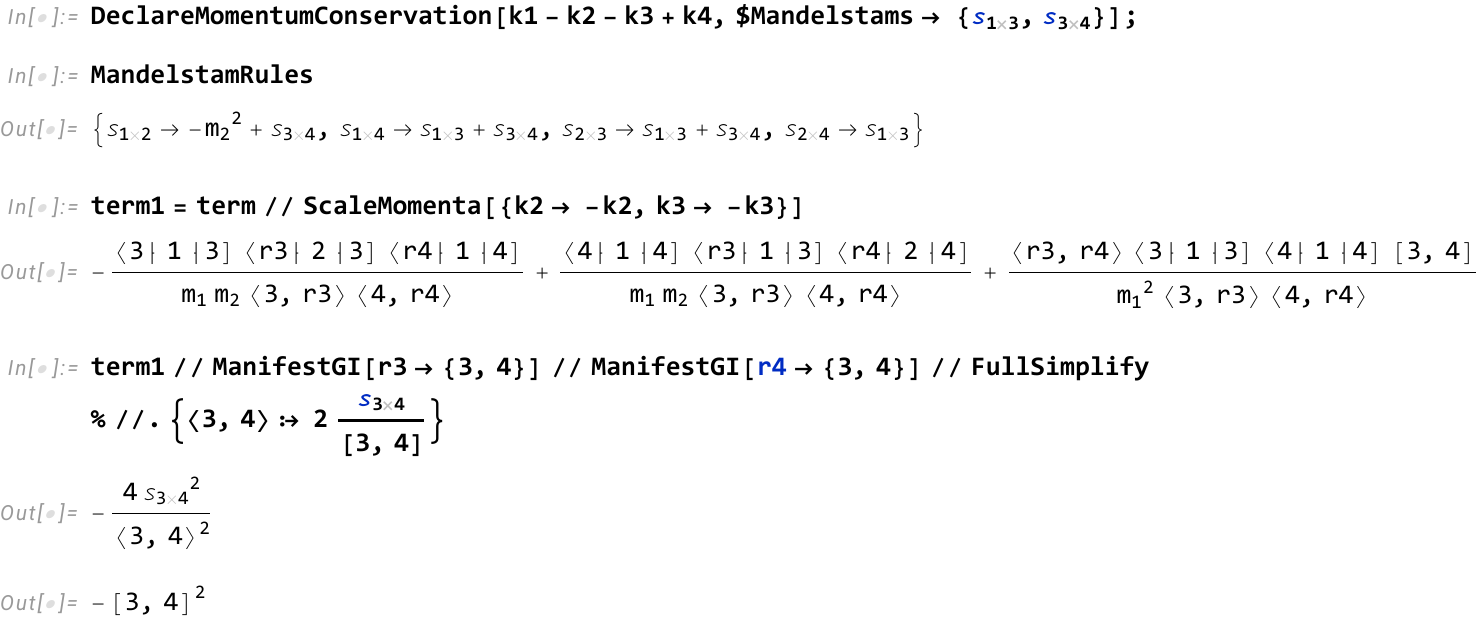}
	}
	
\end{examplebox}

\subsubsection{Contact-term completion}
\begin{documentationbox}{\tt{FindContactTerm}}{\tt{FindContactTerm[leg\_][GaugeTransfExpr\_]}}
	
	The exchange amplitude may not be gauge invariant by itself.  Therefore, a contact
	term is required whose gauge transformation cancels that of the exchange
	amplitude.
	
	The algorithm for finding such contact terms was explained in the
	spinor-helicity formalism in Ref.~\cite{Kumar:2025znu}, and in Lorentz-vector
	form in Ref.~\cite{Kumar:2025juz}.  The package implements this algorithm
	through
	\begin{align}
		\tt{FindContactTerm[leg\_][GaugeTransfExpr\_]}.
	\end{align}
	The first argument is the massless leg label, and the second argument is the
	gauge-transformed expression.
	
\end{documentationbox}

\begin{examplebox}{Finding the contact term}
	
	\redbox{
		\includegraphics[scale=0.45]{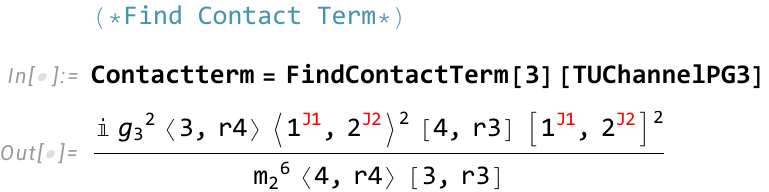}
	}
	
\end{examplebox}

\subsection{A worked out example: Compton amplitude}

\begin{conceptbox}{}
	
	We now demonstrate the computation of a scattering amplitude using the example
	of \(2\to2\) Compton scattering of two spin-\(\mathtt{1}\) particles with
	photons of opposite helicity.  We work in unitary gauge.
	
	The relevant exchange diagrams are the \(t\)- and \(u\)-channel diagrams shown
	below. The \textcolor{red}{red} lines denote massive particles, while the
	\textcolor{violet}{violet} lines denote massless photons.  All momenta are
	taken to be incoming.
	
\end{conceptbox}

\begin{figure}[ht!]
	\begin{center}
		\begin{tikzpicture}[line width=1.5 pt, scale=1]
			
			\begin{scope}[shift={(0,0)}]	
				
				\treefourpointexchange{chspin}{photon}{photon}{chspinbar}{chspin}{0}	
				
				\labeltreefourpointexchange{ $k_2$ }{ $k_3$ }{ $k_4$ }{ $k_1$ }{}{0} 
				
				\draw (0,-2.5) node {$t$-channel};
			\end{scope}

			\begin{scope}[shift={(7,0)}]	
				
				\treefourpointcrossexchange{chspinbar}{photon}{photon}{chspin}{chspinbar}{0}	
				
				\labeltreefourpointexchange{ $k_2$ }{ $k_3$ }{ $k_4$ }{ $k_1$ }{}{0} 
				\draw (0,-2.5) node {$u$-channel};
				
			\end{scope}

		\end{tikzpicture} 
	\end{center}
	\caption{Higher-spin Compton scattering}
	\label{fig:higherspincomptondiag}
\end{figure}
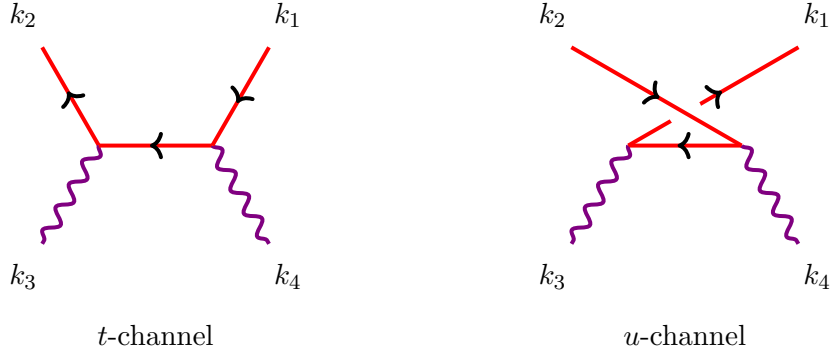

\begin{enumerate}
	\item  {\bf Compton set-up}
	
	We first declare the external legs using
	\begin{align}
		\tt{DeclareLegs[\{1,2\}][\{3,4\}]}.
	\end{align}
	We then declare momentum conservation using
	\begin{align}
		\tt{DeclareMomentumConservation[k1+k2+k3+k4]}.
	\end{align}
	We also impose the equal-mass condition for the massive legs:
	\begin{align}
		\tt{AddAssumptions[\{m[1] == m[2]\}]}.
	\end{align}
	Finally, we choose the independent Mandelstam variables to be the poles in the
	\(t\)- and \(u\)-channels:
	\begin{align}
		\mathtt{\textit{s}_{14}},
		\qquad
		\mathtt{\textit{s}_{13}}.
	\end{align}

	\begin{examplebox}{Compton setup}
		
		\redbox{
			\includegraphics[scale=0.45]{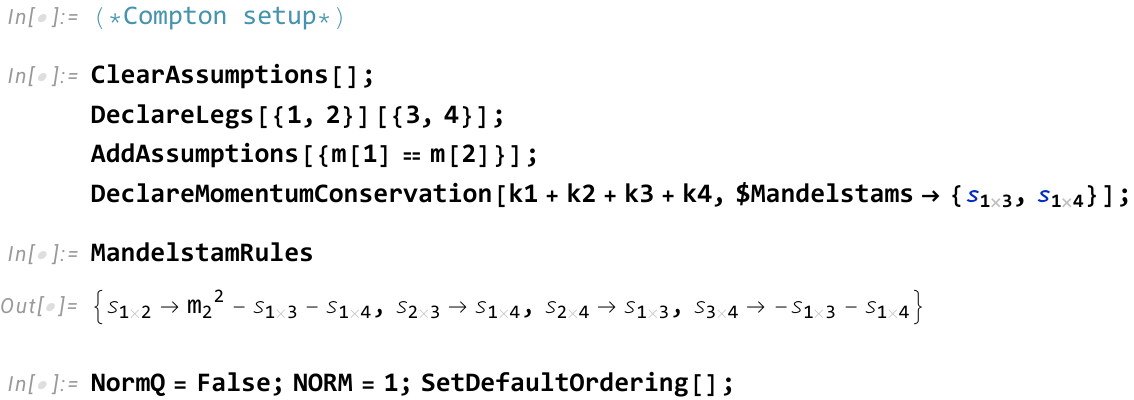}
		}
		
	\end{examplebox}

	\item  {\bf Stripped three-point vertices}
	
	We compute the \(t\)-channel explicitly.  The two stripped vertices in the
	\(t\)-channel are obtained using \(\tt{TwoMassiveOneMasslessAmplitude}\) and \(\tt{StripPol}\).
	
	We choose the Weinberg minimal coupling, which contains the same number of
	angle and square brackets:
	\begin{align}
		\mathtt{\langle 1^{\ttr{J1}}\,2^{\ttr{J2}}\rangle},
		\qquad
		\mathtt{[1^{\ttr{J1}}\,2^{\ttr{J2}}]}.
	\end{align}

	\begin{examplebox}{Stripped vertices}
		
		\redbox{
			\includegraphics[scale=0.45]{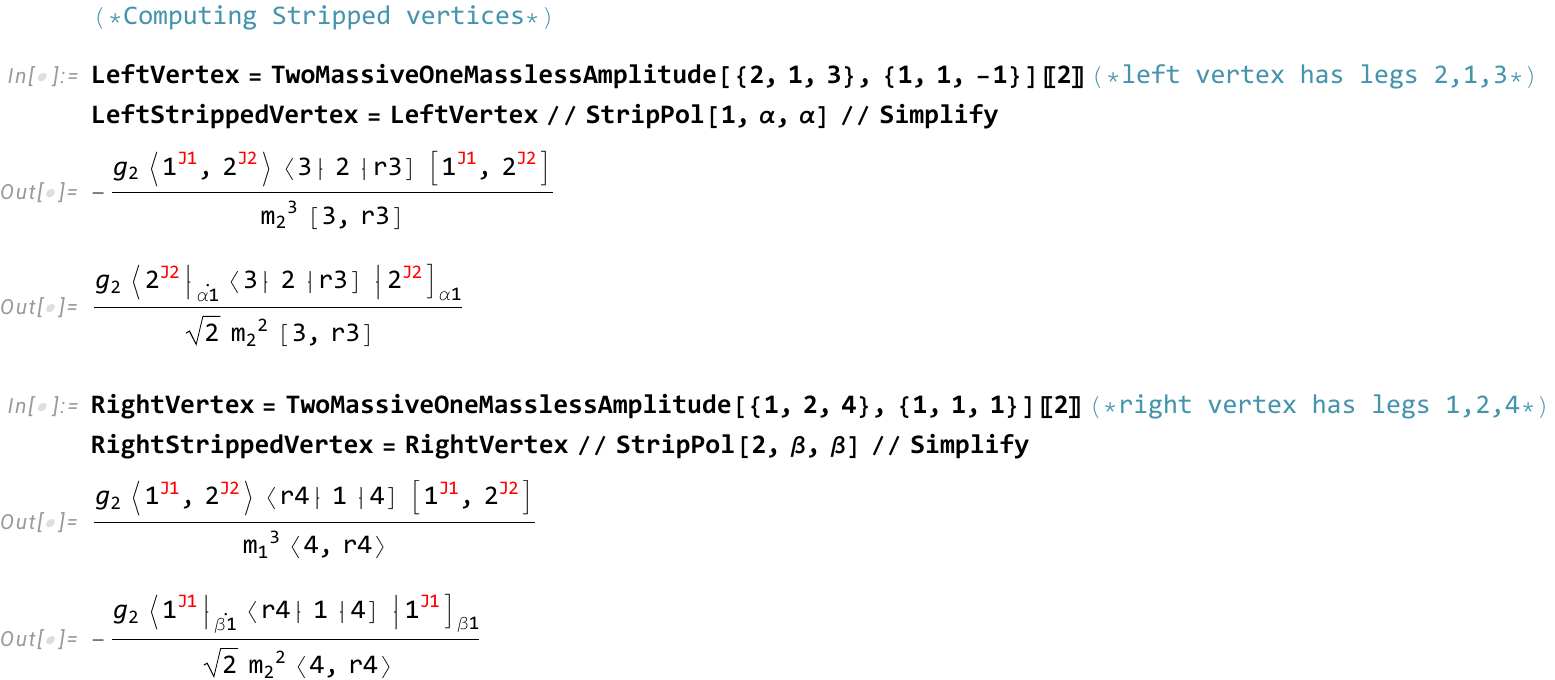}
		}
		
	\end{examplebox}
	
	\item  {\bf Exchange amplitude}
	
	The \(t\)-channel amplitude is obtained by sandwiching the unitary gauge
	propagator between the left and right stripped vertices. The \(u\)-channel amplitude is obtained from the \(t\)-channel amplitude by the
	exchange $\mathtt{1\leftrightarrow 2}$.
	The full exchange amplitude is the sum of the \(t\)- and \(u\)-channel
	amplitudes.

	\begin{examplebox}{\(t\)- and \(u\)-channel exchange amplitudes}
		
		\redbox{
			\includegraphics[scale=0.45]{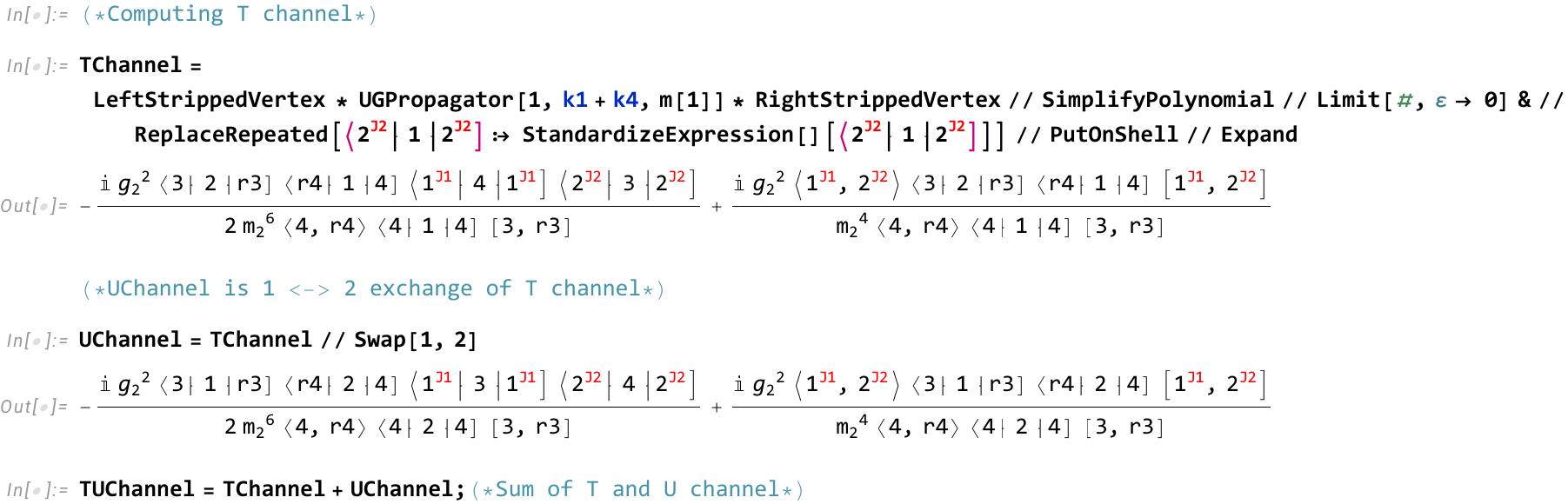}
		}
		
	\end{examplebox}

	\item  {\bf Gauge-invariance check for leg \tt{3}}
	
	Next, we check gauge invariance with respect to the massless leg
	\(\tt{3}\). We often use the command 
	\begin{align}
		\tt{StandardizeExpression[momconsRule\_]}
	\end{align}
	puts an expression into a standard form by replacing a dependent momentum.  The
	default value of the argument is the momentum-conservation rule
	\(\tt{momconsRule}\).

	\begin{examplebox}{Gauge variation of the exchange amplitude for leg \tt{3}}
		
		\redbox{
			\includegraphics[scale=0.35]{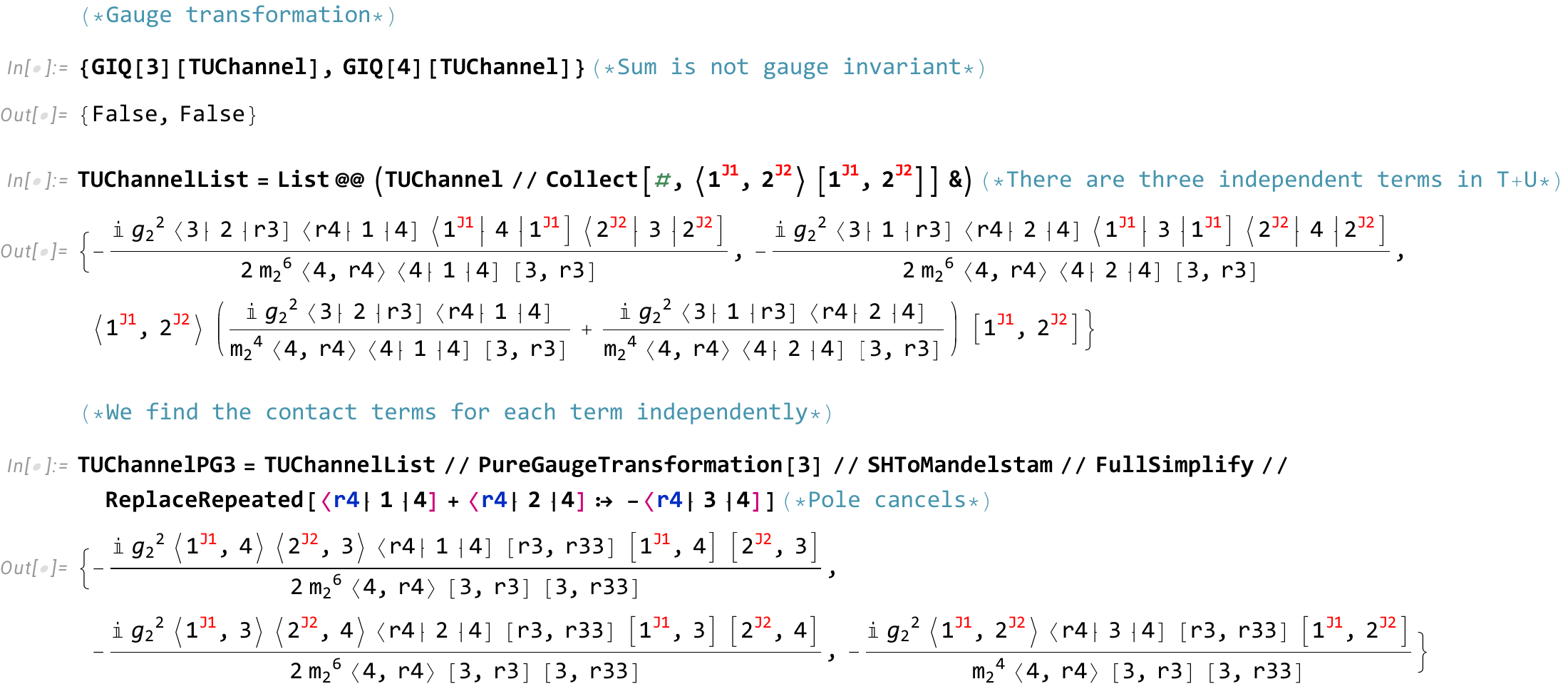}
		}
		
	\end{examplebox}
	
	\item  {\bf Gauge-invariance under leg \tt{3}}
	
	We see that the gauge transformed amplitude \tt{TUChannelPG3}  is non-zero and more importantly it does not have any poles. So the gauge invariance can be restored by introducing extra contact terms\cite{Kumar:2025znu,Kumar:2025juz}.
	
		\begin{examplebox}{Contact term for leg \tt{3}}
		
		\redbox{
			\includegraphics[scale=0.35]{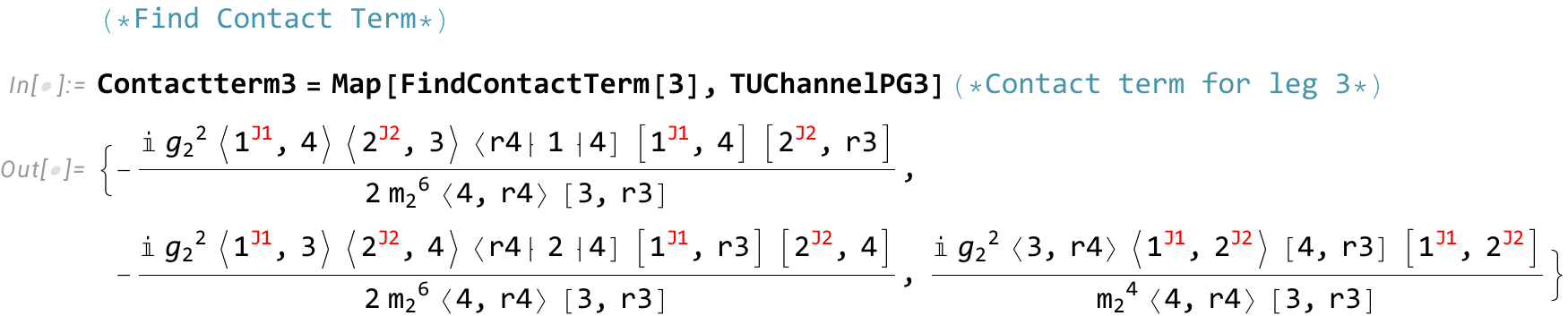}
		}
		
	\end{examplebox}

	Subtracting the contact term \tt{ContactTerm3} from the exchange amplitude gives an
	amplitude which is gauge invariant under leg \tt{3}. The result can then be rewritten in a form which is manifestly independent of \tt{r3}.
	
	\begin{examplebox}{Gauge-invariance under leg \tt{3}}
		
		\redbox{
			\includegraphics[scale=0.35]{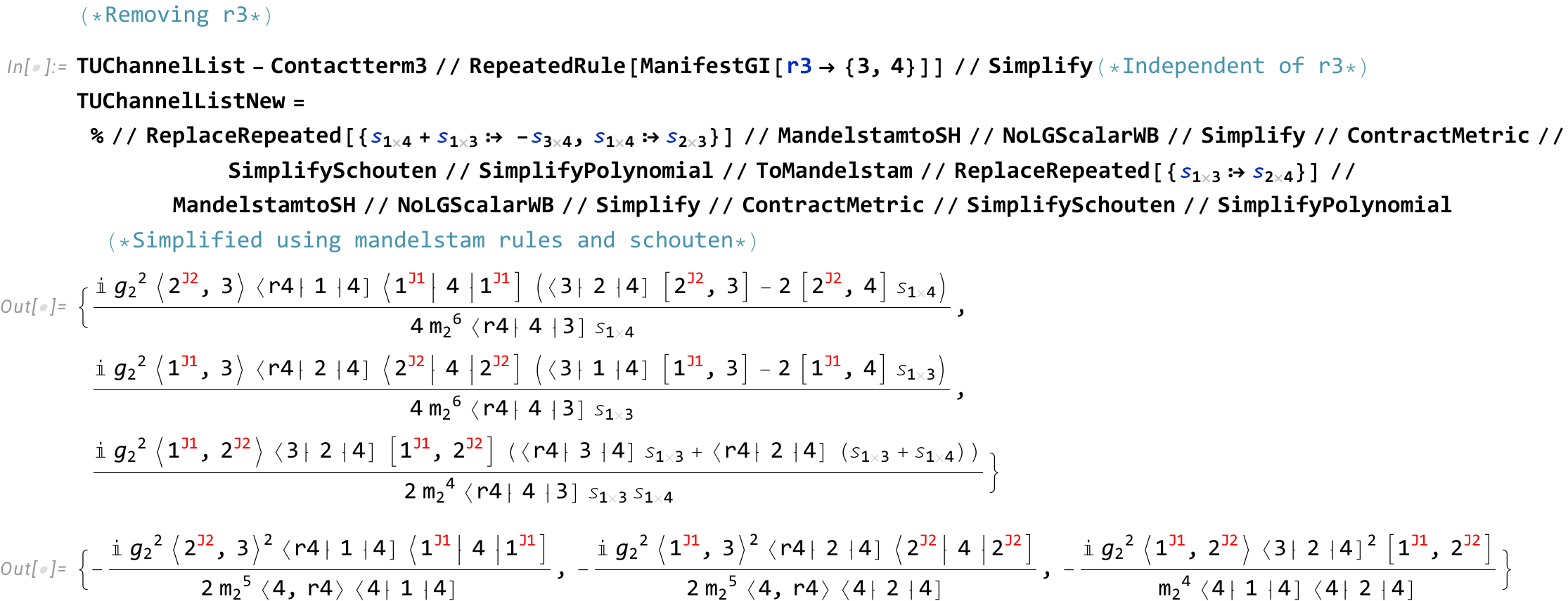}
		}
		
	\end{examplebox}
	
	\item  {\bf Gauge-invariance under leg \tt{4}}
	
	We see that the new gauge transformed amplitude \tt{TUChannelListNew}  is now dependent only on $\tt{r4}$. So we follow the same steps to remove the \tt{r4} dependence.
	
	\begin{examplebox}{Gauge invariant amplitude}
		
		\redbox{
			\includegraphics[scale=0.35]{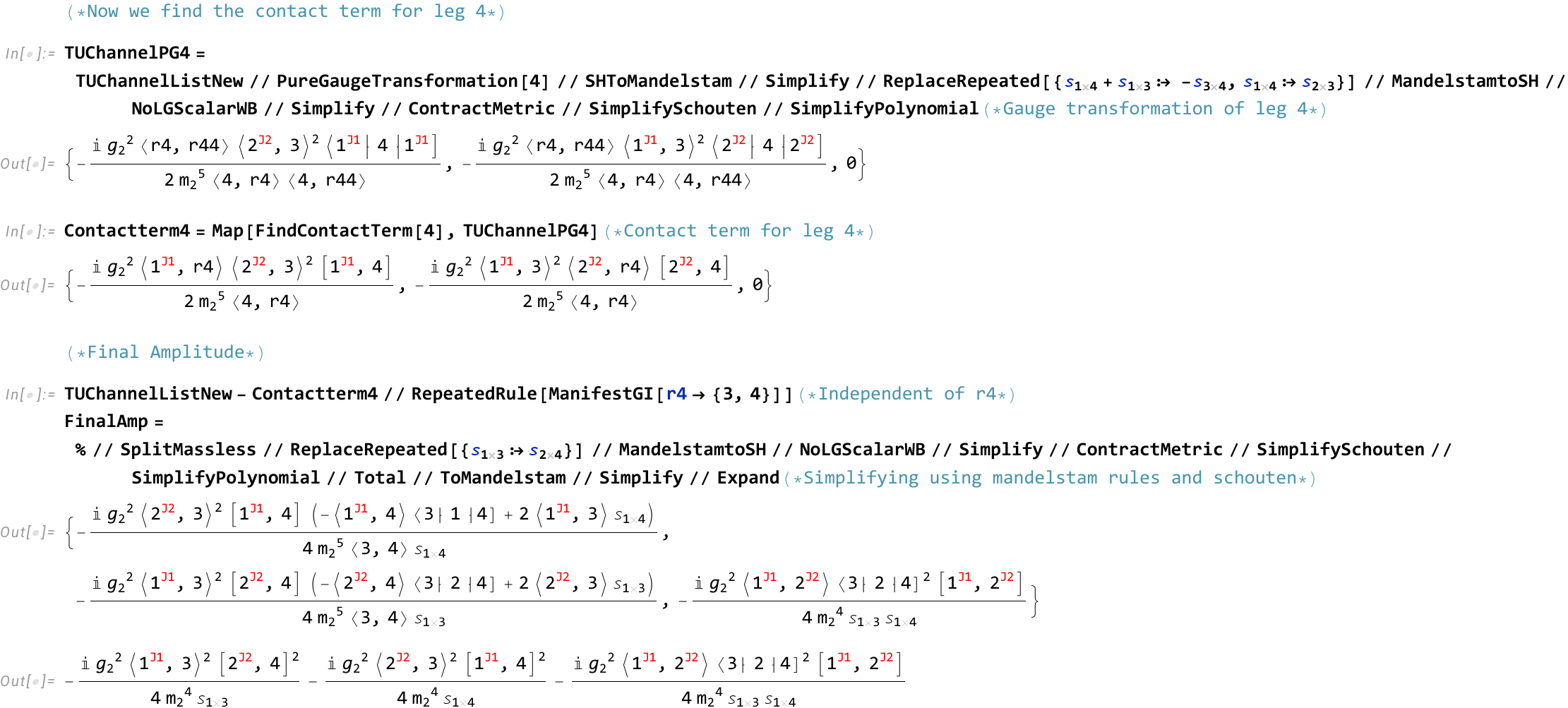}
		}
		
	\end{examplebox}

	Subtracting the contact term \tt{ContactTerm4} gives the final gauge invariant
	amplitude \tt{FinalAmp} which is gauge invariant under both legs \tt{\{3,4\}}.

		\begin{conceptbox}{Amplitude factorization}
			
			The same computation can also be performed using amplitude factorization\cite{Kumar:2025znu,Arkani-Hamed:2017jhn}  In
			this approach, the amplitude is written as a product of two on-shell
			gauge-invariant three-point vertices.  Since each three-point vertex is already
			gauge invariant, no separate contact-term construction is needed.
			
		\end{conceptbox}
		
		\begin{examplebox}{Amplitude factorization}
			
			\redbox{
				\includegraphics[scale=0.4]{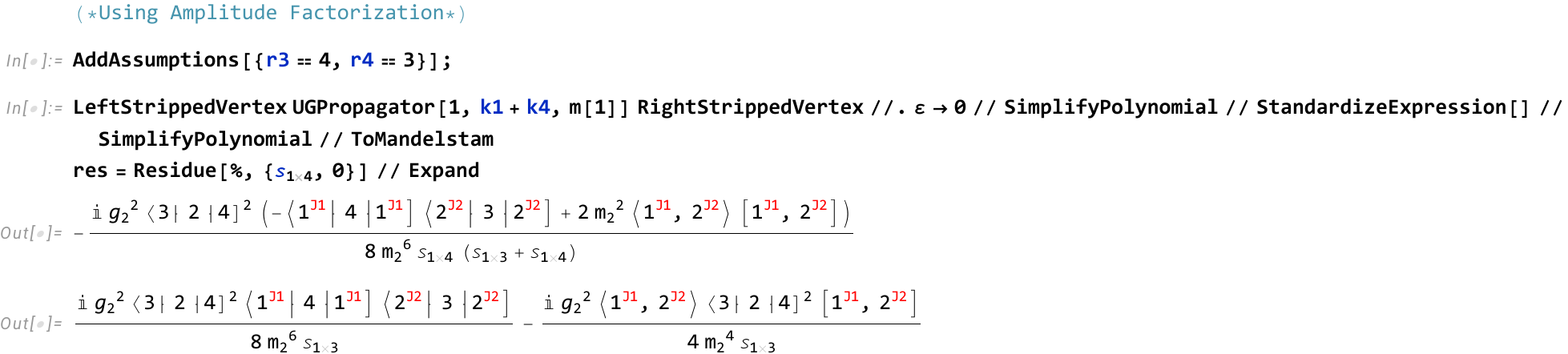}
			}
				\redbox{
				\includegraphics[scale=0.39]{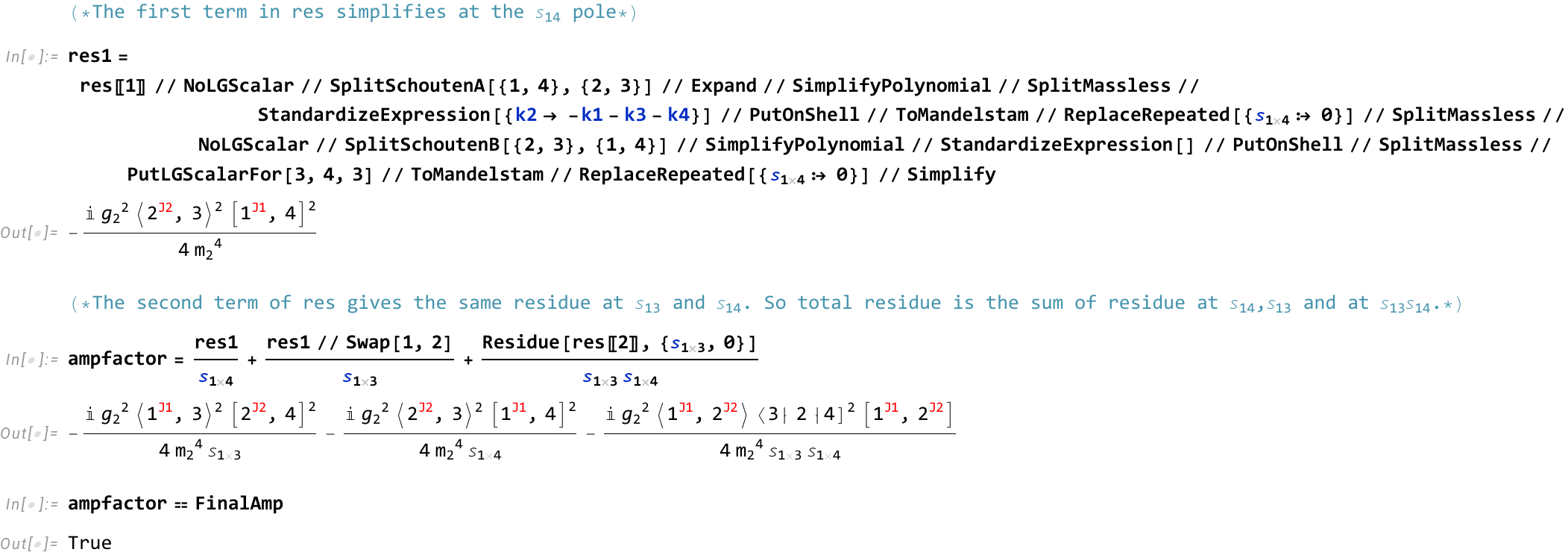}
			}
		\end{examplebox}
		
\end{enumerate}

\newpage

%% file: UsefulTools.tex
\section{A few useful features}
\label{sec:usefultools}

\subsection{Helicity-weight operators}

\begin{conceptbox}{}
	For the massless little group \(U(1)\), there is one generator, a.k.a. the helicity-weight operator, defined by:
	\begin{align}
		\mathtt{\textit{h}_i}
		&=
		\mathtt{\frac{1}{2}}
		\left(
		|\tt{i}\rangle^{\dot{\alpha}}
		\frac{\partial}{\partial |\tt{i}\rangle^{\dot{\alpha}}}
		-
		[\tt{i}|^{\alpha}
		\frac{\partial}{\partial [\tt{i}|^{\alpha}}
		\right).
	\end{align}
	The repeated spinor indices on the right-hand side are summed over.
	
	For the massive little group \(\mathtt{SU(2)}\), there are three traceless generators defined by
	\begin{align}
		\mathtt{\mathcal{J}^{I}{}_{J}}
		&=
		\mathtt{\textit{J}^{I}{}_{J}
			-\frac{1}{2}\delta^{I}{}_{J}\textit{J}^{K}{}_{K}},
		\\
		\mathtt{\textit{J}^{I}{}_{J}}
		&=
		|\tt{i}^{\tt{I}}\rangle^{\dot{\alpha}}
		\frac{\partial}{\partial|\tt{i}^{\tt{J}}\rangle^{\dot{\alpha}}}
		-
		[\tt{i}^{\tt{I}}|^{\alpha}
		\frac{\partial}{\partial[\tt{i}^{\tt{J}}|^{\alpha}} .
	\end{align}
	The three independent components are the raising, lowering, and diagonal
	operators
	\begin{align}
		\mathtt{\mathcal{J}_{+}}
		&=
		\mathtt{\mathcal{J}^{1}{}_{2}},
		&
		\mathtt{\mathcal{J}_{-}}
		&=
		\mathtt{\mathcal{J}^{2}{}_{1}},
		&
		\mathtt{\mathcal{J}_{0}}
		&=
		\mathtt{\mathcal{J}^{1}{}_{1}}
		=
		\mathtt{-\mathcal{J}^{2}{}_{2}} .
	\end{align}
	
\end{conceptbox}

\begin{documentationbox}{\tt{OpMlHW}}{\tt{OpMlHW[leg\_][expr\_]}}

	applies the massless helicity-weight operator to \(\tt{expr}\).
	
\end{documentationbox}

\begin{documentationbox}{\tt{OpMsHW}}{\tt{OpMsHW[leg\_, I\_, J\_][expr\_]}}

	applies the massive little-group generator
	\(\mathtt{\mathcal{J}^{I}{}_{J}}\) to \(\tt{expr}\).   
\end{documentationbox}

\begin{examplebox}{\tt{OpMlHW} and \tt{OpMsHW}}
	
	\redbox{
		\begin{minipage}{0.4\linewidth}
			\includegraphics[scale=0.45]{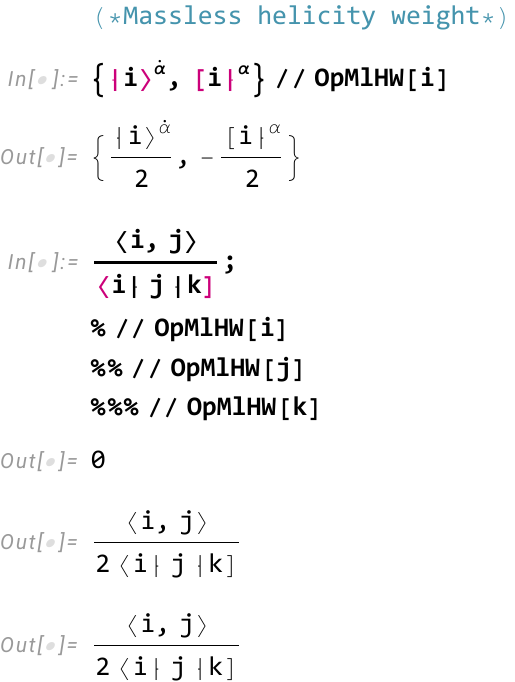}
		\end{minipage}
		\begin{minipage}{0.3\linewidth}
			\includegraphics[scale=0.4]{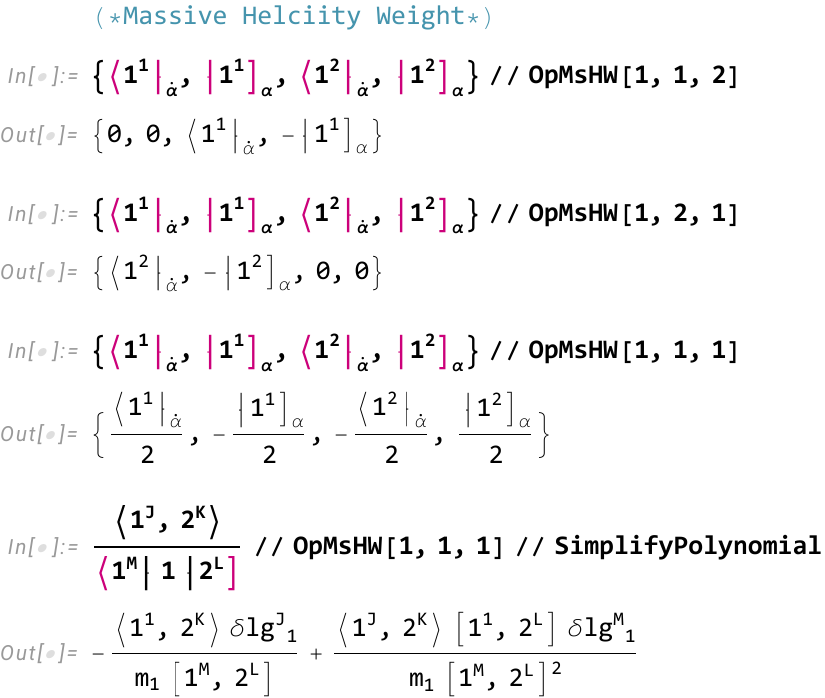}
		\end{minipage}
	}
	
\end{examplebox}

\subsection{Analyzing discrete symmetries: \(\tt{C, P, T}\)}
\label{sec:CPT}

\begin{conceptbox}{Discrete transformations}
	
	The \(\tt{CPT}\) theorem states that any Lorentz-invariant theory admits an
	anti-linear transformation that flips spacetime and leaves every Hermitian
	Lorentz-invariant term invariant.  This transformation is called the
	\(\tt{CPT}\) transformation, where \(\tt{C}\) denotes charge conjugation,
	\(\tt{P}\) denotes parity, and \(\tt{T}\) denotes time reversal.
	
	As discussed in App.~\ref{app:CPT}, the \(\tt{C}\), \(\tt{P}\), and
	\(\tt{T}\) transformations of quantum fields contain possible ambiguities from
	intrinsic parity phases.  By contrast, the transformations of polarizations,
	and hence of helicity spinors, are free of these intrinsic-parity ambiguities.
	They depend only on the basis used to describe the polarizations.  We discuss
	the spin and helicity bases in App.~\ref{app:helicitybasis}.  For
	spinor-helicity computations, the helicity basis is the natural one; the spin
	basis can be obtained from it by rotations, as reviewed in
	App.~\ref{app:spinbasis}.
	
	The full \(\tt{CPT}\) transformation must exist in a Lorentz-invariant theory.
	However, the theorem does not require the individual transformations
	\(\tt{C}\), \(\tt{P}\), or \(\tt{T}\) to exist separately.  For example, in a
	theory of only left-handed fermions, parity is not a valid transformation
	because it maps a left-handed representation to a right-handed one.  In such a
	theory, the valid transformations may instead be \(\tt{CP}\), \(\tt{T}\), and
	\(\tt{CPT}\).
	
\end{conceptbox}

\subsubsection{Complex conjugation}\label{sec:ComplexConjugation}

\begin{conceptbox}{Complex conjugation of spinors}
	
	Complex conjugation maps \(\mathtt{SU(2)_L}\) representations to
	\(\mathtt{SU(2)_R}\) representations and vice versa.  Equivalently, it maps
	undotted and dotted spinor indices into each other.  We denote complex
	conjugation by \(*\), while Hermitian conjugation is denoted by
	\(\dagger\).  In component form, Hermitian conjugation is the same as complex
	conjugation.
	
	For Hermitian momenta,
	\begin{align}
		p^{\mu\dagger}=p^\mu,
	\end{align}
	the angle and square spinors are related by complex conjugation:
	\begin{align}
		\text{Massless:}\qquad
		\big(|\mathtt{p}]_\alpha\big)^*
		&=
		\langle \mathtt{p}|_{\dot{\alpha}},
		\label{ComplexConjugationReal1}
		\\
		\text{Massive:}\qquad
		\big(|\mathtt{p^{J}}]_\alpha\big)^*
		&=
		\langle \mathtt{p_J}|_{\dot{\alpha}},
		&
		\big(|\mathtt{p_{J}}]_\alpha\big)^*
		&=
		-\langle \mathtt{p^J}|_{\dot{\alpha}}.
		\label{ComplexConjugationReal2}
	\end{align}
	
	If the momentum is not Hermitian, the angle and square spinors are independent.
	In that case, the complex conjugate of a left-handed spinor is a new
	right-handed spinor, and the complex conjugate of a right-handed spinor is a
	new left-handed spinor.  We denote these new conjugate spinors using a tilde:
	\begin{align}
		\text{Massless:}\qquad
		\big(|\mathtt{p}]_\alpha\big)^*
		&\equiv
		\langle \mathtt{\tilde p}|_{\dot{\alpha}},
		&
		\big(\langle \mathtt{p}|_{\dot{\alpha}}\big)^*
		&\equiv
		|\mathtt{\tilde p}]_\alpha,
		\label{ComplexConjugationComplex1}
		\\
		\text{Massive:}\qquad
		\big(|\mathtt{p^{J}}]_\alpha\big)^*
		&\equiv
		\langle \mathtt{\tilde p_J}|_{\dot{\alpha}},
		&
		\big(\langle \mathtt{p^J}|_{\dot{\alpha}}\big)^*
		&\equiv
		-|\mathtt{\tilde p_J}]_\alpha.
		\label{ComplexConjugationComplex2}
	\end{align}
	The tilde on the conjugate spinor should not be confused with the usual tilde
	spinor appearing in the momentum matrix
	\(\mathtt{p_{\alpha\dot{\alpha}}=-\lambda_\alpha\tilde\lambda_{\dot{\alpha}}}\).
	
	In this notation, imposing the reality condition on a momentum is equivalent
	to removing the tilde:
	\begin{align}
		\text{Real momenta:}\qquad
		\mathtt{\tilde p=p}.
	\end{align}
	
\end{conceptbox}

\begin{documentationbox}{\tt{ComplexConjugate}}{\tt{ComplexConjugate[expr\_]}}

	implements the complex-conjugation rules for spinors and invariant tensors.
	It is the same as Mathematica's built-in function \(\tt{Conjugate}\).
	
\end{documentationbox}

\begin{morematerialbox}{}
	The rules depend on whether the corresponding leg is massive or not.  The
	criteria for massive and non-massive variables were described in
	sec.~\ref{app:LGcriteria}.  A non-massive leg is either a massless leg declared
	using \(\tt{DeclareMasslessLegs[]}\), or else it is treated as off shell.
	
	For massive legs, complex conjugation includes an additional minus sign when
	acting on spinors with lowered little-group indices.  This sign appears because
	massive spinors transform in a pseudoreal representation of
	\(\mathtt{SU(2)_{LG}}\).	
\end{morematerialbox} 

\begin{examplebox}{\tt{ComplexConjugate} of massless and massive spinors}
	
	\redbox{
		\begin{minipage}{0.5\linewidth}
			\includegraphics[scale=0.5]{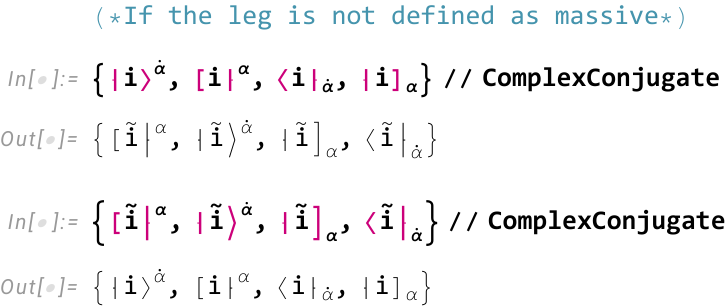}
		\end{minipage}
		\begin{minipage}{0.3\linewidth}
			\includegraphics[scale=0.5]{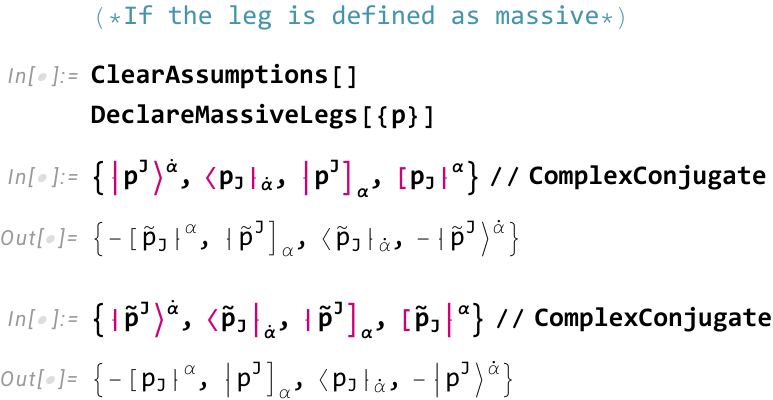}
		\end{minipage}
	}
	
\end{examplebox}

\begin{documentationbox}{\tt{ImposeReality}}{\tt{ImposeReality[expr\_]}}
	
	imposes reality condition on momenta. For a leg \(\tt{p}\), this sets
	\begin{align}
		\mathtt{\tilde p=p}.
	\end{align}
	
\end{documentationbox}

\begin{examplebox}{\tt{ImposeReality} on massless and massive spinors}
	
	\redbox{
		\includegraphics[scale=0.5]{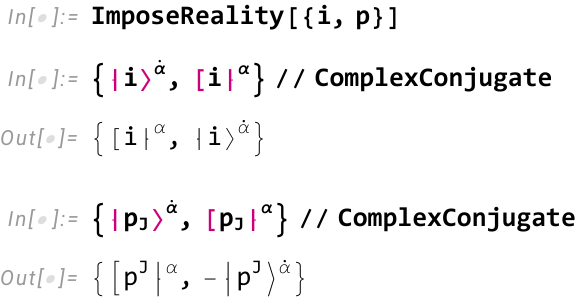}
	}
	
\end{examplebox}

\begin{conceptbox}{Complex conjugation of invariant tensors}
	
	The signs appearing under complex conjugation are independent of the spacetime
	metric convention.  They depend only on the convention chosen for the
	Levi-Civita tensors.  Complex conjugation maps upper little-group indices to
	lower ones and vice versa.  It also maps the
	\(\mathtt{SU(2)_L}\) and \(\mathtt{SU(2)_R}\) invariant tensors into each
	other:
	\begin{align}
		\left(\mathtt{\epsilon lg^{JK}}\right)^*
		&=
		\mathtt{-\epsilon lg_{JK}},
		&
		\left(\mathtt{\delta lg^{J}{}_{K}}\right)^*
		&=
		\mathtt{\delta lg^{K}{}_{J}},
		\\
		\left(\mathtt{\epsilon^{\alpha\beta}}\right)^*
		&=
		\mathtt{\epsilon d}^{\dot{\alpha}\dot{\beta}},
		&
		\left(\mathtt{\delta^{\alpha}{}_{\beta}}\right)^*
		&=
		\mathtt{\delta d}^{\dot{\alpha}}{}_{\dot{\beta}}.
	\end{align}
	
\end{conceptbox}

\begin{examplebox}{Complex conjugation of invariant bilinears}
	
	\redbox{
		\includegraphics[scale=0.5]{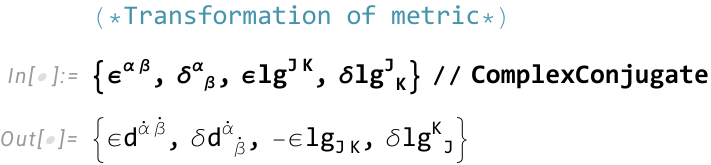}
	}
	
\end{examplebox}

\subsubsection{Charge conjugation}

\begin{conceptbox}{Charge conjugation}
	
	Charge conjugation is an anti-linear transformation on the field space.  It
	maps a particle to its antiparticle.  The \(\tt{C}\) transformations of
	massive and massless, or more generally non-massive, legs are discussed in
	Eqs.~\eqref{ChargeConjugationMassless}--\eqref{ChargeConjugationMassive}.
	
	As described in Eq.~\eqref{ChargeConjugationHelicityPolarizations}, charge
	conjugation maps an angle or box spinor to the conjugate of the corresponding
	box or angle spinor.  This brings the spinor back to the same representation.
	Therefore, for both massive and non-massive legs, charge conjugation maps a
	spinor to its tilde version without changing helicity or chirality:
	\begin{align}
		\tt{C}:\qquad
		\begin{cases}
			\mathtt{|p]}_{\alpha}
			\rightarrow
			\mathtt{|\tilde p]}_{\alpha},
			&
			\mathtt{|\tilde p\rangle}^{\dot{\alpha}}
			\rightarrow
			\mathtt{|p\rangle}^{\dot{\alpha}},
			\\
			\mathtt{|p^{I}]}_{\alpha}
			\rightarrow
			\mathtt{|\tilde p^{I}]}_{\alpha},
			&
			\mathtt{|\tilde p^{I}\rangle}^{\dot{\alpha}}
			\rightarrow
			\mathtt{|p^{I}\rangle}^{\dot{\alpha}}.
		\end{cases}
	\end{align}
	
\end{conceptbox}

\begin{documentationbox}{\tt{ChargeConjugate}}{\tt{ChargeConjugate[expr\_]}}

	implements charge conjugation on spinor-helicity expressions.  
	
	It maps the spinors of a particle to the corresponding spinors of the antiparticle, denoted by tilde variables.
	
\end{documentationbox}

\begin{examplebox}{\tt{ChargeConjugate} of massless and massive spinors}
	
	\redbox{
		\begin{minipage}{0.5\linewidth}
			\includegraphics[scale=0.5]{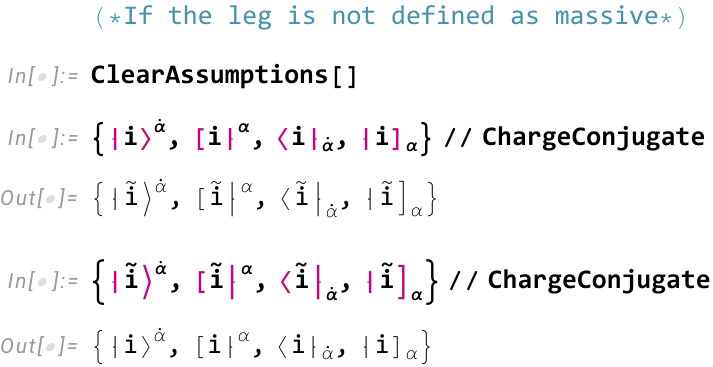}
		\end{minipage}
		\begin{minipage}{0.3\linewidth}
			\includegraphics[scale=0.5]{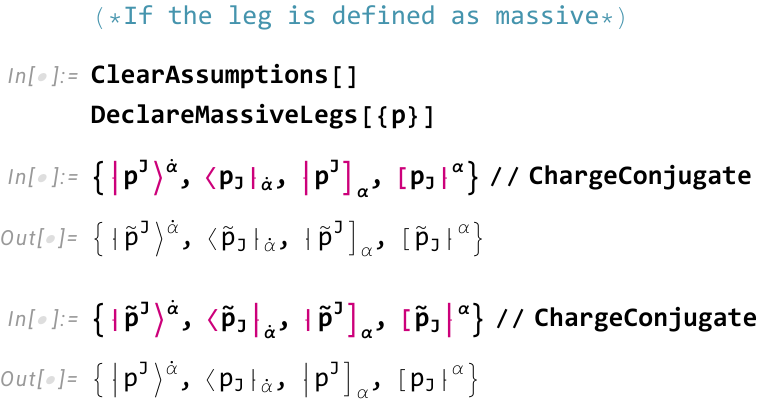}
		\end{minipage}
	}
	
\end{examplebox}

\begin{conceptbox}{Charge conjugation of invariant tensors}
	
	Unlike complex conjugation, charge conjugation leaves the left- and
	right-handed invariant tensors unchanged.  Since charge conjugation commutes
	with the little-group representations, it also leaves the little-group bilinear
	unchanged:
	\begin{align}
		\mathtt{\epsilon^{\alpha\beta}}
		&\rightarrow
		\mathtt{\epsilon^{\alpha\beta}},
		&
		\mathtt{\delta^{\alpha}{}_{\beta}}
		&\rightarrow
		\mathtt{\delta^{\alpha}{}_{\beta}},
		\\
		\mathtt{\epsilon d}^{\dot{\alpha}\dot{\beta}}
		&\rightarrow
		\mathtt{\epsilon d}^{\dot{\alpha}\dot{\beta}},
		&
		\mathtt{\delta d}^{\dot{\alpha}}{}_{\dot{\beta}}
		&\rightarrow
		\mathtt{\delta d}^{\dot{\alpha}}{}_{\dot{\beta}},
		\\
		\mathtt{\epsilon lg^{JK}}
		&\rightarrow
		\mathtt{\epsilon lg^{JK}},
		&
		\mathtt{\delta lg^{J}{}_{K}}
		&\rightarrow
		\mathtt{\delta lg^{J}{}_{K}}.
	\end{align}
	
\end{conceptbox}

\begin{examplebox}{\tt{ChargeConjugate} of scalars and invariant tensors}
	
	\redbox{
		\includegraphics[scale=0.5]{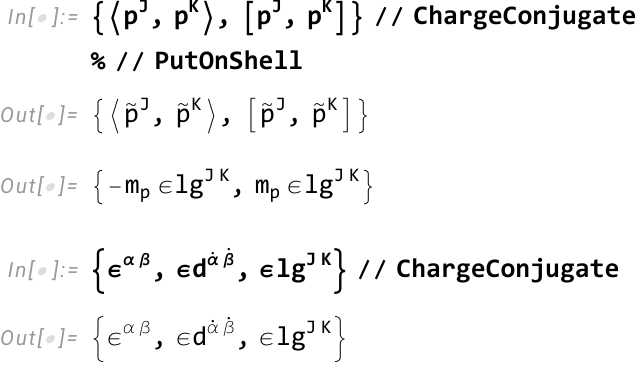}
	}
	
\end{examplebox}

\subsubsection{Parity}

\begin{conceptbox}{Parity}
	
	Parity is a linear operator on both Hilbert space and field space.  Since it
	commutes with rotations but flips momentum, it flips the helicity of particles.
	In angle and box notation, the parity transformations
	\eqref{ParityMasslessAngBox}--\eqref{ParityMassiveAngBox} exchange angle and
	box spinors:
	\begin{align}
		\tt{P}:\qquad
		\begin{cases}
			\mathtt{|p]}_{\alpha}
			\rightarrow
			\mathtt{\iimg |p\rangle}^{\dot{\alpha}},
			&
			\mathtt{|p\rangle}^{\dot{\alpha}}
			\rightarrow
			\mathtt{\iimg |p]}_{\alpha},
			\\
			\mathtt{|p^{I}]}_{\alpha}
			\rightarrow
			\mathtt{\iimg \Sigma^{I}{}_{J}|p^{J}\rangle}^{\dot{\alpha}},
			&
			\mathtt{|p^{I}\rangle}^{\dot{\alpha}}
			\rightarrow
			\mathtt{\iimg \Sigma^{I}{}_{J}|p^{J}]}_{\alpha}.
		\end{cases}
	\end{align}
	The matrix \(\mathtt{\Sigma^{I}{}_{J}}\), defined in
	Eq.~\eqref{SigmaDefinition}, has the same components as the Pauli matrix
	\(\sigma_1\).
	
\end{conceptbox}

\begin{documentationbox}{\tt{Parity}}{\tt{Parity[expr\_]}}

	implements the parity transformation on spinor-helicity expressions.  
\end{documentationbox}

\begin{morematerialbox}
	If a leg
	is declared as massive, the transformation includes the little-group matrix
	\(\mathtt{\Sigma}\).
	
	For a leg label \(\tt{leg}\), this matrix has the syntax
	\begin{align}
		\tt{\(\Sigma\)[leg\_][I\_, J\_]}.
	\end{align}
	We attach the leg label to \(\mathtt{\Sigma}\) because little-group indices
	belonging to different legs should not be identified.  The output form of
	\(\tt{\(\Sigma\)[leg][J,K]}\) is
	\begin{align}
		\mathtt{\left(\Sigma_{\mathrm{leg}}\right){}^{J}{}_{K}}.
	\end{align}
	Putting minus signs in front of the little-group indices places them in the
	subscript.
	
\end{morematerialbox}

\begin{examplebox}{\tt{Parity} transformation of massless and massive spinors}
	
	\redbox{
		\begin{minipage}{0.4\linewidth}
			\includegraphics[scale=0.5]{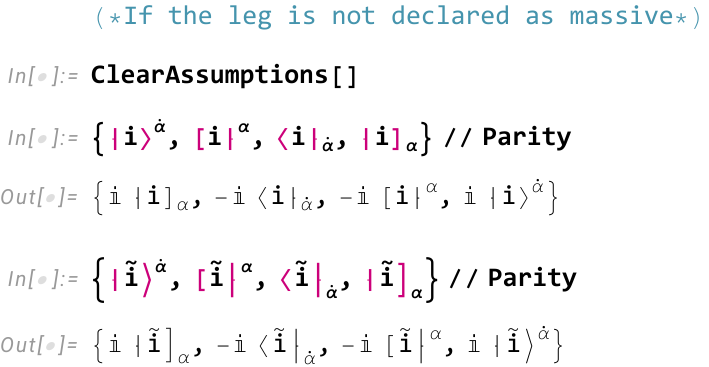}
		\end{minipage}
		\begin{minipage}{0.3\linewidth}
			\includegraphics[scale=0.5]{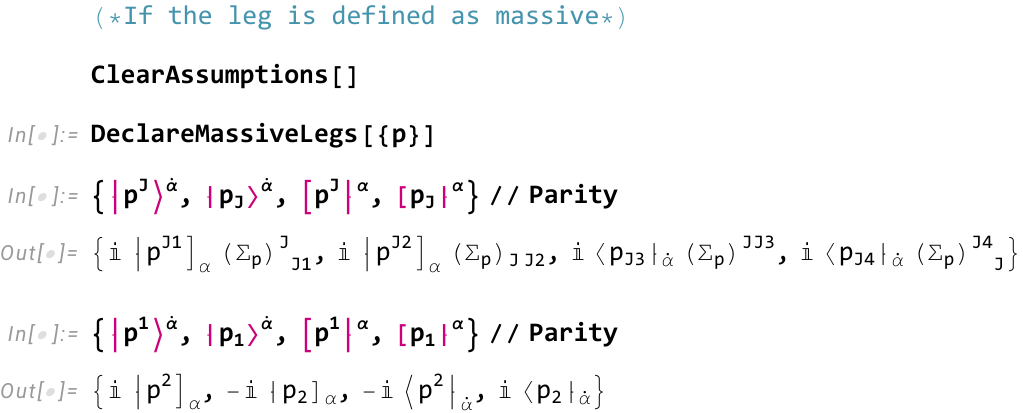}
		\end{minipage}
	}
	
\end{examplebox}

\begin{examplebox}{\tt{Parity} transformation of scalars}
	
	\redbox{
		\includegraphics[scale=0.5]{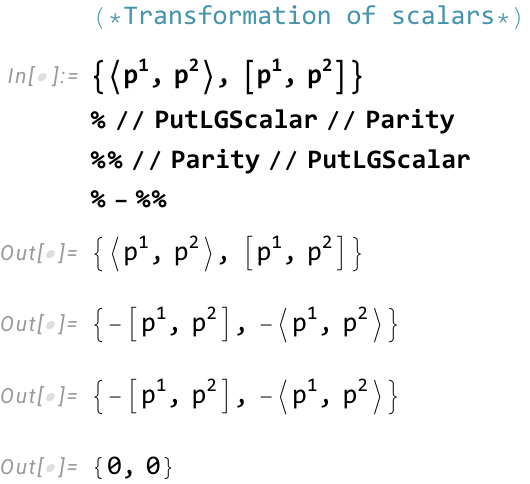}
	}
	
\end{examplebox}

\begin{conceptbox}{Parity transformation of invariant tensors}
	
	Parity maps upper \(\mathtt{SU(2)_L}\) invariant tensors to lower
	\(\mathtt{SU(2)_R}\) invariant tensors, and vice versa.  The little-group
	bilinear is left unchanged:
	\begin{align}
		\tt{P}:\qquad
		\begin{cases}
			\epsilon_{\alpha\beta}
			\leftrightarrow
			-\epsilon\tt{d}^{\dot{\alpha}\dot{\beta}},
			&
			\epsilon\tt{lg}^{\tt{JK}}
			\leftrightarrow
			\epsilon\tt{lg}^{\tt{JK}},
			\\
			\delta^{\beta}{}_{\alpha}
			\leftrightarrow
			\delta\tt{d}^{\dot{\alpha}}{}_{\dot{\beta}},
			&
			\delta\tt{lg}^{\tt{J}}{}_{\tt{K}}
			\leftrightarrow
			\delta\tt{lg}^{\tt{J}}{}_{\tt{K}}.
		\end{cases}
	\end{align}
	
\end{conceptbox}

\begin{examplebox}{Parity and bilinear contractions}
	
	\redbox{
		\includegraphics[scale=0.5]{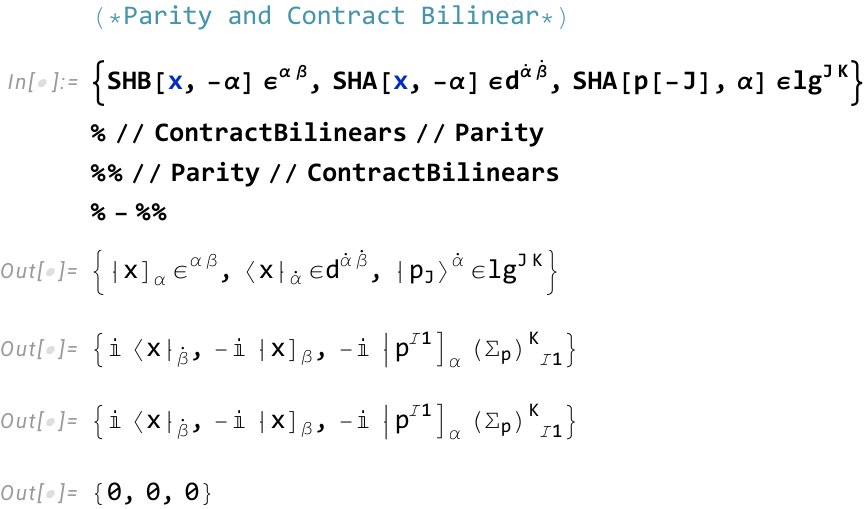}
	}
	
\end{examplebox}

\subsubsection{Time reversal}

\begin{conceptbox}{Time reversal}
	
	Time reversal \(\tt{T}\) is an anti-linear operation on Hilbert space.  It
	anti-commutes with momentum and with rotations, and therefore does not change
	the helicity of a particle.  Its action on angle and box spinors is derived in
	Eqs.~\eqref{TimeReversalMasslessAngBox} and
	\eqref{TimeReversalMassiveAngBox}.  It maps angle and box spinors to angle and
	box spinors, up to the little-group matrix \(\mathtt{\Omega^{I}{}_{J}}\)
	defined in Eq.~\eqref{OmegaDefinition}:
	\begin{align}
		\tt{T}:\qquad
		\begin{cases}
			\mathtt{|p]_{\alpha}}
			\rightarrow
			\mathtt{-\iimg [p|^{\alpha}},
			&
			\mathtt{|p\rangle}^{\dot{\alpha}}
			\rightarrow
			\mathtt{-\iimg \langle p|}_{\dot{\alpha}},
			\\
			\mathtt{|p^{I}]_{\alpha}}
			\rightarrow
			\mathtt{\iimg \Omega^{I}{}_{J}[p^{J}|^{\alpha}},
			&
			\mathtt{|p^{I}\rangle}^{\dot{\alpha}}
			\rightarrow
			\mathtt{-\iimg \Omega^{I}{}_{J}\langle p^{J}|}_{\dot{\alpha}}.
		\end{cases}
	\end{align}
	
\end{conceptbox}

\begin{documentationbox}{\tt{TimeReversal}}{\tt{TimeReversal[expr\_]}}

	implements the time-reversal transformation on spinor-helicity expressions.
	
\end{documentationbox}

\begin{examplebox}{\tt{TimeReversal} of massless and massive spinors}
	
	\redbox{
		\begin{minipage}{0.4\linewidth}
			\includegraphics[scale=0.5]{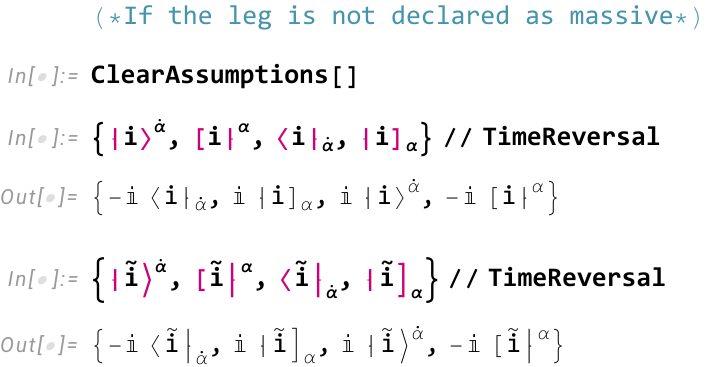}
		\end{minipage}
		\begin{minipage}{0.3\linewidth}
			\includegraphics[scale=0.5]{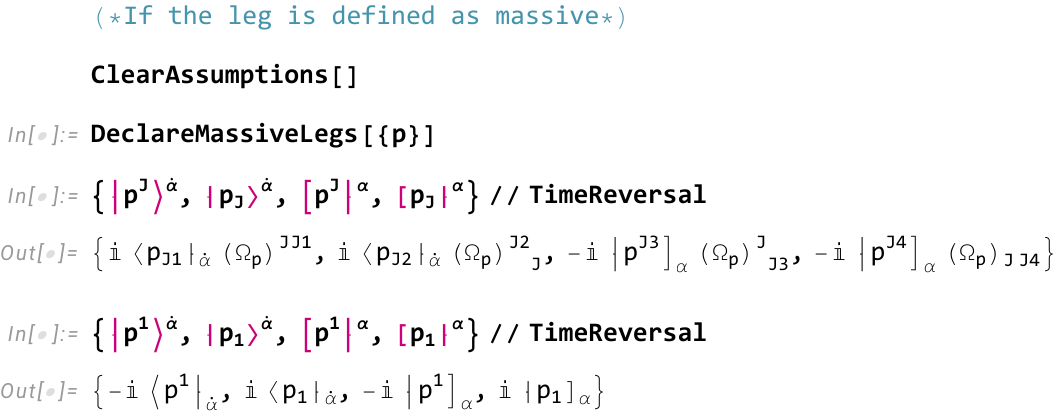}
		\end{minipage}
	}
	
\end{examplebox}

\begin{conceptbox}{Time reversal of invariant tensors}
	
	Time reversal maps upper left- and right-handed bilinears to their corresponding
	lower bilinears.  The little-group bilinear is left unchanged:
	\begin{align}
		\epsilon^{\alpha\beta}
		&\leftrightarrow
		-\epsilon_{\alpha\beta},
		&
		\delta^{\alpha}{}_{\beta}
		&\leftrightarrow
		\delta^{\beta}{}_{\alpha},
		\\
		\epsilon\tt{d}^{\dot{\alpha}\dot{\beta}}
		&\leftrightarrow
		-\epsilon\tt{d}_{\dot{\alpha}\dot{\beta}},
		&
		\delta\tt{d}^{\dot{\alpha}}{}_{\dot{\beta}}
		&\leftrightarrow
		\delta\tt{d}^{\dot{\beta}}{}_{\dot{\alpha}},
		\\
		\epsilon\tt{lg}^{\tt{JK}}
		&\rightarrow
		\epsilon\tt{lg}^{\tt{JK}},
		&
		\delta\tt{lg}^{\tt{J}}{}_{\tt{K}}
		&\leftrightarrow
		\delta\tt{lg}^{\tt{J}}{}_{\tt{K}}.
	\end{align}
	
\end{conceptbox}

\begin{examplebox}{Time reversal and bilinear contractions}
	
	\redbox{
		\includegraphics[scale=0.5]{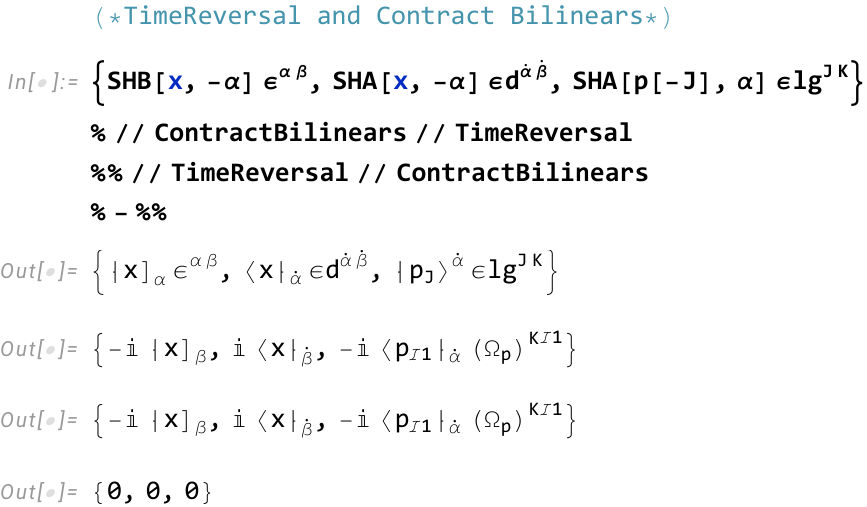}
	}
	
\end{examplebox}

\begin{examplebox}{Time reversal and polynomial rules}
	
	\redbox{
		\includegraphics[scale=0.5]{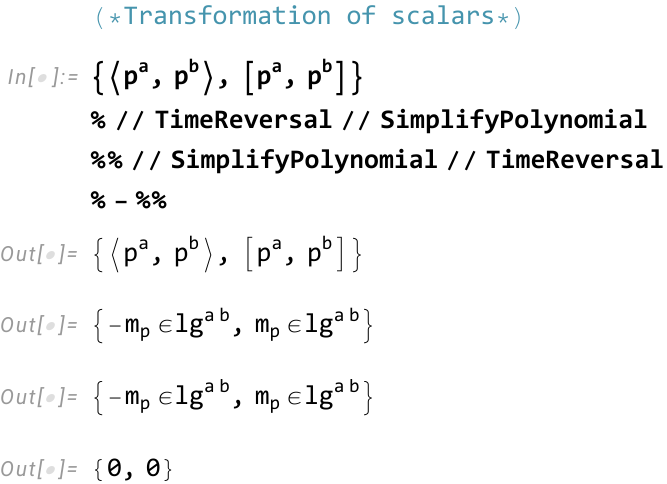}
	}
	
\end{examplebox}

\subsubsection{\(\tt{CPT}\)}

\begin{documentationbox}{\tt{CPT}}{\tt{CPT[expr\_]}}
	
	\(\tt{CPT}\) transformation of massive and massless spinors

\end{documentationbox}
\begin{morematerialbox}
	
	The matrices \(\Sigma\), \(\Omega\), and \(\Delta\), together with their
	properties, are defined in App.~\ref{app:CPT}.
	
\end{morematerialbox}

\begin{examplebox}{\tt{CPT} transformation of spinors}
	
	\redbox{
		\begin{minipage}{0.5\linewidth}
			\includegraphics[scale=0.5]{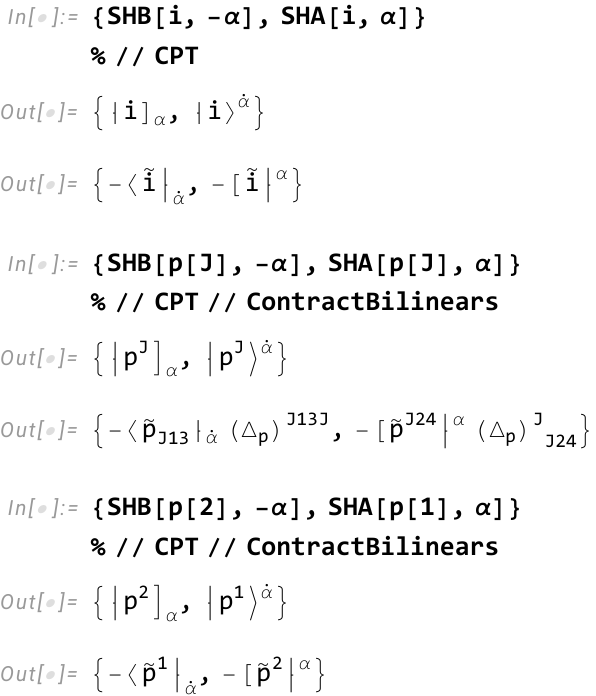}
		\end{minipage}
		\begin{minipage}{0.3\linewidth}
			\includegraphics[scale=0.5]{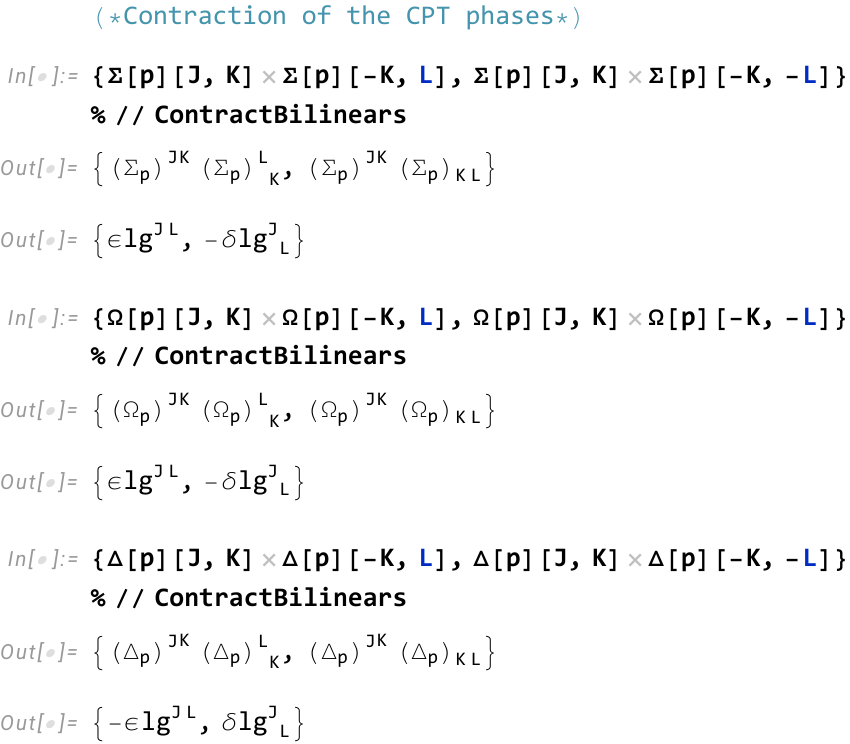}
		\end{minipage}
	}
	
\end{examplebox}

\subsection{Component form}
\label{sec:ComponentForm}

\begin{conceptbox}{Basic idea}
	
	A central goal in scattering computations is to obtain numerical quantities
	such as cross-sections, which can be compared with experimental data.  Since a
	cross-section depends on the explicit configuration of the external particles,
	one often needs to convert spinor-helicity expressions into component form.

	Spinor-helicity angle and box spinors are momentum eigenvectors.  Therefore,
	knowing the components of the corresponding momentum matrix is enough to
	evaluate spinor-helicity expressions numerically.  The package provides commands
	that convert angle and box spinors directly into components.
	
	Once the component forms of the basic spinors are known, the component forms of
	\(\mathtt{SU(2)_L}\), \(\mathtt{SU(2)_R}\), and \(SL(2,\mathbb C)\) scalars
	follow by contraction.  The remaining simple objects are invariant
	antisymmetric bilinears, whose component forms are fixed by convention.
	
\end{conceptbox}

\begin{documentationbox}{\tt{ComponentForm}}{\tt{Direct component conversion}}
	
	Writing the component form using \(\tt{r[A][]}\), \(\tt{l[B][]}\), and related
	objects is sometimes unnecessarily verbose.  Since the package already has the
	simpler objects \(\tt{SHA[]}\) and \(\tt{SHB[]}\), one can directly obtain
	their component forms using
	\begin{align}
		\tt{ComponentForm}.
	\end{align}
	
\end{documentationbox}

\begin{examplebox}{Direct spinor-component conversion}
	
	\redbox{
		\includegraphics[scale=0.5]{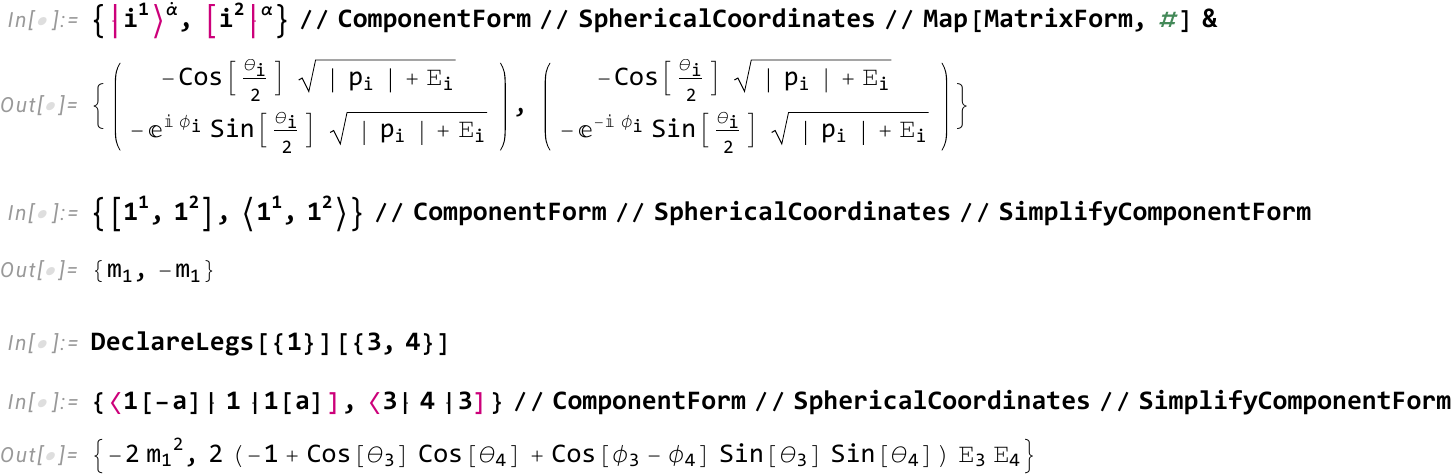}
	}
	
\end{examplebox}

\subsection{Numerics}\label{sec:Numerics}
The component form allows us to expand the spinors and their scalars in terms of actual momentum components. A step forward is to map these components to numerical values. The numerical analysis becomes very important while with the scattering observables like cross-section. It also becomes handy in doing various tests like checking two expressions are same without simplifying them using Schouten identities. The numerical values makes it easy for checking the gauge invariance of a scattering amplitude.

\subsubsection{Generating numerics}
\begin{conceptbox}{}
	The \tt{GenerateNumerics} command generates the numerical values for the momenta of the incoming and outgoing particles. The generated data can be applied to any expression using the commands \tt{PutNumerics} and can be accessed in rules format using \tt{MassNumerics}, \tt{MomentaNumerics} and \tt{ComponentsNumerics}. The generated data can also be used to check the gauge invariance of a scattering amplitude. The \tt{GenerateNumerics} command uses the \tt{RAMBO} engine in the background and so it works only if we provide a non zero threshold energy whose default value is kept as \tt{100}. There are multiple \tt{Options} available for \tt{GenerateNumerics} to get a better control over the numerics.
\end{conceptbox}
\begin{documentationbox}{\tt{GenerateNumerics}}{\tt{GenerateNumerics[label\_,momconsrule\_,OptionsPattern[]]}}
	
	First argument,\tt{label} is any user preferred name(without space), the numberical data would be generated for this \tt{label}. This \tt{label} is for accessing the data later on. Second argument, the momentum conservation rule which is in the form $\mathtt{k_1+\cdots+k_m \longrightarrow k_{m+1}+\cdots+k_{n} }$. The momenta can in general have any arbitrary numerical coefficients. For example, one can also write $\mathtt{k_1+k_2\rightarrow 2k_3-k_4}$. Since the command uses the momentum labels, it needs the corresponding legs to be declared first. Declaring the legs defines the momentum labels as well as their masses. So the user doesn't need to give the masses separately unless they have a specific numeric value.
\end{documentationbox}

\begin{examplebox}{\tt{GenerateNumerics}}
	As an example consider the decay of a pion ($\pi^+$) into a muon ($\mu^+$) and a neutrino ($\nu_\mu$) which is almost massless.
	\begin{align}
		\tt{piondecay}\quad:\quad \pi^+\longrightarrow \mu^++\nu_\mu
	\end{align}
	Let's label the legs by $\tt{\{1,2,3\}}$ for $\{\pi^+,\mu^+,\nu_\mu\}$ respectively. Then the momentum conservation becomes : $\mathtt{k_1\rightarrow k_2+k_3}$. The numerics can be simply generated as follows
	\redbox{\includegraphics[scale=0.4]{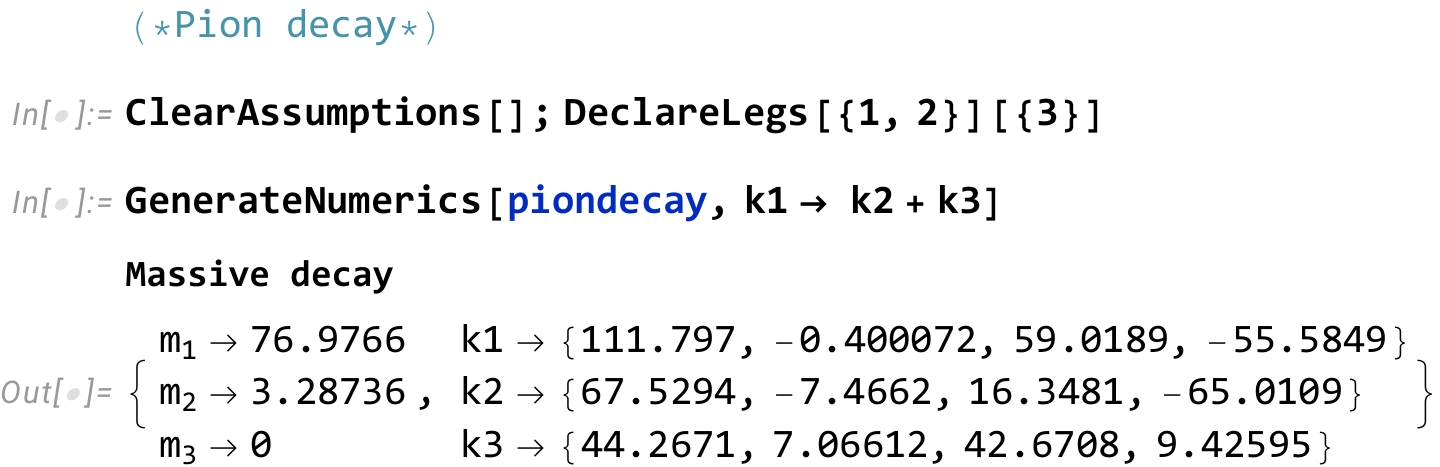}}
	In the example above, \tt{\textcolor{blue}{piondecay}} is only a label and is not a function already defined in the package. It is completely upto the user to choose.
\end{examplebox}

\begin{morematerialbox}{}
	\tt{GenerateNumerics} uses the \tt{RAMBO}\ref{app:RAMBO} engine in the background and so it works only if we provide a non zero threshold energy whose default value is kept as \tt{100}. There are multiple \tt{Options} available for \tt{GenerateNumerics} to get a better control over the numerics.
	
	\begin{table}[ht]
		\centering
		\begin{tabular}{|c|c|}
			\hline
			\tt{\textbf{Options}} & \tt{\textbf{Default}}\\ [2ex]
			\hline
			\tt{COM} & \tt{False}\\
			\hline
			\tt{Energy} & 	\tt{100}\\
			\hline
			\tt{MassConstraints}&	\tt{\$Assumptions} \\
			\hline
			\tt{Seed}& 	\tt{Automatic}\\
			\hline
			\tt{Iterations}& 	\tt{20}\\
			\hline
		\end{tabular}
		\caption{Options for \tt{GenerateNumerics}}
		\label{tab:GenNumOptions}
	\end{table} 
	
	Let's understand them one-by-one.
	\begin{enumerate}
		\item \tt{COM} : Using this option one can choose whether to generate the numerics in the \tt{COM} frame of the incoming particles or any arbitrary lab frame which is randomly generated. If \tt{COM$\rightarrow$True}, the numerics is generated in the COM frame of incoming particles. If it is \tt{False}, it is generated in a randomly chosen lab frame.
		
		\redbox{\includegraphics[scale=0.4]{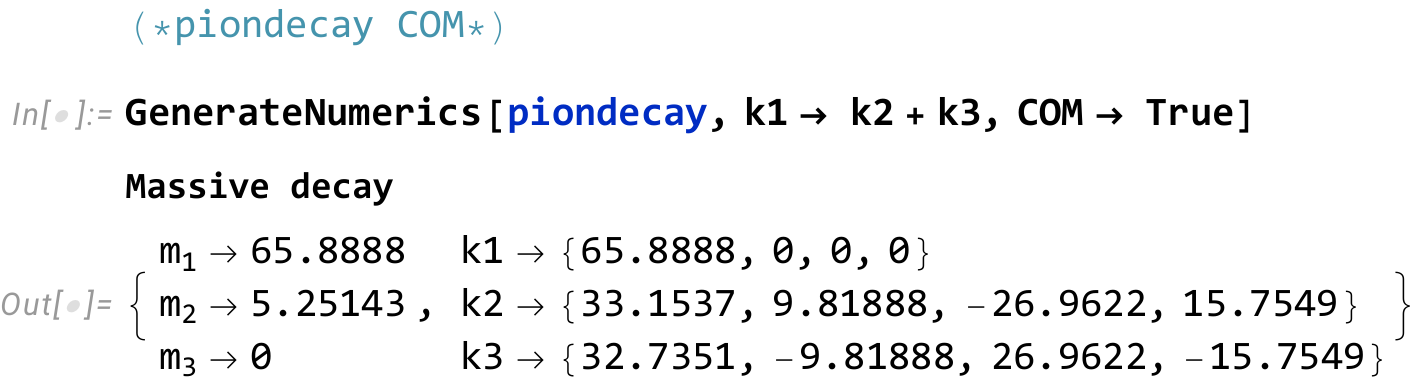}}
		
		\item \tt{Energy} : This option let's the user to threshold energy can be changed. For example, adding \tt{Energy$\rightarrow$200} in the argument sets the threshold energy to \tt{200}.
		
		\item  \tt{MassConstraints} : Using this option one can insert any equality or inequality among the masses. One can also insert exact values of the masses, if known, given that the threshold energy is not crossed. As an example, the mass of the pion is \tt{139.57 MeV}, mass of the muon is \tt{105.658 MeV} and the neutrino is massless. So we simply include in the arguments 
		\begin{center}
			\tt{\tt{MassConstraints}$\rightarrow$ \tt{\{$\mathtt{m_1}$==139.57, $\mathtt{m_2}$==105.658,$\mathtt{m_3}$==0\}}}
		\end{center}
		The mass constraints can also be added globally in the list of assumptions \tt{\$Assumptions} using \tt{AddAssumptions}. This way one-doesn't need to write them again and again.
		
		\redbox{\includegraphics[scale=0.35]{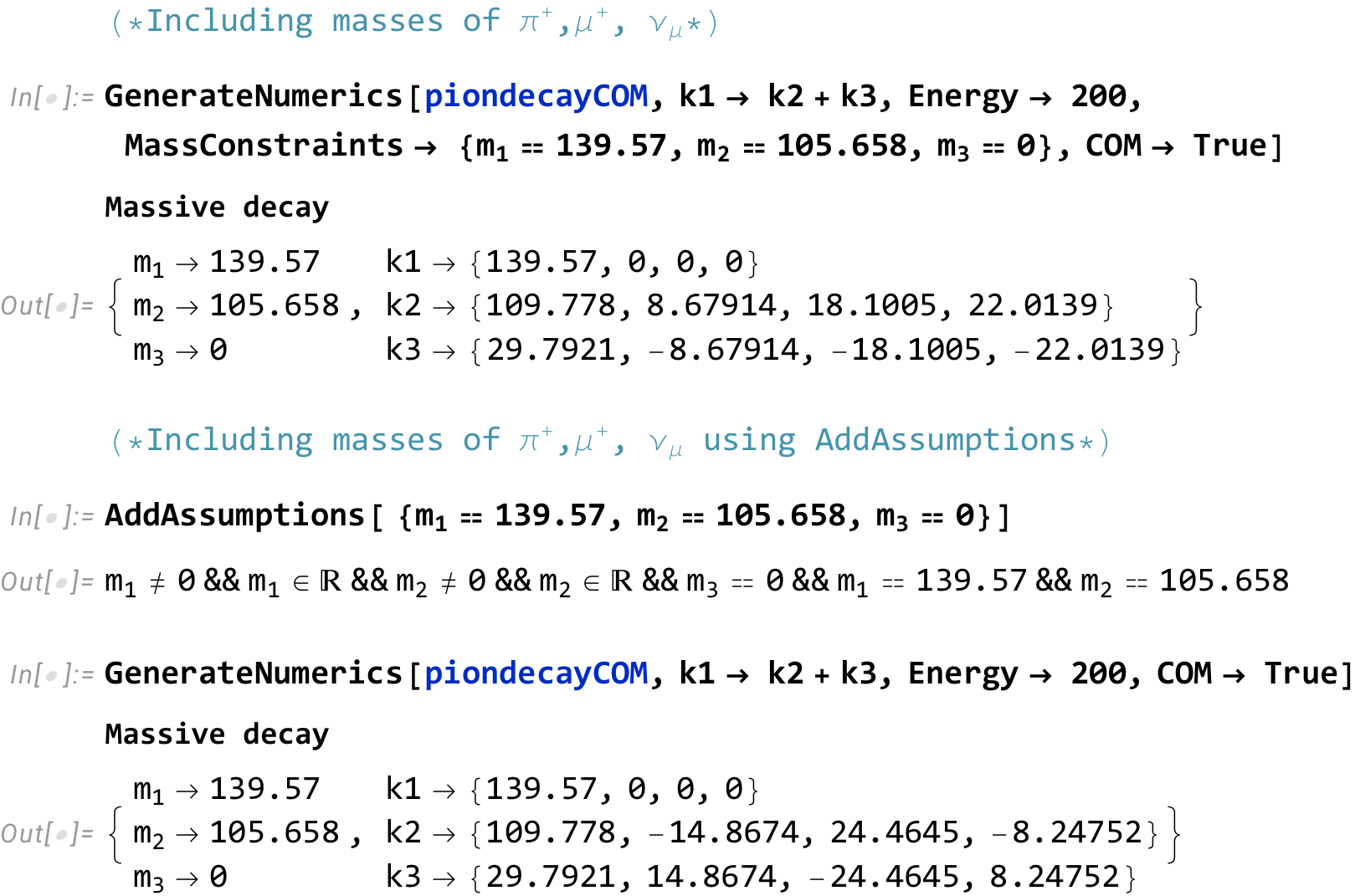}}
		
		In the example above, one can check that the energy of the neutrino is correctly generated as \tt{29.7921 MeV}\cite{Assamagan:1994mb}.
		
		\item \tt{Seed} : The \tt{label} is useful in identifying a particular numerical data if various numerics are generated. To regenerate a data with the same values, one can use \tt{Seed} whose value is either a string or an integer. For the same \tt{Seed} value, the same data can be reproduced. It's default value is \tt{Automatic} for which the generated data is completely random and irreproducible. 
		\redbox{\includegraphics[scale=0.35]{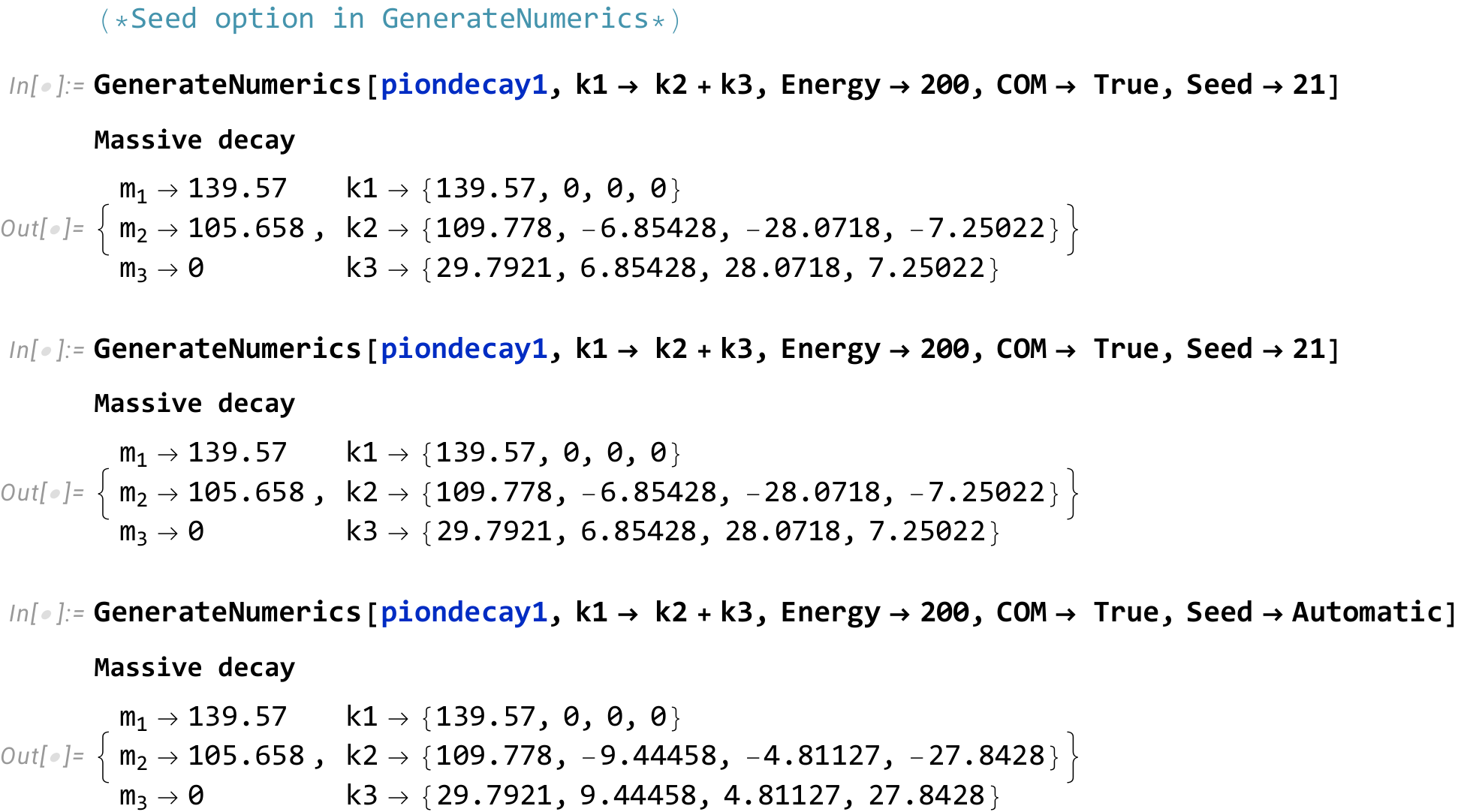}}
		
		\item \tt{Iterations} : This option can be used to do more iterations in the \tt{RAMBO} engine if required. Its default value is set to \tt{20}.
	\end{enumerate}
	All the labels are stored in the list \tt{NumericLabels}. The numerics can be cleared using \tt{ClearNumerics[ labels\_List]}. The default argument in \tt{ClearNumerics} is the list of all the labels \tt{NumericLabels}.
	\redbox{\includegraphics[scale=0.4]{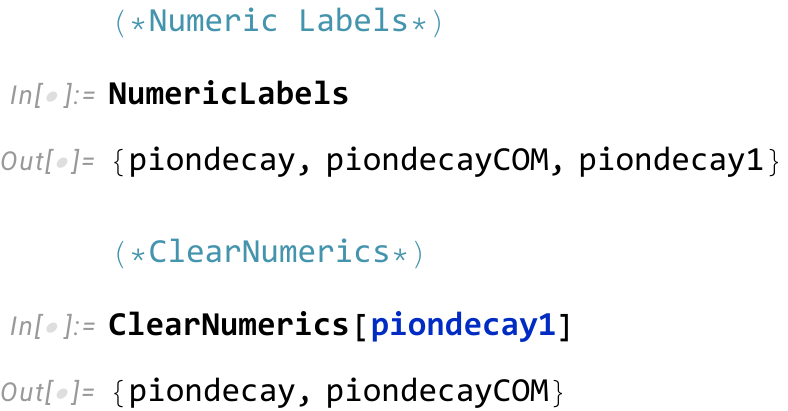}}
	
\end{morematerialbox}
\begin{documentationbox}{\tt{PutNumerics}}{\tt{PutNumerics[label\_][expr\_]}\quad \text{or}\quad \tt{PutNumerics[labels\_List][expr\_]}}
	Using these commands, The numerical data can be used in any expression whether it is in the SH notation or in the component form. 
\end{documentationbox}
\begin{examplebox}{\tt{PutNumerics}}
	We can show that the pion decay numerics are consistent with the on-shell conditions, momentum conservation and the spinor helicity relations.
	\redbox{\begin{minipage}{0.4\linewidth}
			\includegraphics[scale=0.35]{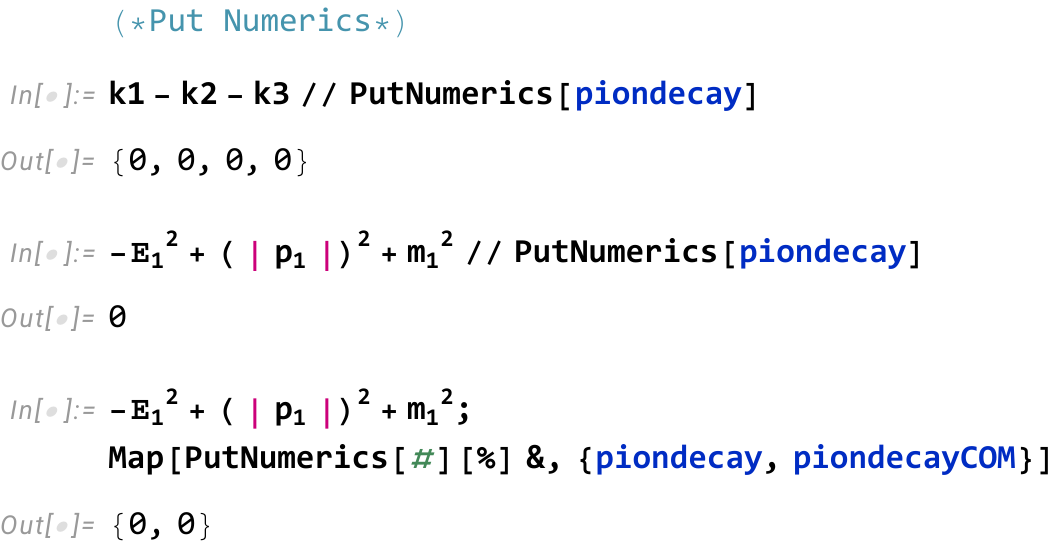}
		\end{minipage}
		\begin{minipage}{0.3\linewidth}
			\includegraphics[scale=0.35]{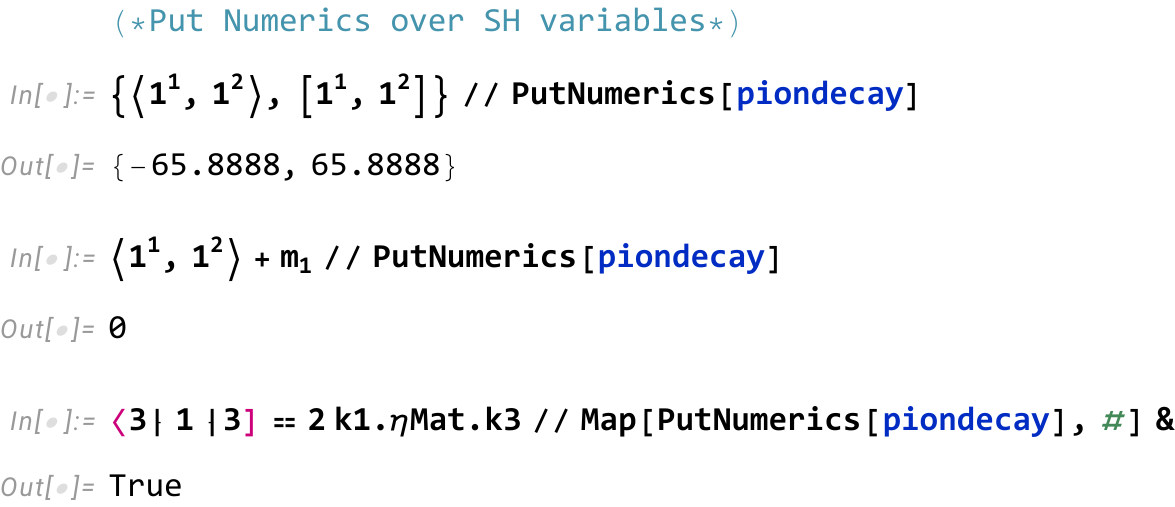}
		\end{minipage}
	}
\end{examplebox}

\begin{documentationbox}{\tt{MassNumerics, MomentaNumerics, ComponentNumerics}}{\tt{MassNumerics[label\_], MomentaNumerics[label\_], ComponentNumerics[label\_]}}
	The generated numerical data is stored in the following commands
	\begin{align}
		\mathtt{MassNumerics[label]}\quad,\quad \mathtt{MomentaNumerics[label]}\quad,\quad \mathtt{ComponentNumerics[label]}
	\end{align}
	As the name suggests, the \tt{MassNumerics} and \tt{MomentaNumerics} has the values for the masses generated against \tt{label}. These values are then passed on to the momenta components $\mathtt{p_0,p_1,p_2,p_3}$ and is stored in \tt{ComponentNumerics} with the same \tt{label}. 
\end{documentationbox}

\begin{examplebox}{\tt{MassNumerics, MomentaNumerics, ComponentNumerics}}
	For example, the numerics generated above for \tt{piondecay}, \tt{piondecayCOM} can be obtained as follows
	\redbox{\begin{minipage}{0.4\linewidth}
			\includegraphics[scale=0.2]{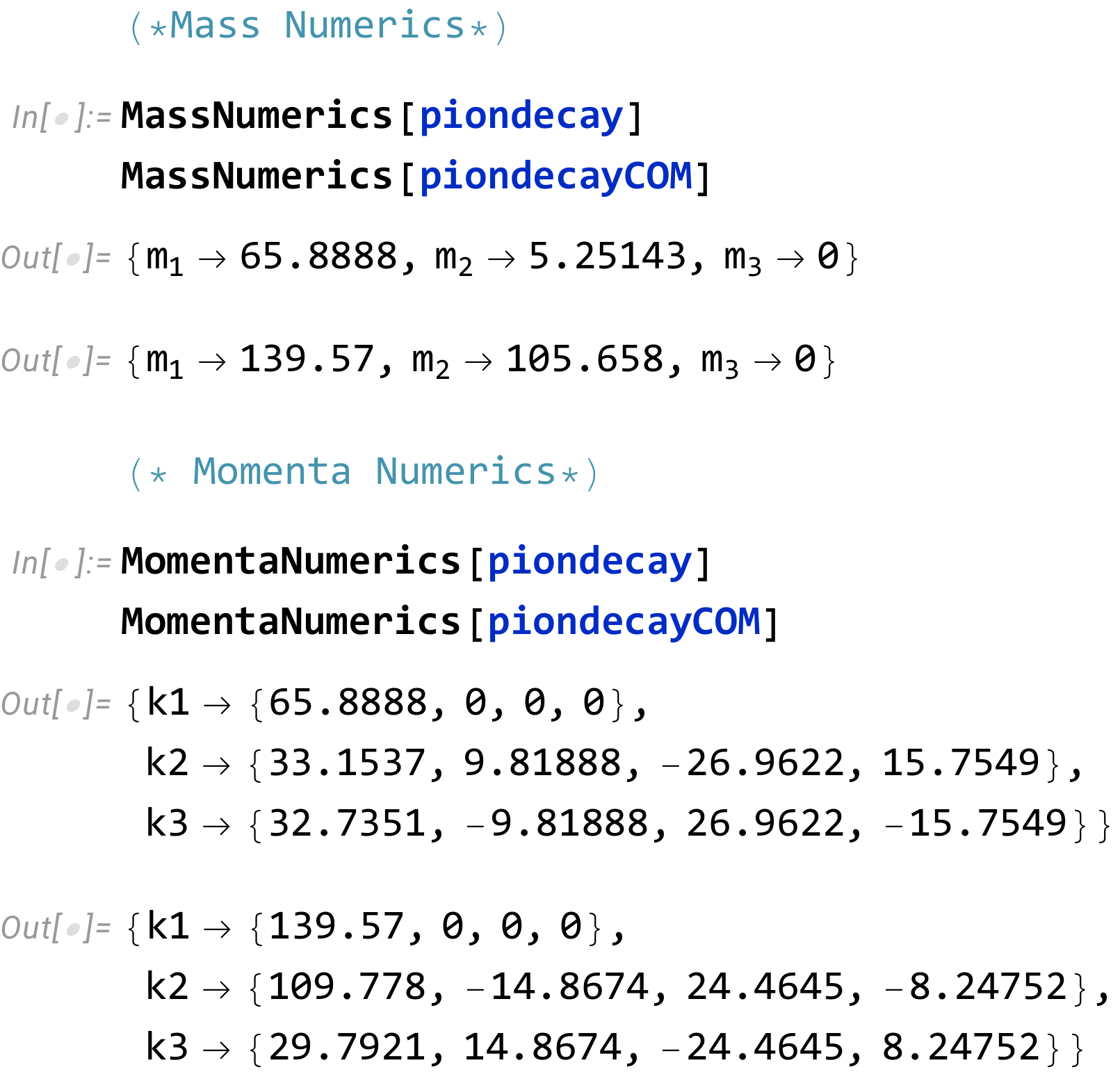}
		\end{minipage}
		\begin{minipage}{0.3\linewidth}
			\includegraphics[scale=0.25]{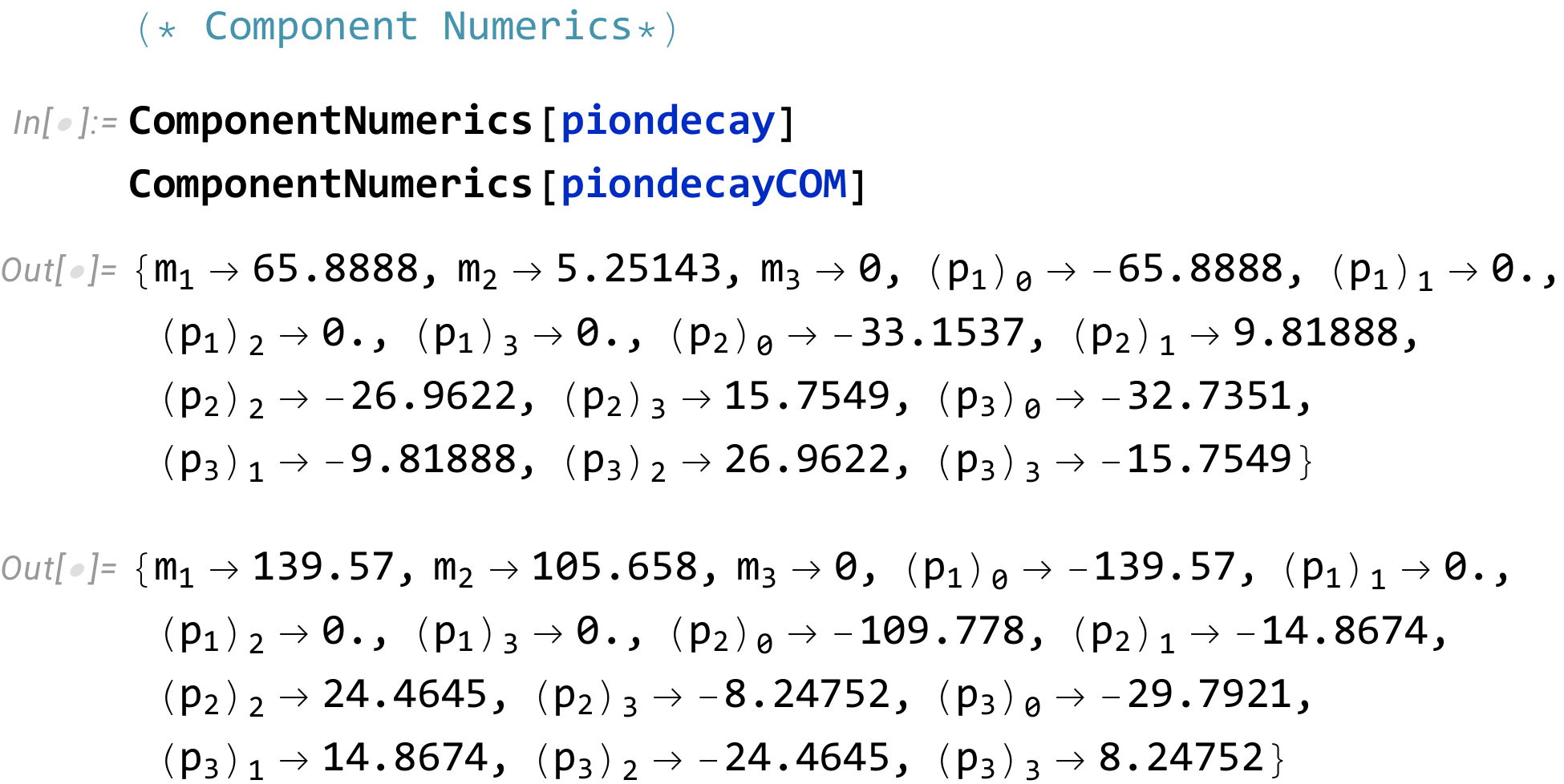}
		\end{minipage}
	}
\end{examplebox}
\textbf{Gauge invariance using numerics:} The numerical data can be used to do various consistency checks. One such beautiful check is the gauge invariance of an amplitude which in SH notation is the invariance under the change of reference spinor. As an example, consider the following expression which can be written independent of the reference spinor $\mathtt{r3}$ using a simple Schouten  identity.
\begin{align}
	\mathtt{term=\frac{\langle 1\,2\rangle  \langle 3\,\text{r3}\rangle -\langle 3\,2\rangle  \langle
			1\,\text{r3}\rangle }{\langle 2\,\text{r3}\rangle }= \langle 1\,3\rangle}
\end{align}
The same can be shown by choosing two different numerics for leg $\mathtt{r3}$ and showing that their difference vanishes. For example
\redbox{\begin{minipage}{0.5\linewidth}
		\includegraphics[scale=0.35]{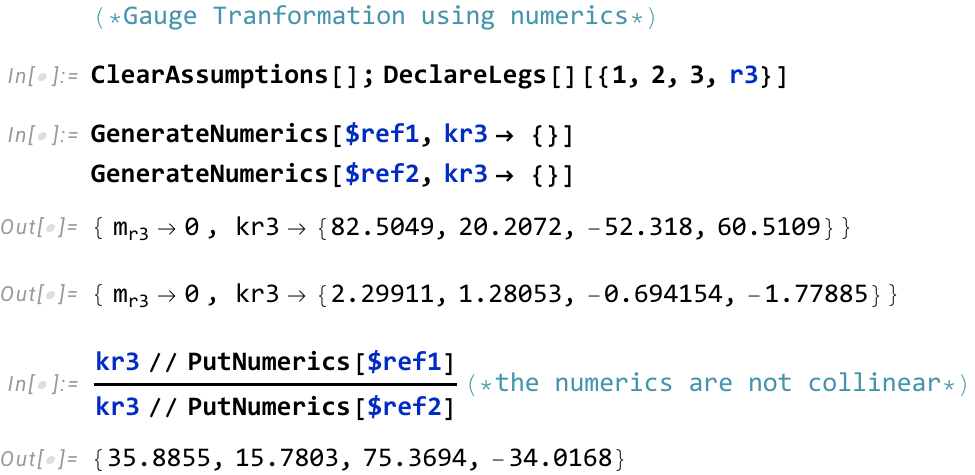}
	\end{minipage}
	\begin{minipage}{0.3\linewidth}
		\includegraphics[scale=0.35]{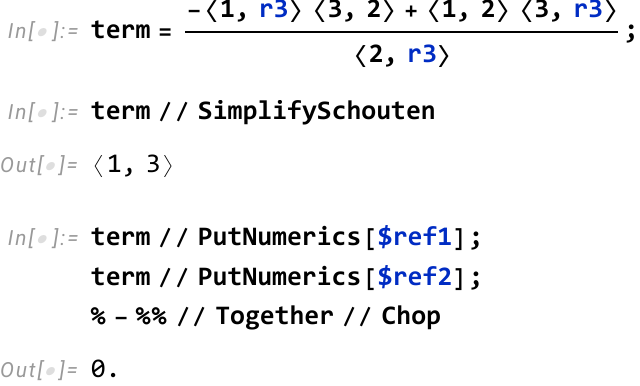}
	\end{minipage}
}

\subsubsection{Complex kinematics}
There are certain kinematic spaces which are unique and more interesting to the amplitudes community. For example, the fusion of massless particles and the three point scattering of a charged massive particles with a massless particle. Consider fusion of $\mathtt{n-1}$ massless particles to another massless particle
\begin{align}
	\mathtt{k_1+\cdots +k_{n-1}}\longrightarrow \mathtt{k_n}\label{ngluonfusion}
\end{align}
Since all $\mathtt{k_i^2=0}$, the momentum conservation gives that any two momenta must satisfy $\mathtt{k_i\cdot k_j=0}$. Since all momenta carry positive energy, it implies that they all must be collinear. (In the amplitudes community, it is a usual trend to take all the particles as incoming with some having negative energies. The amplitudes obtained using the kinematics \eqref{ngluonfusion} are related to the all incoming/outgoing amplitudes by crossing symmetry arguments. The processes where more than one massless particle are outgoing are also related by the crossing symmetry.) Since the momenta are collinear, the phase spaces for such processes collapses to a single line. Any amplitude you write down using the collinear momenta will vanish. If the momenta are real, this is same as all angles and boxes vanishing simultaneously as they are related by complex conjugation.
\begin{align}
	\mathtt{\tt{Real Momenta}\quad:\quad \langle i\, j\rangle =0=[i\, j]}\quad,\quad \forall \mathtt{i,j\in \{1,2,\cdots, n\}}
\end{align}
For example, consider the following massless fusion. 
\redbox{\includegraphics[scale=0.45]{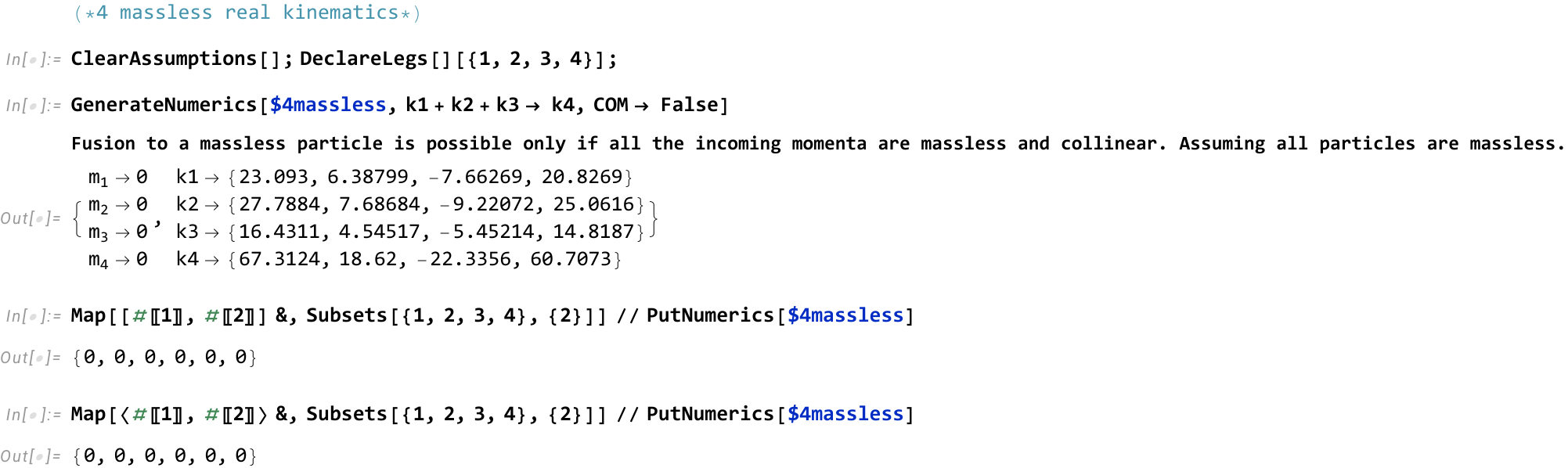}}
The example above clearly demonstrates that the angles and boxes vanish for real kinematics. To write down a non-vanishing massless amplitudes, the trick is to use complex momenta which are not hermitian. In SH notation, this implies that the angles and boxes are no more related by complex conjugation. So either all angles vanish or all boxes vanish.
\begin{align}
	\mathtt{\tt{Complex Momenta}\quad:\quad \langle i\, j\rangle =0 \quad\tt{or} \quad [i\, j]=0}\quad,\quad  \forall \mathtt{i,j\in \{1,2,\cdots, n\}}
\end{align}
In this section, we layout the commands to generate the complex momenta kinematics such that either the angles vanish or the boxes vanish. This is done using the complex shift where the momenta of the $\tt{n}$ legs is shifted by a null light-like momentum $\mathtt{q^\mu=(0,}\vec{\tt{q}}\tt{)}$ which is orthogonal to all other momenta
\begin{align}
	\mathtt{\forall i\quad\quad k_i\rightarrow k_i+\textit{z}_i q}\quad:\quad \mathtt{q^2=0\quad,\quad q\cdot k_i =0}\quad,\quad \mathtt{\sum_{i=1}^{n-1}\textit{z}_i=\textit{z}_{n}} \label{complexshift}
\end{align} 
The complex momenta $\mathtt{q^\mu}$ has two possible solutions : the helicity $\tt{-1}$ solution which can be written as $\tt{q}_{\alpha\dot{\alpha}}\mathtt{\propto|i]\langle q|}$ such that $\mathtt{\langle i\, q\rangle \neq 0} $. Whereas the helicity $\tt{+1}$ solution which can be written as $\tt{q}_{\alpha\dot{\alpha}}\mathtt{\propto|q]\langle i|}$ such that $\mathtt{[ i\, q] \neq 0} $. The helicity $\tt{-1}$ solution leads to a shift in the angles, whereas the boxes remain invariant. On the other hand, helicity $\tt{+1}$ solution leads to the angles being invariant, whereas the boxes are shifted. 
\begin{align}
	\mathtt{\tt{Helicity = -1}\quad:\quad \langle i| \rightarrow \langle i|+z_i\langle q|}\quad, \quad \mathtt{| i] \rightarrow | i]},\\
	\mathtt{\tt{Helicity = +1}\quad:\quad | i] \rightarrow | i]+z_i| q]}\quad, \quad \mathtt{\langle  i| \rightarrow \langle  i|}
\end{align}
After doing the shift it becomes easier to see that either angles vanish or boxes vanish. This shift is realized using the \tt{ComplexifyNumerics} command which has the following syntax
\begin{align}
	\mathtt{ComplexifyNumerics[q\_,label\_,MomConservation\_]}
\end{align}
The first input is the leg index for the light like null momentum \tt{q}. Running the command automatically defines the leg \tt{q} and its momenta. The second input is the \tt{label} used in GenerateNumerics to generate real numerics for the collinear process. The third input is the same momentum conservation rule which was used in \tt{GenerateNumerics}. The command also has two \tt{Options} : 

(1) \tt{Seed} : Setting $\mathtt{Seed\rightarrow}$ as integer or a string generates the same complexified numerics. The default is set to \tt{Automatic}. 

(2) $\mathtt{\$Helicity}$ : Using this option one can generate the numerics such that all angles vanish for $\mathtt{\$Helicity\rightarrow +1}$, whereas all boxes vanish for $\mathtt{\$Helicity\rightarrow -1}$. It's default is set as $\tt{+1}$.

To impose the complex kinematics, we can use the same \tt{PutNumerics} command as before by setting the following Option 
\begin{align}
	\mathtt{\$Kinematics\rightarrow``Complex"}
\end{align}
The default value of this option is $``\tt{Real}"$ which gives the real numerics. The same option can also be used in \tt{MassNumiercs, MomentaNumerics} and \tt{ComponentNumerics}. 

As an example, we can complexify the real numerics generated above for the four massless particle fusion.
\redbox{\includegraphics[scale=0.45]{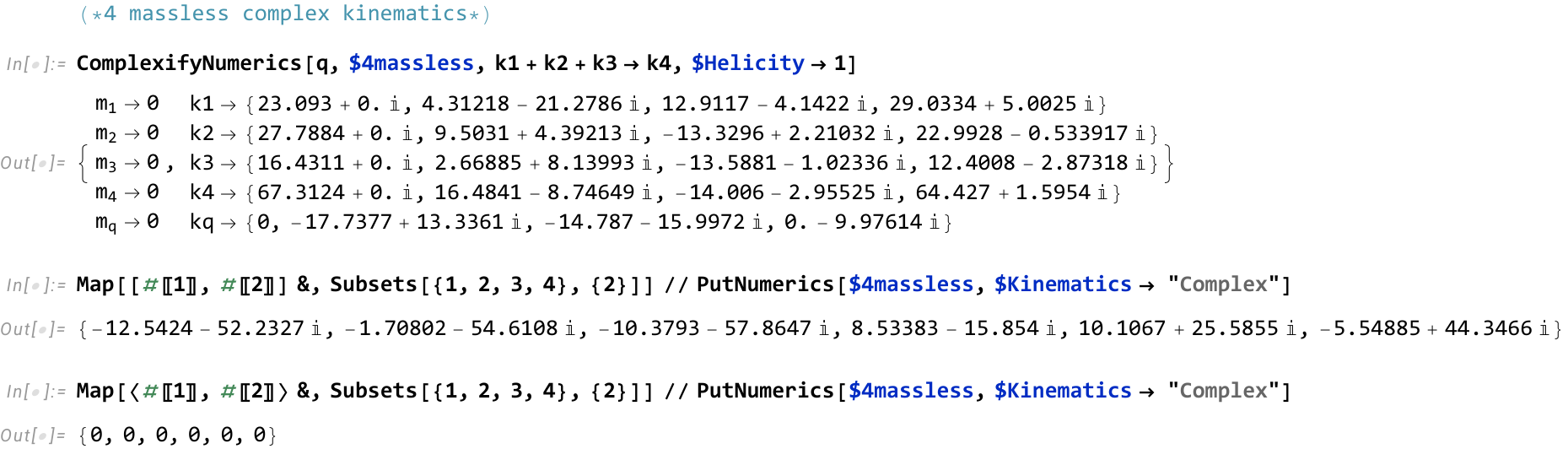}}
Flipping the \tt{\$Helicity} to $\tt{-1}$ makes all the boxes vanish, whereas the angles are non-zero.
\redbox{\includegraphics[scale=0.45]{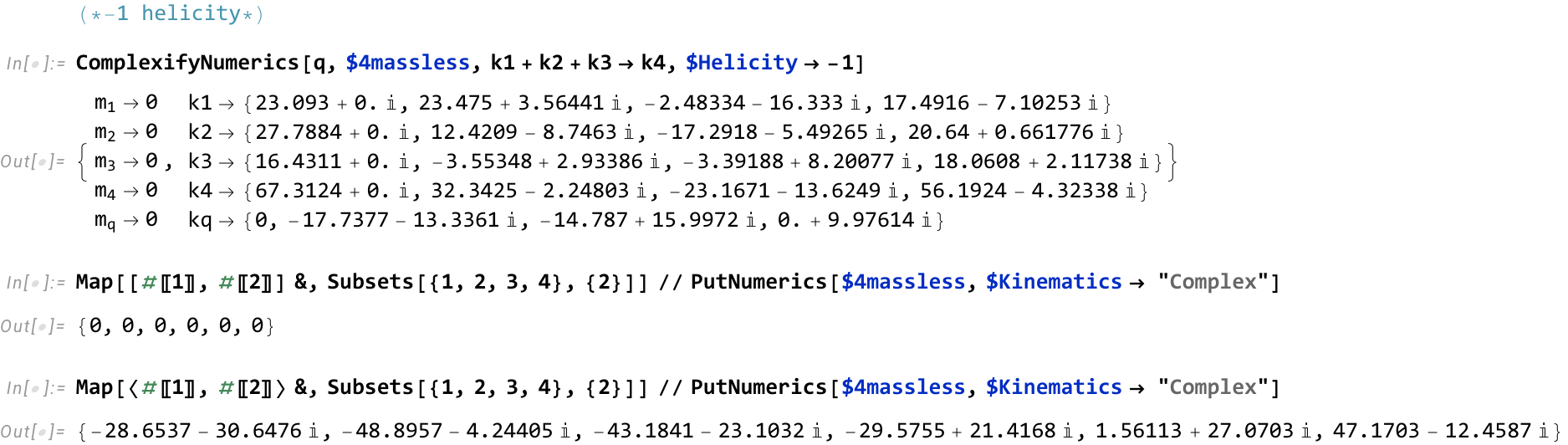}}

\subsection{High energy limit}\label{sec:HighEnergyLimit}
Another important aspect of the massive SH formalism is that it makes it easier to take the higher energy limit of the observables. The high energy limit has been discussed briefly in \cite{Arkani-Hamed:2017jhn} which we have reviewed in appendix \ref{app:BasisDecomposition}. In this section, we lay down the main commands relevant for analyzing the high energy behavior of any massive amplitude. 

\begin{documentationbox}{\tt{ToDimensionfullBasis}}{\tt{ToDimensionfullBasis[Massivelegs\_List]}}
	
This command decomposes the \tt{Massivelegs} in terms of the $\mathtt{SL(2,\mathbb{C})}$ and $\mathtt{SU(2)_{LG}}$ basis discussed in appendix \ref{app:BasisDecomposition}. 

\end{documentationbox}

The dimension full spinors can also be decomposed to the dimensionless spinors using the command \tt{DimFulltoDimLessBasis[Massivelegs\_List]} which we have discussed in appendix \ref{app:MoreonHighEnergy}. More such commands are also discussed in the same appendix.

\begin{examplebox}{\tt{ToDimensionfullBasis}}
	
\redbox{\includegraphics[scale=0.5]{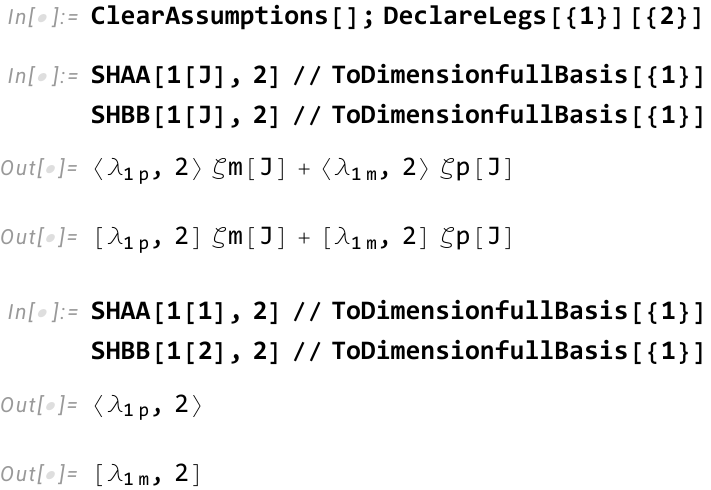}}

\end{examplebox}

\begin{conceptbox}{} 
We define the high energy limit for any particle \tt{i} as the regime where the mass $\mathtt{m_i}$ of the particle is much less than it's energy $\mathtt{E_i}$ (or magnitude of momentum $\mathtt{|p_i|}$).
\begin{align}
	\mathtt{\frac{m_i}{E_i}<< 1 \qquad \text{or} \qquad \frac{m_i}{|p_i|}<< 1}
\end{align}
In this limit, the dimension-full factors which appear in the massive spinors scales as follows
\begin{align}
	\mathtt{\sqrt{E_i+|p_i|}\longrightarrow \sqrt{2E_i} \qquad,\qquad 
		\sqrt{E_i-|p_i|}\longrightarrow \frac{m_i}{\sqrt{2E_i}}}
\end{align}
Therefore, at high energies, the spinors $|\lambda_{\tt{i+}}]_{\alpha}$, $\langle \lambda_{\tt{i-}}|_{\dot{\alpha}}$ scale as $\tt{m}_{\tt{i}}$ and vanish relative to the spinors $|\lambda_{\tt{i-}}]_{\alpha}$, $\langle \lambda_{\tt{i+}}|_{\dot{\alpha}}$. The high energy limit can be imposed over any SH expression involving massive legs using the command \tt{HighEnergyLimit}.
\end{conceptbox}

\begin{documentationbox}{\tt{HighEnergyLimit}}{\tt{HighEnergyLimit[rules\_List][expr\_]}}

Every rule in \tt{rules} is of the form $$\tt{Massiveleg}\rightarrow \tt{MasslesslegatHE}$$ where \tt{Massiveleg} is the massive leg whose high energy massless counterpart is the leg \tt{MasslesslegatHE}.
\end{documentationbox}

\begin{examplebox}{\tt{HighEnergyLimit}}

As an example, consider the scattering of a massive spin $1$ particle with two massless scalars. Let the massive leg be \tt{1} and the massless scalar legs be \tt{2,3}. The corresponding three point amplitude is given by
\begin{align}
\mathtt{\frac{\left\langle 1^a,3\right\rangle  \left\langle 1^b,2\right\rangle +\left\langle 1^a,2\right\rangle  \left\langle 1^b,3\right\rangle }{2
		\langle 2,3\rangle }}
\end{align}
To take the high energy limit we first decompose the amplitude in the ``$\lambda$" basis for different little group components i.e. $\mathtt{(a,b)\in\{(1,1),(1,2),(2,2)\}}$ and then impose the high energy scaling. Let's assume that at high energies the massive leg \tt{1} becomes the massless leg \tt{4}.

	\redbox{\includegraphics[scale=0.5]{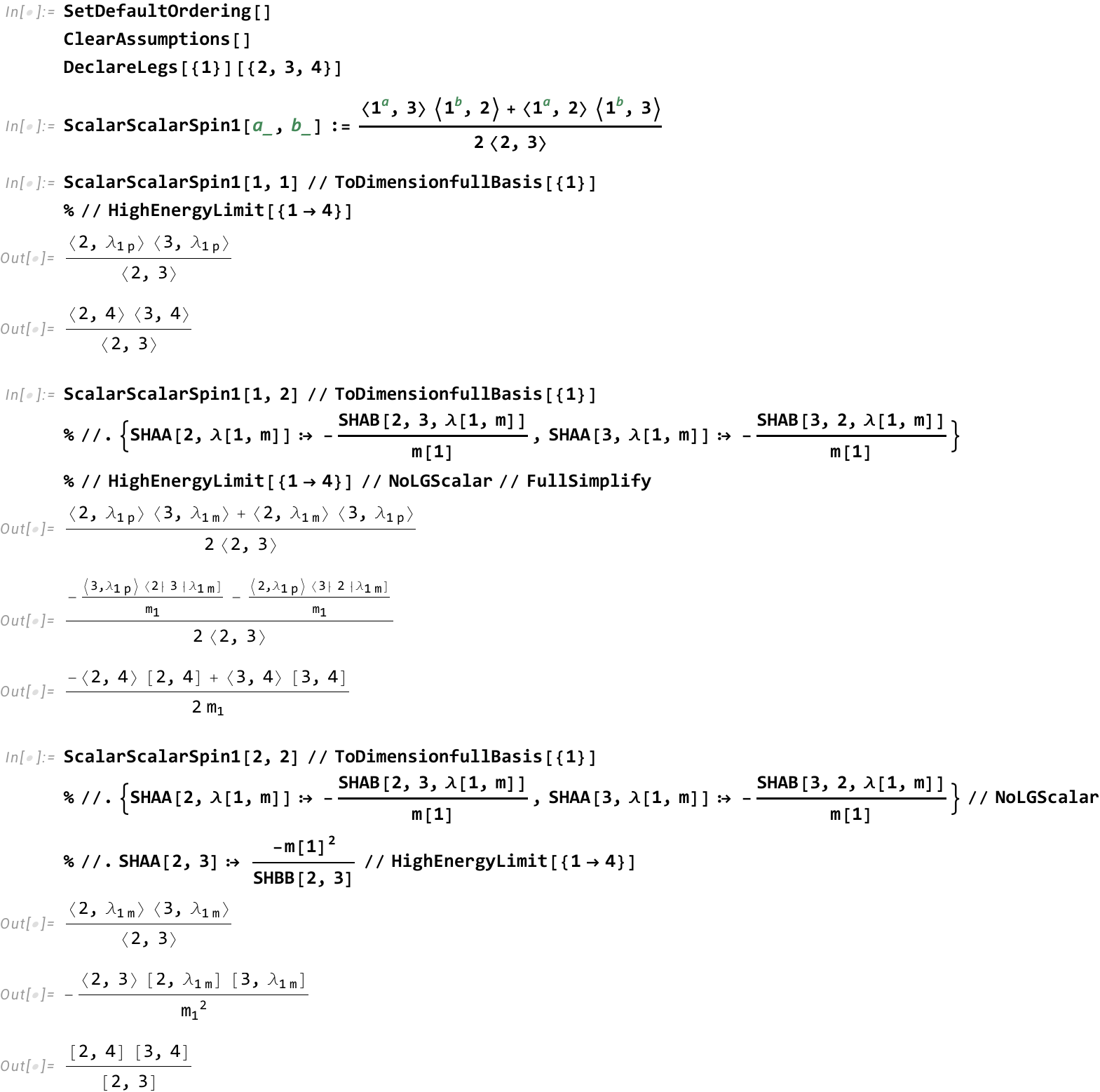}}

	The high energy limit for the little group components $\mathtt{(1,1)}$ is easier to find. It gives the all massless amplitude for negative helicity photon and two massless scalars. However, for $\mathtt{(a,b)\in \{(1,2),(2,2)\}}$, the computation requires a little manipulation using the three point kinematics. The $\mathtt{(2,2)}$ components reduces to all massless positive helicity photon-scalar-scalar amplitude. Whereas, the longitudinal mode $\mathtt{(1,2)}$ is subleading in $\mathtt{m_1}$.

\end{examplebox}

\begin{warningbox}{}

	It should be noted that the high energy limit works after choosing the explicit value of the little group indices within \tt{\{1,2\}}.

\end{warningbox}

\subsection{Lorentz to spinor-helicity conversions}\label{sec:LorentztoSH}

\begin{conceptbox}{Lorentz and spinor-helicity formalisms}
	
	The spinor-helicity formalism often simplifies computations and clarifies many
	physical arguments.  However, compared to the traditional Lorentz-vector
	formalism, it is a relatively new language.  Some computations that are
	difficult in spinor-helicity variables may be more transparent in Lorentz-vector
	notation.  It is therefore useful to convert expressions back and forth between
	the two formalisms.
	
	Conversion from Lorentz notation to spinor-helicity notation is straightforward,
	because every Lorentz tensor can be written in terms of spinor-index
	components, as in Eq.~\eqref{Vectortomatrixrepresentation}.   
	
\end{conceptbox}

\begin{documentationbox}{\tt{ConvertLortoSH}}{\tt{ConvertLortoSH[mslegs\_,mspols\_,msmom\_][mllegs\_,mlpols\_,mlmom\_,hels\_,refs\_,mlFS\_]}}

	converts Lorentz expressions written in
	terms of Lorentz vectors and tensors into spinor-helicity angle and box
	variables.  It can convert the spacetime metric, the four-dimensional
	Levi-Civita tensor, momenta, massive polarizations, massless polarizations, and
	massless field strengths.

	The first input contains the massive variables:
	\begin{itemize}
		\item \(\tt{mslegs}\): list of massive leg labels,
		\item \(\tt{mspols}\): list of massive-polarization heads,
		\item \(\tt{msmom}\): list of massive-momentum heads.
	\end{itemize}
	The second input contains the massless variables:
	\begin{itemize}
		\item \(\tt{mllegs}\): list of massless leg labels,
		\item \(\tt{mlpols}\): list of massless-polarization heads,
		\item \(\tt{mlmom}\): list of massless-momentum heads,
		\item \(\tt{hels}\): list of helicities,
		\item \(\tt{refs}\): list of reference spinors,
		\item \(\tt{mlFS}\): list of massless field-strength heads.
	\end{itemize}
	
\end{documentationbox}

\begin{examplebox}{\tt{ConvertLortoSH}}
	
	\redbox{
		\includegraphics[scale=0.5]{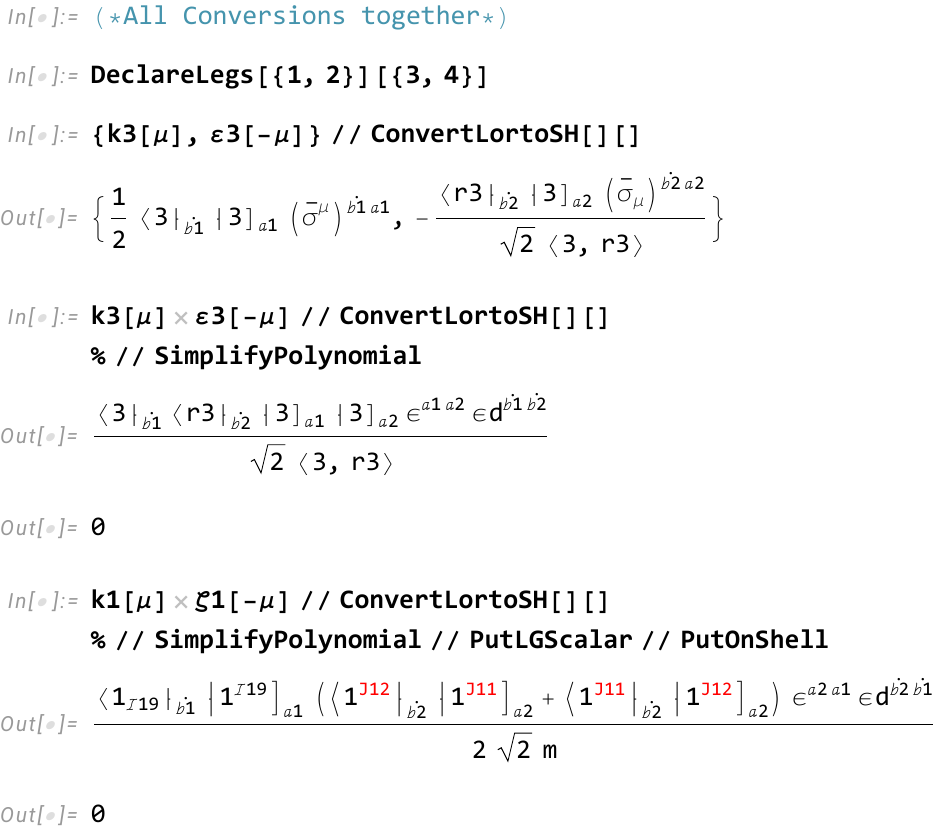}
	}
	
\end{examplebox}

\begin{morematerialbox}{}
	The command \(\tt{ConvertLortoSH}\) combines the following more specialized
	commands:
	\begin{enumerate}
		\item 
		\tt{ConvertMettoSH}: converts the Lorentz spacetime metric and the four-dimensional Levi-Civita tensor to SH expressions
		
		\item 
		\tt{ConvertMomtoSH}: converts momentum (written in terms of Lorentz vector) to SH expressions. 
		
		\item 
		\tt{ConvertMsPoltoSH}: converts massive polarization vector/tensors (written in terms of Lorentz vector) to SH expressions.
		
		\item 
		\tt{ConvertMlPoltoSH}: converts massless polarization (written in terms of Lorentz vector) to SH expressions.
		
		\item 
		\tt{ConvertMlFStoSH}: converts massless field strength (written in terms of Lorentz vector) to SH expressions.
	\end{enumerate}
	See \ref{sec:moreonLorentztoSH} for a more detailed documentation.
	
\end{morematerialbox}

%% file: Conclusion.tex
\section{Conclusion and future directions}
\label{sec:Conclusion}

\texttt{SMaSH} provides a collection of commands for carrying out both basic
and advanced computations in the massless and massive spinor-helicity
formalism. The main motivation behind the package is to provide a practical
platform for massive little-group covariant computations. The package is aimed
at a broad range of users, from beginners learning spinor-helicity methods to
experienced practitioners performing more involved amplitude calculations.

The input and output conventions are chosen to follow the modern notation used
in the spinor-helicity literature. The current version includes tools for
spinor-index and little-group-index manipulations, Lorentz-to-SH conversion,
generation of numerical kinematics, CPT transformations, gauge-invariance
checks, and amplitude computations.

This package sets up the basic platform on which one can build additional niche tools for amplitude computations. We have planned several improvements and extensions, which are as follows:
\begin{enumerate}
	\item Faster algorithm for simplifying expression using Schouten identity and external kinematics. This might require machine learning based implementation like \cite{SchwartzSpinorHelicity}
	
	\item Include spinor-chain simplification algorithms

	\item Generate independent basis for any n-point scattering of massive and massless particles
		
	\item Tools for studying classical limits of amplitudes
	
	\item More systematic tools for constructing and classifying contact terms
	
	\item Include dedicated tools for Standard Model and SMEFT amplitudes
\end{enumerate}

\paragraph{Acknowledgement}{The package has taken this shape over the course of two years. During these years, we have taken inputs from humans, previous packages as well as artificial intelligence tools. We thank Raj Patil for the valuable inputs on the preliminary versions of the package. We thank Prachiti P. Athalye her valuable inputs on this version of the draft. AK would like to thank IISER Bhopal for their institute funding in the first year of this project and providing hospitality for the remaining time. RS gratefully acknowledges support from CSIR. Some parts of this work have been presented by AK during \href{https://asianwinterschool20.github.io}{Asian Winter School 2025} held at IISER Bhopal, India.}

Finally, we are grateful to the people of India for their generous support for research in basic sciences.

%% file: App_Load.tex
\section{Load the package}
\label{sec:load}
The package can be installed using any one of the following ways.
\begin{enumerate}
	\item The easiest method is to download the \tt{.zip} file from the \href{https://github.com/aakash-kmr/SMaSH/archive/refs/heads/main.zip}{\tt{GitRepository}} and extract it in any folder. After extracting, open a \tt{Mathematica} notebook and follow these steps 
	\begin{align*}
		\begin{array}{c}
			\boxed{	\text{\tt{File}}}\\
			\downarrow\\
			\boxed{	\text{\tt{Install...}}}\\
			\downarrow\\
			\text{Choose \boxed{\tt{Packages}} from \boxed{\tt{Type of Item to Install}}}\\
			\downarrow\\
			\text{Choose \boxed{\tt{From File...}} from \boxed{\tt{Source}}}\\
			\downarrow\\
			\text{Choose the folder where the package is extracted, then select the file: "Kernel/init.m"}\\
			\downarrow\\
			\text{ Choose either \boxed{\tt{Install for this user only}} or \boxed{\tt{Install for all users}}}\\
			\downarrow\\
			\text{\boxed{\tt{OK}}}\\
			\downarrow\\
			\text{Run in the mathematica notebook: \tt{<<SMaSH\,$\grave{}$}}
		\end{array}
	\end{align*}
	\item To manually install the package first download the \tt{.zip} file from the \href{https://github.com/aakash-kmr/SMaSH/archive/refs/heads/main.zip}{\tt{GitRepository}}. Get the user base directory by running \tt{\$UserBaseDirectory} and then extract the \tt{.zip} file in the folder \tt{\$UserBaseDirectory$\backslash$Applications}. Then load the package by running \text{\tt{<<SMaSH\,$\grave{}$}}.
	\item To load the package for a particular notebook, download the \tt{.zip} file from the \href{https://github.com/aakash-kmr/SMaSH/archive/refs/heads/main.zip}{\tt{GitRepository}} and extract it in any folder. After extracting, open the \tt{Mathematica} notebook and paste the address of the folder within the command \tt{SetDirectory}. For example, if the package is extracted in the folder \tt{MyPackages} whose complete address is 
	\begin{center}
		\tt{C:Users$\backslash$HP$\backslash$Downloads$\backslash$MyPackages}
	\end{center}
	then simply run
	\begin{align*}
		\begin{array}{c}
			\tt{SetDirectory["C:Users$\backslash$HP$\backslash$Downloads$\backslash$MyPackages"]}\\
			\downarrow\\
			\text{\tt{<<SMaSH\,$\grave{}$}}
		\end{array}
	\end{align*}
\end{enumerate}
After loading the output should show the following
\redbox{\includegraphics[scale=0.5]{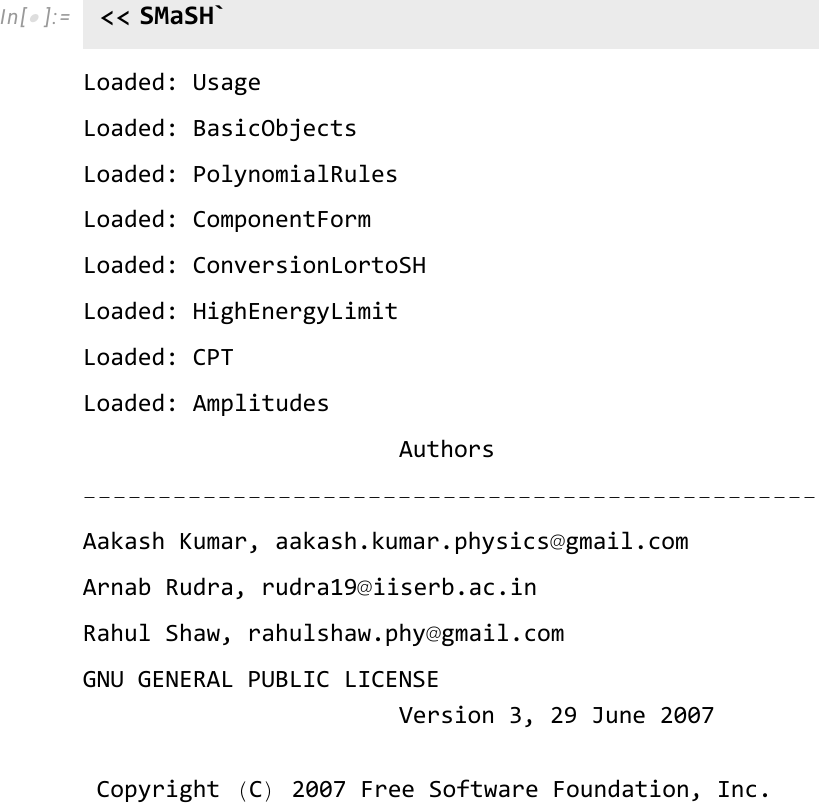}}

%% file: App_CPT.tex
\section{$2$-component helicity basis}\label{app:helicitybasis}

The smallest irreducible representations of the $\mathtt{SU(2)_L}$ and $\mathtt{SU(2)_R}$ are the left and the right handed spinors. These are the $(1/2,0)$ and $(0,1/2)$ Weyl representations of the Lorentz group. One way to understand these representations is to start with a four component Dirac spinor whose Lagrangian is given by
\begin{align}
	\mathcal{L}=\iimg\overline{\Psi}\Gamma^\mu (\partial_\mu \Psi)-m\overline{\Psi}\Psi
\end{align}
where $\overline{\Psi}=  \Psi^\dagger \beta$ is called the Dirac conjugate and $\beta$ is same as $\Gamma^0$ but with the different index structure i.e. $\beta=\begin{pmatrix}
	0&\delta^{\dot{\alpha}}{}_{\dot{\beta}} \\ \delta^{\alpha}{}_{\beta} &0
\end{pmatrix}$. The Dirac spinors satisfy the equation of motion $	(-\iimg \Gamma^\mu\partial_\mu+m)\Psi(x)=0$. The Dirac equation has two plane wave solutions which are given by
\begin{align}
	(\slashed{p}+m)u_\lambda(\vec{p})=0\quad,\quad (\slashed{p}-m)v_\lambda(\vec{p})=0
\end{align}
where $\slashed{p}=p_\mu\Gamma^\mu=\begin{pmatrix}
	0&p_{\alpha\dot{\beta}}\\ p^{\dot{\beta}\alpha}&0
\end{pmatrix}$. 
The subscript $\lambda$ denotes the helicity for a particle moving along the $z$ direction. In other words, the plane wave solutions have eigenvalue $+\frac{\lambda}{2}$ and $-\frac{\lambda}{2}$ under the helicity operator $\frac{\vec{p}\cdot \vec{J}}{|p|}$.
\begin{align}
	J_3 u_\lambda(\hat{z})=\frac{\lambda}{2}u_\lambda(\hat{z})\quad,\quad J_3 v_\lambda(\hat{z})=-\frac{\lambda}{2}v_\lambda(\hat{z})\qquad,\qquad \lambda=\pm\label{helicityoperator}
\end{align}
where $J_3=J^{12}=\frac{1}{2}\begin{pmatrix}
	\sigma_3&0\\ 0&\sigma_3
\end{pmatrix}$ is the angular momentum operator for a particle moving along $z$ direction. The solutions to \eqref{helicityoperator} are the $4$ component spinors which we chose as follows
\footnotesize{\begin{align}
		u_{+}(\hat{z})=\begin{pmatrix}
			-	\sqrt{E-|p|}\\0\\-\sqrt{E+|p|}\\0
		\end{pmatrix}\quad,\quad 
		u_{-}(\hat{z})=\begin{pmatrix}
			0\\\sqrt{E+|p|}\\0\\\sqrt{E-|p|}
		\end{pmatrix}\quad,\quad 
		v_{+}(\hat{z})=\begin{pmatrix}
			0\\\sqrt{E+|p|}\\0\\-\sqrt{E-|p|}
		\end{pmatrix}\quad,\quad 
		v_{-}(\hat{z})=\begin{pmatrix}
			\sqrt{E-|p|}\\0\\-\sqrt{E+|p|}\\0
		\end{pmatrix}	\label{Planewavesolutions}
\end{align}}
\normalsize
The energy $E_p$ of the particle is defined using $p^\mu=(E_p,\vec{p})$. For massive particles, these solutions are different from the choice of plane wave solutions made in \cite{Srednicki:2007qs,Elvang:2015rqa} at zero momentum. However, these are related by a boost along the $z$ axis. The solutions at zero momentum are also called spin polarizations because they are the eigenvalues of the spin operator $\vec{J}$. We will discuss more about this in the next section. For massless particles, the spin and helicity solutions match because there is no rest frame for massless particles and so only helicity is the Lorentz invariant label. One can also check that for massless particles $|p|=E$ implies $u_\lambda(\hat{z})=v_{-\lambda}(\hat{z})$. 

The plane wave solutions satisfy the following inner products
\begin{align}
	\overline{u}_\lambda(\hat{z})\cdot u_{\lambda'}(\hat{z})=2m\delta_{\lambda\lambda'}=- \overline{v}_\lambda(\hat{z})\cdot v_{\lambda'}(\hat{z}) \label{uvinnerproduct}
\end{align}
where $\overline{u}=u^\dagger \beta$ and $\overline{v}=v^\dagger\beta$. Since the inner product is a Lorentz scalar, it holds true for any momenta. The plane wave solutions at any arbitrary momentum $p^\mu$ are obtained by doing the following rotation which aligns the $z$ direction into the direction $\hat{p}$.
\begin{align}
	U(R(\hat{p}))=\exp{(-\iimg \phi J_{3})}\exp{(-\iimg \theta J_{2})}\exp{(\iimg \phi J_{3})}\qquad&,\qquad J_{i}=\epsilon_{ijk}J^{jk}=\frac{1}{2}\begin{pmatrix}
		\sigma_i&0\\ 0&\sigma_i
	\end{pmatrix},\label{rotationsztop}\\
	\exp{(-\iimg \phi\sigma_{3}/2 )}\exp{(-\iimg\theta\sigma_{2}/2)}\exp{(\iimg \phi \sigma_{3}/2)}&=\begin{pmatrix}
		\cos\frac{\theta}{2}  &-\sin\frac{\theta}{2} e^{-\iimg \phi}\\ \sin\frac{\theta}{2} e^{\iimg \phi} & \cos\frac{\theta}{2} 
	\end{pmatrix}\label{rotationsztop1}
\end{align}
One can also use a different rotation matrix with no initial rotation about the third axis. This extra rotation only gives rise to a phase factor $e^{\iimg \phi/2}$ or $e^{-\iimg \phi/2}$ multiplied to the polarizations. With the above defined rotations acting over the polarizations \eqref{Planewavesolutions}, we get the polarizations for massless particles moving in the direction $\hat{p}$.
\begin{align}
	u_+(\vec{p})=\begin{pmatrix}
		0\\ | p\rangle ^{\dot{\alpha }}
	\end{pmatrix}\quad,\quad u_-(\vec{p})=\begin{pmatrix}
		| p]_{\alpha }\\ 0
	\end{pmatrix}\quad,\quad v_+(\vec{p})=u_{-}(\vec{p})\quad,\quad v_-(\vec{p})=u_{+}(\vec{p})\label{MasslessDiractoangbox}
\end{align}
With the component form discussed in \ref{sec:ComponentForm}, it can be shown that both the sides have the same helicity for a momentum pointing along the $\hat{z}$ direction.
In case of massive particles, the polarizations at arbitrary momenta become
\begin{align}
	u_{+}(\vec{p}) =\begin{pmatrix}
		-| p^{1}]_{\alpha }\\ | p^1\rangle ^{\dot{\alpha }}
	\end{pmatrix}\quad,\quad	
	u_{-}(\vec{p}) =\begin{pmatrix}
		| p^{2}]_{\alpha }\\ -| p^2\rangle ^{\dot{\alpha }}
	\end{pmatrix}\quad,\quad 
	v_{+}(\vec{p})=\begin{pmatrix}
		| p^{2}]_{\alpha }\\ | p^2\rangle ^{\dot{\alpha }}
	\end{pmatrix}\quad,\quad 
	v_{-}(\vec{p})=\begin{pmatrix}
		| p^{1}]_{\alpha }\\ | p^1\rangle ^{\dot{\alpha }}
	\end{pmatrix}
\end{align}
It is also useful to write the polarizations at arbitrary momenta $p$ as a linear combination of the spinors $|p^{I}]$ and $|p^{I}\rangle$.
\begin{align}
	u_{\lambda}(\vec{p})= \zeta_{\lambda, I}\begin{pmatrix}
		| p^{I}]_{\alpha }\\ -| p^I\rangle ^{\dot{\alpha }}
	\end{pmatrix}\quad,\quad v_{\lambda}(\vec{p})=\lambda \zeta_{-\lambda, I}\begin{pmatrix}
		| p^{I}]_{\alpha }\\ | p^I\rangle ^{\dot{\alpha }}
	\end{pmatrix}\label{MassiveDiractoangbox}
\end{align}
As discussed before in \ref{sec:HighEnergyLimit}, in the high energy limit, the only non-vanishing scalars are $ | p^1\rangle\rightarrow|p\rangle$ and $ |p^2]\rightarrow|p]$, which matches with the massless spinors in \eqref{MasslessDiractoangbox}.
The $u_{\lambda}$ and $v_{\lambda}$ satisfy the completeness relations (which is a 4$\times$ 4 matrix)
\begin{align}
	\sum_{\lambda=\pm} u_\lambda(\vec{p})\overline{u}_\lambda(\vec{p})=-\slashed{p}+m=\begin{pmatrix}
		m\, \delta^{\beta}{}_{\alpha}&-p_{\alpha\dot{\beta}}\\ -p^{\dot{\beta}\alpha}&m\, \delta^{\dot{\alpha}}{}_{\dot{\beta}}
	\end{pmatrix}\quad,\quad \sum_{\lambda=\pm} v_\lambda(\vec{p})\overline{v}_\lambda(\vec{p})=-\slashed{p}-m=\begin{pmatrix}
		-m\, \delta^{\beta}{}_{\alpha}&-p_{\alpha\dot{\beta}}\\ -p^{\dot{\beta}\alpha}&-m\, \delta^{\dot{\alpha}}{}_{\dot{\beta}}
	\end{pmatrix}\label{uvcompleteness}
\end{align}
For massless particles, we just set $m=0$. The normalization in \eqref{uvinnerproduct} and the completeness relation \eqref{uvcompleteness} matches exactly with \cite{Srednicki:2007qs}.

Since we are working with the Weyl representations, it's easy to classify which of the spinors have positive and negative chirality. The chirality matrix is the Weyl basis is real and diagonal 
\begin{align}
	\Gamma_\star=\begin{pmatrix}
		-	\delta^{\alpha}{}_{\beta} &0\\0&\delta^{\dot{\beta}}{}_{\dot{\alpha}}
	\end{pmatrix} \label{chiralitymatrix}
\end{align}
Therefore, the upper component of a Dirac spinor always has negative chirality, whereas the lower component always has positive chirality.
This implies that for massless particles, the chirality and helicity eigenspaces must coincide i.e. $u_\lambda(\vec{p}),\, v_{-\lambda}(\vec{p})$ have $+\lambda$ helicity as well as $+\lambda$ chirality. For massive particles, the polarizations \eqref{MassiveDiractoangbox} do not have a definite chirality. Rather they are a linear combination of positive and negative chirality components. \textit{In general, it turns out that a box spinor has a negative chirality (left handed), whereas an angle spinor has a positive chirality (right handed).}

The linear combination of the plane wave solutions gives the full Dirac spinor with 4 independent complex components.
\begin{align}
	\Psi(x)&= \sum_{\lambda=\pm}\int \frac{d^3p}{(2\pi)^32E_p}\left(a_\lambda(\vec{p}) u_\lambda(\vec{p}) e^{\iimg p.x} +a^{c\dagger}_\lambda(\vec{p}) v_\lambda(\vec{p})e^{-\iimg p.x}\right),\label{DiracField}\\
	\overline{\Psi}(x)&= \sum_{\lambda=\pm}\int \frac{d^3p}{(2\pi)^32E_p}\left(a^\dagger_\lambda(\vec{p}) \overline{u}_\lambda(\vec{p}) e^{-\iimg p.x} +a^{c}_\lambda(\vec{p}) \overline{v}_\lambda(\vec{p})e^{\iimg p.x}\right)\label{DiracConjugatefield}
\end{align}
where is the Lorentz invariant measure. The operators $a_\lambda(\vec{p})$ and $a^{c\dagger}_\lambda(\vec{p})$ are the annihilation operators for the particles and antiparticles respectively. They satisfy the anti-commutation relations $\{a_\lambda(\vec{p}),a_{\lambda'}^{c\dagger}(\vec{p}')\}=(2\pi)^32E_p\delta^{3}(\vec{p}-\vec{p}')\delta_{\lambda\lambda'}$, whereas the other anti-commutators vanish. The linear combination \eqref{DiracField} is the unique Dirac field which can be used to compute interaction densities which commutes outside the light cone.
\subsection{Relation between spin and helicity basis}\label{app:spinbasis}
The helicity basis polarizations are eigenvectors of the helicity operator $\hat{p}\cdot \vec{J}$. In other words, the helicity is defined as the projection of the angular momentum along the direction of momentum. For massive particles, helicity is not a Lorentz invariant quantum label as it can flip it's sign under boosts\footnote{The same is not true for massless particles because they cannot be boosted such that $\hat{p}\rightarrow -\hat{p}$}. A Lorentz invariant label for massive particles is the spin label $s$ which is the eigenvalue of the spin operator ${\vec{J}}^2$. 
The polarizations in the spin basis (also known as the fixed axis basis) are chosen to be the eigenvectors of the spin matrix $J^{12}$ in the rest frame
\begin{align}
	J^{12}u'_s(0)=\frac{s}{2}u'_s(0)\quad,\quad J^{12}v'_s(0)=\frac{-s}{2}v'_s(0)\qquad,\qquad s=\pm
\end{align}
Let's make the following choice of these $4$ component polarizations (which is same as the choice made in \cite{Srednicki:2007qs,Weinberg:1995mt})
\begin{align}
	u'_{+}(0)=\sqrt{m}\begin{pmatrix}
		1\\0\\1\\0
	\end{pmatrix}\quad,\quad 
	u'_{-}(0)=\sqrt{m}\begin{pmatrix}
		0\\1\\0\\1
	\end{pmatrix}\quad,\quad 
	v'_{+}(0)=\sqrt{m}\begin{pmatrix}
		0\\1\\0\\-1
	\end{pmatrix}\quad,\quad 
	v'_{-}(0)=\sqrt{m}\begin{pmatrix}
		-1\\0\\1\\0
	\end{pmatrix}	\label{PlanewavesolutionsZeroMomenta}
\end{align}
\normalsize
The spin basis at an arbitrary momenta $p^\mu$ is achieved by doing the following boost transformation which takes the particle into the direction $\hat{p}$
\begin{align}
	U(B(\hat{p}))&=\exp(\iimg \eta \hat{p}\cdot \vec{K})=\frac{1}{\sqrt{2} \sqrt{m} \sqrt{E+m}}\begin{pmatrix}
		-p_{\alpha\dot{\alpha}}+m&0\\0&-p^{\dot{\alpha}\alpha}+m
	\end{pmatrix},
\end{align}
where $K^{i}=J^{i0} =\frac{\iimg}{2}\begin{pmatrix}
	\sigma_i&0\\ 0&-\sigma_i
\end{pmatrix}$. The polarizations at arbitrary $p^\mu$ are

\begin{align}
	u'_{+}(\vec{p})=\begin{pmatrix}
		\frac{E+m-|p| \cos\theta}{\sqrt{2} \sqrt{E+m}} \\
		-\frac{|p| e^{i \phi } \sin\theta}{\sqrt{2} \sqrt{E+m}} \\
		\frac{E+m+|p| \cos\theta}{\sqrt{2} \sqrt{E+m}} \\
		\frac{|p| e^{i \phi } \sin\theta}{\sqrt{2} \sqrt{E+m}} 
	\end{pmatrix}
	\quad&,\quad 	
	u'_{-}(\vec{p})=\begin{pmatrix}
		-\frac{|p| e^{-i \phi } \sin\theta}{\sqrt{2} \sqrt{E+m}} \\
		\frac{E+m+|p| \cos\theta}{\sqrt{2} \sqrt{E+m}} \\
		\frac{|p| e^{-i \phi } \sin\theta}{\sqrt{2} \sqrt{E+m}} \\
		\frac{E+m-|p| \cos\theta}{\sqrt{2} \sqrt{E+m}} 
	\end{pmatrix}
	\\
	v'_{+}(\vec{p})=\begin{pmatrix}
		-\frac{|p| e^{-i \phi } \sin\theta}{\sqrt{2} \sqrt{E+m}} \\
		\frac{E+m+|p| \cos\theta}{\sqrt{2} \sqrt{E+m}} \\
		-\frac{|p| e^{-i \phi } \sin\theta}{\sqrt{2} \sqrt{E+m}} \\
		-\frac{E+m-|p| \cos\theta}{\sqrt{2} \sqrt{E+m}} 
	\end{pmatrix}
	\quad&,\quad  
	v'_{-}(\vec{p})=\begin{pmatrix}
		-\frac{E+m-|p| \cos\theta}{\sqrt{2} \sqrt{E+m}} \\
		\frac{|p| e^{i \phi } \sin\theta}{\sqrt{2} \sqrt{E+m}} \\
		\frac{E+m+|p| \cos\theta}{\sqrt{2} \sqrt{E+m}} \\
		\frac{|p| e^{i \phi } \sin\theta}{\sqrt{2} \sqrt{E+m}} 
	\end{pmatrix}
\end{align}
It then becomes easy to find the relation between the spin and helicity basis polarizations which are as follows
\begin{align}
	u'_{s}(\vec{p})=\sum_{\lambda=\pm}X_{s}{}^{\lambda}\, u_{\lambda}(\vec{p})\quad&,\quad v'_{s}(\vec{p})=\sum_{\lambda=\pm}(X^*)_{s}{}^{\lambda}\, v_{\lambda}(\vec{p}),\label{spintohelicity}\\
	u_{\lambda}(\vec{p})=\sum_{s=\pm}(X^\dagger)_{\lambda}{}^{s}\, u'_{s}(\vec{p})\quad&,\quad v_{\lambda}(\vec{p})=\sum_{\lambda=\pm}(X^T)_{\lambda}{}^{s}\, v'_{s}(\vec{p}) \label{helicitytospin}
\end{align}
where $X_{s}{}^{\lambda}$ is the $2\times 2$ unitary matrix given by
\begin{align}
	X_{s}{}^\lambda&=-\frac{\sqrt{E+|p|}+\sqrt{E-|p|}}{\sqrt{2} \sqrt{E+m}}\begin{pmatrix}
		\cos\frac{\theta}{2}  &-\sin\frac{\theta}{2} e^{\iimg \phi}\\ \sin\frac{\theta}{2} e^{-\iimg \phi} & \cos\frac{\theta}{2} 
	\end{pmatrix},\label{Xrotations}\\
	&X_{s}^{\lambda}(X^\dagger)_{\lambda}{}^{s'}=\delta_{s}{}^{s'}\qquad,\qquad 	(X^\dagger)_{\lambda}{}^{s}X_{s}^{\lambda'}=\delta_{\lambda}{}^{\lambda'}
\end{align}
One can always chose a different normalization for the polarizations such that the prefactor in the matrix $X$ is identity. In the matrix notation, the relations are
\begin{align}
	u'=Xu \quad,\quad  v'=X^*v \quad,\quad  u=X^\dagger u' \quad,\quad v=X^T v'
\end{align}
Note that $X^*$ is proportional as the rotation matrix defined in \eqref{rotationsztop1}. The matrix $X$ is actually the $SU(2)$ little group spin half representation which acts on the little group basis spinors $\zeta_{\pm}$ defined in \ref{sec:HighEnergyLimit}. Therefore, the helicity and spin basis are related by redefining the $SU(2)$ basis spinors as
\begin{align}
	\zeta'_{s}=X_{s}{}^{\lambda}\zeta_{\lambda}
\end{align}

The full Dirac spinor $\Psi(x)$ is independent of the choice of basis for it's polarizations because the polarizations are summed over. If we rewrite \eqref{DiracField} in terms of the spin basis polarizations $u_s$ and $v_s$ as
\begin{align}
	\Psi(x)&= \sum_{s=\pm}\int \frac{d^3p}{(2\pi)^32E_p}\left(a'_s(\vec{p}) u'_s(\vec{p}) e^{\iimg p.x} +a'^{c\dagger}_s(\vec{p}) v'_s(\vec{p})e^{-\iimg p.x}\right) \label{DiracFieldSpinBasis}
\end{align}
where we have taken a different set of creation and annihilation operators $a'_s(\vec{p})$ and $a'^{c\dagger}_s(\vec{p})$. Using \eqref{helicitytospin} and then comparing with \eqref{DiracField}, this new set of operators must be related to $a,a^{c\dagger}$ by the same rotations
\begin{align}
	a'_{s}(\vec{p})=\sum_{\lambda=\pm}X_{s}{}^{\lambda}\, a_{\lambda}(\vec{p})\quad&,\quad a'^{c\dagger}_{s}(\vec{p})=\sum_{\lambda=\pm}(X^*)_{s}{}^{\lambda}\, a^{c\dagger}_{\lambda}(\vec{p}),\\
	a_{\lambda}(\vec{p})=\sum_{s=\pm}(X^*)_{\lambda}{}^{s}\, a'_{s}(\vec{p})\quad&,\quad a^{c\dagger}_{\lambda}(\vec{p})=\sum_{s=\pm}X_{\lambda}{}^{s}\, a'^{c\dagger}_{s}(\vec{p})
\end{align}

\subsection{Decomposition to $SU(2)$ basis}\label{app:BasisDecomposition}
A massive helicity spinor can be decomposed to the basis spinors of $\mathtt{SL(2,\mathbb{C})}$ and $\mathtt{SU(2)_{LG}}$. In the angle and box notation, the decomposition is as follows
\begin{align}
	\mathtt{|i^{I}]_\alpha=|\lambda_{i+}]_\alpha \zeta_-^{I}+|\lambda_{i-}]_\alpha\zeta_+^{I}},\\
	\mathtt{\langle i^{I}]}_{\dot{\alpha}}=\mathtt{\langle\lambda_{i-}|}_{\dot{\alpha}} \mathtt{\zeta_+^{I}+\langle\lambda_{i+}|}_{\dot{\alpha}}\zeta_-^{\tt{I}}
\end{align}
The massless spinors $|\lambda_{\tt{i}\pm}]_\alpha$ and $\langle\lambda_{\tt{i}\pm} |^{\dot{\alpha}}$ are dimensionfull spinors which can be further written in terms of the dimensionless spinors defined as $|\chi_{\tt{i}\pm}]_\alpha$ and $\langle\chi_{\tt{i}\pm} |^{\dot{\alpha}}$.
\begin{align}
	\mathtt{|\lambda_{i\pm}]_{\alpha}=\sqrt{E_{i}\mp |p_i|} \, |\chi_{i\pm}]_{\alpha} }\qquad,\qquad 
	\mathtt{\langle\lambda_{i\pm}|}_{\dot{\alpha}}\mathtt{=\sqrt{E_{i}\pm |p_i|} \, \langle\chi_{i\pm}| }_{\dot{\alpha}}
\end{align}
The dimensionless spinors $ |\chi_{\tt{i}\pm}]_{\alpha}$ and $\langle\chi_{\tt{i}\pm}|_{\dot{\alpha}} $ are the $\mathtt{SU(2)_L}$ and $\mathtt{SU(2)_R}$ basis spinors whose component forms are given by
\begin{align}
	\mathtt{
		|\chi_{i+}]_\alpha=\begin{pmatrix}\tt{c}_\tt{i}\\	\tt{s}_\tt{i}\end{pmatrix}\quad,\quad 
		|\chi_{i-}]_\alpha=\begin{pmatrix} \tt{s}^*_\tt{i}\\-\tt{c}_\tt{i}\end{pmatrix}\quad,\quad 	[\chi_{i+}|^\alpha=\begin{pmatrix}\mathtt{s_i}&-\mathtt{c_i}\end{pmatrix}\quad,\quad 
		[\chi_{i-}|^\alpha=\begin{pmatrix}-\tt{c}_\tt{i}&-\tt{s}^*_\tt{i}\end{pmatrix}
	},
	\\
	\mathtt{
		|\chi_{i+}\rangle
	}^{\dot{\alpha}}\mathtt{=\begin{pmatrix} -\mathtt{c_i}\\-\mathtt{s_i}\end{pmatrix}}\quad,\quad  		
	\mathtt{|\chi_{i-}\rangle}^{\dot{\alpha}}\mathtt{=\begin{pmatrix} -\mathtt{s^*_i}\\\mathtt{c_i} \end{pmatrix}}\quad,\quad \mathtt{\langle\chi_{i+}|}_{\dot{\alpha} }\mathtt{=\begin{pmatrix} \mathtt{s_i}& -\mathtt{c_i} 	\end{pmatrix} }\quad,\quad  \mathtt{	\langle\chi_{i-}|}_{\dot{\alpha} }\mathtt{=\begin{pmatrix} -\mathtt{c_i}& -\mathtt{s^*_i} \end{pmatrix} },
\end{align}
where $\mathtt{c_i=\cos\frac{\theta_i}{2}}$ and $\mathtt{s_i=e^{\iimg \phi_i}\sin\frac{\theta_i}{2}}$.
The basis elements for $\mathtt{SU(2)_R}$ and $\mathtt{SU(2)_L}$ are related by the rules of complex conjugation $\mathtt{\langle\chi_{i\pm}|}_{\dot{\alpha}}\mathtt{=\pm (|\chi_{i\mp}]_{\alpha})}^{\dagger}$. The $\zeta_{\pm}^{I}$ are the basis vectors which transform under the fundamental representation of the little group $\mathtt{SU(2)}$.
\begin{align}
	\mathtt{\zeta_-^{I}=\begin{pmatrix}
			1&0
		\end{pmatrix}\quad,\quad 	\zeta_+^{I}=\begin{pmatrix}
			0&1
		\end{pmatrix}\quad,\quad 	\zeta_{-I}=\begin{pmatrix}
			0\\1
		\end{pmatrix}\quad,\quad 	\zeta_{+I}=\begin{pmatrix}
			-1\\0
	\end{pmatrix}}
\end{align}
The basis elements satisfy the following inner product relations
\begin{align}
	\mathtt{\braket{\chi_+\chi_-}=[\chi_+\chi_-]=1\qquad,\qquad \zeta^I_+\zeta_{-I}=1=-\zeta^I_-\zeta_{+I}}\label{BasisInnerPrioduct}
\end{align}

\subsection{$C,P,T$}\label{app:CPT}
Together with the continuous symmetries, it is also important to classify the action of discrete symmetries over the spinors. The $CPT$ theorem states that there exists an anti--linear - anti-unitary transformations which reverses the spacetime direction and keep the hermitian and Lorentz-invariant term in the action invariant. The proof of the $CPT$ theorem is out of scope of this paper. But the reader is encouraged to see the following references\cite{Luders1957,Jost1957,Pauli1955,StreaterWightman2000,Greenberg2003}. In this section, we will give a brief review of charge conjugation $C$, parity (aka space-inversions) $P$ and the time reversal transformations $T$. Their action over the fields have been computed rigorously at various places. However, their realization over the polarizations in the helicity basis hasn't been covered in many standard texts.

\textbf{Charge Conjugation:} Let's start with the charge conjugation transformation which maps a particle to its antiparticle. It is a linear map over the Hilbert space. So it's only job is to flip the charge of the corresponding state. If the state is uncharged (like Majorana states or states of a real photon), it leaves them unchanged. So the action of charge conjugation over the one particle states (in the spin basis) of species $n$ is as follows (see \cite{Srednicki:2007qs, Weinberg:1995mt} for derivation)
\begin{align}
	\tt{C}\quad:\quad |\vec{p},s,n\rangle\rightarrow \eta_{_\mathtt{C}}|\vec{p},s, n^c\rangle
\end{align}
where $n^c$ denotes that it is a state with opposite charge i.e. an antiparticle. The factor $\eta_{_\mathtt{C}}$ is the charge conjugate intrinsic parity with unit magnitude. Similarly, $\eta_{_\mathtt{C}}^{c}$ is the intrinsic parity for a state of species $n^{c}$. If the states are annihilated and created by the operators $a'_s(\vec{p})$ and $a'^{s}_\lambda(\vec{p})$ from the vacuum (which remains invariant under any transformation), then the action of charge conjugation over these operators must be given by
\begin{align}
	\tt{C}\quad:\quad \begin{cases}
		a'_{s}(\vec{p})\rightarrow \eta_{_\mathtt{C}}^{*}a'^c_s(\vec{p})\quad,\quad a'^{\dagger}_{s}(\vec{p})\rightarrow \eta_{_\mathtt{C}}a'^{c\dagger}_s(\vec{p}),\\
		a'^{c}_{s}(\vec{p})\rightarrow \eta_{_\mathtt{C}}^{c*}a'_s(\vec{p})\quad,\quad a'^{c\dagger}_{s}(\vec{p})\rightarrow \eta_{_\mathtt{C}}^{c}a'^{\dagger}_s(\vec{p})
	\end{cases}
\end{align}
This implies that the action over the full Dirac field \eqref{DiracFieldSpinBasis} is given by
\begin{align}
	\tt{C}\quad:\quad \Psi(x)\rightarrow \eta_{_\mathtt{C}}^*\mathcal{B}^{-1}\Psi(x)^*\quad,\quad \eta_{_\mathtt{C}}^c=\eta_{_\mathtt{C}}^* \label{ChargeConjugationDirac}
\end{align}
where $\mathcal{B}$ and it's inverse are given by
\begin{align}
	\mathcal{B}=\sigma_2\otimes \sigma_2= \begin{pmatrix}
		0&&\epsilon^{\dot{\alpha}\dot{\beta}}\\ \epsilon_{\alpha\beta}&&0
	\end{pmatrix}\implies \mathcal{B}^{-1}=\begin{pmatrix}
		0&&\epsilon_{\alpha\beta}\\ \epsilon^{\dot{\alpha}\dot{\beta}}&&0
	\end{pmatrix}\label{Bmatrix}
\end{align}
It is possible to write the action of charge conjugation over the Dirac spinors as \eqref{ChargeConjugationDirac} because the polarizations in the spin basis satisfy the relation
\begin{align}
	u'_s(\vec{p})=\mathcal{B}^{-1}v'_s(\vec{p})^*\quad,\quad v'_s(\vec{p})=\mathcal{B}^{-1}u'_s(\vec{p})^* \label{ChargeConjugationSpinPolarizations}
\end{align}
Note that unlike the fields and the creation \& annihilation operators, the relations which the polarizations satisfy are independent of the intrinsic parities. Using the basis change discussed in \eqref{helicitytospin}, the same transformations are true for the helicity basis polarizations.
\begin{align}
	u_\lambda(\vec{p})=\mathcal{B}^{-1}v_\lambda(\vec{p})^*\quad,\quad v_\lambda(\vec{p})=\mathcal{B}^{-1}u_\lambda(\vec{p})^* \label{ChargeConjugationHelicityPolarizations}
\end{align}
Therefore, the action of charge conjugation over the polarizations can be formally defined by \eqref{ChargeConjugationHelicityPolarizations} as a map. Using \eqref{MasslessDiractoangbox} and \eqref{MassiveDiractoangbox}, it is easy to see that the action of charge-conjugation over the spinor helicity variables $|p]$ and $|p\rangle$ is same as replacing $p$ by $\tilde{p}$
\begin{align}
	\text{Massless}\quad&:\quad \begin{cases}
		|p]_{\alpha}\rightarrow |\tilde{p}]_{\alpha}\quad,\quad 		|p\rangle_{\dot{\alpha}}\rightarrow |\tilde{p}\rangle_{\dot{\alpha}}
	\end{cases},\label{ChargeConjugationMassless}\\
	\text{Massive}\quad&:\quad \begin{cases}
		|p^{I}]_{\alpha}\rightarrow |\tilde{p}^{I}]_{\alpha}\quad,\quad		|p^{I}\rangle_{\dot{\alpha}}\rightarrow |\tilde{p}^{I}\rangle_{\dot{\alpha}}
	\end{cases} \label{ChargeConjugationMassive}
\end{align}
Thus, for the action of charge conjugation to be trivial over real polarizations, we get $p=\tilde{p}$ which is same as the condition for real momenta.
\newline

\textbf{Parity:} Parity $\tt{P}$ is a linear transformation which commutes with the spatial rotations but anticommutes with the boosts. So it's action in the spin basis is such that it doesn't flip the spin of the particle but flips it's direction of motion.
\begin{align}
	\tt{P}\quad:\quad |\vec{p},s,n\rangle \rightarrow \eta_{_\mathtt{P}}|-\vec{p},s,n\rangle 
\end{align}
where $\eta_{_\mathtt{P}}$ is the intrinsic parity with magnitude $1$. The intrinsic parity for the antiparticle state is $\eta_{_\mathtt{P}}^c$. Similar to the action over the states, the action of parity over the creation and annihilation operators is given by
\begin{align}
	\tt{P}\quad:\quad \begin{cases}
		a'_{s}(\vec{p})\rightarrow \eta_{_\mathtt{P}}^{*}a'_s(-\vec{p})\quad,\quad a'^{\dagger}_{s}(\vec{p})\rightarrow \eta_{_\mathtt{P}}a'^{\dagger}_s(-\vec{p}),\\
		a'^{c}_{s}(\vec{p})\rightarrow \eta_{_\mathtt{P}}^{c*}a'^c_s(-\vec{p})\quad,\quad a'^{c\dagger}_{s}(\vec{p})\rightarrow \eta_{_\mathtt{P}}^{c}a'^{c\dagger}_s(-\vec{p})
	\end{cases}
\end{align}
The polarizations in the spin basis satisfy the following inversion properties
\begin{align}
	u'_{s}(-\vec{p})=\beta u'_s(\vec{p})\quad,\quad v'_{s}(-\vec{p})=-\beta v'_s(\vec{p}) \label{ParitySpinPolarizations}
\end{align}
So the action over the full Dirac field is given by
\begin{align}
	\tt{P}\quad:\quad \Psi(x)\rightarrow \eta_{_\mathtt{P}}^* \beta\Psi(\mathcal{P}x)\quad,\quad \eta_{_\mathtt{P}}^c=-\eta_{_\mathtt{P}}^*
\end{align}
In the helicity basis, the polarizations satisfy a little non-simple relation. First we note that the unitary rotations $X(p)$ discussed in \eqref{Xrotations} satisfy the following properties under space-inversions $(\vec{p}\rightarrow -\vec{p}, \theta\rightarrow \pi-\theta, \phi\rightarrow \phi+\pi)$,
\begin{align}
	X(\mathcal{P}p)=X(p)Y(p)\qquad\text{where}\qquad Y(p)=\begin{pmatrix}
		0&e^{\iimg \phi}\\ -e^{-\iimg \phi}&0
	\end{pmatrix}
\end{align}
This implies that the helicity basis polarizations would satisfy the following space inversion properties
\begin{align}
	u_\lambda(-\vec{p})=\beta\,(Y^{\dagger})_{\lambda}{}^{\lambda'}u_{\lambda'}(p) \quad&,\quad  v_\lambda(-\vec{p})=-\beta\,(Y^{T})_{\lambda}{}^{\lambda'}v_{\lambda'}(p) \label{ParityHelicityPolarizations1}\\
	\implies  u_{\lambda}(-\vec{p})=-\lambda\beta e^{\lambda\iimg \phi}u_{-\lambda}(\vec{p}) \quad&,\quad  \implies v_\lambda(-\vec{p})=\lambda\beta e^{-\lambda \iimg \phi}v_{-\lambda}(\vec{p}) \label{ParityHelicityPolarizations2}
\end{align}
Since parity operator anticommutes with the momentum operator, it flips the sign of the helicity operator $\hat{P}\cdot \vec{J}$. Therefore, it takes a particle with helicity $\lambda$ moving in the direction $\hat{p}$ direction to a particle with helicity $-\lambda$ moving in the opposite direction $-\hat{p}$. This is very clear from \eqref{ParityHelicityPolarizations2}. 
Since under parity $e^{\pm \iimg \phi}\rightarrow -e^{\pm \iimg \phi} $, it is easy to show that the parity squares to $+1$. The corresponding mappings for angle and box spinors are as follows
\begin{align}
	\text{Massless}\quad&:\quad |p]_{\alpha}\rightarrow e^{-\iimg \phi}|p\rangle ^{\dot{\alpha}}\quad,\quad |p\rangle^{\dot{\alpha}}\rightarrow -e^{\iimg \phi}|p]_{\alpha},\\
	\text{Massive}\quad&:\quad\begin{cases}
		|p^{1}]_{\alpha}\rightarrow e^{\iimg \phi} |p_1\rangle^{\dot{\alpha}}\quad,\quad |p^{2}]_{\alpha}\rightarrow e^{-\iimg \phi} |p_2\rangle^{\dot{\alpha}},\\
		|p^{1}\rangle^{\dot{\alpha}}\rightarrow e^{\iimg \phi} |p_1]_{\alpha}\quad,\quad |p^{2}\rangle^{\dot{\alpha}}\rightarrow e^{-\iimg \phi} |p_2]_{\alpha}
	\end{cases}
\end{align}

In the first look, the property \eqref{ParityHelicityPolarizations2} doesn't look as simple as \eqref{ParitySpinPolarizations} because of the phase factor $e^{\iimg \phi}$ which is frame dependent. We would be better off with a property where the overall factors are not explicitly momentum dependent. Note that we can always do a phase rotation over the polarizations preserving their equation of motion and completeness relation
\begin{align}
	u_\lambda(\vec{p})\rightarrow e^{\lambda \iimg \phi/2}u_{\lambda}(\vec{p})&\implies u_\lambda(-\vec{p})\rightarrow \lambda \iimg e^{\lambda \iimg \phi/2}u_{\lambda}(-\vec{p}),\label{Helicitybasisphasetransformation1}\\
	v_\lambda(\vec{p})\rightarrow e^{-\lambda \iimg \phi/2}v_{\lambda}(\vec{p})&\implies v_\lambda(-\vec{p})\rightarrow -\lambda \iimg e^{-\lambda \iimg \phi/2}v_{\lambda}(-\vec{p})\label{Helicitybasisphasetransformation2}
\end{align}
In terms of the angles and boxes, this is same as transforming $|p^1]$ and $|p^2]$ with opposite phases. Using complex-conjugation, this results in $|p^1]$ and $|p^{1}\rangle$ with the same phase rotation. After doing these redefinitions, the angle and box spinors transforms as follows
\begin{align}
	\text{Massless}\quad&:\quad  |p]_{\alpha}\rightarrow \iimg |p\rangle ^{\dot{\alpha}}\quad,\quad |p\rangle^{\dot{\alpha}}\rightarrow \iimg |p]_{\alpha},\label{ParityMasslessAngBox}\\
	\text{Massive}\quad&:\quad\begin{cases}
		|p^{1}]_{\alpha}\rightarrow \iimg |p^2\rangle^{\dot{\alpha}}\quad,\quad |p^{2}]_{\alpha}\rightarrow \iimg |p^1\rangle^{\dot{\alpha}},\\
		|p^{1}\rangle^{\dot{\alpha}}\rightarrow \iimg |p^2]_{\alpha}\quad,\quad |p^{2}\rangle^{\dot{\alpha}}\rightarrow \iimg |p^1]_{\alpha}
	\end{cases}\label{ParityMassiveAngBox}
\end{align}
In little group covariant way, the transformations for massive spinors can be written as follows
\begin{align}
	|p^{I}]_{\alpha}\rightarrow \iimg\Sigma^{I}{}_{J}|p^{J}\rangle^{\dot{\alpha}}\quad,\quad |p^{I}\rangle^{\dot{\alpha}}\rightarrow \iimg\Sigma^{I}{}_{J} |p^{J}]_{\alpha}
\end{align}
where $\Sigma^{I}{}_{J}$ is a real and symmetric $2\times 2$ matrix defined as
\begin{align}
	\Sigma^{I}{}_{J}\equiv (\zeta^{I}_{-}\zeta_{-,J}-\zeta^{I}_{+}\zeta_{+,J})&=\begin{pmatrix}
		0&1\\1&0
	\end{pmatrix}=\sigma_1,\label{SigmaDefinition}\\
	\Sigma_{IJ}=\epsilon_{IK}\Sigma^{K}{}_{J}=-\sigma_3\quad&,\quad  \Sigma^{IJ}=\epsilon^{JK}\Sigma^{I}{}_K =\sigma_3,\label{SigmaProperties1}\\
	\Sigma_{I}{}^{J}=\epsilon_{IK}\epsilon^{JL}\Sigma^{K}{}_{L}=\sigma_1\quad&,\quad \Sigma_{IJ}=\Sigma_{JI}\quad,\quad \Sigma_{I}{}^{J}=\Sigma^{J}{}_{I}  \label{SigmaProperties2},\\
	\Sigma^{IK}\Sigma_{KJ}=-\delta^{I}{}_{J}\quad&,\quad \Sigma^{I}{}_K\Sigma^{KJ}=-\epsilon^{IJ},\\
	\Sigma_{IK}\Sigma^{KJ}=-\delta_{I}{}^{J}\quad&,\quad  \Sigma^{I}{}_{K}\Sigma^{K}{}_J=\delta^{I}{}_{J}\quad,\quad  \Sigma_{IK}\Sigma^{K}{}_{J}=\epsilon_{IJ}
\end{align}
The raising and lowering only happens using the $\epsilon_{IJ}$, whereas identity is always given by $\delta^{I}{}_J$. These can also written in terms of the little group basis elements as follows
\begin{align}
	\delta^{I}{}_{J}&= (\zeta^{I}_{+}\zeta_{-,J}-\zeta^{I}_{-}\zeta_{+,J})=\begin{pmatrix}
		1&0\\0&1
	\end{pmatrix},\\
	\epsilon_{IJ}&= (\zeta_{+,I}\zeta_{-,J}-\zeta_{-,I}\zeta_{+,J})=\begin{pmatrix}
		0&-1\\1&0
	\end{pmatrix},
\end{align}

\textbf{Time Reversal:} Time reversal \tt{T} is an anti-linear transformation which commutes with boosts and anti-commutes with the rotations and momentum. Therefore, it reverses the direction of motion of particle. It flips the spin of the particle but preserves it's helicity. It's action over the one-particle state in the spin basis is given by
\begin{align}
	\tt{T}\quad:\quad |\vec{p},s,n\rangle \rightarrow s \,\eta_{_\mathtt{T}}|-\vec{p},-s,n\rangle 
\end{align}
where $\eta_{_\mathtt{T}}$ is the intrinsic parity with unit magnitude. The intrinsic parity for the antiparticle state is $\eta_{_\mathtt{T}}^c$. Similar to parity and charge conjugation, the action of time reversal over the creation and annihilation operators is 
\begin{align}
	\tt{T}\quad:\quad \begin{cases}
		a'_{s}(\vec{p})\rightarrow s \,\eta_{_\mathtt{T}}^{*}a'_{-s}(-\vec{p})\quad,\quad a'^{\dagger}_{s}(\vec{p})\rightarrow s \,\eta_{_\mathtt{T}}a'^{\dagger}_{-s}(-\vec{p}),\\
		a'^{c}_{s}(\vec{p})\rightarrow s \,\eta_{_\mathtt{T}}^{c*}a'^c_{-s}(-\vec{p})\quad,\quad a'^{c\dagger}_{s}(\vec{p})\rightarrow s \,\eta_{_\mathtt{T}}^{c}a'^{c\dagger}_{-s}(-\vec{p})
	\end{cases}
\end{align}
Whereas it's action over the Dirac field \eqref{DiracFieldSpinBasis} is given by
\begin{align}
	\tt{T}\quad:\quad \Psi(x)\rightarrow -\eta_{_\mathtt{T}}^* \mathcal{C}\Gamma_\star\Psi(-\mathcal{P}x)\quad,\quad \eta_{_\mathtt{T}}^c=-\eta_{_\mathtt{T}}^*
\end{align}
where we have used the following property of spin polarizations
\begin{align}
	u'_{-s}(-\vec{p})^*=-s \mathcal{C}\Gamma_\star u'_{s}(\vec{p})\quad,\quad v'_{-s}(-\vec{p})^*=-s \mathcal{C}\Gamma_\star v'_{s}(\vec{p})\label{TimeReversalSpinPolarizations}
\end{align}
The LHS clearly showes that under time-reversal, the spin is flipped and the direction of momentum is reversed. The complex-conjugation is due to the anti-linear nature of time reversal. The matrix $\mathcal{C}$ is called the charge conjugate matrix which maps a representation to it's inverse-transpose. It is given by
\begin{align}
	\mathcal{C}=\begin{pmatrix}
		-\epsilon^{\alpha\beta}&0\\0&-\epsilon_{\dot{\alpha}\dot{\beta}}
	\end{pmatrix}\quad,\quad \mathcal{B}=\mathcal{C}\beta
\end{align}
The property \eqref{TimeReversalSpinPolarizations} is not an independent relation, rather it is a combination of \eqref{ChargeConjugationSpinPolarizations} and \eqref{ParitySpinPolarizations} with the chirality relation
\begin{align}
	\Gamma_\star u'_{s}(\vec{p})=s v'_{-s}(\vec{p})\quad,\quad \Gamma_\star v'_{s}(\vec{p})=-s u'_{-s}(\vec{p})
\end{align}
In the helicity basis, the relation \eqref{TimeReversalSpinPolarizations} translates to
\begin{align}
	u_{\lambda}(-\vec{p})^*= e^{-\lambda \iimg \phi}\mathcal{C}\Gamma_\star u_{\lambda}(\vec{p})\quad,\quad v_{\lambda}(-\vec{p})^*= e^{\lambda \iimg \phi}\mathcal{C}\Gamma_\star v_{\lambda}(\vec{p})
\end{align}
We can again do the phase transformation \eqref{Helicitybasisphasetransformation1}-\eqref{Helicitybasisphasetransformation2} to replace the phase factor $e^{\pm \lambda\iimg \phi}$ with factors which are independent of the momentum. 
\begin{align}
	u_{\lambda}(-\vec{p})^*= \iimg \lambda \mathcal{C}\Gamma_\star u_{\lambda}(\vec{p})\quad,\quad v_{\lambda}(-\vec{p})^*= -\iimg\lambda \mathcal{C}\Gamma_\star v_{\lambda}(\vec{p})
\end{align}
In the angle box notations, we can understand the time reversal transformation as
\begin{align}
	\text{Massless}\quad&:\quad  |p]_{\alpha}\rightarrow -\iimg  [{p}|^{\alpha}\quad,\quad |p\rangle^{\dot{\alpha}}\rightarrow -\iimg  \langle {p}|_{\dot{\alpha}},\label{TimeReversalMasslessAngBox}\\
	\text{Massive}\quad&:\quad\begin{cases}
		|p^1]_{\alpha}\rightarrow \iimg  [{p}^1|^{\alpha}\quad,\quad |p^2]_{\alpha}\rightarrow -\iimg  [{p}^2|^{\alpha},\\
		|p^1\rangle^{\dot{\alpha}}\rightarrow -\iimg  \langle {p}^1|_{\dot{\alpha}}\quad,\quad |p^2\rangle^{\dot{\alpha}}\rightarrow \iimg  \langle {p}^2|_{\dot{\alpha}}
	\end{cases}\label{TimeReversalMassiveAngBox}
\end{align}
In a $\mathtt{SU(2)_{LG}}$ covariant way, the transformations of massive spinors can be written similar to the parity
\begin{align}
	\tt{T}\quad:\quad |p^{I}]_{\alpha}\rightarrow \iimg \Omega^{I}{}_{J}[{p}^{J}|^{\alpha} \quad,\quad |p^{I}\rangle^{\dot{\alpha}}\rightarrow -\iimg \Omega^{I}{}_{J} \langle {p}^{J}|_{\dot{\alpha}}
\end{align}
where $\Omega^{I}{}_{J}$ is the diagonal real and symmetric $2\times 2$ matrix defined as 
\begin{align}
	\Omega^{I}{}_{J}=-(\zeta^{I}_{+}\zeta_{-,J}+\zeta^{I}_{-}\zeta_{+,J})&=\begin{pmatrix}
		1&0\\0&-1
	\end{pmatrix}\label{OmegaDefinition}\\
	\Omega_{IJ}=\epsilon_{IK}\Omega^{K}{}_{J}=\sigma_1\quad&,\quad  \Omega^{IJ}=\epsilon^{JK}\Omega^{I}{}_K =-\sigma_1,\label{OmegaProperties1}\\
	\Omega_{I}{}^{J}=\epsilon_{IK}\epsilon^{JL}\Omega^{K}{}_{L}=\sigma_3\quad&,\quad \Omega_{IJ}=\Omega_{JI}\quad,\quad \Omega_{I}{}^{J}=\Omega^{J}{}_{I}  \label{OmegaProperties2},\\
	\Omega^{IK}\Omega_{KJ}=-\delta^{I}{}_{J}\quad&,\quad \Omega^{I}{}_K\Omega^{KJ}=-\epsilon^{IJ},\\
	\Omega_{IK}\Omega^{KJ}=-\delta_{I}{}^{J}\quad&,\quad  \Omega^{I}{}_{K}\Omega^{K}{}_J=\delta^{I}{}_{J}\quad,\quad \Omega_{IK}\Omega^{K}{}_{J}=\epsilon_{IJ}
\end{align}
This new matrix is different from the off-diagonal matrix $\Sigma^I{}_J$ defined in \eqref{SigmaDefinition} which appeared in the parity transformation. Since $\Omega$ squares to identity and \tt{T} is anti-linear, it squares to $-1$ for both massive and massless polarizations.

\textbf{\tt{CPT}:} The full $\tt{CPT}$ transformation over the spinors in the helicity basis is given by
\begin{align}
	\tt{CPT} \quad:\quad \begin{cases}
		|p]_{\alpha}\rightarrow -\langle \tilde{p}|_{\dot{\alpha}}\quad,\quad |p\rangle^{\dot{\alpha}}\rightarrow [\tilde{p}|^{\alpha},\\
		|p^{I}]_{\alpha}\rightarrow \Delta^{I}{}_{J}\langle \tilde{p}^{J}|_{\dot{\alpha}}\quad,\quad |p^{I}\rangle^{\dot{\alpha}}\rightarrow- \Delta^{I}{}_{J} [\tilde{p}^{J}|^{\alpha}
	\end{cases}
\end{align}
where $\Delta^{I}{}_{J}$ is defined as the product of $\Sigma$ and $\Omega$.
\begin{align}
	\Delta^{I}{}_{J}\equiv \Sigma^{IK}\Omega_{KJ}&=(\zeta^{I}_{+}\zeta_{+,J}+\zeta^{I}_{-}\zeta_{-,J})=\begin{pmatrix}
		0&1\\-1&0
	\end{pmatrix}\label{DeltaDefinition}\\
	\Delta_{I}{}^{J}=\epsilon_{IK}\epsilon^{JL}\Delta^{K}{}_{L}=\sigma_3\quad&,\quad \Delta_{IJ}=\Delta_{JI}\quad,\quad \Delta_{I}{}^{J}=\Delta^{J}{}_{I} \\
	\Delta^{IK}\Delta_{KJ}=\delta^{I}{}_{J}\quad&,\quad \Delta^{I}{}_K\Delta^{KJ}=\epsilon^{IJ},\\
	\Delta_{IK}\Delta^{KJ}=\delta_{I}{}^{J}\quad&,\quad  \Delta^{I}{}_{K}\Delta^{K}{}_J=-\delta^{I}{}_{J}\quad,\quad \Delta_{IK}\Delta^{K}{}_{J}=-\epsilon_{IJ}
\end{align}

%% file: App_RAMBO.tex
\section{RAMBO Algorithm}
\label{app:RAMBO}
In this section, we discuss the physics behind generating the numerical values for the kinematics of any $\mathtt{n}$-point incoming momenta. The method used by the command \tt{GenerateNumerics} is known as \tt{RAMBO}\footnote{The full form is ``Random Momenta Beautifully Organized".} derived by Kleiss, Stirling and Ellis\cite{RAMBO1} in the year 1985. Given the masses and the energy scale, the algorithm can generate numerical values for any $\mathtt{n-}$particle scattering by randomly choosing the $4n$ components. If the masses are randomly generated, this algorithm generates the pseudo-random numerical values for all the kinematic variables. The algorithm was further improved in 2013 by Simon Pl\"atzer in his work titled as \tt{RAMBO on diet}\cite{RAMBO2}. The new algorithm could generate the random kinematics using only for the $\mathtt{3n-4}$ independent variables. We review the \tt{RAMBO} which is simple and easy to implement for any number of incoming momenta.

The algorithm has the following steps : (1) Generate a random set of $\tt{n}$ number of four momenta. (2) Apply the Lorentz transformation over the sum of these momenta to the rest frame which gives a set which satisfies three momentum conservation (3) Scale the momenta such that they satisfy the energy bound. If the particles are massless, the algorithm terminates here. (4) If there are massive particles, the mass-shell condition must be attained for all of them. This is done by introducing a scaling factor $\mathtt{\xi}$ which is tuned using the Newton-Raphson method to land up on the correct mass-shell.

Let's consider that the randomly generated momenta for any leg $\tt{i}$ is given by $\mathtt{p_i^\mu}$ such that they satisfy the on-shell relations
\begin{align}
	\mathtt{p_i^2=-m_i^2}
\end{align}
Since each component is randomly generated, these momenta do not necessarily satisfy the energy-momentum conservation\footnote{The sum of incoming spatial momenta must be same as the sum of outgoing spatial momenta. The total energy of the incoming particle is equal to $\tt{E}$ which must be same as that of the outgoing particles.}. The first step is to arrive at a set of momenta whose spatial momentum is conserved. Let's call the sum of $\mathtt{p_i}'s$ as a new momenta $\mathtt{Q}$ such that its square gives an effective mass which we call as $\mathtt{M_Q}$.
\begin{align}
	\mathtt{Q^\mu=p_1^\mu+p_2^\mu +\cdots +p_n^\mu}\quad, \quad \mathtt{Q^2=-M_Q^2} 
\end{align}
Now let's do a Lorentz boost transformation $\mathtt{\Lambda^\mu{}_\nu}$ to the rest frame of $\mathtt{Q}$ where its spatial momenta is zero i.e.
\begin{align}
	\mathtt{Q'^\mu=\Lambda^\mu{}_\nu Q^\nu } \implies \vec{\mathtt{Q'}}=0
\end{align}
Using the linearity of Lorentz transformations, we can conclude that the same boost transformation acting on the $\mathtt{p_i}s$ gives a new set of momenta which satisfy spatial momentum conservation. Let's call these new set of momenta as $\mathtt{p_i'}$. Then we have
\begin{align}
	\mathtt{p'^\mu=\Lambda^\mu{}_\nu p^\nu }\quad,\quad \mathtt{\sum_{i=1}^{n}}\vec{\mathtt{p}}'\mathtt{_i=0}
\end{align}
The total energy of these momenta is same as the energy in the rest frame of $\mathtt{Q}$ i.e. $\mathtt{M_Q}$. Since the sum of the energies of these new momenta cannot cross the threshold energy $\mathtt{E}$, we scale them appropriately by a conformal transformation. The new set of momenta (call them $\mathtt{k_i}$) have the energies which sum upto total available energy $\mathtt{E}$.
\begin{align}
	\mathtt{k_i^\mu=\frac{E}{M_Q}p'^\mu_i}\implies \mathtt{\sum_{i=1}^{n}k_i^0=\frac{E}{M_Q}Q'^0=E}\quad,\quad \mathtt{(k_i)^2=\frac{E^2}{M_Q^2}(p'_i)^2}
\end{align}
The algorithm stops here if the initial momenta were massless because of the last equality. If there are massive momenta, we need to go a step further to tune in the correct massive on-shell condition which would be $\mathtt{k_i^2=-m_i^2}$. To do that, we rescale the spatial momentum of each $\mathtt{k_i}$ obtained in the last step by the same factor $\xi$ such that the new momenta satisfies on-shell condition.
\begin{align}
	\vec{\tt{k}}_\tt{i}\rightarrow \xi\vec{\tt{k}}_\tt{i}\implies \mathtt{k_i^2=-E_i^2+\xi^2 }(\vec{\tt{k}}_\tt{i})^2\mathtt{=-m_i^2}
\end{align}
Now look at the last equation carefully. We have modified the spatial components uniformally such that the on-shell conditions for all momenta are satisfied. This implies that the total energy has now gone bad and most probably shot up the available threshold $\mathtt{E}$. To fix this trade-off between the on-shell properties and the energy conservation, we tune the parameter $\xi$ such that the total energy is back to $\tt{E}$. So we define the following function $\mathtt{f(\xi)}$ 
\begin{align}
	\mathtt{f(\xi)=\sum_{i=1}^{n}}\sqrt{\mathtt{m_i^2+\xi^2 }(\vec{\tt{k}}_\tt{i})^2}\mathtt{-E}=0
\end{align}
Notice that this function is monotonically increasing. So the value of $\xi$ can be easily found out using the Newton-Raphson method. 
\begin{align}
	\mathtt{\xi_{a+1}=\xi_a-\frac{f(\xi_a)}{f'(\xi_a)}}
\end{align}
Once $\xi$ has converged, the final momenta becomes $\mathtt{k_i=(E_i,}\xi \vec{\tt{k}}_{\mathtt{i}})$.

%% file: App_MoreCommands.tex
\section{A few more useful commands beside the mains ones}\label{app:MoreCommands}
\subsection{More on Declaring massive and massless legs}
\label{subsec:moreondeclaring massive and massless legs}

\begin{documentationbox}{\tt{DeclareMassiveLegs}}{\tt{DeclareMassiveLegs[Legs\_List]}}
	
	Declares the massive legs. The argument is a list of leg labels. A leg label can be either a
	\href{https://reference.wolfram.com/language/ref/Symbol.html?q=Symbol}{\tt{Symbol}}
	or an element of
	\href{https://reference.wolfram.com/language/ref/PositiveIntegers.html}{\tt{PositiveIntegers}}.
	Negative labels are reserved for negative momenta; see
	\eqref{negativespinors}.
\end{documentationbox}

\begin{morematerialbox}
	These default values can be edited, as discussed later. The declared
	associations can be accessed using	\tt{MassiveData},
\end{morematerialbox}

\begin{documentationbox}{\tt{DeclareMasslessLegs}}{\tt{DeclareMasslessLegs[Legs\_List]} }

	Declares the massless legs. The argument is a list of leg labels. A leg label can be either a
	\href{https://reference.wolfram.com/language/ref/Symbol.html?q=Symbol}{\tt{Symbol}}
	or an element of
	\href{https://reference.wolfram.com/language/ref/PositiveIntegers.html}{\tt{PositiveIntegers}}.
	Negative labels are reserved for negative momenta; see
	\eqref{negativespinors}.

\end{documentationbox}

\begin{morematerialbox}
	These default values can be edited, as discussed later. The declared
	associations can be accessed using	\tt{MasslessData}.
	
\end{morematerialbox}

\begin{examplebox}{Declaring massive and massless legs separately}
	
	\redbox{\includegraphics[scale=0.5]{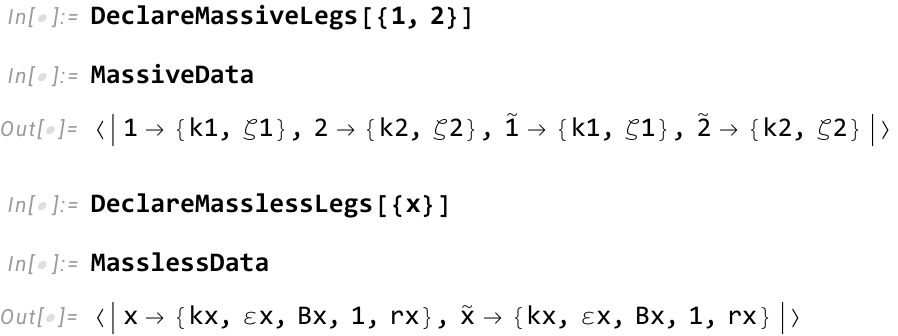}}
	
\end{examplebox}

\begin{morematerialbox}{} 
	
	The declared massive kinematic data can be accessed using
	\begin{align}
		\tt{AllMassiveLegs[]},
		\qquad
		\tt{AllMassiveMomenta[]},
		\qquad
		\tt{AllMassivePolarizations[]} .
	\end{align}
	Similarly, the declared massless kinematic data can be accessed using
	\begin{align}
		\tt{AllMasslessLegs[]},
		\qquad
		\tt{AllMasslessMomenta[]},
		\qquad
		\tt{AllMasslessPolarizations[]},
	\end{align}
	and
	\begin{align}
		\tt{AllMasslessFieldStrengths[]},
		\qquad
		\tt{AllMasslessHelicities[]},
		\qquad
		\tt{AllMasslessReferenceSpinors[]} .
	\end{align}
	
\end{morematerialbox}

\subsubsection{Edit kinematical data}\label{sec:EditData}

\begin{documentationbox}{\tt{EditMassiveData}, \tt{EditMasslessData}}{\tt{Editing declared kinematic data}}
	
	The declared kinematic data can be edited using the commands
	\begin{align}
		\tt{EditMassiveData[]},
		\qquad
		\tt{EditMasslessData[]} .
	\end{align}
	Running either command with empty input opens an editable palette containing
	the currently declared kinematic data.
	
\end{documentationbox}

\begin{examplebox}{Editing kinematic data}
	
	\redbox{\includegraphics[scale=0.5]{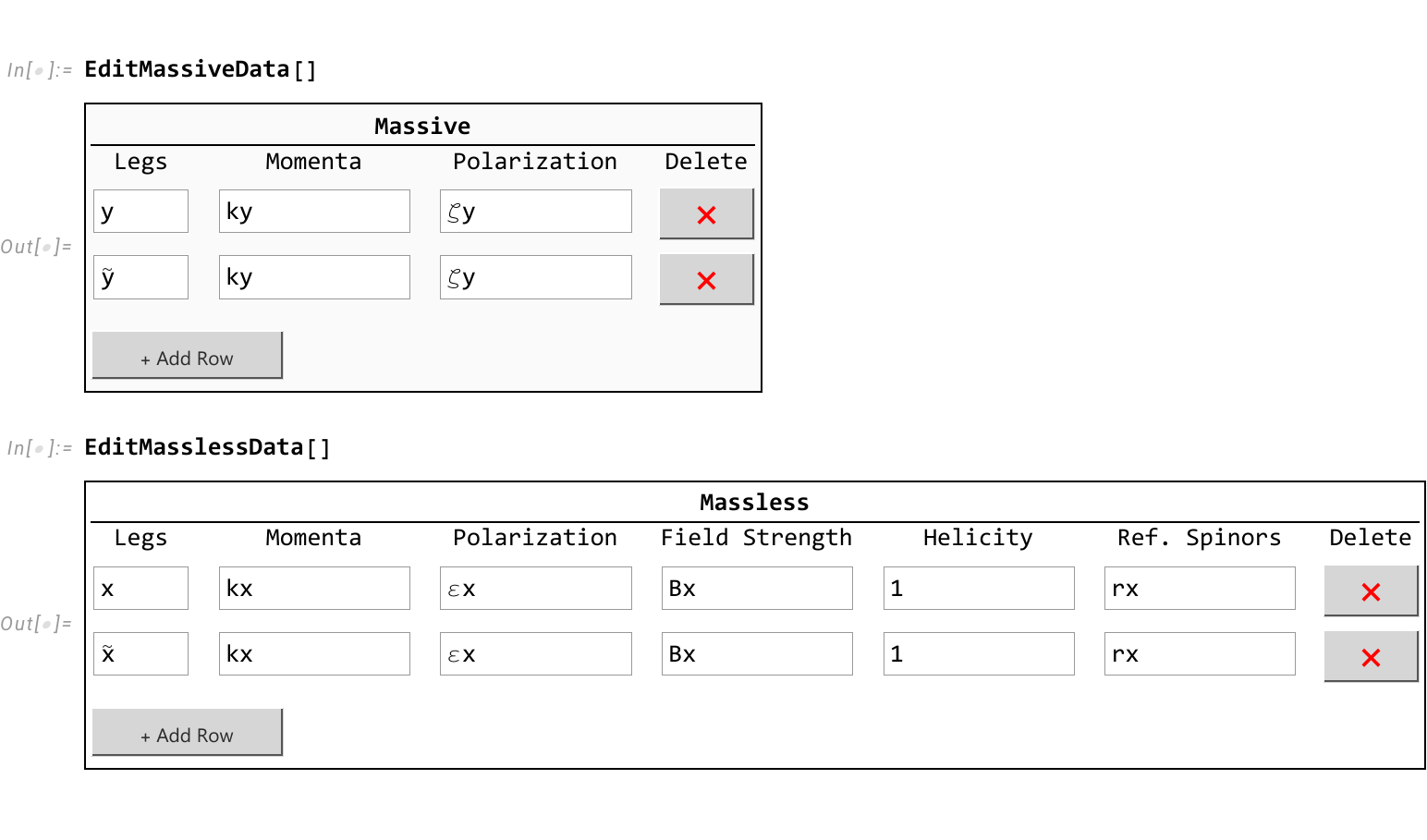}}
	
	\textit{Editing rows in the palette}: A new entry can be added using the \tt{+Add Row} button. Existing leg data can
	be deleted by clicking the \textcolor{red}{\(\times\)} button at the end of the
	corresponding row.
	
\end{examplebox}

\subsubsection{Little group indices}\label{app:LGcriteria}
\begin{conceptbox}{}
As discussed earlier, a \tt{particlelabel} can be any positive integer or a symbol that has been declared as massive or massless. Any variable that has not been declared in this way is treated as off-shell.

Once the particle labels have been declared, the next question is how to incorporate the massive little group indices. 
\begin{center}
	\textit{A massive \tt{leg} is represented by its \tt{particlelabel} with a little group index i.e. }\tt{particlelabel[LGIndex]}.
\end{center}

Whenever a label is declared as massive, any expression of the form \tt{particlelabel[LGIndex]} is interpreted as massive particle carrying little group index \tt{LGIndex}. More generally, for a compound object of arbitrary depth \tt{particlelabel[~][~]$\cdots$[LGIndex]}, the last argument is always interpreted as the little group index\footnote{In \tt{Mathematica} notation, the last variable is accessed using the \href{https://reference.wolfram.com/language/ref/Part.html}{\tt{Part}} \tt{[[1]]}, for example \tt{particlelabel[~][~]...[Index][[1]]=Index.}}. This convention allows additional labels such as charge, flavor, or species to be included alongside the little group index without any ambiguity.

A similar convention applies to the massless particles. Once a \tt{particlelabel} is declared as massless, the corresponding variable may contain any number of additional labels, for example \tt{particlelabel[charge][flavor]$\cdots$[species]}. In this case, no special interpretation is associated with the last argument since there is no little group index to be specified.

In practice, we will mostly be working with the simplest cases, where no additional labels are present. Thus, massive particles are represented as \tt{particlelabel[LGIndex]}, while massless particles are represented simply by \tt{particlelabel}.
\end{conceptbox}

\subsection{More on Three point function}
\label{subsec:moreonthreepointfunctions}

\begin{conceptbox}{Unequal masses}
	
	For unequal masses, \(\mathtt{m_1\neq m_2}\), the independent stripped
	structures are built from the spinors
	\begin{align}
		\mathtt{(k_1[3|)}_{\dot{\alpha}},
		\qquad
		\mathtt{\langle 3|}_{\dot{\alpha}} .
	\end{align}
	Schematically, the stripped amplitude takes the form
	\begin{align}
		\mathtt{
			m_1\neq m_2
			\quad:\quad
			\sum_i g_i
			\frac{1}{(m_1m_2)^{3\frac{S_1+S_2}{2}-\frac{h}{2}}}
			\left(
			\langle 3|^{S_1+S_2+h}
			(k_1[3|)^{S_1+S_2-h}
			\right)^i
		}_{\{\dot{\alpha}_1\cdots \dot{\alpha}_{2S_1}\};
			\{\dot{\beta}_1\cdots \dot{\beta}_{2S_2}\}} .
	\end{align}
	Here the constants \(\mathtt{g_i}\) label the independent couplings.

\begin{documentationbox}{\tt{Coupling}}{\tt{Coupling[g\_, i\_]}}
	
	gives independent couplings are denoted by \(\mathtt{g_i}\).  
	\paragraph{Keyboard shortcut}	\boxed{\tt{Esc}}\ \tt{gc}\ \boxed{\tt{Esc}} .
	
\end{documentationbox}
	
	With no loss of generality, take \(\mathtt{S_1\geq S_2}\). Then the full
	unequal-mass amplitude can be written as
	\begin{align}
		\mathtt{
			m_1\neq m_2
			\quad:\quad
			\sum_i g_i
			\frac{1}{(m_1m_2)^{S_1+S_2}}
			[1^{a}\,3]^i
			[2^{b}\,3]^{S_1+S_2+h-i}
			\langle 1^{a}\,3\rangle^{2S_1-i}
			\langle 2^{b}\,3\rangle^{S_2-S_1-h+i}
		}.
	\end{align}
	The allowed range of \(\mathtt{i}\) is
	\begin{align}
		\mathtt{0\leq h\leq S_1+S_2}
		&\implies
		\mathtt{
			S_1-S_2+h\leq i\leq Min[S_1+S_2+h,2S_1]
		},
		\\
		\mathtt{-S_1-S_2\leq h\leq 0}
		&\implies
		\mathtt{
			Max[0,S_1-S_2+h]\leq i\leq S_1+S_2+h
		}.
	\end{align}
	For all other values of \(\mathtt{h}\), the amplitude vanishes.
	
\end{conceptbox}

\begin{conceptbox}{Equal masses and the \(x\)-factor}
	
	The equal-mass case is more subtle. When
	\(\mathtt{m_1=m_2}\), three-point kinematics implies that the spinors
	\(\mathtt{(k_1[3|)}_{\dot{\alpha}}\) and
	\(\mathtt{\langle 3|}_{\dot{\alpha}}\) are proportional. The proportionality
	constant is the \(x\)-factor:
	\begin{align}
		\mathtt{(k_1)}_{\alpha\dot{\alpha}}\mathtt{[3|^{\alpha}}
		=
		\mathtt{m_1 x_{31}\langle 3|}_{\dot{\alpha}},
		\qquad
		\mathtt{
			x_{31}
			=
			\frac{\langle r_3|1|3]}{m\langle 3\,r_3\rangle}
		}.
	\end{align}
	Here \(r_3\) is a null reference spinor satisfying
	\begin{align}
		\mathtt{\langle 3\,r_3\rangle\neq 0}.
	\end{align}
	
	Therefore, in the equal-mass case, the stripped amplitude can be constructed
	using only the \(SL(2,\mathbb C)\) spinor
	\(\mathtt{\langle 3|}_{\dot{\alpha}}\) and the invariant tensor
	\(\mathtt{\epsilon_{\dot{\alpha}\dot{\beta}}}\). For any helicity
	\(\mathtt{h}\), the amplitude is spanned by
	\begin{align}
		\mathtt{2Min[S_1,S_2]+1}
	\end{align}
	independent structures.
	
\end{conceptbox}

\begin{conceptbox}{Equal-mass amplitude basis}
	
	With the massive spin indices stripped, the equal-mass amplitude is
	\begin{align}
		\mathtt{
			m_1=m_2
			\quad:\quad
			\sum_{i=|S_1-S_2|}^{S_1+S_2}
			g_i
			\frac{1}{m_1^i}
			x^{h+i}
			\left(
			\langle 3|^{2i}
			\epsilon^{S_1+S_2-i}
			\right)
		}_{\{\dot{\alpha}_1\cdots \dot{\alpha}_{2S_1}\};
			\{\dot{\beta}_1\cdots \dot{\beta}_{2S_2}\}}
		\\
		=
		\mathtt{
			\sum_{i=|S_1-S_2|}^{S_1+S_2}
			g_i
			\frac{1}{m_1^i}
			x^h
			\left(
			\langle 3|^i
			\left(
			\frac{k_1[3|}{m_1}
			\right)^i
			\epsilon^{S_1+S_2-i}
			\right)
		}_{\{\dot{\alpha}_1\cdots \dot{\alpha}_{2S_1}\};
			\{\dot{\beta}_1\cdots \dot{\beta}_{2S_2}\}} .
	\end{align}
	
	where $\tt{x}$ is the x-factor which can be accessed using the command
	
\begin{documentationbox}{\tt{Xfactor}}{\tt{Xfactor[masslessleg,massiveleg,h]}}
	
	The \(x\)-factor appearing in equal-mass three-point kinematics. 
\end{documentationbox}

	Because the invariant tensor \(\epsilon\) can now appear explicitly, the
	equal-mass case contains additional scalar structures such as
	\begin{align}
		\mathtt{\langle 1^a\,2^b\rangle},
		\qquad
		\mathtt{[1^a\,2^b]} ,
	\end{align}
	which need not appear in the unequal-mass case.
	
	For \(\mathtt{S_1\geq S_2}\), the full equal-mass amplitude can be written as
	\begin{align}
		\mathtt{
			m_1=m_2
			\quad:\quad
			\frac{x^h}{m_1^{2(S_1+S_2)}}
			\langle 1^a|3|1^a]^{S_1-S_2}
			\sum_{i=0}^{2S_2}
			g_i
			\langle 1^a\,2^b\rangle^i
			[1^a\,2^b]^{2S_2-i}
		}.
	\end{align}
	The amplitude for \(\mathtt{S_1\leq S_2}\) is obtained by exchanging
	\(\mathtt{1\leftrightarrow 2}\).
	
\end{conceptbox}

\subsection{More on propagator}
\label{subsec:moreonpropagator}

\begin{conceptbox}{Propagator}
	
	The unitary gauge propagator \tt{UG$\Theta$} was ad-hocly upgraded to a general $\xi$ propagator by simply promoting the spin $\tt{1}$ projector \tt{UG$\Theta$} to $\mathtt{\Theta}$.
		\begin{align}
		\mathtt{\mathcal{P}}^{\{\alpha\dot{\alpha}\};\{\beta\dot{\beta}\}}
		\mathtt{(k,\xi)}
		&\equiv
		\frac{-\iimg}{\mathtt{k^2+m^2+\iimg\epsilon}}
		\sum_{a=0}^{\lfloor S/2\rfloor}
		\mathtt{A(S,a)}
		\bigg[
		\mathtt{\Theta}^{\alpha_{\tt 1}\dot{\alpha}_{\tt 1},
			\alpha_{\tt 2}\dot{\alpha}_{\tt 2}}
		\cdots
		\mathtt{\Theta}^{\alpha_{\tt{2a-1}}\dot{\alpha}_{\tt{2a-1}},
			\alpha_{\tt{2a}}\dot{\alpha}_{\tt{2a}}}
		\nonumber\\
		&\hspace{2.8cm}\times
		\mathtt{\Theta}^{\beta_{\tt 1}\dot{\beta}_{\tt 1},
			\beta_{\tt 2}\dot{\beta}_{\tt 2}}
		\cdots
		\mathtt{\Theta}^{\beta_{\tt{2a-1}}\dot{\beta}_{\tt{2a-1}},
			\beta_{\tt{2a}}\dot{\beta}_{\tt{2a}}}
		\nonumber\\
		&\hspace{2.8cm}\times
		\mathtt{\Theta}^{\alpha_{\tt{2a+1}}\dot{\alpha}_{\tt{2a+1}},
			\beta_{\tt{2a+1}}\dot{\beta}_{\tt{2a+1}}}
		\cdots
		\mathtt{\Theta}^{\alpha_{\tt{2S}}\dot{\alpha}_{\tt{2S}},
			\beta_{\tt{2S}}\dot{\beta}_{\tt{2S}}}
		\bigg]_{\tt{sym}(\alpha,\dot{\alpha},\beta,\dot{\beta})}.
	\end{align}
	
	 This new propagator at $\xi\rightarrow \infty$ and $\xi\rightarrow1$ gives the already known unitary gauge and the Feynman gauge analogues of the higher spin propagators. 
	
\end{conceptbox}

\begin{documentationbox}{\tt{Propagator}}{\tt{Propagator[S\_, k\_, m\_, LeftPrefix\_, RightPrefix\_,OptionsPattern[]]}}
	
	Gives the general propagator for higher spin \tt{S}, where the input variables are same as the \tt{UGPropagator} (defined in \eqref{UGPropagator}) with an extra option of \(\mathtt{\xi}\) which is the gauge parameter option whose default value is $\xi$ itself. 	
\end{documentationbox}

\begin{documentationbox}{\tt{FGPropagator}}{\tt{FGPropagator[S\_, k\_, m\_, LeftPrefix\_, RightPrefix\_,OptionsPattern[]]}}
	
	This provides the Feynman-gauge propagator.  It corresponds to the choice
	\(\mathtt{\xi=1}\). The options include \tt{LeftSymm} and \tt{RightSymm} with the same defualt values as the \tt{Propagator}.
	
\end{documentationbox}

\begin{examplebox}{\tt{Propagator} and \tt{FGPropagator}}
	
	\redbox{\includegraphics[scale=0.4]{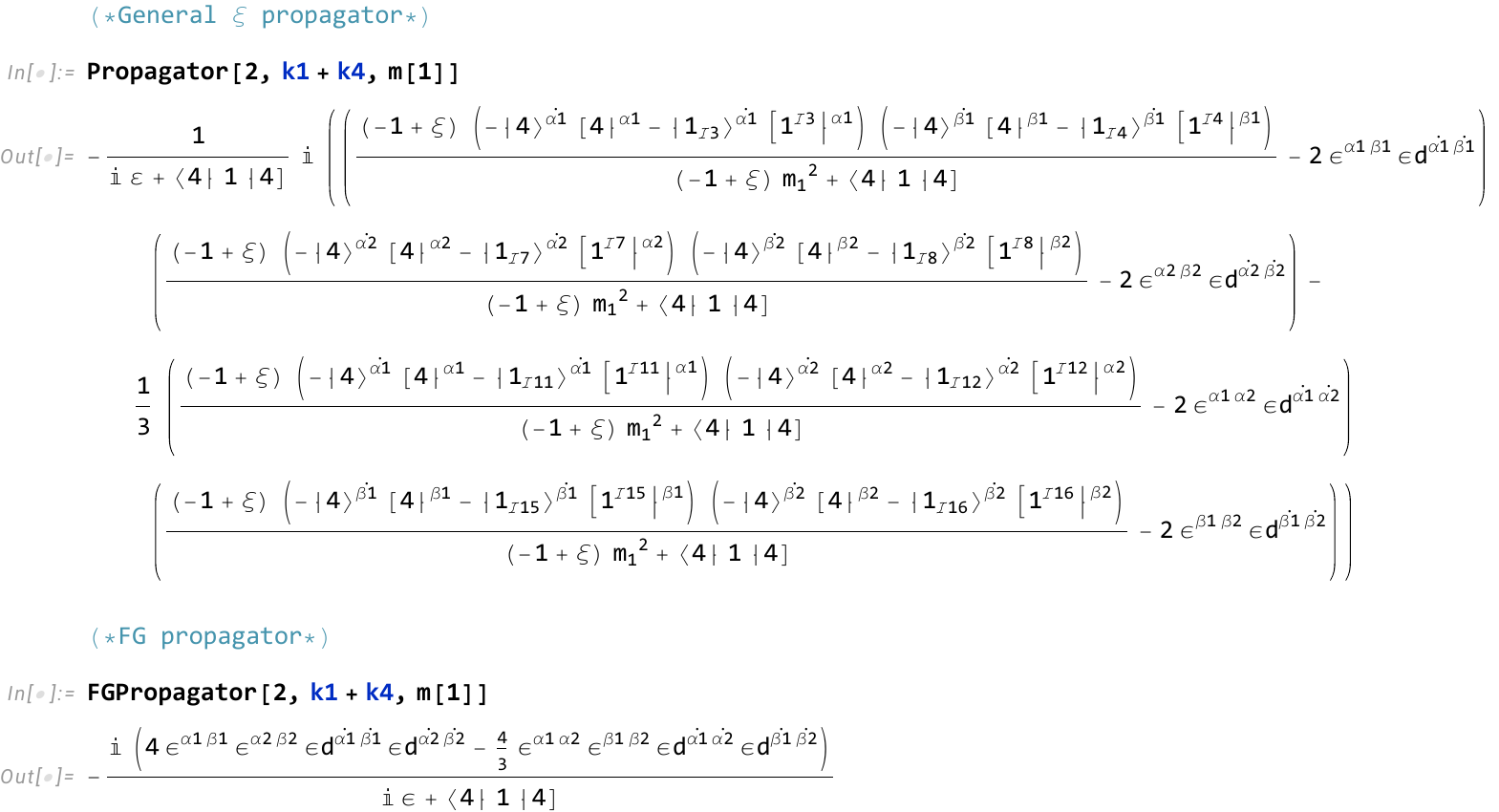}}
	
\end{examplebox}

\subsection{More on Mandelstam variables}
\label{subsec:moreonmandelstamvar}

\begin{documentationbox}{\tt{MandelstamtoSH} and \tt{SHToMandesltam}}{\tt{Conversion between Mandelstam and spinor-helicity notation}}
	
	Mandelstam variables can be converted to spinor-helicity notation using
	\begin{align}
		\tt{MandelstamtoSH}.
	\end{align}
	Conversely, spinor-helicity expressions can be converted to Mandelstam
	variables using
	\begin{align}
		\tt{SHToMandesltam}.
	\end{align}
	
\end{documentationbox}

\begin{documentationbox}{\tt{MandelstamtoLor} and \tt{LorToMandesltam}}{\tt{Conversion between Mandelstam and Lorentz notation}}
	
	Mandelstam variables can be converted to Lorentz-vector notation using
	\begin{align}
		\tt{MandelstamtoLor}.
	\end{align}
	Conversely, Lorentz-vector expressions can be converted to Mandelstam
	variables using
	\begin{align}
		\tt{LorToMandesltam}.
	\end{align}
	
\end{documentationbox}

\begin{examplebox}{\tt{MandelstamtoSH}, \tt{SHToMandesltam}, \tt{MandelstamtoLor}, \tt{LorToMandesltam}}
	
	\redbox{
		\begin{minipage}{0.5\linewidth}
			\includegraphics[scale=0.5]{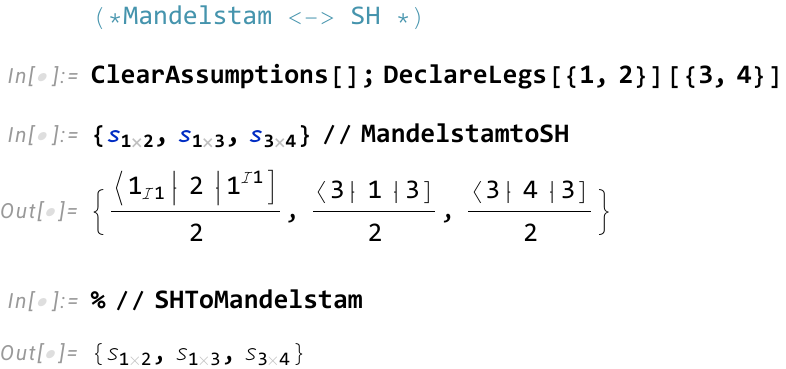}
		\end{minipage}
		\begin{minipage}{0.5\linewidth}
			\includegraphics[scale=0.5]{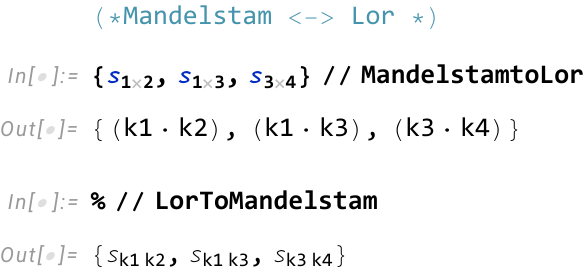}
		\end{minipage}
	}
	
\end{examplebox}

\subsection{More on Declaring momentum conservation}
\label{subsec:moreondeclaringmomconv}

\begin{documentationbox}{\tt{momconsRule}}{\tt{momconsRule[]}}

	returns the momentum-conservation replacement rule generated by
	\(\tt{DeclareMomentumConservation}\).  This rule can be used to replace the
	leading momentum in terms of the remaining momenta.
	
\end{documentationbox}

\begin{documentationbox}{\tt{MandelstamRules}}{\tt{MandelstamRules[]}}
	
	returns the rules that express dependent Mandelstam variables in terms of the
	independent Mandelstam variables determined by momentum conservation.

\end{documentationbox}

\begin{examplebox}{\tt{DeclareMomentumConservation}}
	
	\redbox{
		\begin{minipage}{0.37\linewidth}
			\includegraphics[scale=0.225]{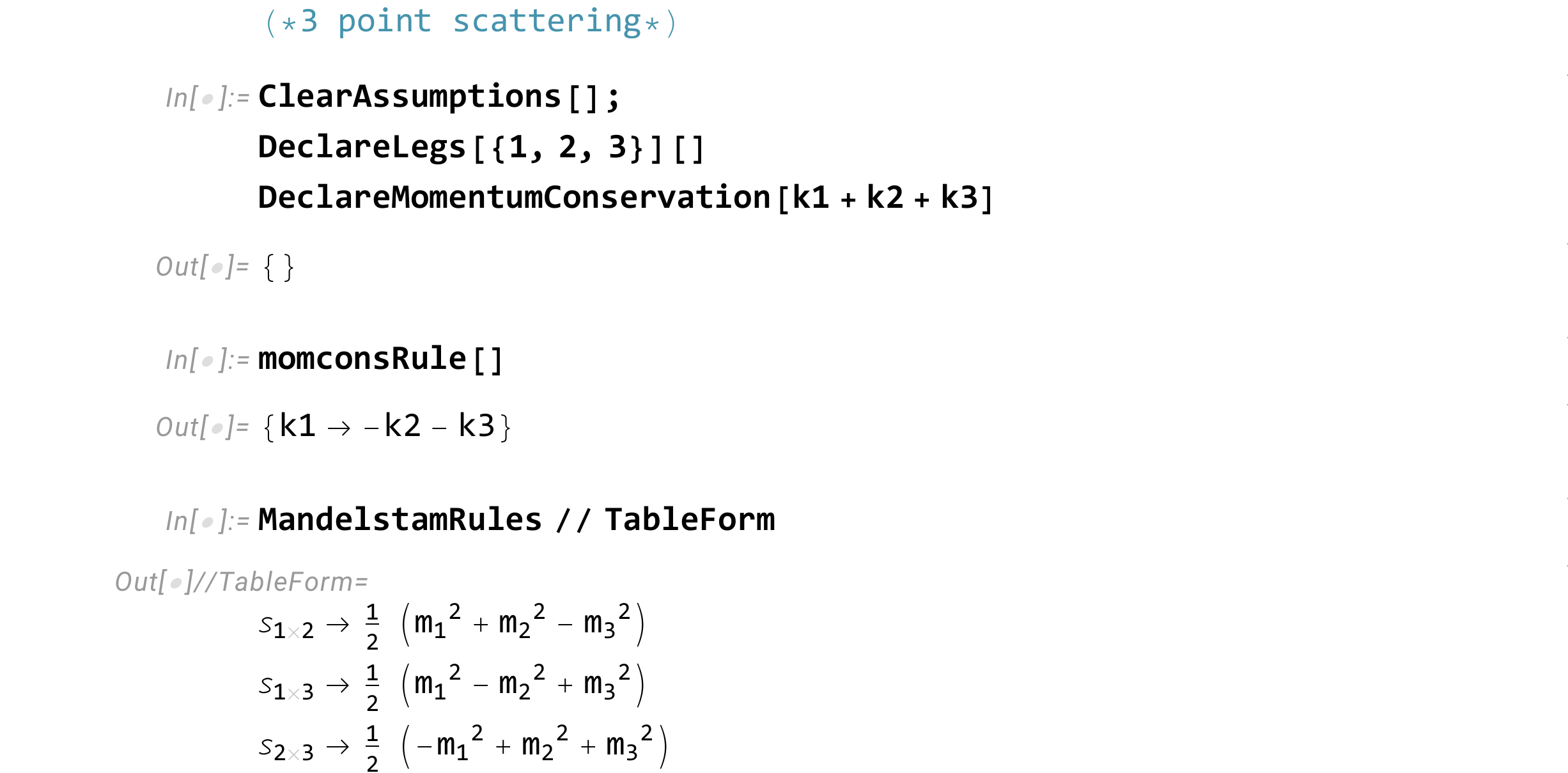}
		\end{minipage}
		\begin{minipage}{0.5\linewidth}
			\includegraphics[scale=0.225]{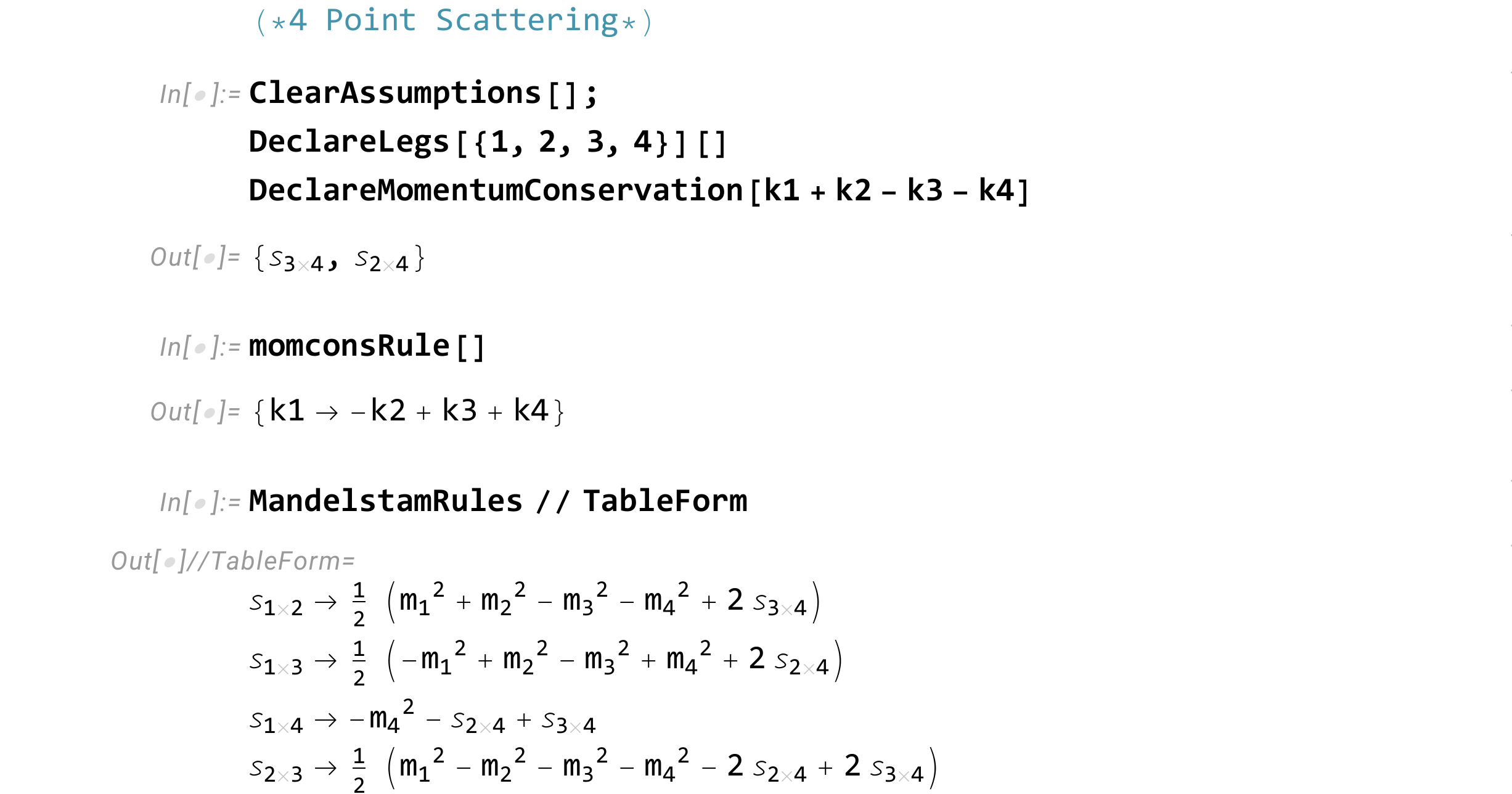}
		\end{minipage}
	}
	
\end{examplebox}

\begin{documentationbox}{\tt{ClearMomentumConservation}}{\tt{ClearMomentumConservation[]}}

	clears the momentum-conservation rules.  It resets the defined momentum
	replacement rules and Mandelstam simplification rules to
	\tt{Null}.
	
\end{documentationbox}

\begin{examplebox}{\tt{ClearMomentumConservation}}
	
	\redbox{
		\includegraphics[scale=0.45]{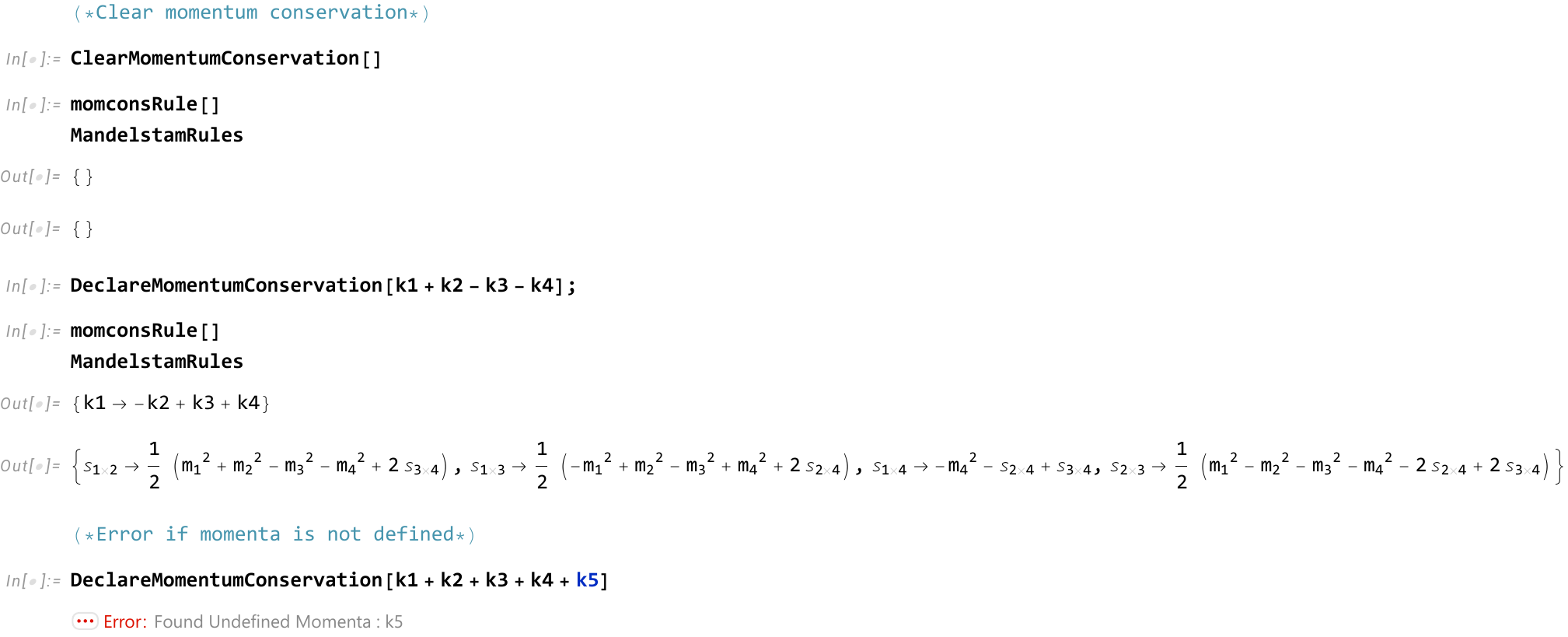}
	}
	
\end{examplebox}

\begin{documentationboxWL}{\tt{\$Mandelstams -> list}}{\tt{Choosing independent Mandelstam variables}}
	\label{dcmnt:mandelstamrules}
	
	By default, \(\tt{DeclareMomentumConservation}\) chooses the independent
	Mandelstam variables automatically using Mathematica's
	\href{https://reference.wolfram.com/language/ref/Solve.html}{\tt{Solve}}.
	
	One can specify a preferred list of independent Mandelstam variables using the
	option
	\begin{align}
		\tt{\$Mandelstams -> list}.
	\end{align}
	With this option, all other Mandelstam variables are solved in terms of the
	variables provided in \(\tt{list}\).  The corresponding rules are then stored
	in
	\begin{align}
		\tt{MandelstamRules[]}.
	\end{align}
	
\end{documentationboxWL}

\begin{examplebox}{Choosing independent Mandelstam variables}
	
	\redbox{
		\includegraphics[scale=0.45]{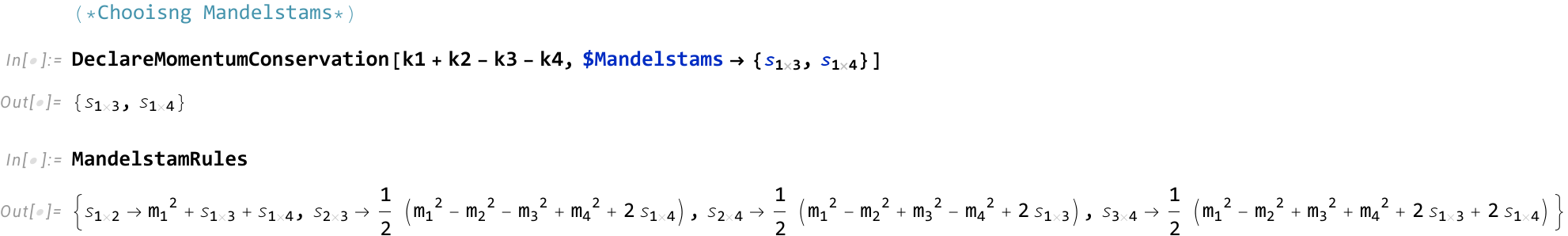}
	}
	
\end{examplebox}

\subsection{More on Contract Bilinears}
\label{subsec:Moreoncontractmetric}

\subsubsection{\tt{ContractLBilinear}}

\begin{documentationbox}{\tt{ContractLBilinear}}{\tt{ContractLBilinear[expr\_]}}
	
	The command \(\tt{ContractLBilinear}\) contracts repeated
	\(\mathtt{SU(2)_L}\) spinor indices using the corresponding Levi-Civita
	bilinear.
	
\end{documentationbox}

\begin{examplebox}{\tt{ContractLBilinear}}

	\redbox{
		\includegraphics[scale=0.5]{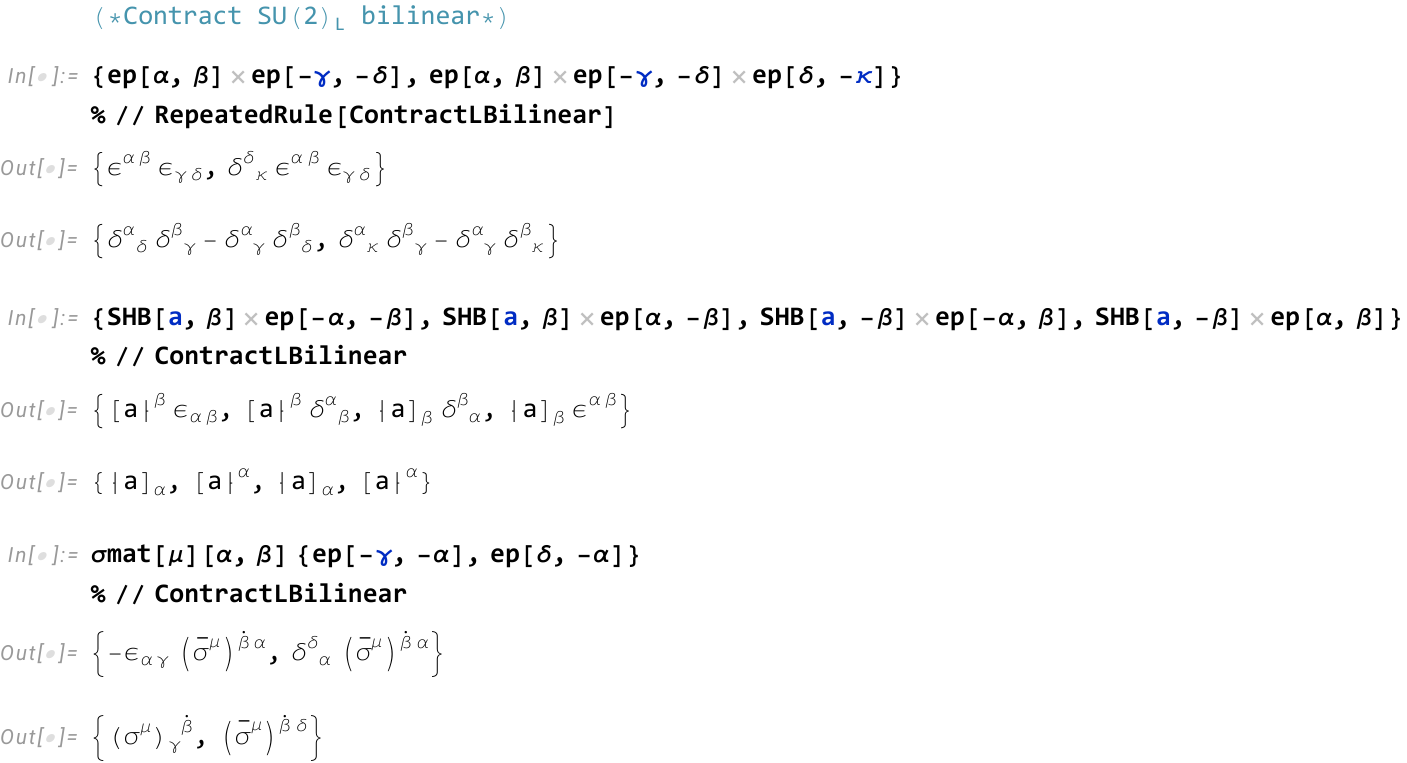}
	}

\end{examplebox}

\subsubsection{\tt{ContractRBilinear}}

\begin{documentationbox}{\tt{ContractRBilinear}}{\tt{ContractRBilinear[expr\_]}}
	
	The command \(\tt{ContractRBilinear}\) contracts repeated
	\(\mathtt{SU(2)_R}\) spinor indices using the corresponding Levi-Civita
	bilinear.
	
\end{documentationbox}

\begin{examplebox}{\tt{ContractRBilinear}}

	\redbox{
		\includegraphics[scale=0.5]{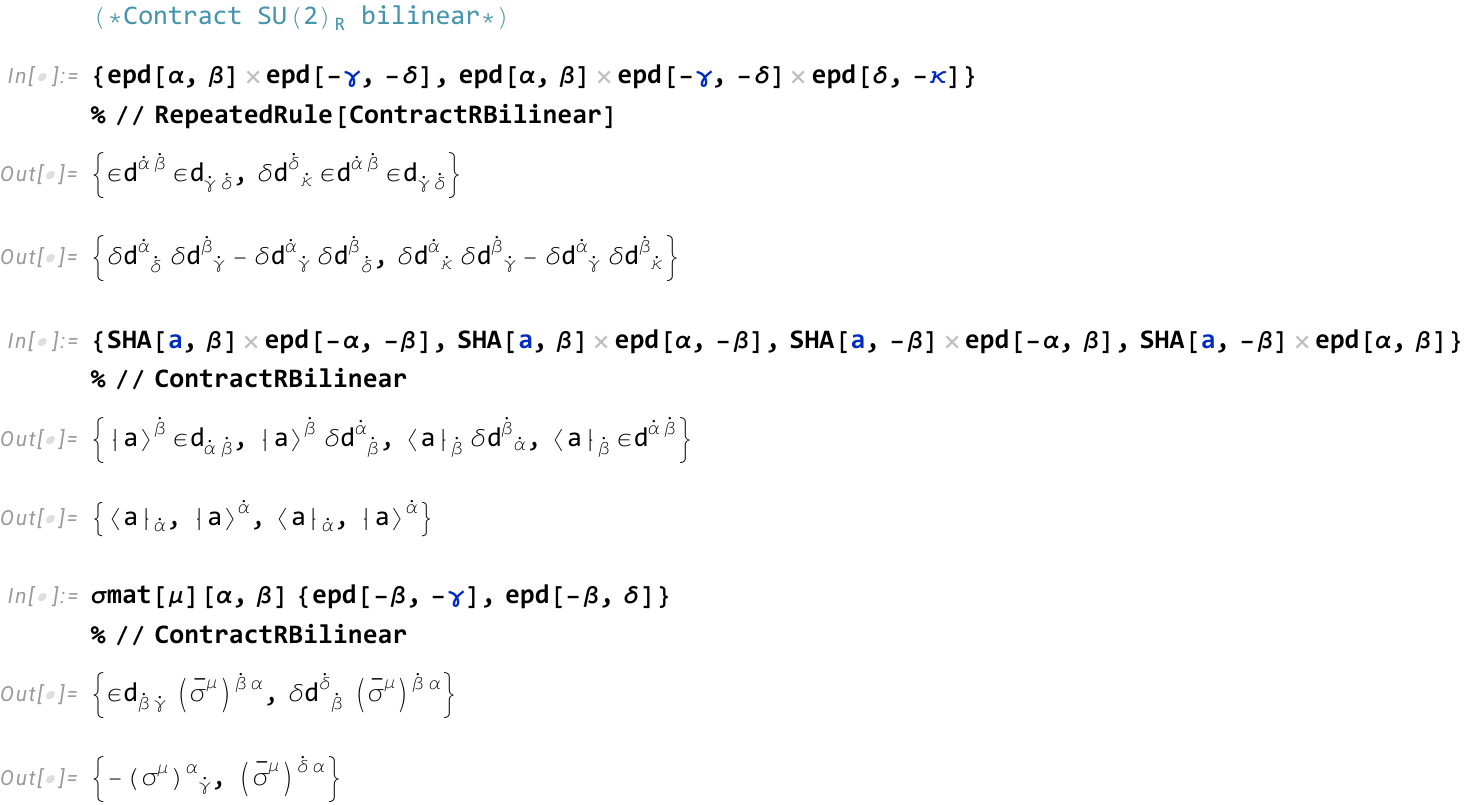}
	}

\end{examplebox}

\subsubsection{\tt{ContractLGBilinear}}

\begin{documentationbox}{\tt{ContractLGBilinear}}{\tt{ContractLGBilinear[expr\_]}}
	
	The command \(\tt{ContractLGBilinear}\) contracts repeated massive
	little-group indices using the \(\mathtt{SU(2)_{LG}}\) Levi-Civita bilinear.
	
\end{documentationbox}

\begin{examplebox}{\tt{ContractLGBilinear}}

	\redbox{
		\includegraphics[scale=0.475]{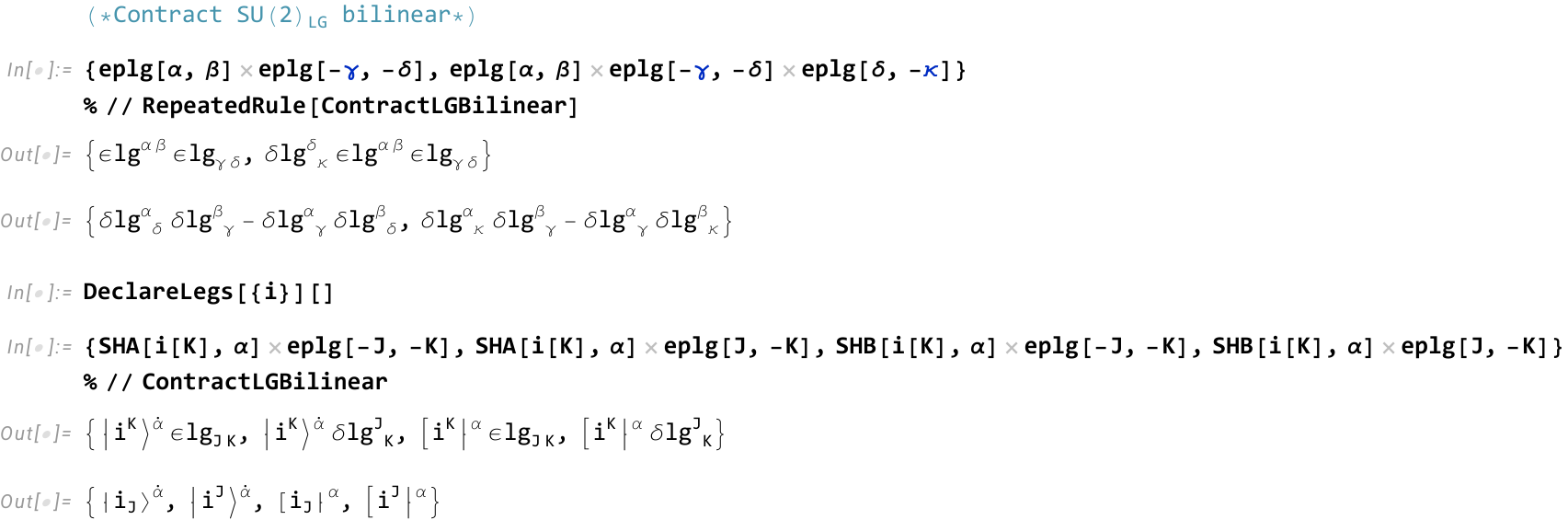}
	}

\end{examplebox}

\subsubsection{\tt{ContractLorMetric}}

\begin{documentationbox}{\tt{ContractLorMetric}}{\tt{ContractLorMetric[expr\_]}}
	
	The command \(\tt{ContractLorMetric}\) contracts repeated Lorentz-vector
	indices using the Lorentz metric \(\eta^{\mu\nu}\).
	
\end{documentationbox}

\begin{examplebox}{\tt{ContractLorMetric}}

	\redbox{
		\includegraphics[scale=0.5]{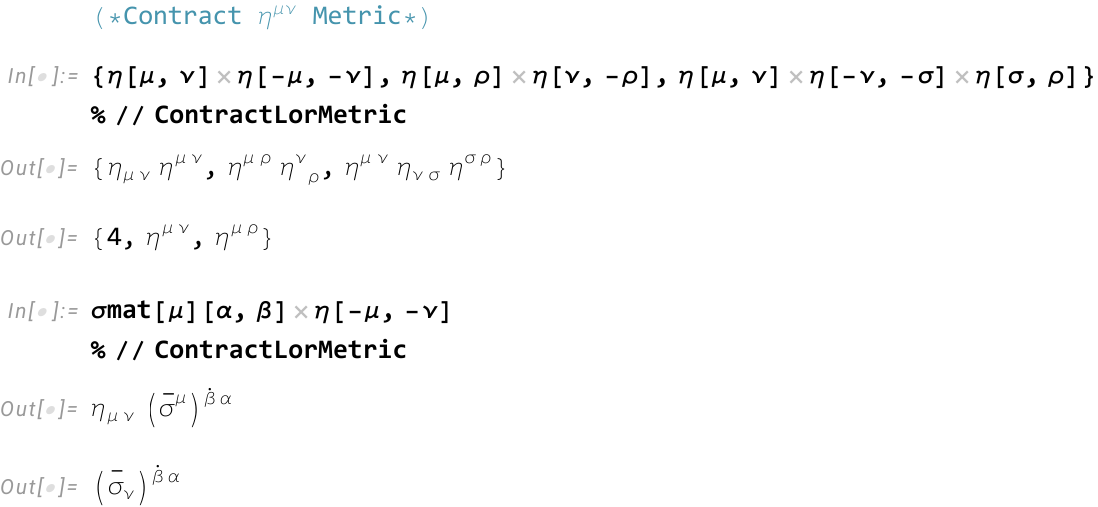}
	}

\end{examplebox}

\subsection{More on \texorpdfstring{\(SL(2,\mathbb C)\)}{SL(2,C)} Scalars}
\label{subsec:Moreonputsltwocscalar}

\begin{documentationbox}{\tt{PutSU2LScalar}}{\tt{PutSU2LScalar[expr\_]}}
	
	The command \(\tt{PutSU2LScalar}\) replaces a product of two spinors with
	contracted \(\mathtt{SU(2)_L}\) indices by the corresponding scalar bracket.
	It acts on contractions of the form
	\begin{equation}
		[p|^{\alpha}|q]_{\alpha}
		=
		[p,q].
	\end{equation}

\end{documentationbox}

\begin{examplebox}{\tt{PutSU2LScalar}}
	\redbox{
		\includegraphics[scale=0.5]{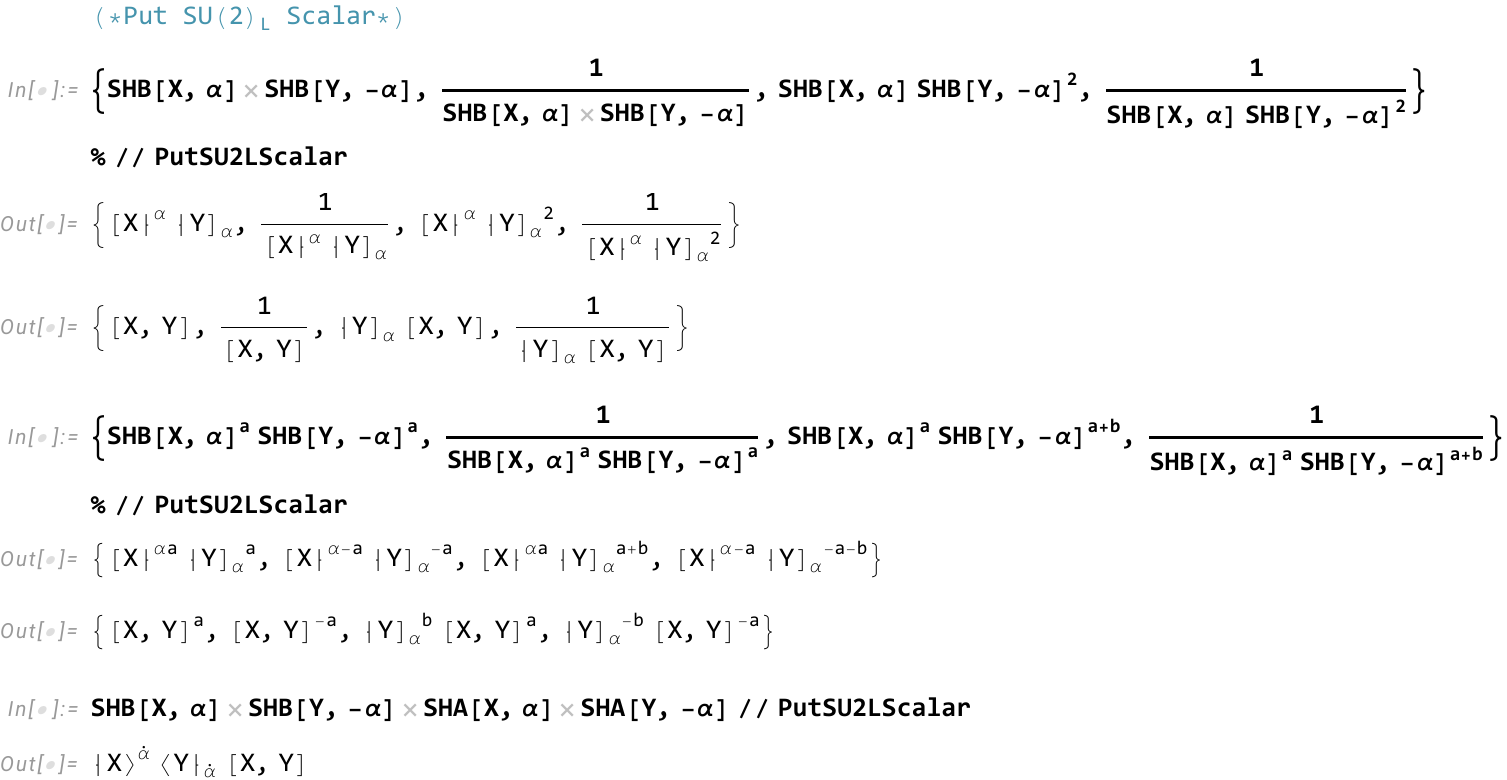}
	}
	
\end{examplebox}

\begin{documentationbox}{\tt{PutSU2RScalar}}{\tt{PutSU2RScalar[expr\_]}}
	
	The command \(\tt{PutSU2RScalar}\) replaces a product of two spinors with
	contracted \(\mathtt{SU(2)_R}\) indices by the corresponding scalar bracket.
	It acts on contractions of the form
	\begin{equation}
		\langle p|_{\dot\alpha}|q\rangle^{\dot\alpha}
		=
		\langle p,q\rangle .
	\end{equation}

\end{documentationbox}

\begin{examplebox}{\tt{PutSU2RScalar}}
	\redbox{
		\includegraphics[scale=0.5]{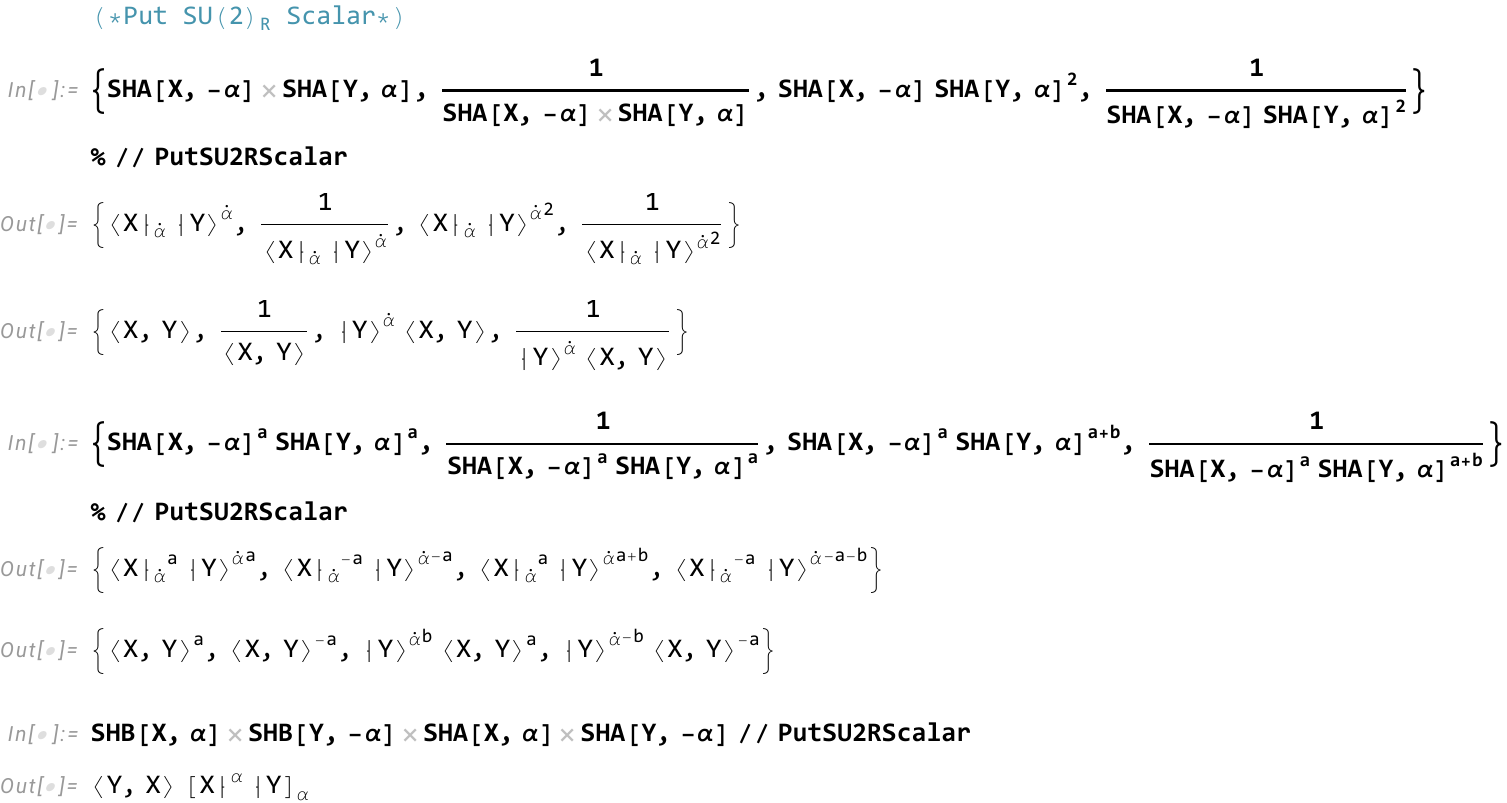}
	}
\end{examplebox}

\subsubsection{More on \tt{NoSL2CScalar}}
\label{subsec:Moreonnosltwocscalar}

\begin{documentationbox}{\tt{NoSU2LScalar}}{\tt{NoSU2LScalar[expr\_]}}
	
	The command \(\tt{NoSU2LScalar}\) performs the inverse operation of
	\(\tt{PutSU2LScalar}\).  It expands a box scalar bracket into a product of two
	spinors with contracted \(\mathtt{SU(2)_L}\) indices:
	\begin{equation}
		[p,q]
		\longrightarrow
		[p|^{\alpha}|q]_{\alpha}.
	\end{equation}

\end{documentationbox}

\begin{examplebox}{\tt{NoSU2LScalar}}
	\redbox{
		\includegraphics[scale=0.5]{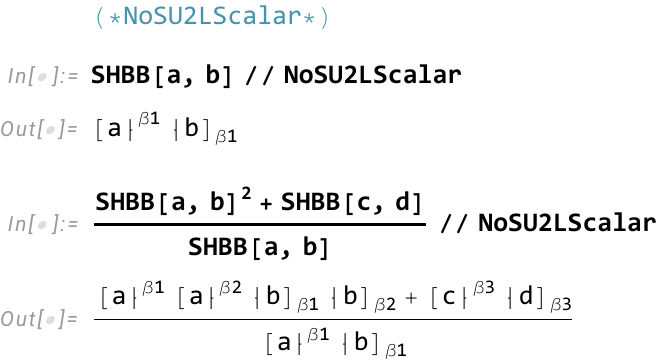}
	}
	
\end{examplebox}

\begin{documentationbox}{\tt{NoSU2RScalar}}{\tt{NoSU2RScalar[expr\_]}}
	
	The command \(\tt{NoSU2RScalar}\) performs the inverse operation of
	\(\tt{PutSU2RScalar}\).  It expands an angle scalar bracket into a product of
	two spinors with contracted \(\mathtt{SU(2)_R}\) indices:
	\begin{equation}
		\langle p,q\rangle
		\longrightarrow
		\langle p|_{\dot\alpha}|q\rangle^{\dot\alpha}.
	\end{equation}
	
\end{documentationbox}

\begin{examplebox}{\tt{NoSU2RScalar}}
	\redbox{
		\includegraphics[scale=0.5]{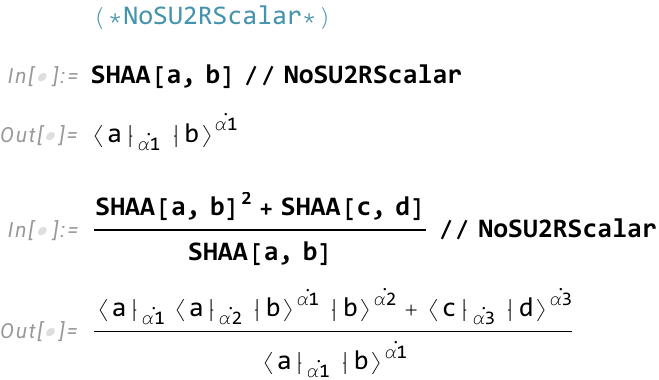}
	}
	
\end{examplebox}

\subsection{More Little-group Scalars}
\label{Subsec:moreonputlgscalar}

\begin{documentationbox}{\tt{PutMasslessLGScalar}}{\tt{PutMasslessLGScalar[expr\_]}}
	
	The command \(\tt{PutMasslessLGScalar}\) replaces a product of
	\(\tt{SHAA}\) and \(\tt{SHBB}\) by \(\tt{SHAB}\) when the common leg is
	not declared massive.  Schematically,
	\begin{equation}
		\langle a,b\rangle [c,b]
		\quad \longrightarrow \quad
		\langle a|b|c].
	\end{equation}
	This command does not require the user to declare massless legs using
	\(\tt{DeclareMasslessLegs}\).  It only checks that the common leg is not
	massive, namely that \(\tt{MsQ[leg]}\) is not \(\tt{True}\).

\end{documentationbox}

\begin{examplebox}{\tt{PutMasslessLGScalar}}
	
	\redbox{
		\includegraphics[scale=0.5]{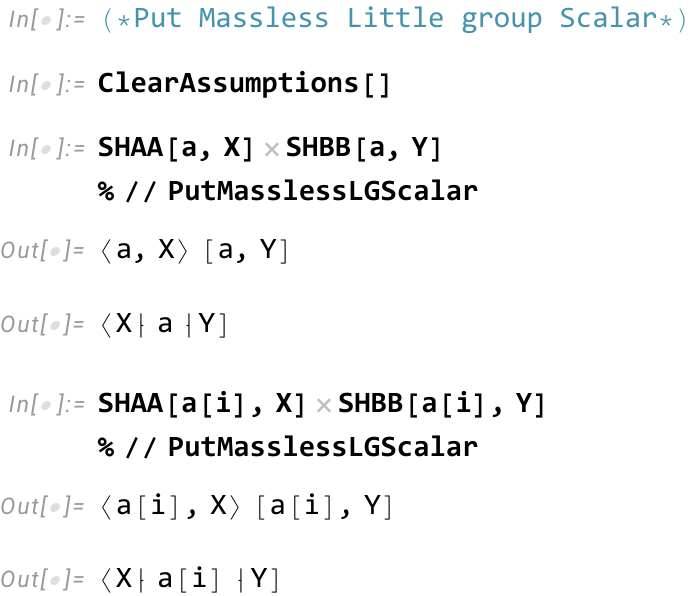}
	}
\end{examplebox}

\begin{documentationbox}{\tt{PutMassiveLGScalar}}{\tt{PutMassiveLGScalar[expr\_]}}
	
	The command \(\tt{PutMassiveLGScalar}\) replaces a product of
	\(\tt{SHAA}\) and \(\tt{SHBB}\) by \(\tt{SHAB}\) when the common leg is
	declared massive and its massive little-group indices are contracted.
	Schematically,
	\begin{equation}
		\langle a,b_J\rangle [c,b^J]
		\quad \longrightarrow \quad
		\langle a|b|c].
	\end{equation}
	This command requires the relevant base leg to be declared massive using
	\(\tt{DeclareMassiveLegs}\).  If no contracted massive little-group indices
	are present, the command leaves the expression unchanged.

\end{documentationbox}

\begin{examplebox}{\tt{PutMassiveLGScalar}}
	
	\redbox{
		\includegraphics[scale=0.5]{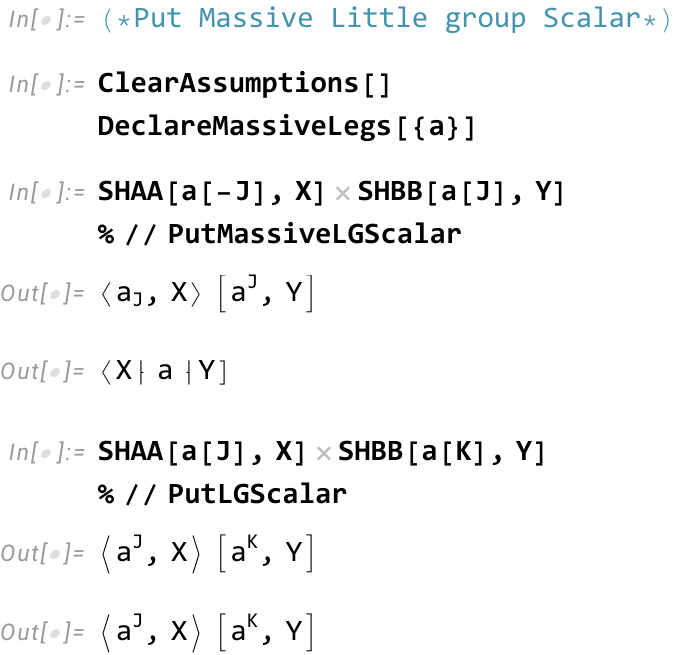}
	}
\end{examplebox}

\subsubsection{More on \tt{NoLGScalar}}
\label{Subsec:moreonnolgscalar}

\begin{documentationbox}{\tt{NoMasslessLGScalar}}{\tt{NoMasslessLGScalar[expr\_]}}
	
	The command \(\tt{NoMasslessLGScalar}\) is the inverse of
	\(\tt{PutMasslessLGScalar}\).  It replaces a massless little-group scalar
	\(\tt{SHAB}\) by a product of \(\tt{SHAA}\) and \(\tt{SHBB}\):
	\begin{equation}
		\langle a|b|c]
		\quad \longrightarrow \quad
		\langle a,b\rangle [c,b],
	\end{equation}
	provided the middle leg \(b\) is not declared massive.

\end{documentationbox}

\begin{examplebox}{\tt{NoMasslessLGScalar}}
	\redbox{
		\includegraphics[scale=0.5]{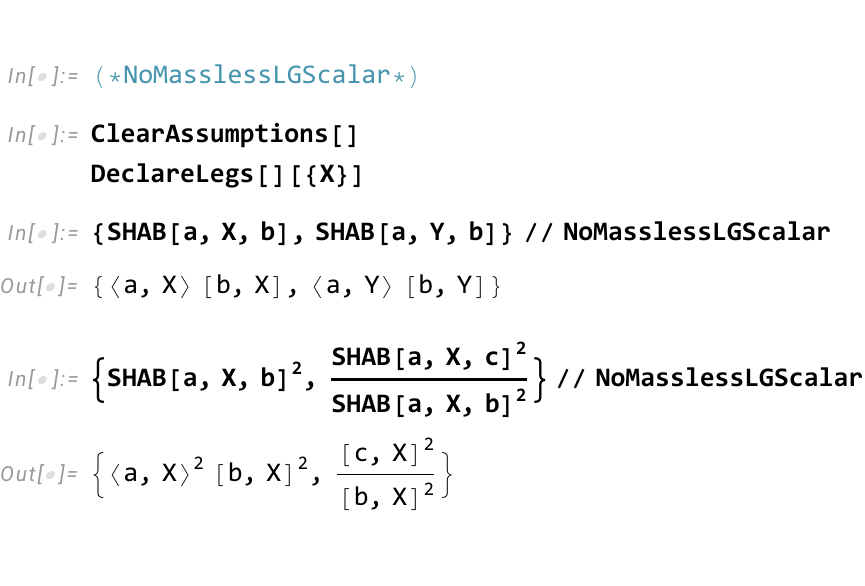}
	}
	If the middle leg is declared massive, the command acts trivially.
	
	\redbox{
		\includegraphics[scale=0.5]{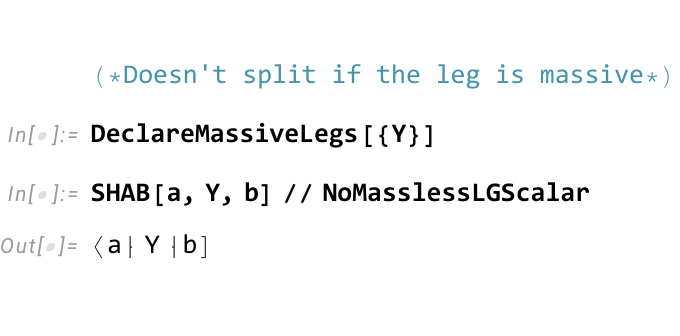}
	}
	
\end{examplebox}

\begin{documentationbox}{\tt{NoMassiveLGScalar}}{\tt{NoMassiveLGScalar[expr\_]}}
	
	The command \(\tt{NoMassiveLGScalar}\) is the inverse of
	\(\tt{PutMassiveLGScalar}\).  If \(X\) is declared massive, it expands
	\(\tt{SHAB[a,X,b]}\) into a product of \(\tt{SHAA[]}\) and \(\tt{SHBB[]}\)
	with contracted massive little-group indices:
	\begin{equation}
		\langle a|X|b]
		\quad \longrightarrow \quad
		-\langle X_{J},a\rangle [X^{J},b].
	\end{equation}
	If \(X\) is not declared massive, the command leaves \(\tt{SHAB[a,X,b]}\)
	unchanged.  When several such scalars are present, the command chooses fresh
	little-group indices to avoid accidental repetitions.

\end{documentationbox}

\begin{examplebox}{\tt{NoMassiveLGScalar}}
	\redbox{
		\includegraphics[scale=0.5]{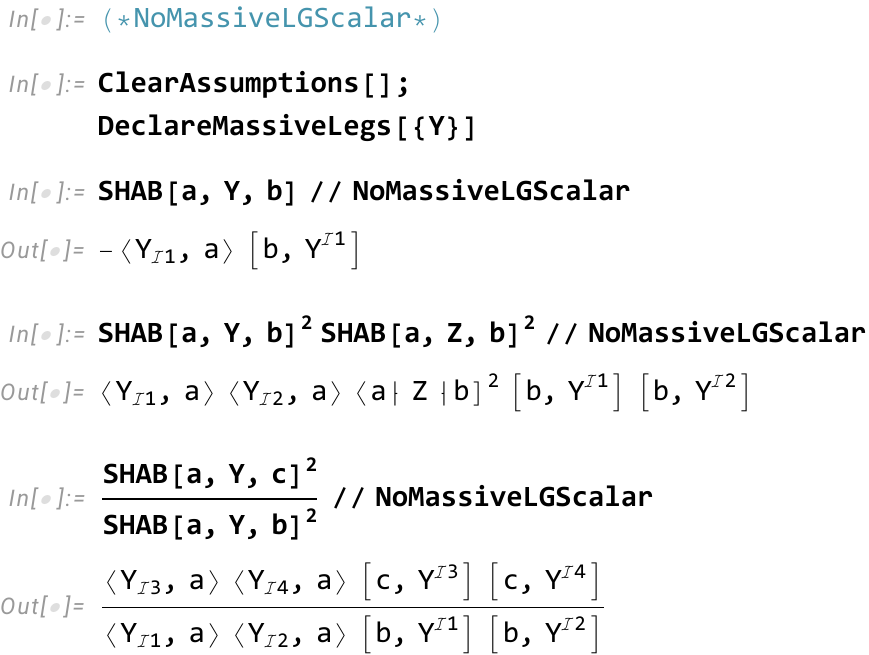}
	}
	
\end{examplebox}

\begin{documentationbox}{\tt{PutMassiveLGScalarFor}}{\tt{PutMassiveLGScalarFor[a\_, b\_, c\_][expr\_]}}
	
	The command \(\tt{PutMassiveLGScalarFor}\) gives explicit control over which
	massive little-group scalar is formed.  It contracts the specified middle leg
	\(b\):
	\begin{equation}
		\langle a,b_J\rangle [c,b^J]
		\quad \longrightarrow \quad
		\langle a|b|c].
	\end{equation}
	This is useful when several possible massive little-group contractions are
	available and the default \(\tt{PutLGScalar}\) choice is not the desired one.

\end{documentationbox}

\subsubsection{\tt{PutLGScalarFor}}

\begin{examplebox}{\tt{PutMassiveLGScalarFor}}
	\redbox{
		\includegraphics[scale=0.5]{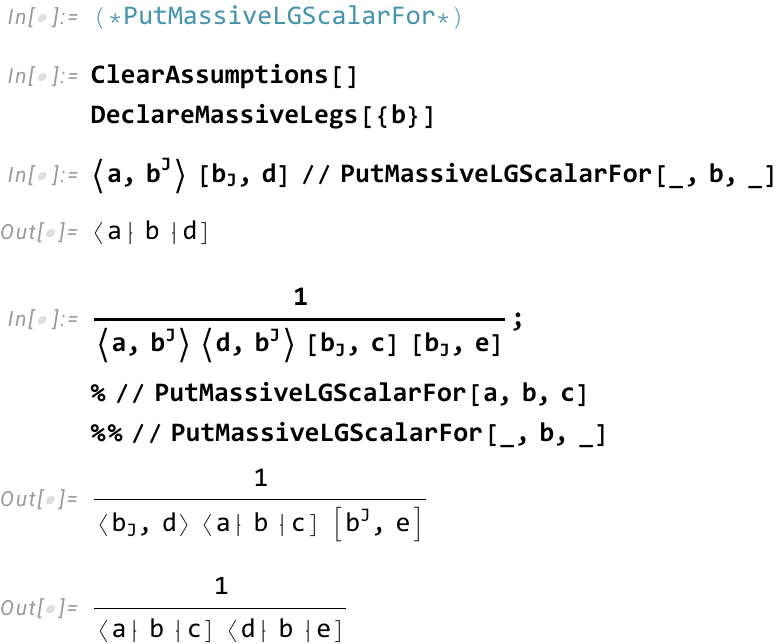}
	}
	
\end{examplebox}

\begin{documentationbox}{\tt{PutMasslessLGScalarFor}}{\tt{PutMasslessLGScalarFor[a\_, b\_, c\_][expr\_]}}
	
	The command \(\tt{PutMasslessLGScalarFor}\) gives explicit control over which
	massless little-group scalar is formed.  It contracts the specified middle leg
	\(b\):
	\begin{equation}
		\langle a,b\rangle [c,b]
		\quad \longrightarrow \quad
		\langle a|b|c].
	\end{equation}
	This is useful when several possible massless contractions are present and one
	wants to choose a particular \(\tt{SHAB[]}\) structure.

\end{documentationbox}

\begin{examplebox}{\tt{PutMasslessLGScalarFor}}
	
	\redbox{
		\includegraphics[scale=0.5]{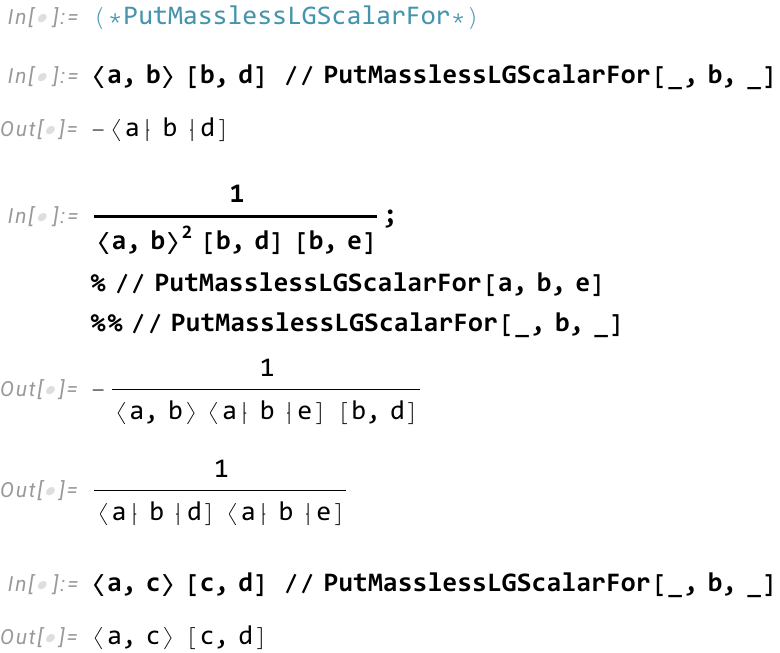}
	}
	
\end{examplebox}

\begin{documentationbox}{\tt{PutLGScalarFor}}{\tt{PutLGScalarFor[a\_, b\_, c\_][expr\_]}}
	
	The command \(\tt{PutLGScalarFor}\) combines
	\(\tt{PutMassiveLGScalarFor}\) and \(\tt{PutMasslessLGScalarFor}\).  It forms
	the little-group scalar with the specified middle leg \(b\), whether \(b\) is
	massive or massless:
	\begin{equation}
		\langle a,b\rangle [c,b]
		\quad \textrm{or} \quad
		\langle a,b_J\rangle [c,b^J]
		\quad \longrightarrow \quad
		\langle a|b|c].
	\end{equation}
	This command should be used when one needs explicit control over the scalar
	formed inside a larger expression.

\end{documentationbox}

\begin{examplebox}{\tt{PutLGScalarFor}}
	\redbox{
		\includegraphics[scale=0.5]{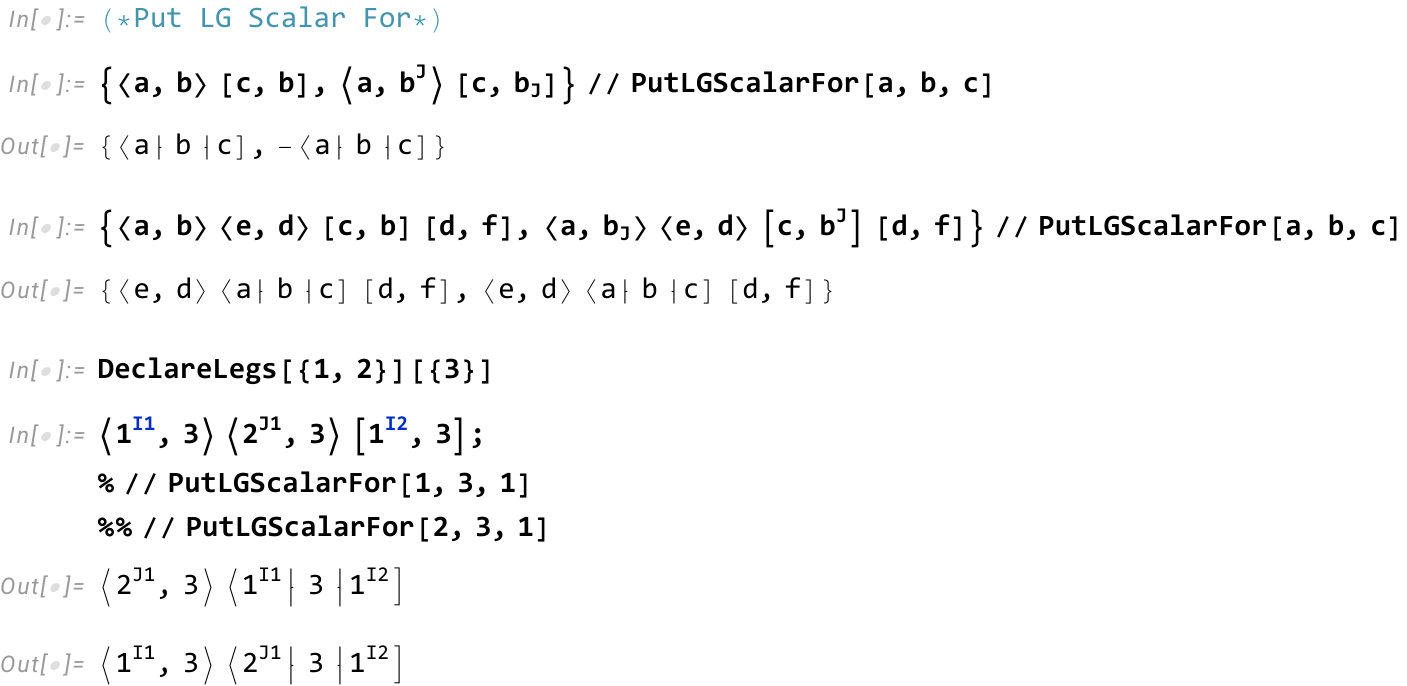}
	}
	
\end{examplebox}

\begin{documentationbox}{\tt{NoMassiveLGScalarWB}}{\tt{NoMassiveLGScalarWB[expr\_]}}
	
	The command \(\tt{NoMassiveLGScalarWB}\) is a variant of
	\(\tt{NoMassiveLGScalar}\).  The suffix \(\tt{WB}\) stands for ``with
	bilinear.''  It expands a massive little-group scalar into a product of
	\(\tt{SHAA[]}\) and \(\tt{SHBB[]}\), with the massive little-group indices
	contracted through an explicit invariant bilinear.
	
	Thus, unlike \(\tt{NoMassiveLGScalar}\), the little-group contraction is kept
	visible in the output.

\end{documentationbox}

\subsubsection{\tt{NoLGScalarWB}}

\begin{examplebox}{\tt{NoMassiveLGScalarWB}}
	
	\redbox{
		\includegraphics[scale=0.5]{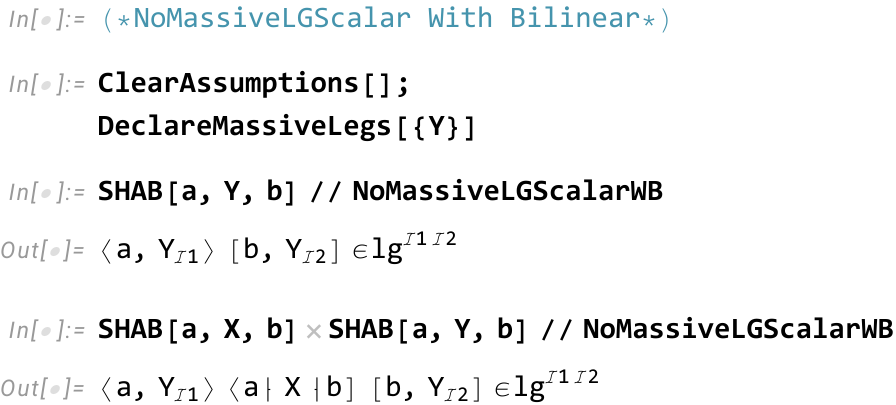}
	}
\end{examplebox}

\begin{documentationbox}{\tt{NoMasslessLGScalarWB}}{\tt{NoMasslessLGScalarWB[expr\_]}}
	
	The command \(\tt{NoMasslessLGScalarWB}\) is the massless counterpart of
	\(\tt{NoMassiveLGScalarWB}\).  Since there is no massive little-group bilinear
	in the massless case, its action is the same as \(\tt{NoMasslessLGScalar}\).
	
\end{documentationbox}

\begin{examplebox}{\tt{NoMasslessLGScalarWB}}
\end{examplebox}

\begin{documentationbox}{\tt{NoLGScalarWB}}{\tt{NoLGScalarWB[expr\_]}}
	
	The command \(\tt{NoLGScalarWB}\) combines
	\(\tt{NoMasslessLGScalarWB}\) and \(\tt{NoMassiveLGScalarWB}\).  It expands
	little-group scalars while keeping massive little-group contractions explicit
	through invariant bilinears whenever the middle leg is massive.
	
	The suffix \(\tt{WB}\) stands for ``with bilinear.''
	
\end{documentationbox}

\begin{examplebox}{\tt{NoSL2CScalar}}

\end{examplebox}

\subsection{More on \tt{SplitSchouten}}
\label{subsec:MoreonSplitSchouten}

\subsubsection{\tt{SplitSchoutenA(ngle)}}
\begin{documentationbox}{\tt{SplitSchoutenA}}{\tt{SplitSchoutenA[\{a\_, b\_\},\{c\_, d\_\}]}}
	
	Here \(a,b,c,d\) may denote either particle labels or spinor indices,
	depending on the object on which the command acts.
	
	\(\tt{SplitSchoutenA}\) applies Schouten identities in the
	angle sector.  It can split products of
	\begin{itemize}
		\item two \(\tt{SHAA}\)'s,
		\item two \(\tt{SHA}\)'s,
		\item two \(\mathtt{SU(2)_L}\) invariant bilinears.
	\end{itemize}
	
\end{documentationbox}

\begin{examplebox}{\tt{SplitSchoutenA}}
	
	\redbox{
		\includegraphics[scale=0.5]{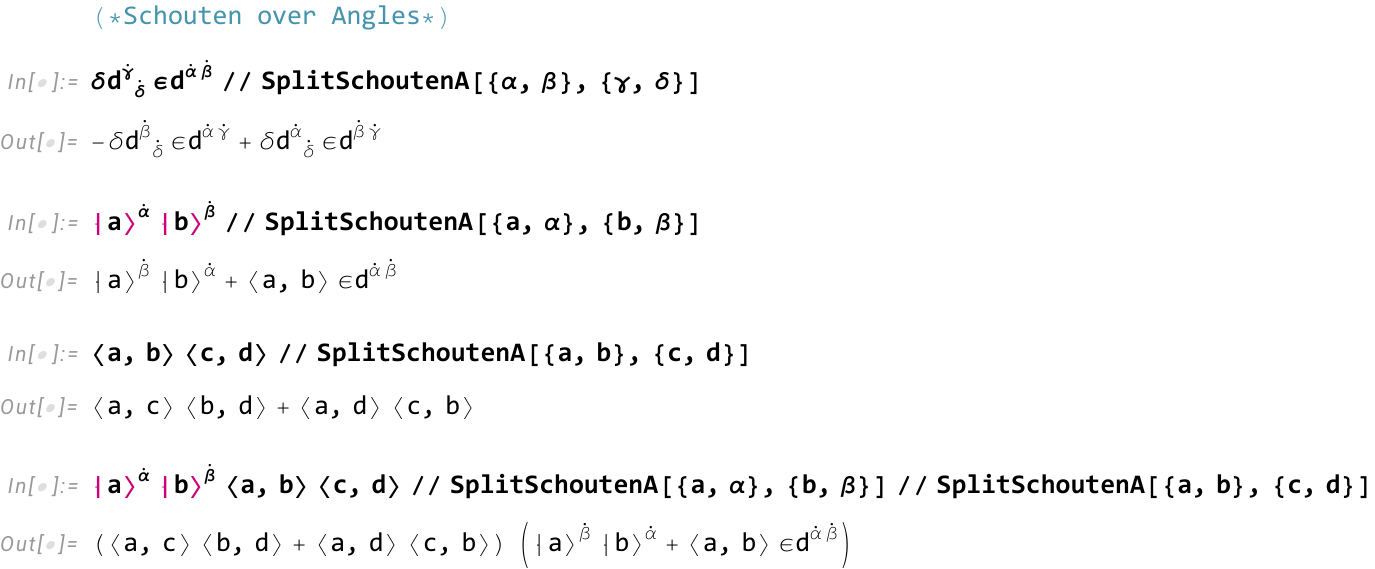}
	}
	
\end{examplebox}

\subsubsection{\tt{SplitSchoutenB(ox)}}

\begin{documentationbox}{\tt{SplitSchoutenB}}{\tt{SplitSchoutenB[\{a\_, b\_\},\{c\_, d\_\}]}}
	
	Here \(a,b,c,d\) may denote either particle labels or spinor indices,
	depending on the object on which the command acts.

	\(\tt{SplitSchoutenB}\) applies Schouten identities in the box
	sector.  It can split products of
	\begin{itemize}
		\item two \(\tt{SHBB}\)'s,
		\item two \(\tt{SHB}\)'s,
		\item two \(\mathtt{SU(2)_R}\) invariant bilinears.
	\end{itemize}

\end{documentationbox}

\begin{examplebox}{\tt{SplitSchoutenB}}

	\redbox{
		\includegraphics[scale=0.5]{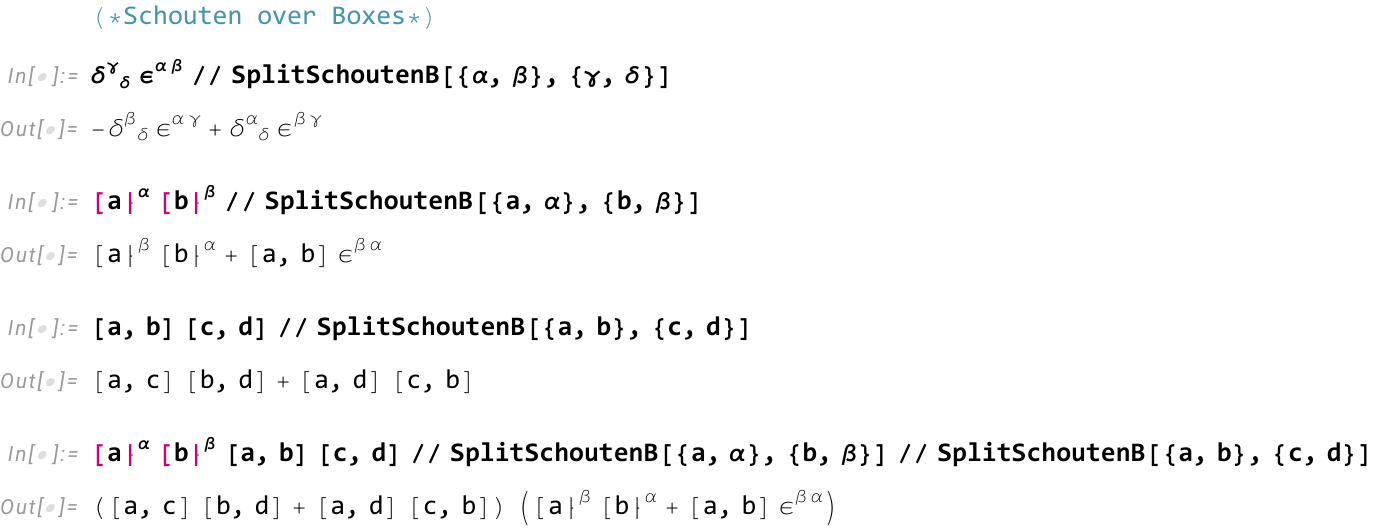}
	}
	
\end{examplebox}

\subsection{More on Canonicalize Indices}\label{app:moreoncanonicalizeindices}
\begin{conceptbox}{}
	The \tt{CanonicalizeIndices} command puts uniform indices throughout, keeping the left and the right handed scalars at the same footing. In some cases a user might need to have better control on canonicalization.  For example, consider an expression with massive legs $\mathtt{1,2}$ and massless legs $\mathtt{3,4}$ of the following schematic
form 
\begin{align}
	\tt{\(\langle\)3|1|3]\(\langle\)3|2|4]
		-\(\langle\)3|1|4]\(\langle\)3|2|3]} 
\end{align}
Since there is a common \(\mathtt{SU(2)_R}\) scalar structure in the two
terms, the expression can be simplified by first extracting the common factor
and then applying a Schouten identity. A similar expression where $\mathtt{SU(2)_L}$ scalars are common would be
\begin{align}
	\tt{\(\langle\)3|2|4]\(\langle\)4|1|4]
		-\(\langle\)3|1|4]\(\langle\)4|2|4]} 
\end{align}	
To put the same indices on the same \(\tt{SHAA[]}\) or
\(\tt{SHBB[]}\) factors, one can use
\begin{align}
	\tt{CanonicalizeSU2RScalars}
	\qquad\text{or}\qquad
	\tt{CanonicalizeSU2LScalars},
\end{align}
respectively.

\end{conceptbox}

\begin{documentationbox}{\tt{CanonicalizeSU2RScalars} and \tt{CanonicalizeSU2LScalars}}{\tt{CanonicalizeSU2RScalars[expr\_]} and \tt{CanonicalizeSU2LScalars[expr\_]}}

The command \tt{CanonicalizeSU2RScalars}
canonicalizes the \(\mathtt{SU(2)_R}\) scalar factors, while \tt{CanonicalizeSU2LScalars} canonicalizes the \(\mathtt{SU(2)_L}\) scalar factors. These commands are useful before applying Schouten identities or simplifying
expressions using momentum conservation.

\end{documentationbox}

\begin{examplebox}{\tt{CanonicalizeSU2RScalars} and \tt{CanonicalizeSU2LScalars}}

\redbox{\includegraphics[scale=0.5]{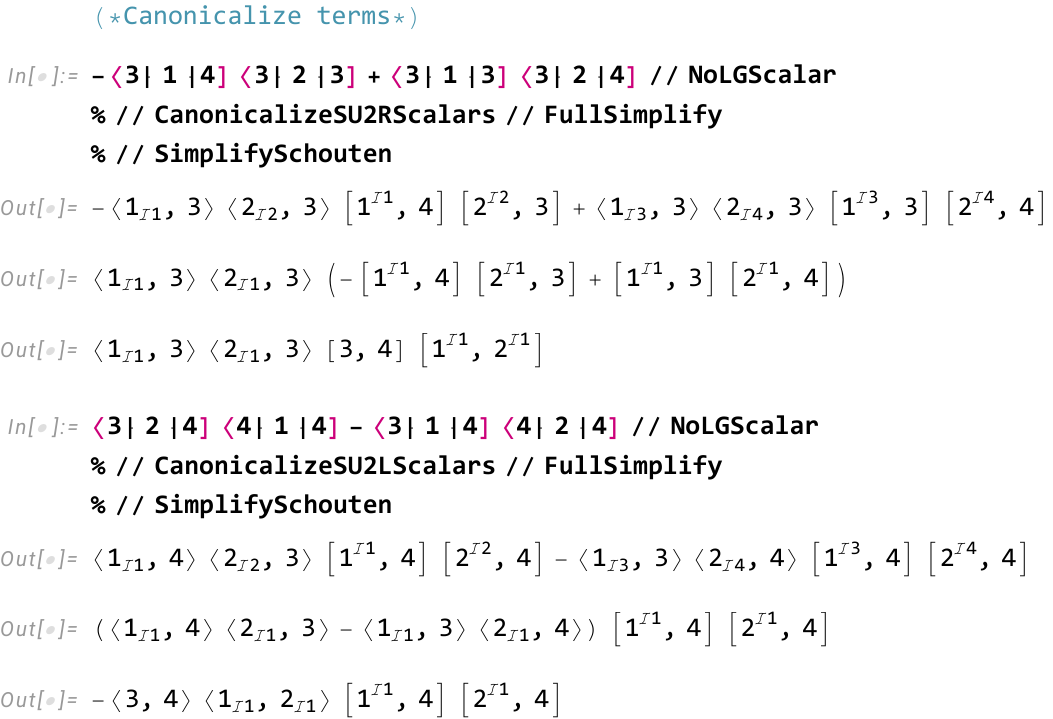}}

The command \(\tt{NoLGScalar}\) splits \(\tt{SHAB[]}\) into
\(\tt{SHAA[]SHBB[]}\), with unique \(SL(2,\mathbb C)\) indices attached to
the massive legs. 
\end{examplebox}

\subsection{More on Simplify Polynomial}
\label{subsec:moreonsimplifypolynomial}

\begin{documentationbox}{\tt{SimplifyPolynomialFunction}}{\tt{SimplifyPolynomialFunction[expr\_]}}
	
	The command \(\tt{SimplifyPolynomialFunction}\) applies one pass of the basic
	spinor-helicity polynomial simplification rules.  It combines the following
	operations \(\tt{ContractBilinears}\), \(\tt{PutSL2CScalar}\), \(\tt{PutOnShell}\),  \(\tt{PutLGScalar}\)

\end{documentationbox}

\subsection{More on Component Form}\label{sec:MoreonComponentForm}

\subsubsection{Bilinear components}

\begin{conceptbox}{Metric and invariant tensors}
	
	The most basic component objects are the Lorentz metric and the left- and
	right-handed Levi-Civita tensors.  The component matrix for the Lorentz metric
	is
	\begin{align}
		\eta\tt{Mat}.
	\end{align}
	The same matrix is used for both \(\eta_{\mu\nu}\) and \(\eta^{\mu\nu}\).
	
	The component forms of the antisymmetric Levi-Civita tensors
	\(\epsilon^{\alpha\beta}\) and \(\epsilon_{\alpha\beta}\) are given by
	\begin{align}
		\epsilon\tt{MatUpper},
		\qquad
		\epsilon\tt{MatLower}.
	\end{align}
	Similarly, the corresponding \(\mathtt{SU(2)_R}\) and
	\(\mathtt{SU(2)_{LG}}\) objects are available through the dotted and
	little-group versions of these commands.
	
	The package also provides component forms for the sigma matrices
	\begin{align}
		\sigma^\mu_{\alpha\dot{\beta}},
		\qquad
		\sigma_{\mu\,\alpha\dot{\beta}},
		\qquad
		\bar\sigma^{\mu\,\dot{\beta}\alpha},
		\qquad
		\bar\sigma_{\mu}^{\dot{\beta}\alpha}.
	\end{align}
	
\end{conceptbox}

\begin{documentationboxWL}{\(\eta\)\tt{Mat}, \(\epsilon\)\tt{Mat}, \(\delta\)\tt{Mat}, and sigma matrices}{Component matrices}
	\label{dcmnt:componentmatrices}
	
	The Lorentz metric, spinor Levi-Civita tensors, little-group invariant tensors,
	and sigma matrices can be accessed through the following component matrices:
	\begin{align}
		&\eta\tt{Mat},
		\\
		&\epsilon\tt{MatUpper},
		\qquad
		\epsilon\tt{MatLower},
		\qquad
		\delta\tt{Mat},
		\\
		&\epsilond\tt{MatUpper},
		\qquad
		\epsilond\tt{MatLower},
		\qquad
		\deltad\tt{Mat},
		\\
		&\epsilonlg\tt{MatUpper},
		\qquad
		\epsilonlg\tt{MatLower},
		\qquad
		\deltalg\tt{Mat},
		\\
		&\sigma\tt{VecUpper},
		\qquad
		\sigma\tt{VecLower},
		\qquad
		\sigma\tt{barVecUpper},
		\qquad
		\sigma\tt{barVecLower}.
	\end{align}
	
\end{documentationboxWL}

\begin{examplebox}{Metric, bilinears and sigma-matrix components}
	
	\redbox{
		\begin{minipage}{0.6\linewidth}
			\includegraphics[scale=0.3]{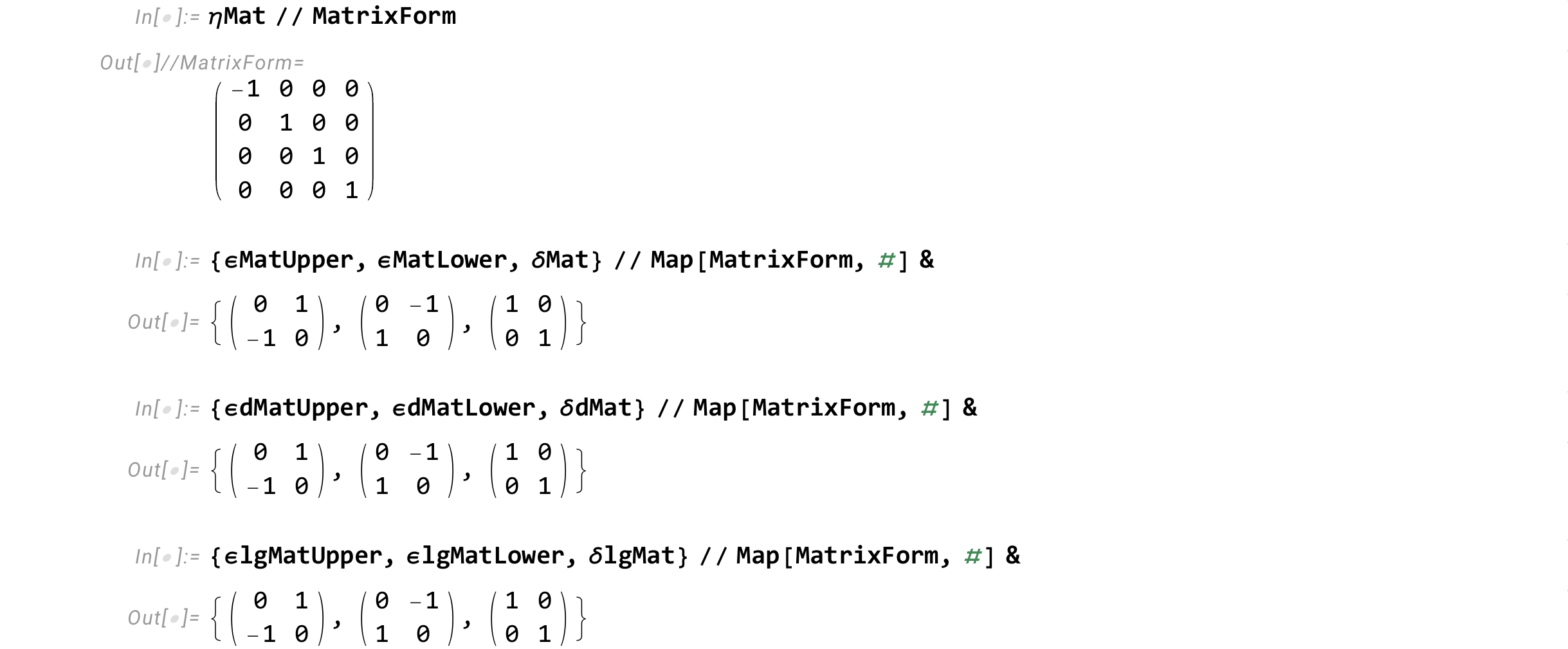}
		\end{minipage}
		\begin{minipage}{0.5\linewidth}
			\includegraphics[scale=0.5]{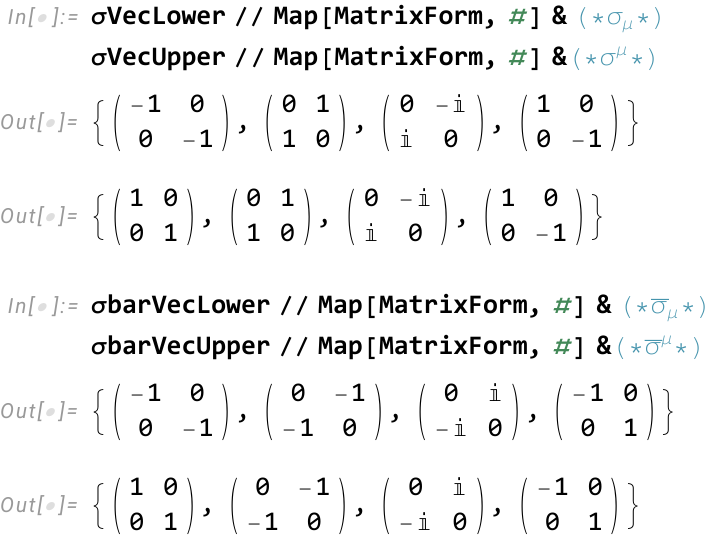}
		\end{minipage}
	}
	
\end{examplebox}

\subsubsection{Momentum components}

\begin{documentationbox}{\tt{pVecLower} and \tt{pVecUpper}}{\tt{pVecLower[leg\_]} and \tt{pVecUpper[leg\_]}}
	
	provides the four-momentum vectors \(p_\mu\) and \(p^\mu\) for any particle leg are  in Cartesian coordinates.

\end{documentationbox}

\begin{documentationbox}{\tt{pMatLower} and \tt{pMatUpper}}{\tt{pMatLower[leg\_]} and \tt{pMatUpper[leg\_]}}
	
	provides spinor-helicity momentum matrix \(p_{\alpha\dot{\beta}}\) and \(p^{\dot{\beta}\alpha}\)  respectively. 
	
\end{documentationbox}

\begin{conceptbox}{Momentum-component notation}
	
	The following notation is used for momentum components, masses, energies, and
	spherical angles.
	
	\begin{center}
		\begin{tabular}{|c|c|c|}
			\hline
			\tt{InputForm} & \tt{OutputForm} & \tt{Keyboard Shortcut} \\[1ex]
			\hline
			\(\mathtt{p[i][comp]}\) & \(\mathtt{(p_i)_{comp}}\) &
			\(\boxed{\tt{Esc}}\tt{p}\boxed{\tt{Esc}}\) \\
			\hline
			\(\mathtt{m[i]}\) & \(\mathtt{m_i}\) &
			\(\boxed{\tt{Esc}}\tt{ms}\boxed{\tt{Esc}}\) \\
			\hline
			\(\mathtt{E[i]}\) & \(\mathtt{E_i}\) &
			\(\boxed{\tt{Esc}}\tt{En}\boxed{\tt{Esc}}\) \\
			\hline
			\(\mathtt{\theta[i]}\) & \(\mathtt{\theta_i}\) &
			\(\boxed{\tt{Esc}}\theta\boxed{\tt{Esc}}\) \\
			\hline
			\(\mathtt{\phi[i]}\) & \(\mathtt{\phi_i}\) &
			\(\boxed{\tt{Esc}}\phi\boxed{\tt{Esc}}\) \\
			\hline
			\(\mathtt{Modp[i]}\) & \(\mathtt{|p_i|}\) &
			\(\boxed{\tt{Esc}}\tt{mp}\boxed{\tt{Esc}}\) \\
			\hline
		\end{tabular}
	\end{center}
	
\end{conceptbox}

\begin{documentationbox}{\tt{SphericalCoordinates} and \tt{CartesianCoordinates}}{\tt{SphericalCoordinates[expr\_]} and \tt{CartesianCoordinates[expr\_]}}
	
	The command	\tt{SphericalCoordinates} converts the momentum components to spherical coordinates.  The Cartesian coordinate representation can be accessed using	\tt{CartesianCoordinates}.
	
\end{documentationbox}

\begin{examplebox}{Momentum vectors and momentum matrices}
	
	\redbox{
		\begin{minipage}{0.4\linewidth}
			\includegraphics[scale=0.4]{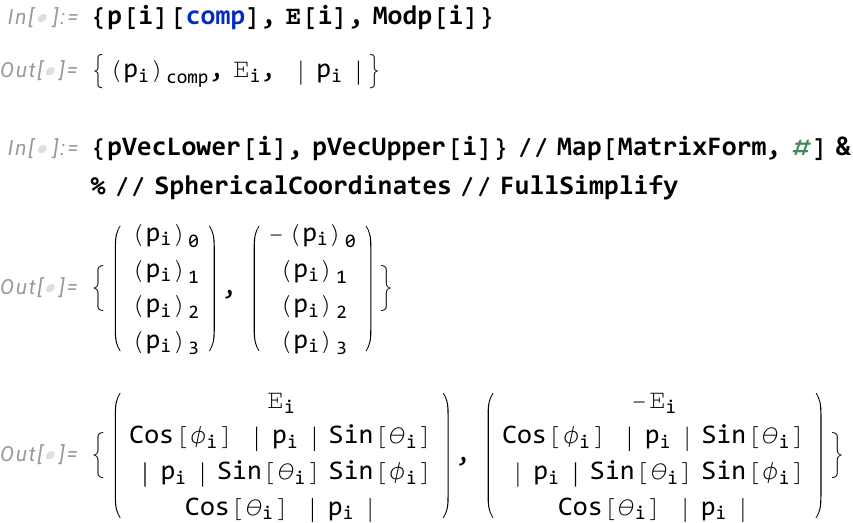}
		\end{minipage}
		\begin{minipage}{0.3\linewidth}
			\includegraphics[scale=0.2]{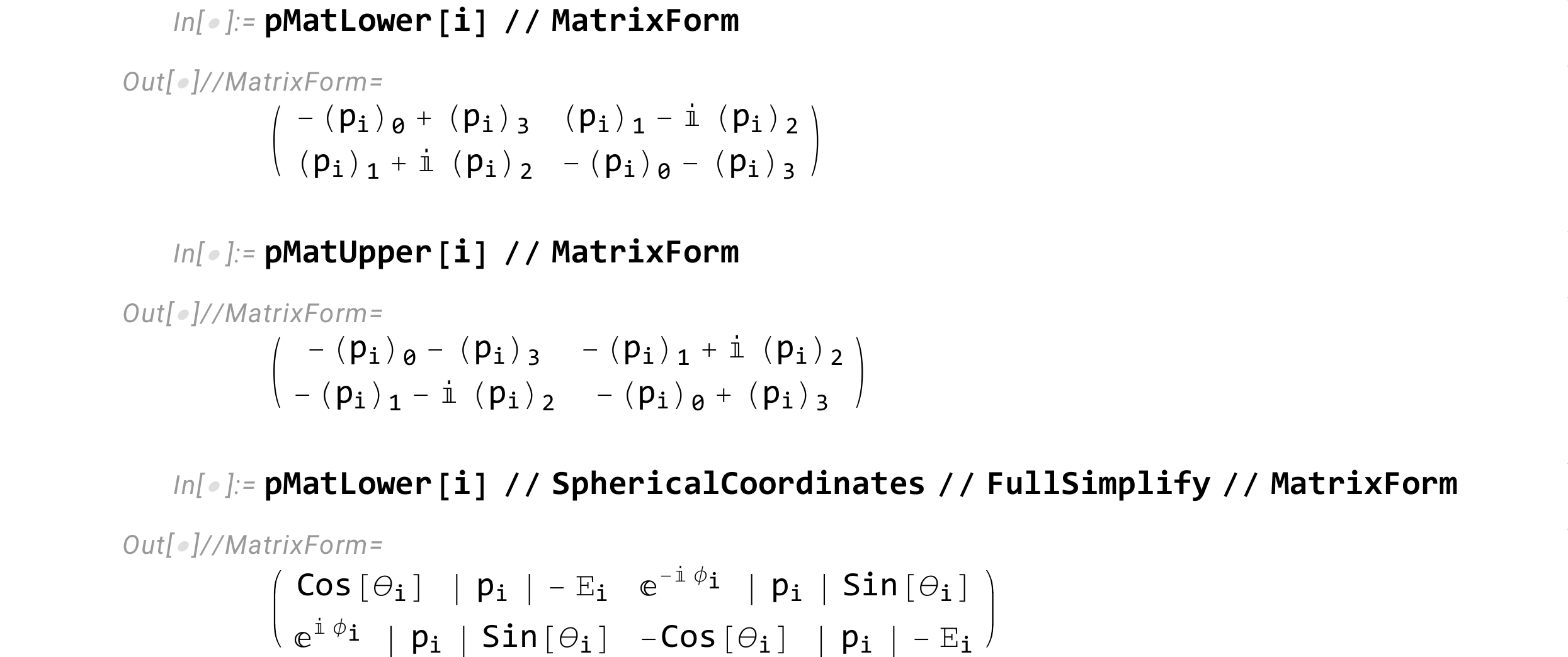}
		\end{minipage}
	}
	
\end{examplebox}

\begin{conceptbox}{Component-form assumptions}
	
	The angles \(\mathtt{\theta_i,\phi_i}\) and momentum components
	\(\mathtt{p_i}\) are assumed to be real.  The mass \(\mathtt{m_i}\), energy
	\(\mathtt{E_i}\), and momentum magnitude \(\mathtt{|p_i|}\) are assumed to be
	positive and real.
	
	These assumptions are appropriate for positive-energy particles.  In an
	all-incoming or all-outgoing convention, negative-energy momenta must first be
	flipped.  This can be done using \(\tt{ScaleSpinors}\) or
	\(\tt{ScaleMomenta}\).
	
\end{conceptbox}

\begin{documentationbox}{\tt{SetComponentFormAssumption} and \tt{UnsetComponentFormAssumption}}{\tt{SetComponentFormAssumption[]} and \tt{UnsetComponentFormAssumption[]}}
	
	The assumptions used in component form can be added to, or removed from, \(\tt{\$Assumptions}\) for declared legs using these commands.
	
\end{documentationbox}

\begin{examplebox}{Component-form assumptions for a massive leg}
	
	\redbox{
		\includegraphics[scale=0.5]{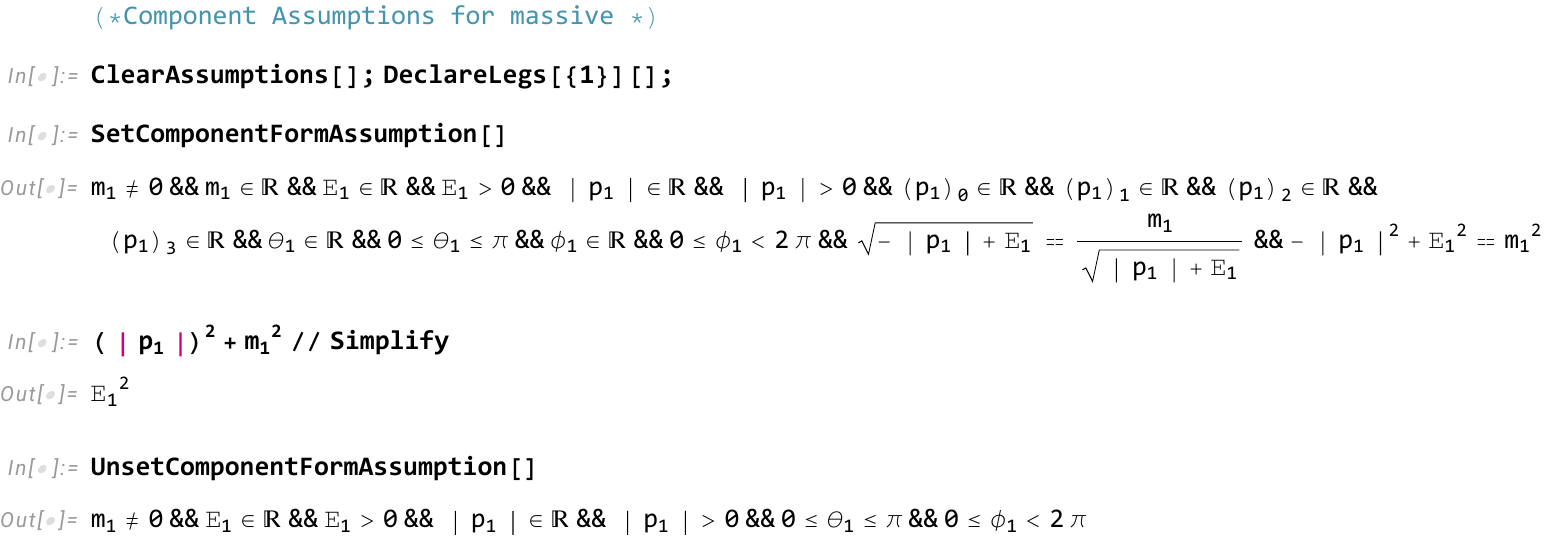}
	}
	
\end{examplebox}

\begin{examplebox}{Component-form assumptions for a massless leg}
	
	\redbox{
		\includegraphics[scale=0.5]{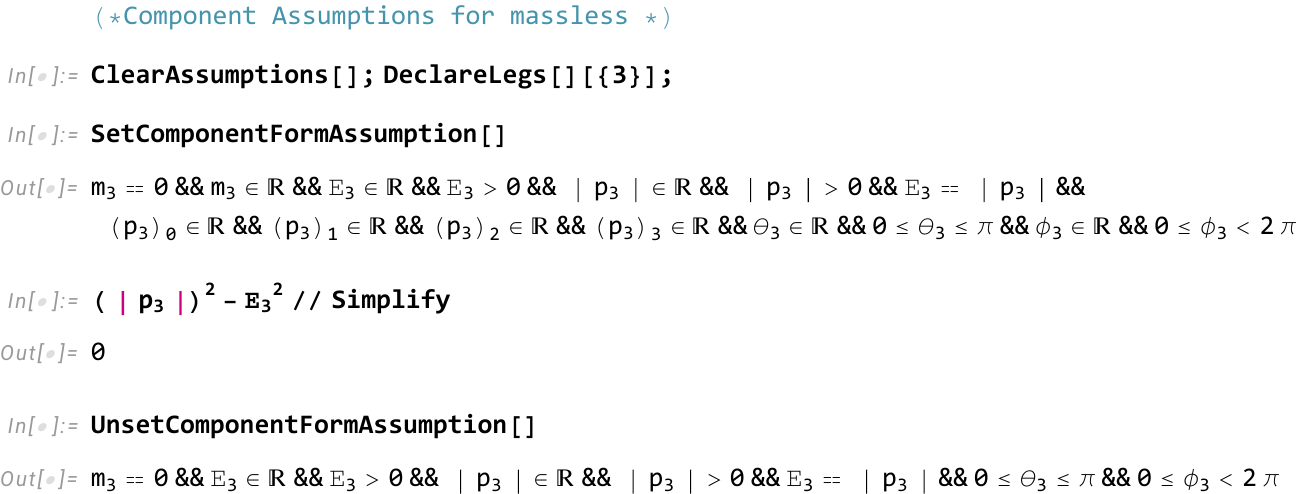}
	}
	
\end{examplebox}

\subsubsection{\(\mathtt{SU(2)_L}\) and \(\mathtt{SU(2)_R}\) components}

\begin{conceptbox}{Spinor components}
	
	The conversion of angle and box spinors depends on whether the corresponding
	leg carries little-group indices.  For a massive leg, the spinors are written as
	\begin{align}
		\tt{SHB[i[J],\(\alpha\)]},
		\qquad
		\tt{SHA[i[J],\(\alpha\)]},
	\end{align}
	where \(\tt{J}\in\{\pm1,\pm2\}\).  The sign of the spinor index determines
	whether the spinor is left-facing or right-facing.
	
	The component form is encoded by the following notation:
	\begin{align}
		\tt{l[B][i[J]]}
		&\longrightarrow
		\tt{SHB[i[J],\(\alpha\)]},
		&
		\tt{r[B][i[J]]}
		&\longrightarrow
		\tt{SHB[i[J],-\(\alpha\)]},
		\\
		\tt{l[A][i[J]]}
		&\longrightarrow
		\tt{SHA[i[J],-\(\alpha\)]},
		&
		\tt{r[A][i[J]]}
		&\longrightarrow
		\tt{SHA[i[J],\(\alpha\)]}.
	\end{align}
	Here \(\tt{l[]}\) denotes a left-facing spinor, and \(\tt{r[]}\) denotes a
	right-facing spinor.
	
\end{conceptbox}

\begin{documentationbox}{\tt{l[B][]}, \tt{r[B][]}, \tt{l[A][]}, and \tt{r[A][]}}{\tt{Spinor components}}
	
	The commands
	\begin{align}
		\tt{l[B][]},
		\qquad
		\tt{r[B][]},
		\qquad
		\tt{l[A][]},
		\qquad
		\tt{r[A][]}
	\end{align}
	give the component forms of box and angle spinors.
	
\end{documentationbox}

\begin{examplebox}{Spinor components}
	
	\redbox{
		\includegraphics[scale=0.5]{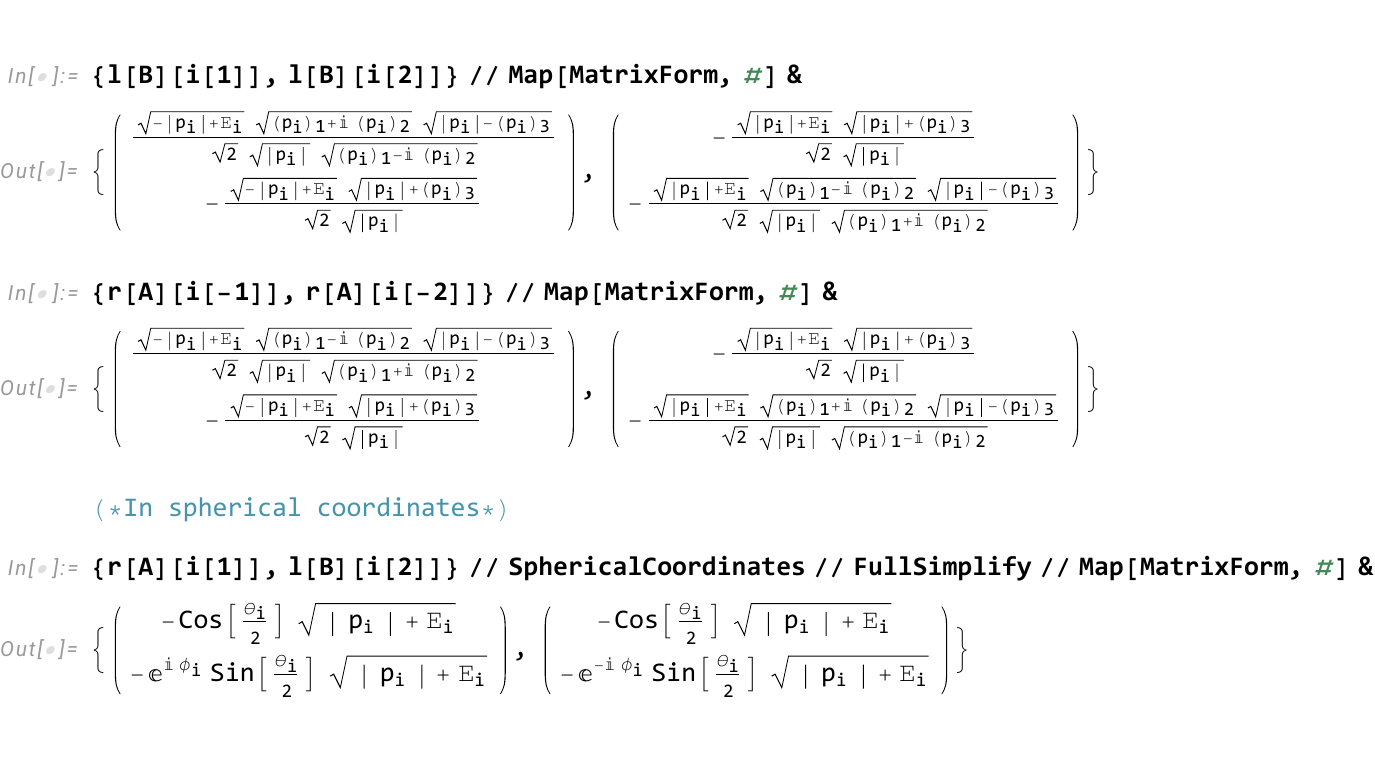}
	}
	
\end{examplebox}

\begin{documentationbox}{\tt{PutOnShellComponent}}{\tt{PutOnShellComponent[expr\_]}}
	
	imposes	$\mathtt{E_i^2-|p_i|^2=m_i^2}$ for each leg.  
	
	
\end{documentationbox}

\begin{examplebox}{Putting component expressions on shell}
	
	\redbox{
		\includegraphics[scale=0.5]{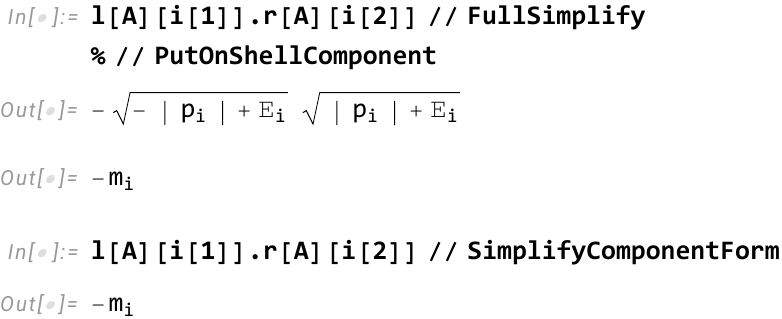}
	}
	
\end{examplebox}

\begin{documentationbox}{\tt{SimplifyComponentForm}}{\tt{SimplifyComponentForm[expr\_]}}

	Use this to simplify polynomials involving these square roots.

\end{documentationbox}

\begin{morematerialbox}
	This command repeatedly applies
	\(\tt{PowerExpand}\), \(\tt{PutOnShellComponent}\), and
	\(\tt{PowerContract}\), together with \(\tt{FullSimplify}\).  This may slow
	down the evaluation, but it is often necessary because component expressions
	with square roots can be difficult to simplify.	
\end{morematerialbox}

\begin{examplebox}{On-shell component simplification}
	
	\redbox{
		\includegraphics[scale=0.5]{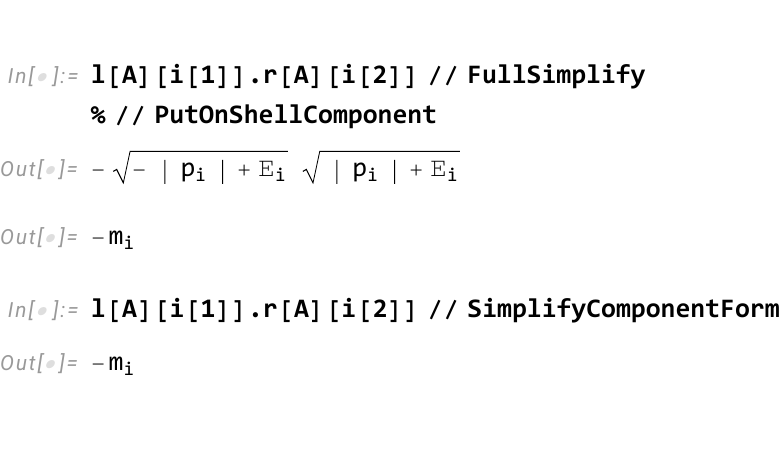}
	}
	
\end{examplebox}

\subsection{More on Numerics}
\begin{documentationbox}{\tt{RAMBO}}{\tt{RAMBO[n\_,Energy\_,massesList\_,iterations\_]}}

	The \tt{RAMBO} command takes as input the number of particles \tt{n}, the energy of the scattering process\footnote{This is the total energy available for the scattering.} and the numerical values of the masses of the \tt{n} particles. The algorithm uses the Newton-Raphson method to satisfy the mass-shell, so the last argument is reserved for the number of iterations used in the method. It's default value is \tt{20}. The output is a list of randomly generated momenta which are on-shell and also satisfy momentum conservation i.e. the sum of all the spatial momenta is zero whereas the sum of their energies is equal to the total energy \tt{E}. If the sum of the masses exceeds the threshold energy $\tt{E}$, it throws an error.
\end{documentationbox}
\begin{examplebox}{\tt{RAMBO}}
	For example,
	\redbox{\includegraphics[scale=0.5]{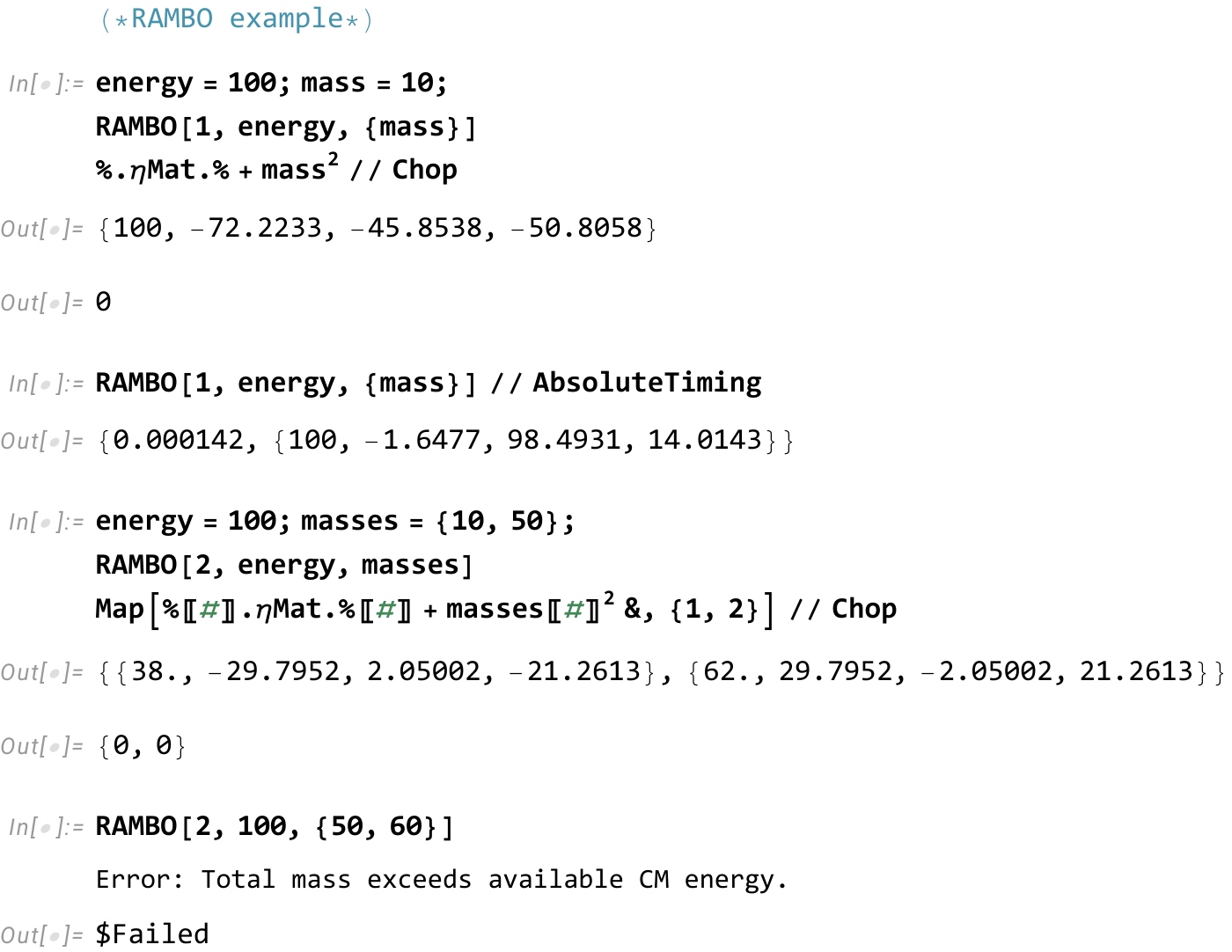}}
\end{examplebox}

\subsection{More on High Energy Limit}
\label{app:MoreonHighEnergy}

\begin{conceptbox}{Decompose to $\mathtt{SU(2)}$ fundamental basis}
	
The basis elements for any leg $\tt{i}$ can be accessed using the commands \tt{SU2LBasis[i]}, \tt{SU2RBasis[i]} and \tt{SU2LGBasis}. One can also extract the components of the little group basis elements using the command \tt{SU2LGBasisComponent}. Following the discussion in \ref{app:BasisDecomposition}, we have fixed the notation for the basis elements as follows : $\chi_{\tt{i}+} \rightarrow \chi \tt{[i,p]}$, $\mathtt{\chi_{i-} \rightarrow \chi \tt{[i,m]}}$, $\mathtt{\lambda_{i+} \rightarrow \lambda \tt{[i,p]}}$ and $\mathtt{\lambda_{i-} \rightarrow \lambda \tt{[i,m]}}$. The \tt{Format} for \tt{$\lambda$[i\_,p\_]} is \tt{$\lambda_{\tt{ip}}$}, and for \tt{$\chi$[i\_,p\_]} is \tt{$\chi_{\tt{ip}}$}. 
	
\end{conceptbox}

\begin{documentationbox}{\tt{SU2LBasis}, \tt{SU2RBasis}, \tt{SU2LGBasis} and \tt{SU2LGBasisComponent}}{\tt{SU2LBasis[leg\_]}, \tt{SU2RBasis[leg\_]}, \tt{SU2LGBasis} and \tt{SU2LGBasisComponent}}

These commands gives the component matrix of the $\mathtt{SU(2)_{L}}$, $\mathtt{SU(2)_{R}}$ and $\mathtt{SU(2)_{LG}}$ spinors for a massive \tt{leg}. The \tt{SU2LGBasisComponent} specifically picks the compponents of the little group basis spinors.
	
\end{documentationbox}

\begin{examplebox}{\tt{SU2LBasis}, \tt{SU2RBasis}, \tt{SU2LGBasis} and \tt{SU2LGBasisComponent}}
	
\redbox{\includegraphics[scale=0.5]{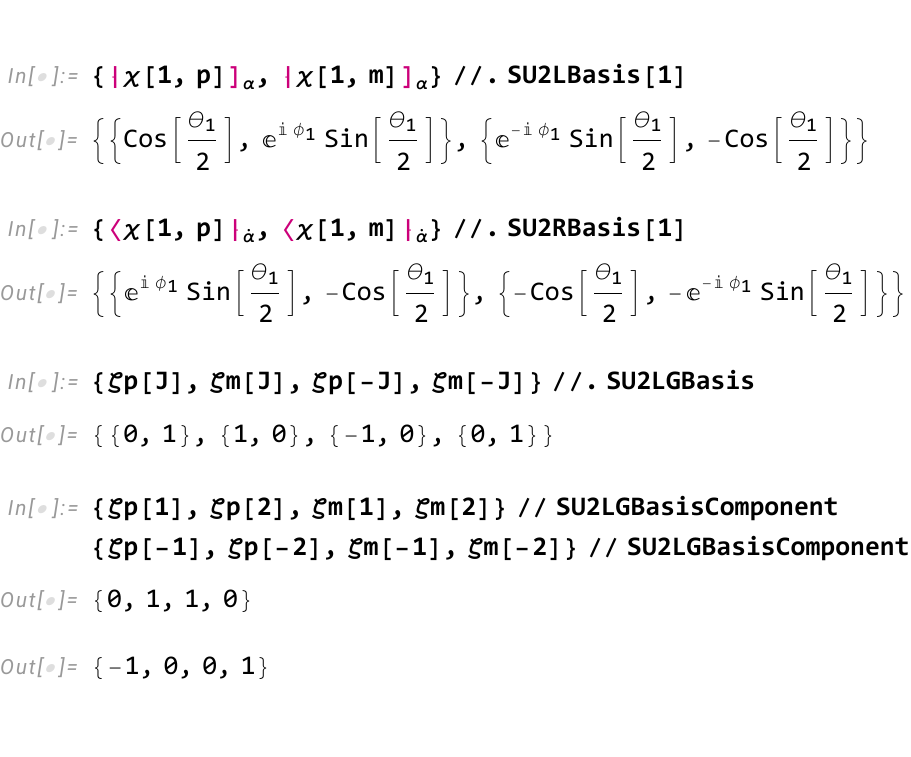}}

\end{examplebox}

\begin{documentationbox}{\tt{SU2LInnerProduct}, \tt{SU2RInnerProduct} and \tt{SU2LGInnerProduct}}{\tt{SU2LInnerProduct[leg]}, \tt{SU2RInnerProduct[leg]} and \tt{SU2LGInnerProduct}}
	
Imposes the inner product rules \eqref{BasisInnerPrioduct} over the basis spinors for a massive \tt{leg}.

\end{documentationbox}

\begin{examplebox}{\tt{SU2LInnerProduct}, \tt{SU2RInnerProduct} and \tt{SU2LGInnerProduct}}
	
\redbox{\includegraphics[scale=0.5]{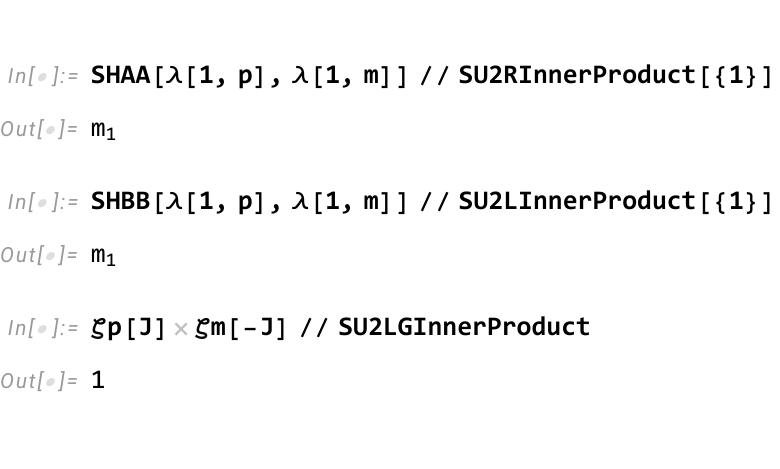}}

\end{examplebox}

\begin{documentationbox}{\tt{DimFulltoDimLessBasis}}{\tt{DimFulltoDimLessBasis[Massivelegs\_List]}}

This command decomposes the dimensionfull basis spinors to the dimensionless basis spinors for the \tt{Massivelegs}.
	
\end{documentationbox}

\begin{examplebox}{\tt{DimFulltoDimLessBasis}}

\redbox{\includegraphics[scale=0.5]{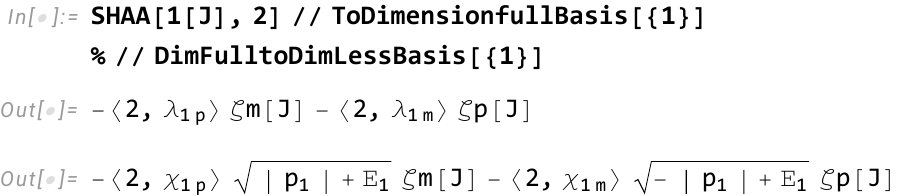}}

\end{examplebox}

\begin{conceptbox}{Decompose to massless}

	It is sometimes useful to get away from the little group covariance and rather write the spinor components $\mathtt{\lambda_{i\pm}}$ in terms of the massless spinor $\mathtt{\lambda}$ and its reference spinor $\mathtt{\eta}$.
 \begin{align}
\mathtt{|\lambda_{i-}] = |\lambda] \quad,\quad \langle \lambda_{i+}] = \langle \lambda| \quad,\quad |\lambda_{i+}] = |\eta]\quad,\quad \langle \lambda_{i-} |=\langle \eta|}
 \end{align}
  It is easier to impose all these rules using a single command \tt{ToMassless}.
\end{conceptbox}

\begin{documentationbox}{\tt{ToMassless}}{\tt{ToMassless[rules\_List][expr\_]}}
	
Decomposes the massive spinors in terms of massless spinors. Every rule in the argument is of the following form $$\mathtt{Massiveleg\rightarrow \{MasslessLeg1,MasslessLeg2\}}$$ where a \tt{Massiveleg} goes to a pair of massless legs \tt{\{MasslessLeg1,MasslessLeg2\}}.

\end{documentationbox}

\begin{examplebox}{\tt{ToMassless}}
	
\redbox{\includegraphics[scale=0.5]{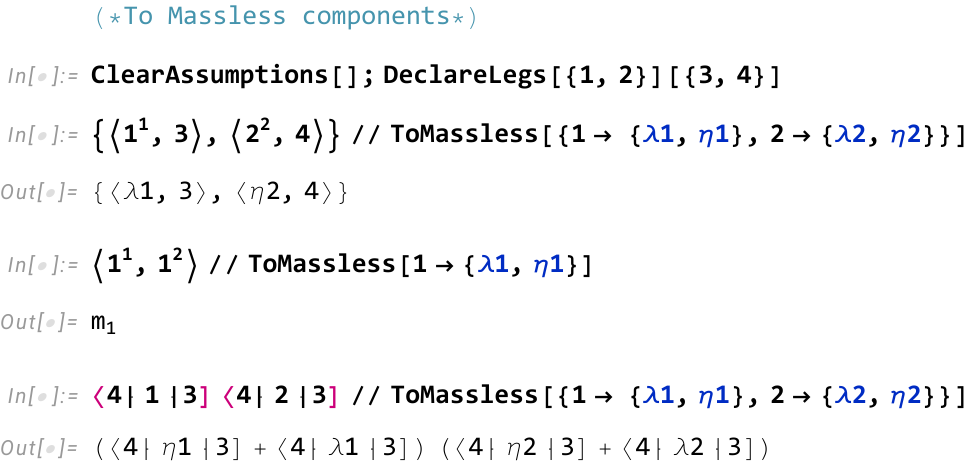}}

\end{examplebox}

\subsection{More on Lorentz to spinor-helicity conversions}
\label{sec:moreonLorentztoSH}

\begin{documentationbox}{\tt{ConvertMettoSH}}{\tt{ConvertMettoSH[expr\_]}}
	
	converts the Lorentz spacetime metric	\tt{\(\eta\)[\(\mu\),\(\nu\)]} and the four-dimensional Levi-Civita tensor into the corresponding
	\(\mathtt{SU(2)_L}\) and \(\mathtt{SU(2)_R}\) Levi-Civita tensors.
	
\end{documentationbox}

\begin{examplebox}{\tt{ConvertLortoSH}}
	
	\redbox{
		\includegraphics[scale=0.5]{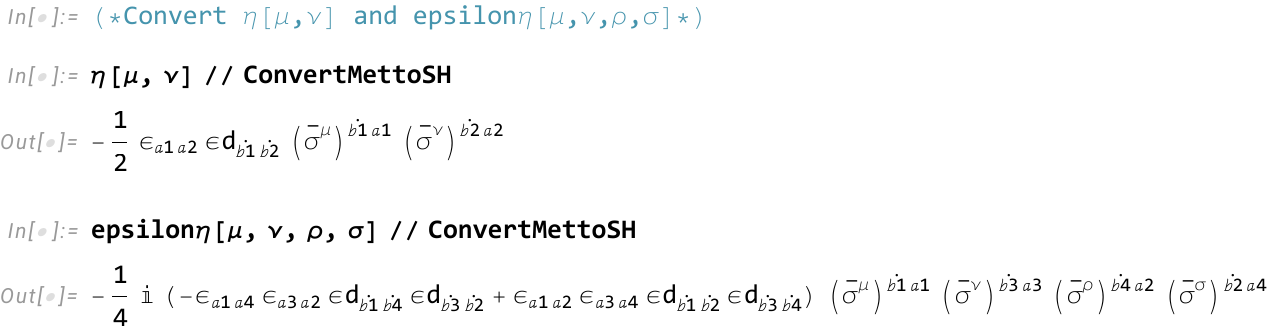}
	}
	
\end{examplebox}

\begin{examplebox}{\tt{ConvertLortoSH} II}
	
	\redbox{
		\includegraphics[scale=0.5]{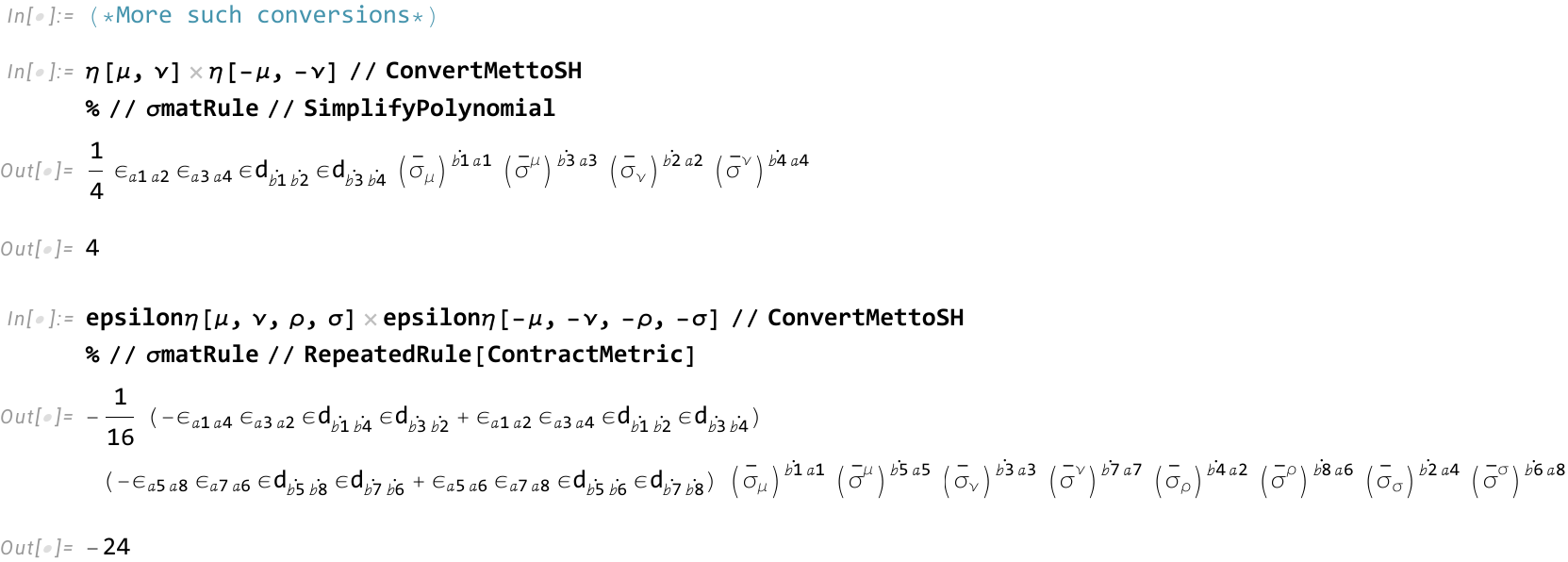}
	}
	
\end{examplebox}

\begin{documentationbox}{\tt{ConvertMomtoSH}}{\tt{ConvertMomtoSH[legs\_List,momheads\_List]}}
	
	converts Lorentz momenta into spinor-helicity variables.
	
	It takes two arguments, both of which are lists:
	\begin{itemize}
		\item \(\tt{legs}\): the list of leg labels,
		\item \(\tt{momheads}\): the corresponding list of momentum heads.
	\end{itemize}
\end{documentationbox}
For example, to convert an expression involving the Lorentz momenta
\(\mathtt{k1^\mu}\) and \(\mathtt{k2^\mu}\) for legs \(\tt{1}\) and \(\tt{2}\), one
uses
\begin{align}
	\mathtt{ConvertMomtoSH[\{1,2\},\{k1,k2\}]} .
\end{align}
\begin{examplebox}{\tt{ConvertMomtoSH}}
	
	\redbox{
		\includegraphics[scale=0.5]{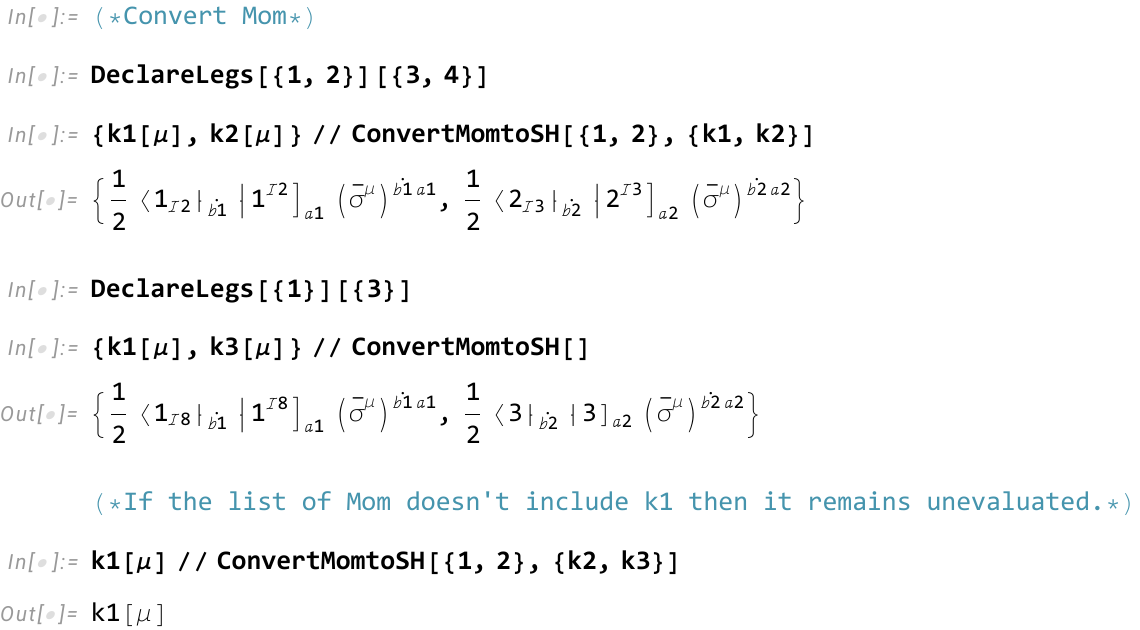}
	}
	
\end{examplebox}

The command \(\tt{ConvertMomtoSH}\) can also be used to convert Mandelstam
variables.  For example, in \(2\to 2\) Compton scattering with masses satisfying
\begin{align}
	\tt{m[1]\(\neq\)0},
	\qquad
	\tt{m[2]\(\neq\)0},
	\qquad
	\tt{m[3]==0},
	\qquad
	\tt{m[4]==0},
\end{align}
one may define
\begin{align}
	\mathtt{s}
	&=
	\mathtt{2(k_3\cdot k_4)},
	&
	\mathtt{t+m^2}
	&=
	\mathtt{2(k_1\cdot k_4)},
	&
	\mathtt{u+m^2}
	&=
	\mathtt{2(k_1\cdot k_3)} .
\end{align}
The corresponding spinor-helicity forms can then be obtained using
\(\tt{ConvertMomtoSH}\).

\begin{examplebox}{Mandelstam variables in spinor-helicity notation}
	
	\redbox{
		\includegraphics[scale=0.5]{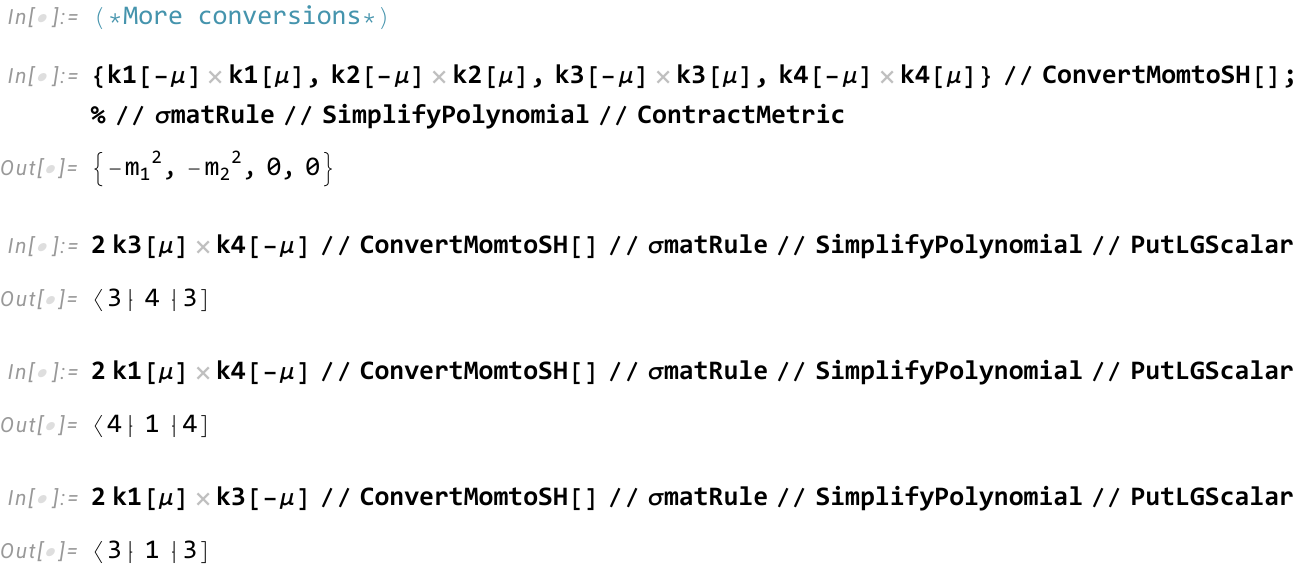}
	}
	
\end{examplebox}

\begin{documentationbox}{\tt{ConvertMsPoltoSH}}{\tt{ConvertMsPoltoSH[legs\_List,polheads\_List]}}

	converts massive Lorentz polarizations into spinor-helicity variables.
	
	It takes two arguments:
	\begin{itemize}
		\item \(\tt{legs}\): the list of leg labels,
		\item \(\tt{polheads}\): the list of polarization heads, ordered in the
		same way as \(\tt{legs}\).
	\end{itemize}
	The conversion automatically assigns little-group indices based on the leg
	label.  
\end{documentationbox}

\begin{examplebox}{\tt{ConvertMsPoltoSH}}
	
	\redbox{
		\includegraphics[scale=0.5]{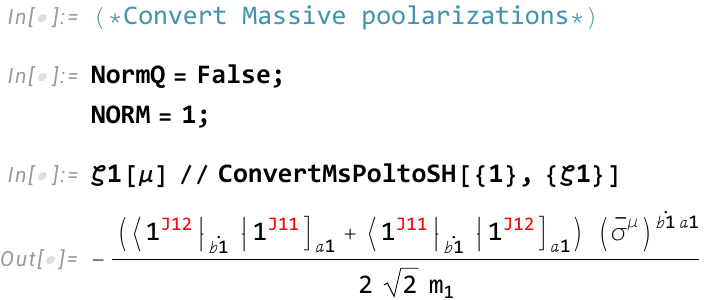}
	}
	
\end{examplebox}

\begin{conceptbox}{An example: three-point spin-\(1\)-scalar couplings}
	
	A useful application of \(\tt{ConvertMsPoltoSH}\) is the decomposition of
	three-point on-shell amplitudes involving a charged massive spin-\(1\) particle
	and a scalar.
	
	By little-group covariance, the space of three-point couplings is spanned by
	degree-two polynomials in
	\begin{align}
		\left\langle
		\tt{1}^{\ttr{J1}}\,
		\tt{2}^{\ttr{J2}}
		\right\rangle,
		\qquad
		\left[
		\tt{1}^{\ttr{J1}}\,
		\tt{2}^{\ttr{J2}}
		\right].
	\end{align}
	There are three independent Lorentz-invariant couplings for a massive charged
	spin-\(1\) field and a scalar:
	\begin{align}
		\zeta_1\cdot\zeta_2,
		\qquad
		\frac{1}{m^2}
		(\zeta_1\cdot k_2)(\zeta_2\cdot k_1),
		\qquad
		\frac{\iimg}{m^2}
		\epsilon^{\mu\nu\rho\sigma}
		\zeta_{1\mu}\zeta_{2\nu}k_{1\rho}k_{2\sigma}.
	\end{align}
	Using the conversion commands, these Lorentz structures can be decomposed into
	linear combinations of
	\begin{align}
		\left\langle
		\tt{1}^{\ttr{J1}}\,
		\tt{2}^{\ttr{J2}}
		\right\rangle^2,
		\qquad
		\left[
		\tt{1}^{\ttr{J1}}\,
		\tt{2}^{\ttr{J2}}
		\right]^2,
		\qquad
		\left\langle
		\tt{1}^{\ttr{J1}}\,
		\tt{2}^{\ttr{J2}}
		\right\rangle
		\left[
		\tt{1}^{\ttr{J1}}\,
		\tt{2}^{\ttr{J2}}
		\right].
	\end{align}
	
\end{conceptbox}

\begin{examplebox}{Spin-\(1\)-scalar three-point structures}
	
	\redbox{
		\includegraphics[scale=0.445]{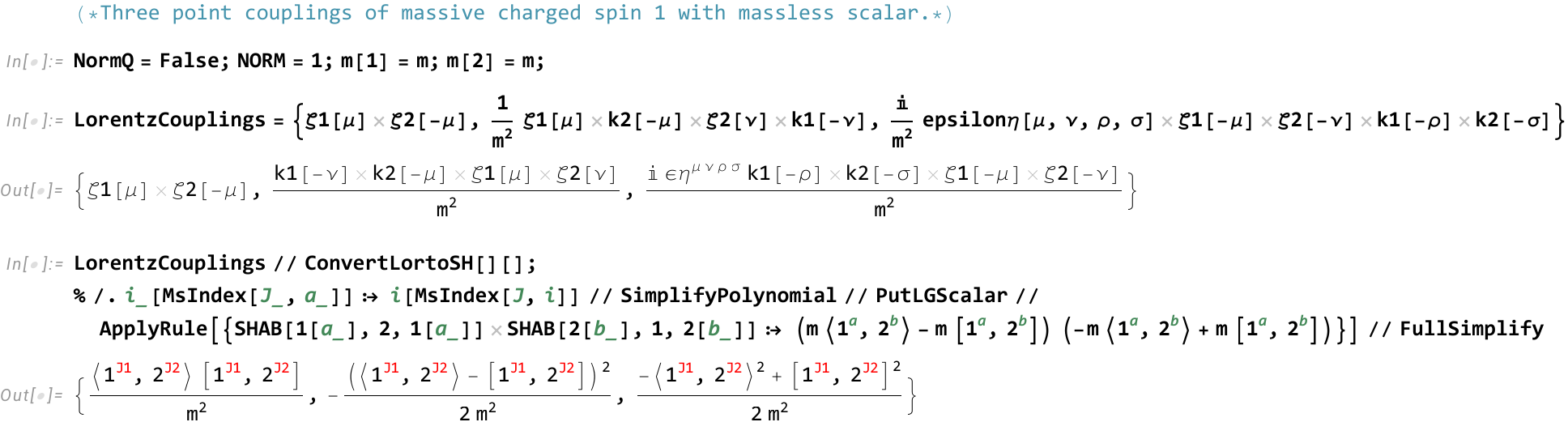}
	}
	
\end{examplebox}

\begin{documentationbox}{\tt{ConvertMlPoltoSH}}{\tt{ConvertMlPoltoSH[legs\_List,polheads\_List,hels\_List,refs\_List]}}

	converts spin-\(1\) massless Lorentz polarizations into spinor-helicity
	variables.
	
	The four arguments are:
	\begin{itemize}
		\item \(\tt{legs}\): the list of leg labels,
		\item \(\tt{polheads}\): the list of polarization heads,
		\item \(\tt{hels}\): the list of helicities,
		\item \(\tt{refs}\): the list of reference spinors.
	\end{itemize}
	All four lists should be given in the same order.
	
\end{documentationbox}

\begin{examplebox}{\tt{ConvertMlPoltoSH}}
	
	\redbox{
		\includegraphics[scale=0.5]{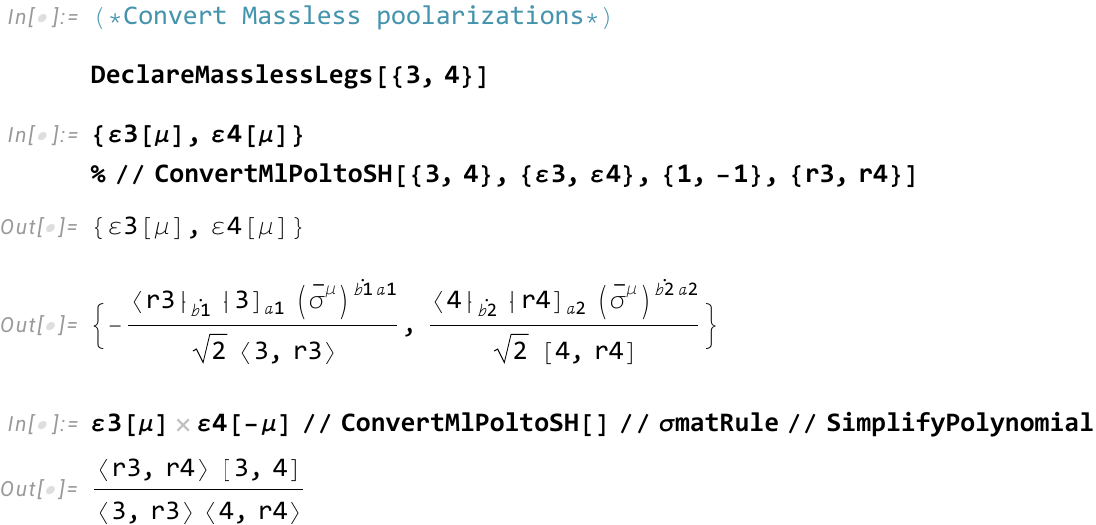}
	}
	
\end{examplebox}

\begin{documentationbox}{\tt{ConvertMlFStoSH}}{\tt{ConvertMlFStoSH[legs\_List,fsheads\_List,hels\_List]}}

	converts massless field strengths into spinor-helicity variables in a way that
	is manifestly independent of reference spinors.

\end{documentationbox}

\begin{examplebox}{\tt{ConvertMlFStoSH}}
	
	\redbox{
		\includegraphics[scale=0.5]{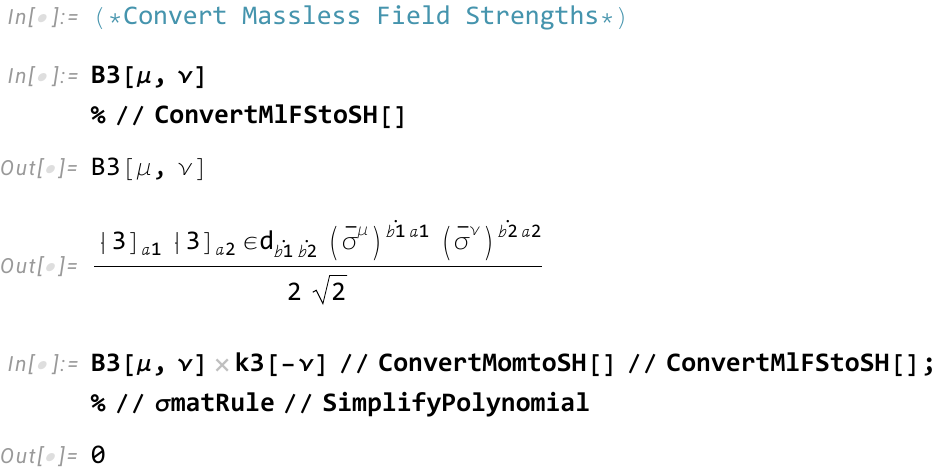}
	}
	
\end{examplebox}
\subsection{Replace spinors and momenta}\label{ReplaceSpinors}

\begin{conceptbox}{Replacing and scaling spinors}
	
	In amplitude computations, it is often necessary to replace spinors by linear
	combinations of spinors.  For example, in BCFW shifts one replaces the spinors
	of two external legs by shifted spinors.  The commands in this subsection
	provide direct control over such operations.
	
\end{conceptbox}

\begin{documentationbox}{\tt{ReplaceSpinors}}{\tt{ReplaceSpinors[ruleList\_]}}

	replaces spinors according to the delayed replacement rules in
	\(\tt{ruleList}\).  The left-hand side of each rule is the spinor to be
	replaced, and the right-hand side is a linear combination of spinors.

	The left-hand side may also involve patterns.
	
\end{documentationbox}

\begin{examplebox}{Replacing spinors}
	
	\redbox{
		\begin{minipage}{0.45\linewidth}
			\includegraphics[scale=0.45]{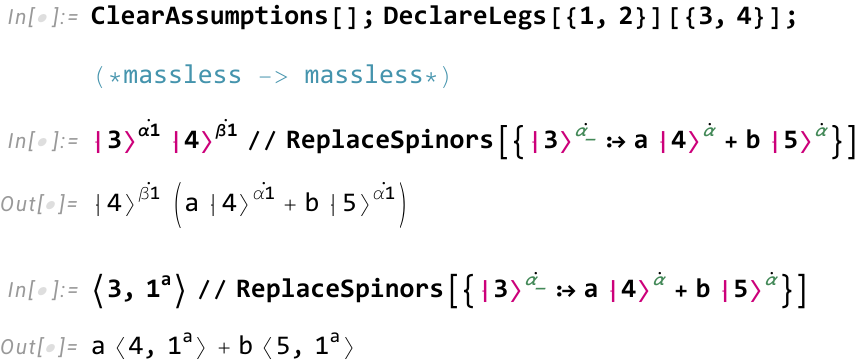}
		\end{minipage}
		\begin{minipage}{0.5\linewidth}
			\includegraphics[scale=0.45]{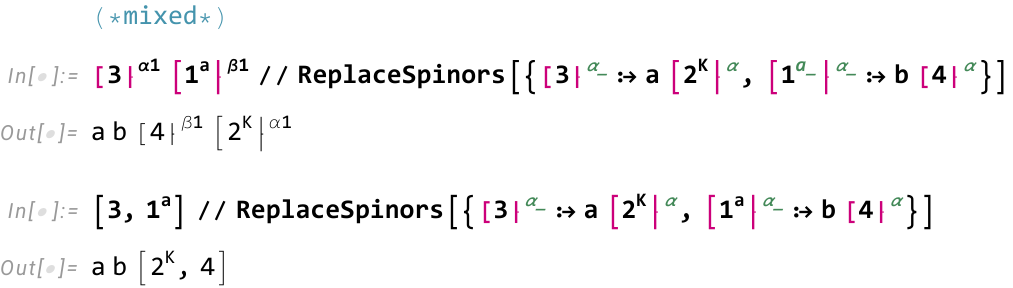}
		\end{minipage}
	}
	
\end{examplebox}

\begin{examplebox}{Replacing angle and box spinors}
	
	\redbox{\includegraphics[scale=0.5]{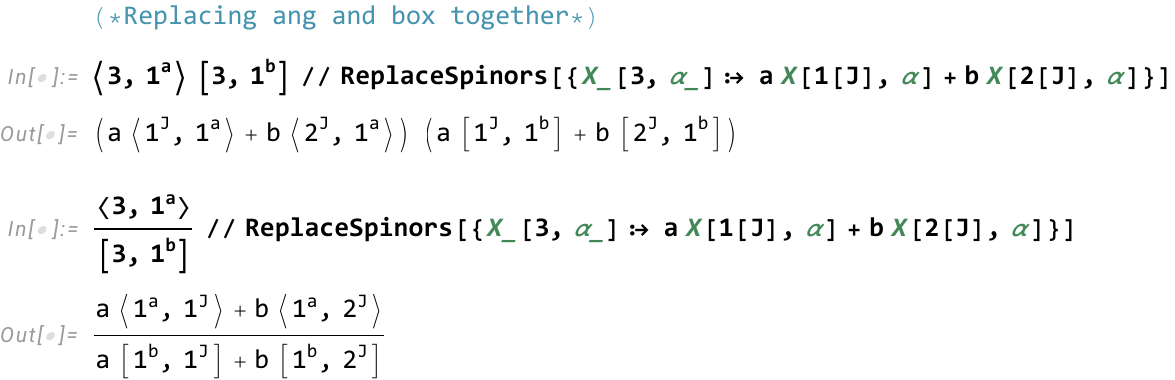}}
	
\end{examplebox}

\begin{documentationbox}{\tt{ScaleSpinors}}{\tt{ScaleSpinors[ruleList\_]}}
	
	used to rescale spinors by arbitrary factors or to shift them by new spinors.

	It has the same syntax as \(\tt{ReplaceSpinors}\).
\end{documentationbox}

\begin{examplebox}{Scaling spinors}
	
	\redbox{\includegraphics[scale=0.5]{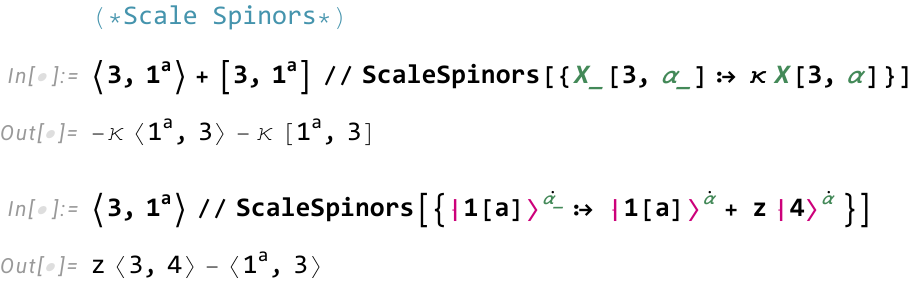}}
	
\end{examplebox}
\begin{conceptbox}{}
	
	Since spinors are two dimensional objects, they can always be written in terms of two independent basis vectors. It is often useful to rewrite a spinor in a fixed basis using the Schouten identity. \tt{ToBasis} does this. \begin{align}
		\mathtt{
			\label{eqToBasis}
			|i\rangle
			=
			\frac{\langle i\,b\rangle}{\langle a\,b\rangle}
			|a\rangle
			-
			\frac{\langle i\,a\rangle}{\langle a\,b\rangle}
			|b\rangle
		}.
	\end{align}
	The analogous formula holds for box spinors.
\end{conceptbox}
\begin{documentationbox}{\tt{ToBasis}}{\tt{ToBasis[rule\_]}}

	This command does the basis transformation given in \ref{eqToBasis} and the corresponding analog for box spinors. The is given as below
	\begin{align}
		\tt{i -> \{a,b\}} .
	\end{align}
	where \tt{i,a,b} are leg indices.
	
\end{documentationbox}

\begin{examplebox}{Spinor basis transformation}
	
	\redbox{
		\begin{minipage}{0.5\linewidth}
			\includegraphics[scale=0.5]{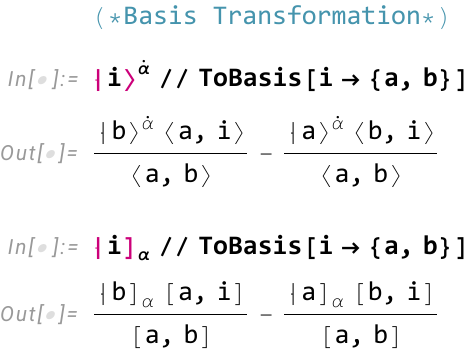}
		\end{minipage}
		\begin{minipage}{0.5\linewidth}
			\includegraphics[scale=0.5]{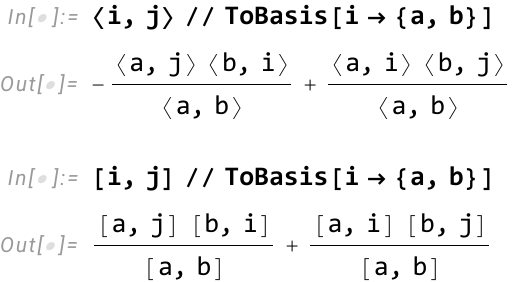}
		\end{minipage}
	}
	
\end{examplebox}

\begin{conceptbox}{Replacing momenta versus replacing spinors}
	
	Replacing spinors is not the same as replacing momenta.  Replacing spinors
	acts separately on angle and box spinors, and therefore produces cross terms
	between different spinors.  Replacing momenta instead replaces products of
	angle and box spinors in such a way that no such cross terms are generated.
	
\end{conceptbox}

\begin{documentationbox}{\tt{ReplaceMomenta}}{\tt{ReplaceMomenta[momentumReplaceRulesList\_]}}
	
	replaces momentum factors.  The argument is a list of replacement rules such
	that the left-hand side is the head of the momentum to be replaced, while the
	right-hand side is a linear combination of other momentum heads.

\end{documentationbox}

Since the linear combination may be arbitrary, all momenta appearing on the
right-hand side must be declared first.  The evaluation gives a warning if an
undeclared momentum appears.

\begin{examplebox}{Replacing momenta}
	
	\redbox{
		\begin{minipage}{0.5\linewidth}
			\includegraphics[scale=0.5]{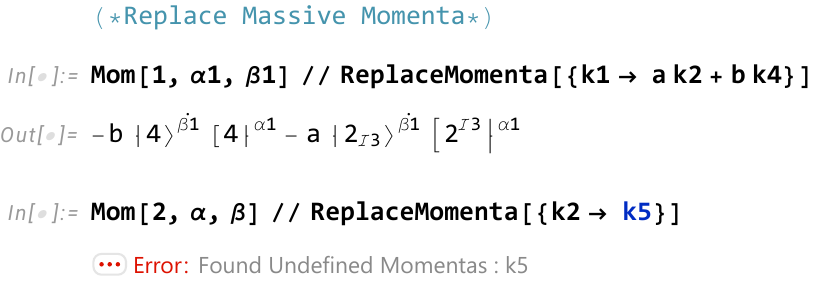}
		\end{minipage}
		\begin{minipage}{0.5\linewidth}
			\includegraphics[scale=0.5]{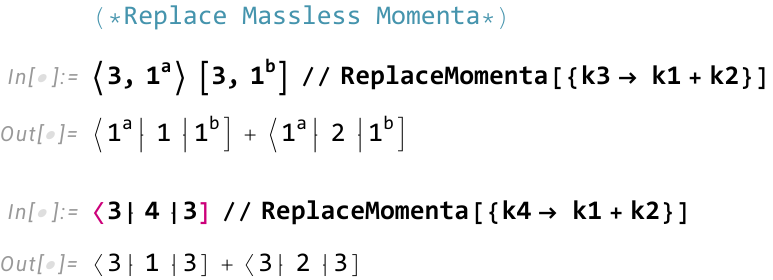}
		\end{minipage}
	}
	
\end{examplebox}

\begin{documentationbox}{\tt{ShiftMomenta}}{\tt{ShiftMomenta[shiftedMomentaRulesList\_]}}
	
	The command \tt{ShiftMomenta} should be used when a momentum is replaced by a
	linear combination that contains the same momentum.

\end{documentationbox}

\begin{morematerialbox}
	The command \(\tt{ReplaceMomenta}\) repeatedly replaces a momentum by a linear
	combination of other momenta.  Therefore, if the linear combination contains
	the same momentum being replaced, the evaluation may enter an infinite loop.
\end{morematerialbox}

\begin{examplebox}{Shifting momenta}
	
	\redbox{\includegraphics[scale=0.5]{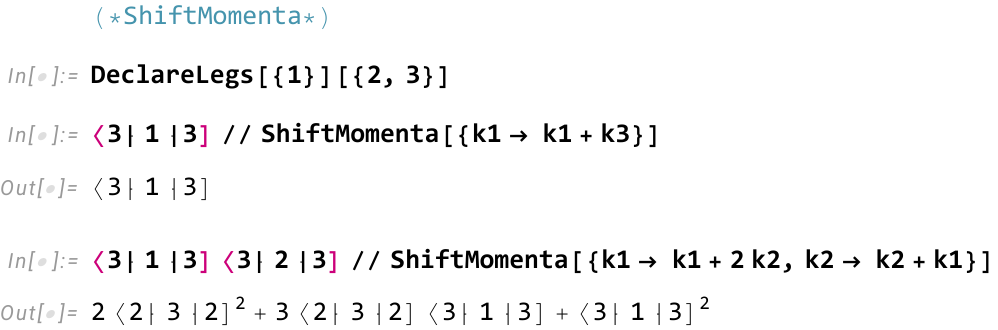}}
	
\end{examplebox}

\begin{documentationbox}{\tt{ScaleMomenta}}{\tt{ScaleMomenta[scaledMomentaRulesList\_]}}

	scales momenta by arbitrary factors.  This command is kept separate from
	\(\tt{ShiftMomenta}\), because scaling a momentum requires scaling the
	corresponding angle and box spinors by appropriate factors.
	
	With the convention used here, if a momentum is scaled by a factor
	\(\mathtt{\alpha}\), the corresponding angle spinor is multiplied by
	\(-1\), while the box spinor is multiplied by \(-\mathtt{\alpha}\).
	
\end{documentationbox}

\begin{examplebox}{Scaling momenta}
	
	\redbox{\includegraphics[scale=0.5]{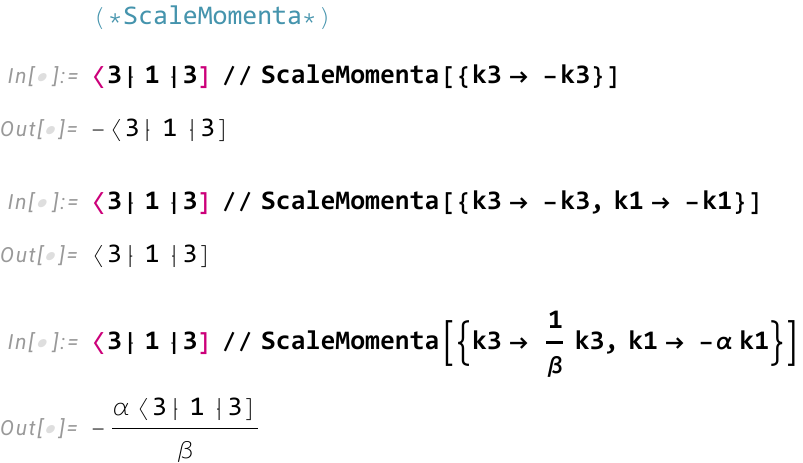}}
	
\end{examplebox}

\subsection{Symmetrize}

\begin{documentationbox}{\tt{Symmetrized}}{\tt{Symmetrized[list\_][expr\_]}}

	symmetrizes an expression with respect to the objects in \(\tt{list}\).
	
	These objects may be \(SL(2,\mathbb C)\) spinor indices,
	\(\mathtt{SU(2)_{LG}}\) little-group indices, or particle labels.
	
\end{documentationbox}

\begin{examplebox}{Symmetrization}
	
	\redbox{
		\begin{minipage}{0.5\linewidth}
			\includegraphics[scale=0.5]{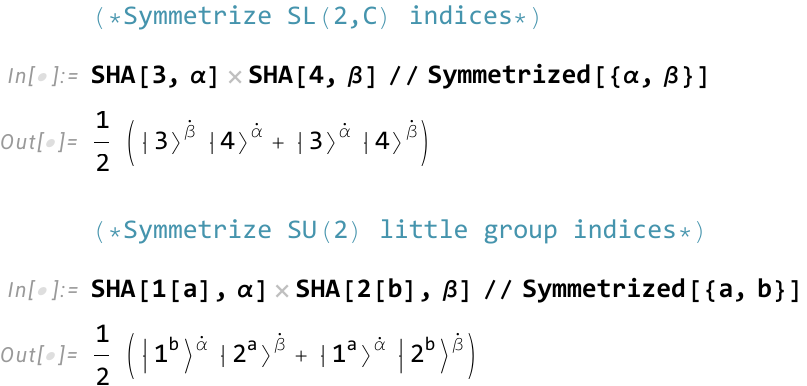}
		\end{minipage}
		\begin{minipage}{0.5\linewidth}
			\includegraphics[scale=0.5]{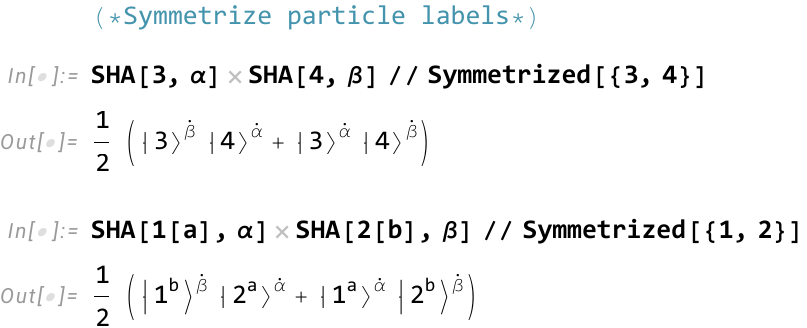}
		\end{minipage}
	}
	
\end{examplebox}
\subsection{Differentiate}\label{sec:Differentiate}
\begin{documentationbox}{\tt{Differentiate}}{\tt{Differentiate[X\_][expr\_]}, \tt{Differentiate[X\_List][expr\_]}}
	
	The \tt{Differentiate} command is used to obtain the derivative of any expression with respect to any spinor or a scalar or a list of spinors or scalars.
	
	\textbf{Keyboard Shortcut:} \boxed{\tt{Esc}}\tt{diff}\boxed{\tt{Esc}}
\end{documentationbox}

\begin{examplebox}{\tt{Differentiate}}
	\redbox{\begin{minipage}{0.2\linewidth}
			\includegraphics[scale=0.45
			]{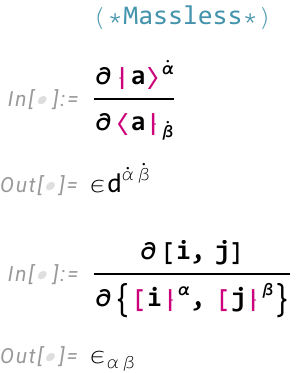}
		\end{minipage}\begin{minipage}{0.4\linewidth}
			\includegraphics[scale=0.45
			]{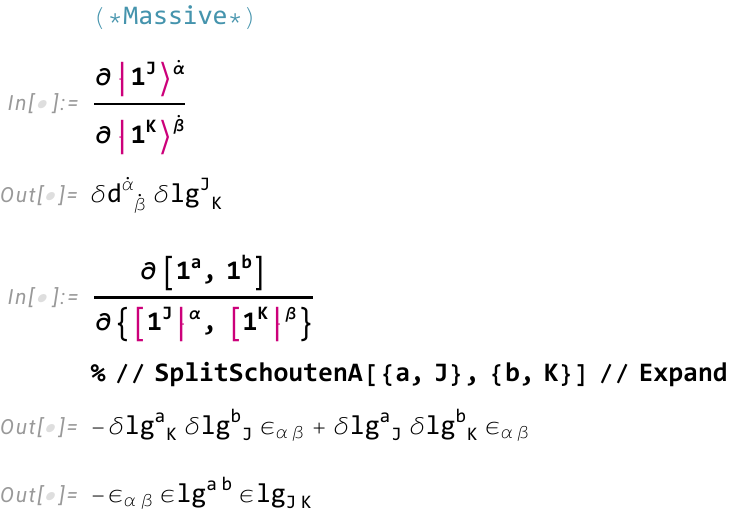}
		\end{minipage}\begin{minipage}{0.3\linewidth}
			\includegraphics[scale=0.45
			]{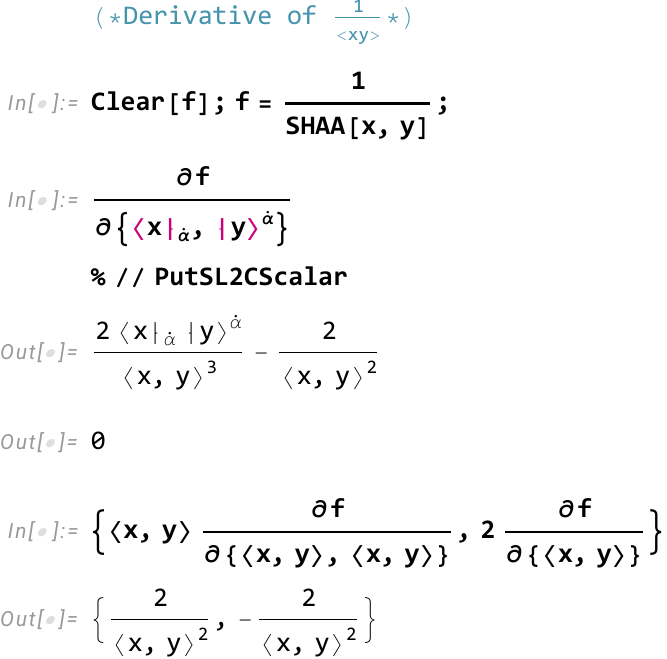}
	\end{minipage}}
	As shown in the third column of the example, one has to be careful while using the differentiate command because of the following identities
	\begin{align}
		\mathtt{\frac{	\partial^2f([x,y])}{\partial x^{\alpha}\partial{y_{\alpha}}}\neq \frac{\partial f([x,y])}{\partial [x,y]}} 
	\end{align}
	similarly for angles. Instead we have,
	\begin{align}
		\mathtt{\frac{	\partial^2f([x,y])}{\partial x^{\alpha}\partial{y_{\alpha}}}=[x,y]\frac{\partial^2f([x,y])}{\partial^2 [x,y]}+2 \frac{\partial f([x,y])}{\partial[x,y]}} 
	\end{align}
\end{examplebox}

\begin{documentationbox}{\tt{HigherSpinDiffA}, \tt{HigherSpinDiffB} and \tt{HigherSpinDiff}}{\tt{HigherSpinDiffA[leg\_,Spin\_,SpinIndexA\_]}\\ \tt{HigherSpinDiffB[leg\_,Spin\_,SpinIndexB\_]} \\ \tt{HigherSpinDiff[leg\_,Spin\_,SpinIndexB\_,SpinIndexA\_]}}
	
	To take the derivative wrt a higher spin leg (or a spinor with multiple copies), one can use the \tt{HigherSpinDiffA}, \tt{HigherSpinDiffB} or the \tt{HigherSpinDiff} commands with the following syntax. The first argument is the leg label, the second is the spin, and the third/fourth argument takes the prefix of the spinor dotted or undotted indices.
\end{documentationbox}
\begin{examplebox}{\tt{HigherSpinDiffA}, \tt{HigherSpinDiffB} and \tt{HigherSpinDiff}}
	\redbox{\begin{minipage}{0.5\linewidth}
			\includegraphics[scale=0.5
			]{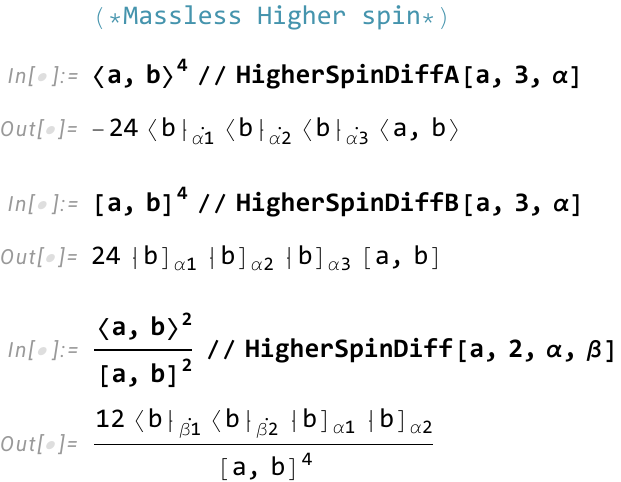}
		\end{minipage}\begin{minipage}{0.5\linewidth}
			\includegraphics[scale=0.5
			]{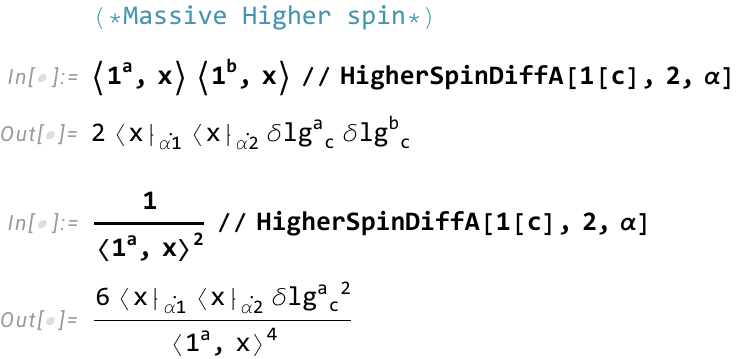}
	\end{minipage}}
\end{examplebox}

%% file: App_SectionWiseSummary.tex
\pagebreak
\section{Section-wise Summary of Commands}
\label{app:sectionwisesummary}

\subsection{List of commands for declaring and editing legs}
\begin{commandtable}
\texttt{DeclareLegs} & Declare massive and massless legs for further computation & \pageref{dcmnt:DeclareLegs} \\
\texttt{UndeclareLegs} & Remove massive or massless declaration & \pageref{dcmnt:UndeclareLegs} \\
\texttt{DeclareMassiveLegs} & Declare masslive legs  & \pageref{dcmnt:DeclareMassiveLegs} \\
\texttt{DeclareMasslessLegs} & Declare massless legs & \pageref{dcmnt:DeclareMasslessLegs} \\
\texttt{EditMassiveData} & Opens interactive palette for editing massive legs & \pageref{dcmnt:EditMassiveData} \\
\texttt{EditMasslessData} & Opens interactive palette for editing massless legs & \pageref{dcmnt:EditMassiveData} \\
\texttt{MsQ} & True if leg is declared as massive, otherwise False. & \pageref{dcmnt:MsQ} \\
\texttt{MlQ} & Same as MasslessQ. & \pageref{dcmnt:MlQ}
\end{commandtable}
\subsection{List of commands for Basic spinor helicity objects}
\label{sec:Summaryofbasicobjects}
\begin{basicobjectstable}
\texttt{SHA} & $<$ and $>$ & Angle spinor for any leg and spinor index $\alpha$ & \pageref{dcmnt:SHA} \\
\texttt{SHB} & $[$ and $]$ & Box spinor for any leg and spinor index $\alpha$ & \pageref{dcmnt:SHB} \\
\texttt{SHAA} & $<>$ & Angle-Angle $SL(2, \mathbb{C})$ scalar & \pageref{dcmnt:SHAA} \\
\texttt{SHBB} & $[]$ & Box-Box $SL(2, \mathbb{C})$ scalar & \pageref{dcmnt:SHAA} \\
\texttt{SHAB} & $<]$ & Angle-Box $SL(2, \mathbb{C})$ and little group scalar & \pageref{dcmnt:SHAB} \\
\texttt{SHBA} & $[>$ & Box-Angle $SL(2, \mathbb{C})$ and little group scalar & \pageref{dcmnt:SHAB} \\
\texttt{ep} & $\epsilon$ & $SU(2)_L$ invariant bilinear, sign of indices decides symmetric or antisymmetric  & \pageref{dcmnt:ep} \\
\texttt{epd} & $\epsilond$ & $SU(2)_R$ invariant bilinear, sign of indices decides symmetric or antisymmetric  & \pageref{dcmnt:ep} \\
\texttt{eplg} & $\epsilonlg$ & $SU(2)_{\text{LG}}$ invariant bilinear, sign of indices decides symmetric or antisymmetric & \pageref{dcmnt:ep} \\
\texttt{$\eta$} & \ & Lorentz metric tensor & \pageref{dcmnt:eta}\\
\texttt{$\sigma$mat} &\ & Clifford algebra matrix elements & \pageref{dcmnt:sigma} \\
\texttt{SM} &\ & Mandelstam variables of $(k_i + k_j + \dots)^2$ kind & \pageref{dcmnt:SM} \\
\texttt{sm} &\tt{sm} & Mandelstam variables of $(k_i \cdot k_j)$ kind & \pageref{dcmnt:sm} 
\end{basicobjectstable}
\pagebreak
\subsection{List of commands for simplifying spinor helicity polynomials}
\label{sec:SummaryofrulesforSimplifyingSpinorHelicityPolynomials}
\begin{commandtable}
\texttt{PutOnShell} & Applies the on-shell inner product rules for massive spinors. & \pageref{dcmnt:PutOnShell} \\
\texttt{PutCanonicalOrder} & Places canonical ordering in SH scalars. & \pageref{dcmnt:PutCanonicalOrder} \\
\texttt{SetDefaultOrdering} & Resets the canonical ordering of SH scalars to default. & \pageref{dcmnt:SetDefaultOrdering} \\
\hline
\texttt{ContractBilinears} & Contracts all possible metrics. & \pageref{dcmnt:ContractBilinears} \\
\texttt{ContractLBilinear} & Contracts the left-handed antisymmetric bilinear. & \pageref{dcmnt:ContractLBilinear} \\
\texttt{ContractRBilinear} & Contracts the right-handed antisymmetric bilinear. & \pageref{dcmnt:ContractRBilinear} \\
\texttt{ContractLGBilinear} & Contracts the little group antisymmetric bilinear. & \pageref{dcmnt:ContractLGBilinear} \\
\texttt{ContractLorMetric} & Contracts the Lorentz spacetime metric. & \pageref{dcmnt:ContractLorMetric} \\
\hline
\texttt{ToMandelstam} & Converts contracted momenta factors into Mandelstam variables. & \pageref{dcmnt:ToMandelstam} \\
\texttt{MandelstamtoSH} & Converts mandelstam variables \tt{sm[i,j]} to the SH expression \tt{SHAB[i, j, i]} & \pageref{dcmnt:MandelstamtoSH} \\
\texttt{SHToMandelstam} & Converts the spinor helicity scalars to Mandelstam variables, \tt{sm[i,j]} wherever possible  & \pageref{dcmnt:MandelstamtoSH} \\
\texttt{MandelstamtoLor} & Writes the Mandelstam variable \tt{sm[i,j]} in terms of the contracted Lorentz momenta $(ki . kj)$. & \pageref{dcmnt:MandelstamtoLor} \\
\texttt{MandelstamRules} & Gives the list of all Mandelstam variables in terms of all independent mandelstam variables(\tt{AllIndependentMandelstams[]}). & \pageref{dcmnt:MandelstamRules} \\
\hline
\texttt{PutSL2CScalar} & Contracts all possible $SL(2,\mathbb{C})$ indices to write expression in terms of \tt{SHAA} and \tt{SHBB} scalars & \pageref{dcmnt:PutSL2CScalar} \\
\texttt{NoSL2CScalar} & Breaks SHAA and SHBB into individual spinors. & \pageref{dcmnt:NoSL2CScalar} \\
\texttt{PutSU2LScalar} & Contracts the $SU(2)_L$ indices and constructs the \tt{SHBB} scalars. & \pageref{dcmnt:PutSU2LScalar} \\
\texttt{PutSU2RScalar} & Contracts the $SU(2)_R$ indices and constructs the \tt{SHAA} scalars. & \pageref{dcmnt:PutSU2RScalar} \\
\texttt{NoSU2LScalar} & Breaks only the \tt{SHBB} scalars into individual spinors. & \pageref{dcmnt:NoSU2LScalar} \\
\texttt{NoSU2RScalar} & Breaks only the \tt{SHAA} scalars into individual spinors. & \pageref{dcmnt:NoSU2RScalar} \\
\hline
\texttt{PutLGScalar} & Construct the little group scalars, \tt{SHAB} wherever possible by contracting \tt{SHAA} and \tt{SHBB} & \pageref{dcmnt:PutLGScalar} \\
\texttt{NoLGScalar} & Breaks \tt{SHAB} and \tt{SHBA} into \tt{SHAA} and SHBB. & \pageref{dcmnt:NoLGScalar} \\
\texttt{PutMasslessLGScalar} & Construct the massless little group scalars, \tt{SHAB} wherever possible by contracting \tt{SHAA} and \tt{SHBB} & \pageref{dcmnt:PutMasslessLGScalar} \\
\texttt{PutMassiveLGScalar} & Construct the massive little group scalars, \tt{SHAB} wherever possible by contracting \tt{SHAA} and \tt{SHBB} & \pageref{dcmnt:PutMassiveLGScalar} \\
\texttt{NoMasslessLGScalar} & Breaks \tt{SHAB} and \tt{SHBA} into \tt{SHAA} and \tt{SHBB} only if the middle leg is massless. & \pageref{dcmnt:NoMasslessLGScalar} \\
\texttt{NoMassiveLGScalar} & Breaks \tt{SHAB} and \tt{SHBA} into \tt{SHAA} and \tt{SHBB} only if the middle leg is massive. & \pageref{dcmnt:NoMassiveLGScalar} \\
\texttt{PutMassiveLGScalarFor} & Construct the massive leg specific little group scalars, \tt{SHAB} wherever possible by contracting \tt{SHAA} and \tt{SHBB}. Arguments can also take patterns. & \pageref{dcmnt:PutMassiveLGScalarFor} \\
\texttt{PutMasslessLGScalarFor} & Construct the massless leg specific little group scalars, \tt{SHAB} wherever possible by contracting \tt{SHAA} and \tt{SHBB}. Arguments can also take patterns. & \pageref{dcmnt:PutMasslessLGScalarFor} \\
\texttt{PutLGScalarFor} & Construct leg specific little group scalars, \tt{SHAB} wherever possible by contracting \tt{SHAA} and \tt{SHBB} Arguments can also take patterns. & \pageref{dcmnt:PutLGScalarFor} \\
\texttt{NoMassiveLGScalarWB} & Breaks \tt{SHAB} and \tt{SHBA} into \tt{SHAA} and \tt{SHBB} only if the middle leg is massive, such that all massive little group indices are lowered and contracted with invariant bilinears. & \pageref{dcmnt:NoMassiveLGScalarWB} \\
\texttt{NoMasslessLGScalarWB} & Same as NoMasslessLGScalar. & \pageref{dcmnt:NoMasslessLGScalarWB} \\
\texttt{NoLGScalarWB} & Breaks \tt{SHAB} and \tt{SHBA} into \tt{SHAA} and \tt{SHBB} with all massive little group indices lowered and contracted with invariant bilinears. WB denotes With Bilinear. & \pageref{dcmnt:NoLGScalarWB} \\
\hline
\texttt{CanonicalizeSU2RScalars} & Canonicalizes the right-handed spinor helicity, \tt{SHAA} scalars in an expression. & \pageref{dcmnt:CanonicalizeSU2RScalars} \\
\texttt{CanonicalizeSU2LScalars} & Canonicalizes the left-handed spinor helicity, \tt{SHBB} scalars in an expression. & \pageref{dcmnt:CanonicalizeSU2RScalars} \\
\texttt{CanonicalizeIndices} & Canonicalizes the contracted little group and spinor indices in an expression. & \pageref{dcmnt:CanonicalizeIndices} \\
\hline
\texttt{SimplifySchouten} & Simplifies an expression by applying the Schouten identities recursively & \pageref{dcmnt:SimplifySchouten} \\
\texttt{SplitSchouten} & Applies the Schouten rule over the pairs specific set of legs wherever possible. The pairs can consist of angle or box or both spinors, spinor indices, or a mixture of both. & \pageref{dcmnt:SplitSchouten} \\
\texttt{SplitSchoutenA} & Applies the Schouten rule over the pairs specific set of legs wherever possible. The pairs can consist of angle spinors, dotted spinor indices, or a mixture of both. & \pageref{dcmnt:SplitSchoutenA} \\
\texttt{SplitSchoutenB} & Applies the Schouten rule over the pairs specific set of legs wherever possible. The pairs can consist of box spinors, undotted spinor indices, or a mixture of both. & \pageref{dcmnt:SplitSchoutenB} \\
\texttt{SimplifyPolynomial} & Applies the \tt{PutOnShell}, \tt{PutScalar}, and \tt{ContractBilinears} rules using \tt{RepeatedRule}. & \pageref{dcmnt:SimplifyPolynomial} \\
\texttt{SimplifyPolynomialFunction} & Applies the \tt{PutOnShell}, \tt{PutScalar}, and \tt{ContractBilinears} rules. & \pageref{dcmnt:SimplifyPolynomialFunction}
\end{commandtable}

\subsection{List of commands for calculating amplitudes in spinor helicity formalism}
\label{sec:SummaryOfSACommands}
\begin{commandtable}
\texttt{UGPropagator} & Unitary gauge propagator for integer spin particle & \pageref{dcmnt:UGPropagator} \\
\texttt{Propagator} & $R_\xi$ analog gauge propagator for integer spin particle & \pageref{dcmnt:Propagator} \\
\texttt{FGPropagator} & Feynman analog gauge propagator for integer spin particle & \pageref{dcmnt:FGPropagator} \\
\texttt{PropCoeff} & The binomial like combinatorial factor appearing in the terms of integer spin propagator. & \pageref{dcmnt:PropCoeff} \\
\hline
\texttt{AddAssumptions} & Adds the list of assumptions over mass etc for scattering process. & \pageref{dcmnt:AddAssumptions} \\
\texttt{RemoveAssumptions} & Removes assumptions & \pageref{dcmnt:AddAssumptions} \\
\texttt{ClearAssumptions} & Clears all assumptions, any declared data, and the generated numerics. & \pageref{dcmnt:ClearAssumptions} \\
\hline
\texttt{Mom} & Gives the momentum for the leg with dotted and undotted indices & \pageref{dcmnt:Mom} \\
\texttt{DressMom} & Places the spinor indices over the linear combination of momentas. & \pageref{dcmnt:DressMom} \\
\texttt{MsPol} & Gives the massive spin-1 polarization for particular leg. & \pageref{dcmnt:MsPol} \\
\texttt{MsPolNonSymm} & Gives the massive spin-1 polarization for a particular leg & \pageref{dcmnt:MsPolNonSymm} \\
\texttt{HsPol} & Gives the higher-spin polarization for a particular leg & \pageref{dcmnt:HsPol} \\
\texttt{MlPol} & Gives the massless polarization  for a particular leg & \pageref{dcmnt:MlPol} \\
\texttt{DeclareMomentumConservation} & Defines the Mandelstam variables for the given momentum conservation equation & \pageref{dcmnt:DeclareMomentumConservation} \\
\texttt{momconsRule} & Momentum conservation rule defined using DeclareMomentumConservation & \pageref{dcmnt:momconsRule} \\
\texttt{ClearMomentumConservation} & Clears the declared momentum conservation. Also clears the Mandelstam rules & \pageref{dcmnt:ClearMomentumConservation} \\
\hline
\texttt{ThreeMasslessAmplitude} & Gives the all massless three-point amplitude & \pageref{dcmnt:ThreeMasslessAmplitude} \\
\texttt{TwoMasslessOneMassiveAmplitude} & Gives the two massless and one massive three point amplitude & \pageref{dcmnt:TwoMasslessOneMassiveAmplitude} \\
\texttt{TwoMassiveOneMasslessAmplitude} & Gives the two massive and one massless three point amplitude & \pageref{dcmnt:TwoMassiveOneMasslessAmplitude} \\
\texttt{Coupling} & Represents coupling coefficient & \pageref{dcmnt:Coupling} \\
\texttt{Xfactor} & Gives the spinor helicity expression for the Lorentz scalar matching the contraction of the massless polarization vector and the massive momentum vector & \pageref{dcmnt:Xfactor} \\
\texttt{StripPol} & Strips both, angle and box polarizations for a leg from an expression(of amplitude) & \pageref{dcmnt:StripPol} \\
\hline
\texttt{PureGaugeTransformation} & Applies the gauge transformation for the massless leg & \pageref{dcmnt:PureGaugeTransformation} \\
\texttt{GIQ} & When acted on an expression, gives True if the expression is gauge invariant, otherwise gives False & \pageref{dcmnt:GIQ} \\
\texttt{ManifestGI} & If the expression is independent of ReferenceSpinor. Uses basis transformation to write expression in gauge invariant form. & \pageref{dcmnt:ManifestGI} \\
\texttt{FindContactTerm} & Finds the contact term for a massless leg given a gauge-transformed expression. & \pageref{dcmnt:FindContactTerm} \\
\end{commandtable}

\subsection{List of commands for various other useful operations in spinor helicity formalism}
\label{sec:SummaryOfUsefulTools}
\begin{commandtable}
\texttt{OpMlHW} & Helicity Weight operator for the massless leg. & \pageref{dcmnt:OpMlHW} \\
\texttt{OpMsHW} & Helicity Weight operator for the massive leg. & \pageref{dcmnt:OpMsHW} \\
\texttt{Differentiate} & Differentiates an expression w.r.t spinor & \pageref{dcmnt:Differentiate} \\
\texttt{HigherSpinDiffA} & Takes the derivative with respect to angled spinor of given leg. & \pageref{dcmnt:HigherSpinDiffA} \\
\texttt{HigherSpinDiffB} & Takes the derivative with respect to boxed spinor of given leg. & \pageref{dcmnt:HigherSpinDiffA} \\
\texttt{HigherSpinDiff} & Takes the derivative with respect to angled and boxed both spinors of given leg. & \pageref{dcmnt:HigherSpinDiffA} \\
\hline
\texttt{ChargeConjugate} & Applies a charge conjugation transformation. & \pageref{dcmnt:ChargeConjugate} \\
\texttt{ComplexConjugate} & Applies complex conjugation over a SH expression. & \pageref{dcmnt:ComplexConjugate} \\
\texttt{ImposeReality} & Imposes reality over a list of declared legs. & \pageref{dcmnt:ImposeReality} \\
\texttt{Parity} & Applies a parity transformation over the SH expression. Shortcut: pt & \pageref{dcmnt:Parity} \\
\texttt{TimeReversal} & Applies time reversal transformations over angles and boxes. Shortcut: tr & \pageref{dcmnt:TimeReversal} \\
\texttt{CPT} & Applies CPT transformations. & \pageref{dcmnt:CPT} \\
\hline

\texttt{ComponentForm} & Writes the SH expression in terms of momentum components in Cartesian coordinates. & \pageref{dcmnt:ComponentForm} \\
\texttt{SphericalCoordinates} & Converts the Cartesian momentum components to spherical components. & \pageref{dcmnt:SphericalCoordinates} \\
\texttt{CartesianCoordinates} & Converts spherical coordinates to Cartesian coordinates. & \pageref{dcmnt:SphericalCoordinates} \\
\texttt{SetComponentFormAssumption} & Adds $E$[i] $\in$ PositiveReals, Modp[i] $\in$ PositiveReals, p[i][0] $\in$ Reals, p[i][1] $\in$ Reals, p[i][2] $\in$ Reals, p[i][3] $\in$ Reals, $\theta$[i] $\in$ Reals, and $\phi$[i] $\in$ Reals to \$Assumptions for all declared legs i. & \pageref{dcmnt:SetComponentFormAssumption} \\
\texttt{UnsetComponentFormAssumption} & Removes the energy, spatial momentum magnitude, Cartesian component, and angular assumptions from \$Assumptions for all declared legs i. & \pageref{dcmnt:SetComponentFormAssumption} \\
\texttt{PutOnShellComponent} & Applies the on-shell relations between momentum components. & \pageref{dcmnt:PutOnShellComponent} \\
\texttt{SimplifyComponentForm} & Simplifies the component form using RepeatedRule. & \pageref{dcmnt:SimplifyComponentForm} \\
\texttt{$\eta$Mat} & Matrix form of the Lorentz metric tensor. & \pageref{dcmnt:componentmatrices} \\
\texttt{$\epsilon$Mat} & Matrix components of the upper $SU(2)_L$ antisymmetric bilinear matrix. & \pageref{dcmnt:componentmatrices} \\
\texttt{$\delta$Mat} & Matrix components of the $SU(2)_L$ identity & \pageref{dcmnt:componentmatrices} \\
\texttt{$\sigma$VecUpper} &  $\sigma$-matrix of the upper lorentz index & \pageref{dcmnt:componentmatrices} \\
\texttt{$\sigma$VecLower} & $\sigma$-matrix of the lower lorentz index  & \pageref{dcmnt:componentmatrices} \\
\texttt{$\sigma$barVecUpper} & $\bar{\sigma}$-matrix of the upper lorentz index  & \pageref{dcmnt:componentmatrices} \\
\texttt{$\sigma$barVecLower} & $\bar{\sigma}$-matrix of the lower lorentz index  & \pageref{dcmnt:componentmatrices} \\
\texttt{pVecLower} & Gives the momentum vector components for any leg. & \pageref{dcmnt:pVecLower} \\
\texttt{pVecUpper} & Gives the momentum vector components for any leg. & \pageref{dcmnt:pVecLower} \\
\texttt{pMatLower} & Lower component matrix for a leg. & \pageref{dcmnt:pMatLower} \\
\texttt{pMatUpper} & Upper component matrix for a leg.
 & \pageref{dcmnt:pMatLower} \\
\hline
\texttt{GenerateNumerics} & Generates numerical kinematics against the chosen label and momentum conservation. & \pageref{dcmnt:GenerateNumerics} \\
\texttt{PutNumerics} & Converts an SH expression into numerics using numerical kinematics generated for the label via \tt{GenerateNumerics}. & \pageref{dcmnt:PutNumerics} \\
\texttt{MassNumerics} & Gives the randomly generated mass values created against the label using \tt{GenerateNumerics}. & \pageref{dcmnt:MassNumerics, MomentaNumerics, ComponentNumerics} \\
\texttt{MomentaNumerics} & Gives the list of numerical values for the momenta generated using \tt{GenerateNumerics} against the label. & \pageref{dcmnt:MassNumerics, MomentaNumerics, ComponentNumerics} \\
\texttt{ComponentNumerics} & Stores the momentum component numerical values generated against the label. & \pageref{dcmnt:MassNumerics, MomentaNumerics, ComponentNumerics} \\
\texttt{RAMBO} & Randomly generates numerical values of momenta such that they satisfy mass-shell conditions, their spatial momenta sum to zero, and their total energy equals Energy. & \pageref{dcmnt:RAMBO} \\
\hline
\texttt{ToDimensionfullBasis} & Decompposes a massive spinor to the $\mathtt{SU(2)}$ fundamental basis.& \pageref{dcmnt:ToDimensionfullBasis}\\
\texttt{HighEnergyLimit} & Takes the high energy limit over the massive legs.& \pageref{dcmnt:HighEnergyLimit}\\
\texttt{SU2LBasis} & Gives the fundamental left handed basis spinors. & \pageref{dcmnt:SU2LBasis}\\
\texttt{SU2RBasis} & Gives the fundamental right handed basis spinors. & \pageref{dcmnt:SU2LBasis}\\
\texttt{SU2LGBasis} & Gives the fundamental little group basis spinors. & \pageref{dcmnt:SU2LBasis}\\
\texttt{SU2LGBasisComponent} & Gives the fundamental little group basis spinor components. & \pageref{dcmnt:SU2LBasis}\\
\texttt{DimFulltoDimLessBasis} & Decomposes the dimension-full basis spinors to dimension-less spinors. & \pageref{dcmnt:DimFulltoDimLessBasis}\\
\texttt{ToMassless} & Decomposes the massive spinors to a pair of massless spinors. & \pageref{dcmnt:ToMassless}\\
\hline
\texttt{ConvertLortoSH} & Converts a Lorentz vector expression to spinor helicity. & \pageref{dcmnt:ConvertLortoSH} \\
\texttt{ConvertMettoSH} & Converts the Lorentz metric to spinor helicity. & \pageref{dcmnt:ConvertMettoSH} \\
\texttt{ConvertMomtoSH} & Converts Lorentz momentum to spinor helicity. & \pageref{dcmnt:ConvertMomtoSH} \\
\texttt{ConvertMsPoltoSH} & Converts massive polarizations to spinor helicity. & \pageref{dcmnt:ConvertMsPoltoSH} \\
\texttt{ConvertMlPoltoSH} & Converts massless polarizations to spinor helicity. & \pageref{dcmnt:ConvertMlPoltoSH} \\
\texttt{ConvertMlFStoSH} & Converts massless field strengths to spinor helicity. & \pageref{dcmnt:ConvertMlFStoSH} \\
\texttt{LorToMandelstam} & Writes the contracted momenta $(k_i \cdot k_j)$ in terms of the Mandelstam variable \tt{sm[i,j]}. & \pageref{dcmnt:MandelstamtoLor} \\
\hline
\texttt{ReplaceSpinors} & Replaces spinors in the list of rules where each rule is a delayed rule of a spinor in terms of other spinors. & \pageref{dcmnt:ReplaceSpinors} \\
\texttt{ScaleSpinors} & Scales the spinors. & \pageref{dcmnt:ScaleSpinors} \\
\texttt{ToBasis} & Performs a basis transformation. & \pageref{dcmnt:ToBasis} \\
\texttt{ReplaceMomenta} & Replaces the momenta in the list of rules. & \pageref{dcmnt:ReplaceMomenta} \\
\texttt{ShiftMomenta} & Shifts the momenta up to a linear combination of other momenta. & \pageref{dcmnt:ShiftMomenta} \\
\texttt{ScaleMomenta} & Scales the momenta in the list of rules. & \pageref{dcmnt:ScaleMomenta} \\
\hline
\texttt{Symmetrized} & Symmetrizes the objects (indices or leg labels) in the list. & \pageref{dcmnt:Symmetrized} \\
\end{commandtable}

%% file: SMaSHbiblio.bib
@article{WeylvanderWaerden,
author="Hermann Weyl",
title="{The Theory of Groups and Quantum Mechanics}",
journal="Dover Books on Mathematics",
pages="464",
year="1950"
}

@article{Dittmaier_1998,
   title={Weyl–van der Waerden formalism for helicity amplitudes of massive particles},
   volume={59},
   ISSN={1089-4918},
   url={http://dx.doi.org/10.1103/PhysRevD.59.016007},
   DOI={10.1103/physrevd.59.016007},
   number={1},
   journal={Physical Review D},
   publisher={American Physical Society (APS)},
   author={Dittmaier, Stefan},
   year={1998},
   month=Dec }

@article{Berends:1981rb,
    author = "Berends, Frits A. and Kleiss, R. and De Causmaecker, P. and Gastmans, R. and Wu, Tai Tsun",
    title = "{Single Bremsstrahlung Processes in Gauge Theories}",
    reportNumber = "Print-81-0356 (LEIDEN)",
    doi = "10.1016/0370-2693(81)90685-7",
    journal = "Phys. Lett. B",
    volume = "103",
    pages = "124--128",
    year = "1981"
}

@article{Berends:1981uq,
    author = "Berends, Frits A. and Kleiss, R. and De Causmaecker, P. and Gastmans, R. and Troost, W. and Wu, Tai Tsun",
    title = "{Multiple Bremsstrahlung in Gauge Theories at High-Energies. 2. Single Bremsstrahlung}",
    reportNumber = "KUL-TF-81/18",
    doi = "10.1016/0550-3213(82)90489-8",
    journal = "Nucl. Phys. B",
    volume = "206",
    pages = "61--89",
    year = "1982"
}

@article{Xu:1986xb,
    author = "Xu, Zhan and Zhang, Da-Hua and Chang, Lee",
    title = "{Helicity Amplitudes for Multiple Bremsstrahlung in Massless Nonabelian Gauge Theories}",
    reportNumber = "TUTP-86/9a, TUTP-84-3, TUTP-84-4, TUTP-84-5",
    doi = "10.1016/0550-3213(87)90479-2",
    journal = "Nucl. Phys. B",
    volume = "291",
    pages = "392--428",
    year = "1987"
}

@article{Parke:1986gb,
    author = "Parke, Stephen J. and Taylor, T. R.",
    title = "{An Amplitude for $n$ Gluon Scattering}",
    reportNumber = "FERMILAB-PUB-86-042-T",
    doi = "10.1103/PhysRevLett.56.2459",
    journal = "Phys. Rev. Lett.",
    volume = "56",
    pages = "2459",
    year = "1986"
}

@article{S@M,
   title={S@M, a mathematica implementation of the spinor-helicity formalism},
   volume={179},
   ISSN={0010-4655},
   url={http://dx.doi.org/10.1016/j.cpc.2008.05.002},
   DOI={10.1016/j.cpc.2008.05.002},
   number={7},
   journal={Computer Physics Communications},
   publisher={Elsevier BV},
   author={Maître, D. and Mastrolia, P.},
   year={2008},
   month=Oct, pages={501–534} }

@misc{spinorhelicity4d,
      title={SpinorHelicity4D: a Mathematica toolbox for the four-dimensional spinor-helicity formalism}, 
      author={Manuel Accettulli Huber},
      year={2023},
      eprint={2304.01589},
      archivePrefix={arXiv},
      primaryClass={hep-th},
      url={https://arxiv.org/abs/2304.01589}, 
}

@misc{spinorsextras,
      title={SpinorsExtras - Mathematica implementation of massive spinor-helicity formalism}, 
      author={Jakub Kuczmarski},
      year={2014},
      eprint={1406.5612},
      archivePrefix={arXiv},
      primaryClass={hep-ph},
      url={https://arxiv.org/abs/1406.5612}, 
}

@article{SchwartzSpinorHelicity,
   title={Learning the simplicity of scattering amplitudes},
   volume={18},
   ISSN={2542-4653},
   url={http://dx.doi.org/10.21468/SciPostPhys.18.2.040},
   DOI={10.21468/scipostphys.18.2.040},
   number={2},
   journal={SciPost Physics},
   publisher={Stichting SciPost},
   author={Cheung, Clifford and Dersy, Aurélien and Schwartz, Matthew},
   year={2025},
   month=Feb }

@article{RAMBO1,
    author = "Kleiss, R. and Stirling, W. James and Ellis, S. D.",
    title = "{A New Monte Carlo Treatment of Multiparticle Phase Space at High-energies}",
    reportNumber = "CERN-TH-4299/85",
    doi = "10.1016/0010-4655(86)90119-0",
    journal = "Comput. Phys. Commun.",
    volume = "40",
    pages = "359",
    year = "1986"
}

@article{RAMBO2,
    author = {Pl{\"a}tzer, Simon},
    title = "{RAMBO on diet}",
    eprint = "1308.2922",
    archivePrefix = "arXiv",
    primaryClass = "hep-ph",
    reportNumber = "DESY-13-145, MCNET-13-10",
    month = "8",
    year = "2013"
}

@article{Conde_2016,
   title={Lorentz constraints on massive three-point amplitudes},
   volume={2016},
   ISSN={1029-8479},
   url={http://dx.doi.org/10.1007/JHEP09(2016)041},
   DOI={10.1007/jhep09(2016)041},
   number={9},
   journal={Journal of High Energy Physics},
   publisher={Springer Science and Business Media LLC},
   author={Conde, Eduardo and Marzolla, Andrea},
   year={2016},
   month=Sept }

@article{Assamagan:1994mb,
    author = "Assamagan, K. and others",
    title = "{Measurement of the muon momentum in pion decay at rest using a surface muon beam}",
    reportNumber = "PSI-PR-94-19",
    doi = "10.1016/0370-2693(94)91419-2",
    journal = "Phys. Lett. B",
    volume = "335",
    pages = "231--236",
    year = "1994"
}

@article{Kumar:2025znu,
    author = "Kumar, Aakash and Rudra, Arnab and Shaw, Rahul",
    title = "{Compton amplitude and Contact term(s) in the Spinor Helicity formalism}",
    eprint = "2506.12431",
    archivePrefix = "arXiv",
    primaryClass = "hep-th",
    month = "6",
    year = "2025"
}

@article{Luders1957,
  author  = {Gerhart L{\"u}ders},
  title   = {Proof of the TCP Theorem},
  journal = {Annals of Physics},
  volume   = {2},
  number   = {1},
  pages    = {1--15},
  year     = {1957},
  doi      = {10.1016/0003-4916(57)90032-5}
}

@article{Jost1957,
  author  = {Res Jost},
  title   = {Eine Bemerkung zum CTP-Theorem},
  journal = {Helvetica Physica Acta},
  volume   = {30},
  pages    = {409--416},
  year     = {1957}
}

@incollection{Pauli1955,
  author    = {Wolfgang Pauli},
  title     = {Exclusion Principle, Lorentz Group and Reflection of Space-Time and Charge},
  booktitle = {Niels Bohr and the Development of Physics},
  publisher = {Pergamon Press},
  address   = {London},
  year      = {1955},
  pages     = {30--51}
}

@book{StreaterWightman2000,
  author    = {Raymond F. Streater and Arthur S. Wightman},
  title     = {PCT, Spin and Statistics, and All That},
  edition   = {Corrected Third Printing},
  publisher = {Princeton University Press},
  address   = {Princeton},
  year      = {2000}
}

@article{Greenberg2003,
  author        = {O. W. Greenberg},
  title         = {Why is CPT Fundamental?},
  eprint        = {hep-ph/0309309},
  archivePrefix = {arXiv},
  primaryClass  = {hep-ph},
  year          = {2003}
}

@misc{cheung2017tasilecturesscatteringamplitudes,
      title={TASI Lectures on Scattering Amplitudes}, 
      author={Clifford Cheung},
      year={2017},
      eprint={1708.03872},
      archivePrefix={arXiv},
      primaryClass={hep-ph},
      url={https://arxiv.org/abs/1708.03872}, 
}

@misc{LDixonSH,
  doi = {10.5170/CERN-2014-008.31},
  
  url = {https://cds.cern.ch/record/1613349},
  
  author = {Dixon, Lance J},
  
  language = {en},
  
  title = {A brief introduction to modern amplitude methods},
  
  publisher = {CERN},
  
  year = {2014}
}

@article{Aoude:2020onz,
    author = "Aoude, Rafael and Haddad, Kays and Helset, Andreas",
    title = "{On-shell heavy particle effective theories}",
    eprint = "2001.09164",
    archivePrefix = "arXiv",
    primaryClass = "hep-th",
    reportNumber = "SAGEX-20-02-E, MITP/20-002",
    doi = "10.1007/JHEP05(2020)051",
    journal = "JHEP",
    volume = "05",
    pages = "051",
    year = "2020"
}

@article{Vines:2018gqi,
	title        = {{Spinning-black-hole scattering and the test-black-hole limit at second post-Minkowskian order}},
	author       = {Vines, Justin and Steinhoff, Jan and Buonanno, Alessandra},
	year         = 2019,
	journal      = {Phys. Rev. D},
	volume       = 99,
	number       = 6,
	pages        = {064054},
	doi          = {10.1103/PhysRevD.99.064054},
	eprint       = {1812.00956},
	archiveprefix = {arXiv},
	primaryclass = {gr-qc}
}

@article{Ochirov:2018uyq,
	title        = {{Helicity amplitudes for QCD with massive quarks}},
	author       = {Ochirov, Alexander},
	year         = 2018,
	journal      = {JHEP},
	volume       = {04},
	pages        = {089},
	doi          = {10.1007/JHEP04(2018)089},
	eprint       = {1802.06730},
	archiveprefix = {arXiv},
	primaryclass = {hep-ph}
}

@article{Guevara:2017csg,
	title        = {{Holomorphic Classical Limit for Spin Effects in Gravitational and Electromagnetic Scattering}},
	author       = {Guevara, Alfredo},
	year         = 2019,
	journal      = {JHEP},
	volume       = {04},
	pages        = {033},
	doi          = {10.1007/JHEP04(2019)033},
	eprint       = {1706.02314},
	archiveprefix = {arXiv},
	primaryclass = {hep-th}
}

@article{Guevara:2018wpp,
	title        = {{Scattering of Spinning Black Holes from Exponentiated Soft Factors}},
	author       = {Guevara, Alfredo and Ochirov, Alexander and Vines, Justin},
	year         = 2019,
	journal      = {JHEP},
	volume       = {09},
	pages        = {056},
	doi          = {10.1007/JHEP09(2019)056},
	eprint       = {1812.06895},
	archiveprefix = {arXiv},
	primaryclass = {hep-th}
}

@article{Maybee:2019jus,
	title        = {{Observables and amplitudes for spinning particles and black holes}},
	author       = {Maybee, Ben and O'Connell, Donal and Vines, Justin},
	year         = 2019,
	journal      = {JHEP},
	volume       = 12,
	pages        = 156,
	doi          = {10.1007/JHEP12(2019)156},
	eprint       = {1906.09260},
	archiveprefix = {arXiv},
	primaryclass = {hep-th}
}

@article{Bautista:2022wjf,
	title        = {{Scattering in black hole backgrounds and higher-spin amplitudes. Part II}},
	author       = {Bautista, Yilber Fabian and Guevara, Alfredo and Kavanagh, Chris and Vines, Justin},
	year         = 2023,
	journal      = {JHEP},
	volume       = {05},
	pages        = 211,
	doi          = {10.1007/JHEP05(2023)211},
	eprint       = {2212.07965},
	archiveprefix = {arXiv},
	primaryclass = {hep-th}
}

@article{Bautista:2021wfy,
	title        = {{Scattering in black hole backgrounds and higher-spin amplitudes. Part I}},
	author       = {Bautista, Yilber Fabian and Guevara, Alfredo and Kavanagh, Chris and Vines, Justin},
	year         = 2023,
	journal      = {JHEP},
	volume       = {03},
	pages        = 136,
	doi          = {10.1007/JHEP03(2023)136},
	eprint       = {2107.10179},
	archiveprefix = {arXiv},
	primaryclass = {hep-th}
}

@article{Johansson:2019dnu,
	title        = {{Double copy for massive quantum particles with spin}},
	author       = {Johansson, Henrik and Ochirov, Alexander},
	year         = 2019,
	journal      = {JHEP},
	volume       = {09},
	pages        = {040},
	doi          = {10.1007/JHEP09(2019)040},
	eprint       = {1906.12292},
	archiveprefix = {arXiv},
	primaryclass = {hep-th},
	reportnumber = {UUITP-24/19, NORDITA 2019-070}
}

@article{Guevara:2019fsj,
	title        = {{Black-hole scattering with general spin directions from minimal-coupling amplitudes}},
	author       = {Guevara, Alfredo and Ochirov, Alexander and Vines, Justin},
	year         = 2019,
	journal      = {Phys. Rev. D},
	volume       = 100,
	number       = 10,
	pages        = 104024,
	doi          = {10.1103/PhysRevD.100.104024},
	eprint       = {1906.10071},
	archiveprefix = {arXiv},
	primaryclass = {hep-th}
}

@article{Chung:2018kqs,
	title        = {{The simplest massive S-matrix: from minimal coupling to Black Holes}},
	author       = {Chung, Ming-Zhi and Huang, Yu-Tin and Kim, Jung-Wook and Lee, Sangmin},
	year         = 2019,
	journal      = {JHEP},
	volume       = {04},
	pages        = 156,
	doi          = {10.1007/JHEP04(2019)156},
	eprint       = {1812.08752},
	archiveprefix = {arXiv},
	primaryclass = {hep-th},
	reportnumber = {NCTS-TH/1817}
}

@article{Vines:2017hyw,
	title        = {{Scattering of two spinning black holes in post-Minkowskian gravity, to all orders in spin, and effective-one-body mappings}},
	author       = {Vines, Justin},
	year         = 2018,
	journal      = {Class. Quant. Grav.},
	volume       = 35,
	number       = 8,
	pages        = {084002},
	doi          = {10.1088/1361-6382/aaa3a8},
	eprint       = {1709.06016},
	archiveprefix = {arXiv},
	primaryclass = {gr-qc}
}

@article{Kumar:2025juz,
	title        = {{Compton amplitude for massive bosons of arbitrary spin}},
	author       = {Kumar, Aakash and Rudra, Arnab and Shah, Manav and Shaw, Rahul},
	year         = 2025,
	month        = 4,
	eprint       = {2504.06343},
	archiveprefix = {arXiv},
	primaryclass = {hep-th}
}

@article{Arkani-Hamed:2019ymq,
	title        = {{Kerr black holes as elementary particles}},
	author       = {Arkani-Hamed, Nima and Huang, Yu-tin and O'Connell, Donal},
	year         = 2020,
	journal      = {JHEP},
	volume       = {01},
	pages        = {046},
	doi          = {10.1007/JHEP01(2020)046},
	eprint       = {1906.10100},
	archiveprefix = {arXiv},
	primaryclass = {hep-th},
	reportnumber = {NCTS-TH/1905}
}

@article{Chiodaroli:2021eug,
	title        = {{Compton black-hole scattering for s \ensuremath{\leq} 5/2}},
	author       = {Chiodaroli, Marco and Johansson, Henrik and Pichini, Paolo},
	year         = 2022,
	journal      = {JHEP},
	volume       = {02},
	pages        = 156,
	doi          = {10.1007/JHEP02(2022)156},
	eprint       = {2107.14779},
	archiveprefix = {arXiv},
	primaryclass = {hep-th},
	reportnumber = {UUITP-34/21, NORDITA 2021-013}
}

@article{Ingraham:1974un,
	title        = {{Covariant propagators and vertices for higher spin bosons}},
	author       = {Ingraham, R. L.},
	year         = 1974,
	journal      = {Prog. Theor. Phys.},
	volume       = 51,
	pages        = {249--261},
	doi          = {10.1143/PTP.51.249}
}

@book{Elvang:2015rqa,
	title        = {{Scattering Amplitudes in Gauge Theory and Gravity}},
	author       = {Elvang, Henriette and Huang, Yu-tin},
	year         = 2015,
	publisher    = {Cambridge University Press},
	isbn         = {9781316191422, 9781107069251},
	url          = {http://www.cambridge.org/mw/academic/subjects/physics/theoretical-physics-and-mathematical-physics/scattering-amplitudes-gauge-theory-and-gravity?format=AR},
	slaccitation = {%%CITATION = INSPIRE-1384881;%%}
}

@book{Srednicki:2007qs,
	title        = {{Quantum field theory}},
	author       = {Srednicki, M.},
	year         = 2007,
	publisher    = {Cambridge University Press},
	isbn         = {9780521864497, 9780511267208},
	slaccitation = {%%CITATION = INSPIRE-752478;%%}
}

@book{Weinberg:1995mt,
	title        = {{The Quantum theory of fields. Vol. 1: Foundations}},
	author       = {Weinberg, Steven},
	year         = 2005,
	publisher    = {Cambridge University Press},
	isbn         = {9780521670531, 9780511252044},
	slaccitation = {%%CITATION = INSPIRE-406190;%%}
}

@book{Schwartz:2013pla,
	title        = {{Quantum Field Theory and the Standard Model}},
	author       = {Schwartz, Matthew D.},
	year         = 2014,
	publisher    = {Cambridge University Press},
	isbn         = {1107034736, 9781107034730},
	url          = {http://www.cambridge.org/us/academic/subjects/physics/theoretical-physics-and-mathematical-physics/quantum-field-theory-and-standard-model},
	reportnumber = {ISBN-9781107034730},
	slaccitation = {%%CITATION = ISBN-9781107034730;%%}
}

@article{Arkani-Hamed:2017jhn,
	title        = {{Scattering amplitudes for all masses and spins}},
	author       = {Arkani-Hamed, Nima and Huang, Tzu-Chen and Huang, Yu-tin},
	year         = 2021,
	journal      = {JHEP},
	volume       = 11,
	pages        = {070},
	doi          = {10.1007/JHEP11(2021)070},
	eprint       = {1709.04891},
	archiveprefix = {arXiv},
	primaryclass = {hep-th},
	reportnumber = {NCTS-TH/1714, NCTS-TH-1714}
}
